%% file: main.tex
\documentclass[%
 preprint,
 twocolumn,
 10pt,
 amsmath,amssymb,
 aps,
prb,
]{revtex4-1}

\pdfoutput=1
\usepackage{graphicx,xcolor}
\usepackage{mathtools}
\usepackage{hyperref}
\usepackage{epstopdf}
\usepackage{comment}

\usepackage{feynmp}
\usepackage{pgfplots}
\input{./defs}

\begin{document}

\title{Review: Systematic Quantum Cluster Typical Medium Method For the Study of Localization in Strongly Disordered Electronic Systems}


\author{Hanna Terletska$^1$, Yi Zhang$^{2,3}$, Ka Ming Tam$^{2,3}$,
Tom Berlijn$^{4,5}$, L. Chioncel$^{6,7}$, N.\ S.\ Vidhyadhiraja and Mark Jarrell$^{2,3}$} 
\affiliation{$^1$ Department of Physics and Astronomy, Middle Tennessee State University, Computational Science Program, Murfreesboro, TN 37132, USA}
\affiliation{$^2$
Department of Physics and Astronomy, Louisiana State University, Baton Rouge, LA 70803, USA
}
\affiliation{$^3$
Center for Computation and Technology, Louisiana State University, Baton Rouge, LA 70803, USA
}
\affiliation{$^4$Center for Nanophase Materials Sciences, Oak Ridge National Laboratory, Oak Ridge, TN 37831, USA}
\affiliation{$^5$Computer Science and Mathematics Division, Oak Ridge National Laboratory, Oak Ridge, Tennessee 37831, USA}
\affiliation{$^6$Augsburg Center for Innovative Technologies, University of Augsburg,
D-86135 Augsburg, Germany}
\affiliation{$^7$Theoretical Physics III, Center for Electronic
Correlations and Magnetism, Institute of Physics, University of
Augsburg, D-86135 Augsburg, Germany}
\affiliation{$^8$Theoretical Sciences Unit, Jawaharlal Nehru Center for Advanced Scientific Research,
Bengaluru 560064, India
}

\date{\today}

\begin{abstract}
Great progress has been made in the last several years towards understanding the properties of disordered electronic systems. In part, this is made possible by recent advances in quantum effective medium methods which enable the study of disorder and electron-electronic interactions on equal footing.  They include dynamical mean field theory and the coherent potential approximation, and their cluster extension, the dynamical cluster approximation.  Despite their successes, these methods do not enable the first-principles study of the strongly disordered regime, including the effects of electronic localization.  The main focus of this review is the recently developed typical medium dynamical cluster approximation for disordered electronic systems. This method has been constructed to capture disorder-induced localization, and is based on a mapping of a lattice onto a quantum cluster embedded in an effective typical medium, which is determined self-consistently. Unlike the average effective medium based methods mentioned above,  typical medium based methods properly capture the states localized by disorder. The typical medium dynamical cluster approximation not only provides the proper order parameter for Anderson localized states but it can also incorporate the full complexity of DFT-derived potentials into the analysis, including the effect of multiple bands, non-local disorder, and electron-electron interactions. After a brief historical review of other numerical methods for disordered systems, we discuss coarse-graining as a unifying principle for the development of translationally invariant quantum cluster methods.  Together, the Coherent Potential Approximation, the Dynamical Mean Field Theory and the Dynamical Cluster Approximation may be viewed as a single class of approximations with a much needed small parameter of the inverse cluster size which may be used to control the approximation.  We then present an overview of various recent applications of the typical medium dynamical cluster approximation to a variety of models and systems, including single and multi-band Anderson model, and models with local and off-diagonal disorder.  We then present the application of the method to realistic systems in the framework of the density functional theory. and demonstrate that the resulting method is able to provide a systematic first principles method validated by experiment and capable of making experimentally relevant predictions.  We also discuss the application of the typical medium dynamical cluster approximation to systems with disorder and electron-electron interactions. Most significantly, we show that in the limits of strong disorder and weak interactions treated perturbatively, that the phenomena of 3D localization, including a mobility edge, remains intact.  However, the metal-insulator transition is pushed to larger disorder values by the local interactions.  We also study the limits of strong disorder and strong interactions capable of producing moment formation and screening, with a non-perturbative local approximation.  Here, we find that the Anderson localization quantum phase transition is accompanied by a quantum-critical fan in the energy-disorder phase diagram.  
\end{abstract}

\maketitle
\section*{Keywords:} Disordered electrons, Anderson localization, metal-insulator transition, coarse-graining, typical medium, quantum cluster methods, first principles.

\tableofcontents


\section{Introduction} 
\label{sec:introduction}

The metal-to-insulator transition (MIT) is one of the most spectacular effects in condensed matter physics and materials science. The dramatic change in electrical properties of materials undergoing such a transition is exploited in electronic devices that are components of data storage and memory technology\cite{newyorktimes,greenpeace}. It is generally recognized that the underlying mechanism of MITs are the interplay of electron correlation effects (Mott type) and disorder effects (Anderson type) ~\cite{imada_mit,n_mott_68,d_belitz_94,f_evers_08, dobrosavljevic_book}. Recent developments in many-body physics make it possible to study these phenomena on equal footing rather than having to disentangle the two.

The purpose of this review is to bring together the various developments and applications of such a new method, namely the Typical Medium Dynamical Cluster Approach (TMDCA)~\cite{m_jarrell_01a,v_dobrosavljevic_03,c_ekuma_15b, c_ekuma_15c,y_zhang_15a}, for investigating interacting disordered quantum systems.  

The organization of this article is as follows: 
Sec.~\ref{sec:intro} is dedicated to a few basic aspects of modeling disorder in solids.  We discuss a couple of examples of materials that are believed to have relevant technological applications connected to the problem of localization. The corresponding subsections deal with theoretical modeling. We then follow with a review of the Anderson and Mott mechanisms leading to electronic localization, as well as their interplay.

In Sec.~\ref{sec:DirectMethods} we review three alternative numerical methods for solving the Anderson model and discuss their advantages and limitations in chemically-specific modeling.  These methods are employed in Sec.~\ref{sec:Applications} to validate the developed formalism.

In Sec.~\ref{sec:CGmethods} we shift our focus to the discussion of the effective medium methods. First, we present the concept of coarse-graining. The coarse-graining procedure allows us to draw similarities present in infinite dimension between the Dynamical Mean Field Theory (DMFT) ~\cite{w_metzner_89a,e_mullerhartmann_89a,e_mullerhartmann_89b,a_georges_92a,m_jarrell_92a,t_pruschke_95,a_georges_96a} of interacting electrons and the Coherent potential Approximation (CPA) ~\cite{soven_cpa,velicky_cpa,r_elliott_74} of non-interacting electrons in disordered external potentials.  We then provide a detailed discussion of the Dynamical Cluster Approximation\cite{m_hettler_98a,m_hettler_00a,m_jarrell_01a}, a non-local effective medium approximation, which systematically incorporates the non-local correlation effects missing in the DMFT and CPA by refining the course graining.

The central focus of this review, is the typical medium theories of Anderson localization, which are discussed in Sec.~\ref{sec:tmdca}.  We show how this method is used to study disorder-induced electron localization. Starting from the single-site typical medium theory, we present its natural cluster extension, discussing several algorithms for the self-consistent embedding of periodic clusters fulfilling the original symmetries of the lattice in addition to other desirable properties. We present details of how this method can be used to incorporate the full chemical complexity of various systems, including off-diagonal disorder and multi-band nature, along with the interplay of disorder and electron-electron interactions.  

In Sec.~\ref{sec:edhm} we discuss how the developed typical medium methods can be practically applied to real materials. This is done in a three-step process in which DFT results are used to generate an effective disordered  Hamiltonian, which is passed to the typical medium cluster/single-site solver to compute spectral densities and estimate the degree of localization. Section Sec.\ref{sec:Applications} reviews the application of the TMDCA from single-band three dimensional models to more complex cases such as off-diagonal disorder,  multi-orbital cases and electronic interactions. Finally the concluding remarks are presented in Sec.~\ref{sec:Conclusion}.

\section{Background: electron localization in disordered medium } 
\label{sec:intro}

\begin{figure}
\includegraphics[width=1.1\columnwidth,clip=true]{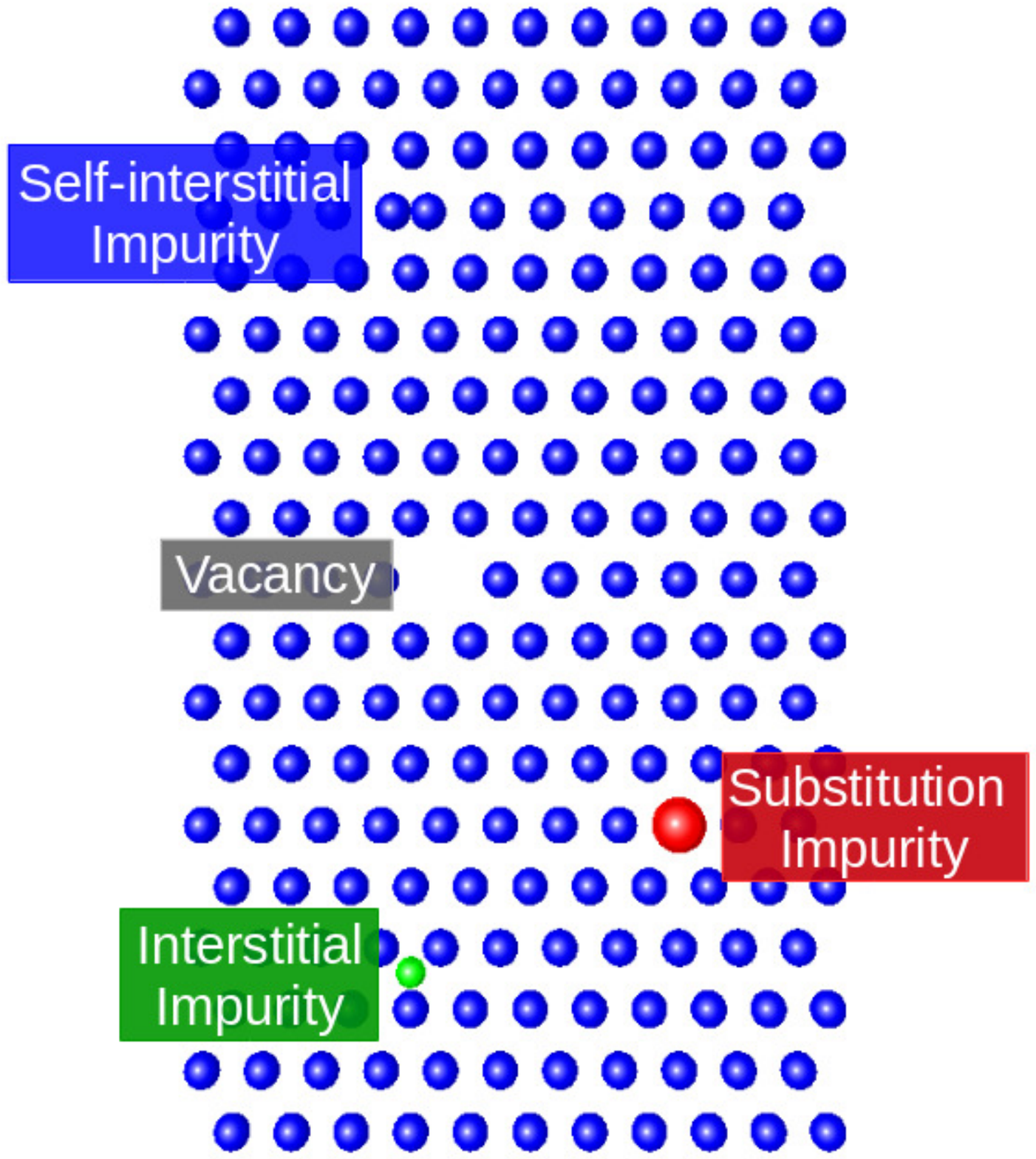}
\caption{Examples of various types of disorder, including substitution and interstitial impurities, and
vacancies. In addition (not shown), disorder can originate from other ways of breaking the translational symmetry, including the external disorder potentials,amorphous systems, random arrangement of spins, etc.}
\label{fig:typesofdisorer}
\end{figure}

Disorder is a common feature of many materials and often plays a key role in changing and controlling their properties. As a ubiquitous feature of real systems it can arise in varying degrees in the crystalline host for a number of reasons.  As shown in Figure~\ref{fig:typesofdisorer}, disorder may range from a few impurities or defects in perfect crystals, (vacancies, dislocations, interstitial atoms, etc), chemical substitutions in alloys and random arrangements of electron spins or glassy systems. 

One of the most important effects of disorder is that it can induce spatial localization of electrons and lead to a metal-insulator transition, which is known as  Anderson localization. Anderson predicted~\cite{p_anderson_58} that in a disordered medium, electrons scattered off randomly distributed impurities can become localized in certain regions of space due to interference between multiple-scattering paths. 

Besides being a fundamental solid-state physics phenomena,  Anderson localization has a profound consequences on many functional properties of materials.  For example, the substitution of P or B for Si may be used to dope holes or particles into Si increasing its functionality.  Disorder appears to play a crucial role also in formation of inhomogeneities in commercially important CMR materials ~\cite{e_dagotto_05}.
At the same time, in dilute magnetic semiconductors such as GaMnAs, there is a subtle interplay between magnetism and Anderson localization~\cite{l_rokhinson_07,Dobrowolska12,m_sawicki_10a,m_flatte_11,n_samarth_12}. Intermediate band semiconductors are another type of material where disorder may play an important role in manipulating their properties.  These materials hold the promise to significantly improve solar cell efficiency, but only if the electrons in the impurity band are extended~\cite{a_luque_01,y_okada_15,j_zhang_15}. Also recently, Anderson localization of phonons has been suggested as the basis of relaxor behavior~\cite{m_manley_14}. These examples show that Anderson localization has profound consequences for functional materials that we need to understand and try to control for a positive outcome.

In 1977 P. W. Anderson and N. Mott shared one third each of the Nobel prize~\cite{1977Nobel}. Both were, at least in part, for rather different perspectives on the localization of electrons. In Mott's picture, localization is driven by interactions, albeit originally only at the level of Thomas-Fermi screening of impurities~\cite{n_mott_68}. The transition is first order, with the finite temperature second order terminus. In Anderson's picture, localization is a quantum phase transition driven by disorder.  Despite more than five decades of intense research~\cite{a_lagendijk_09,e_abrahams_10}, a completely satisfactory picture of Anderson localization does not exist, especially when applied to real materials. 

\begin{figure}[t]
\includegraphics[width=0.5\textwidth,clip=true]{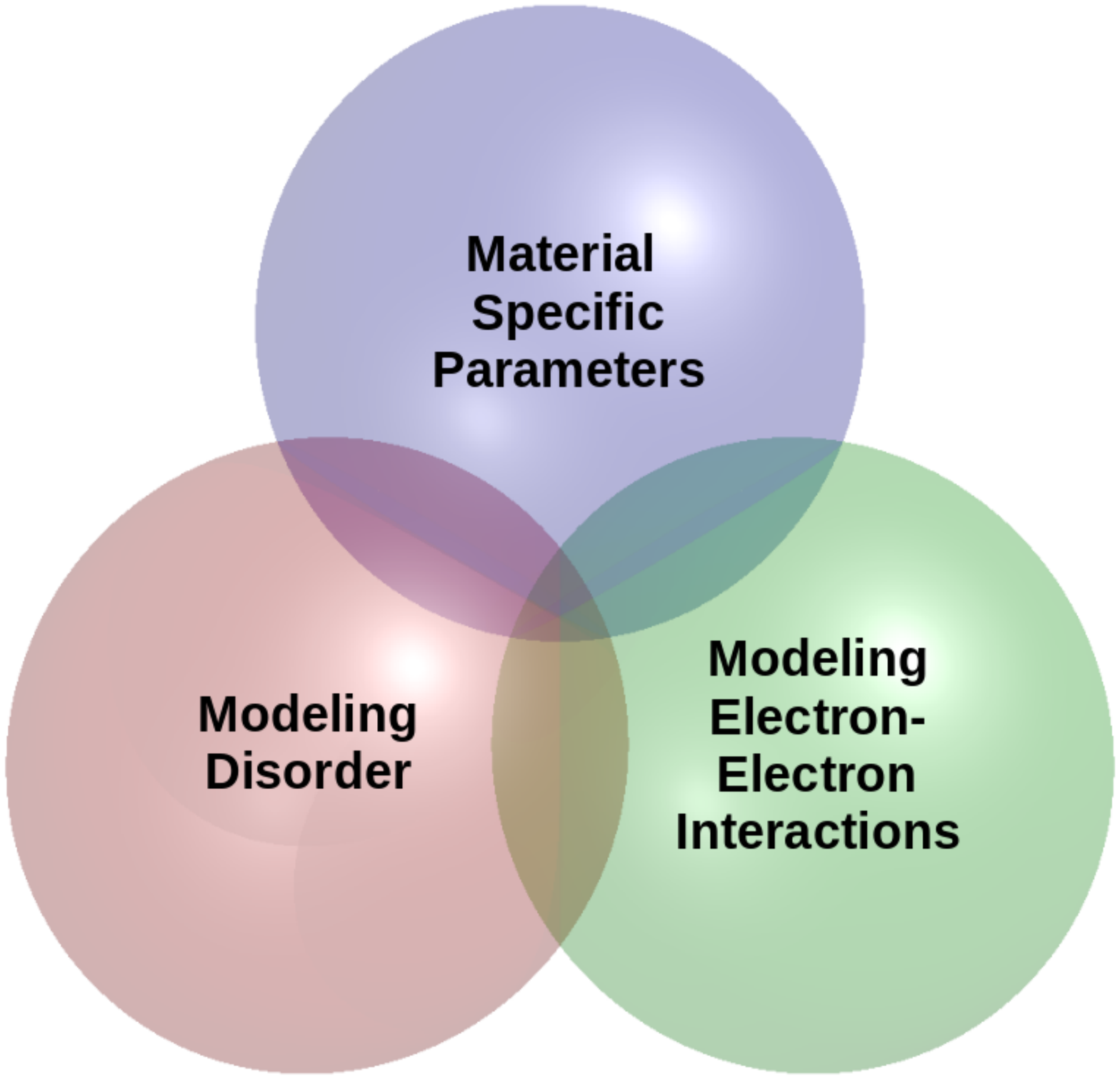}
\caption{Simultaneous treatment of the material specific parameters, modeling disorder and electron-electron interactions present one of the major challenges for theoretical studies of electron localization in real materials. }
\label{fig:complexity}
\end{figure}

Several standard computationally exact numerical techniques including exact diagonalization, transfer matrix method~\cite{Kramer_etal_2010,Markos_2006,Kramer_MacKinnon_1993}, and kernel polynomial method~\cite{a_weisse_06} have been developed. They are extensively applied to study the Anderson model (a tight binding model with a random local potential). While these are very robust methods for the Anderson model, their application to real modern materials is highly non-trivial.This is due to the computational difficulty in treating simultaneously the effects of multiple orbitals and complex real disorder potentials (Figure~\ref{fig:complexity}) for large system sizes. In particular, it is very challenging to include the electron-electron interaction. Practical calculations are limited to rather small systems. Also the effects from the long range disorder potential which happens in real materials, such as semi-conductors, are completely absent. This, perhaps, is  not surprising, as direct numerical calculations on interacting systems even in the clean limit often come with various challenges. Reliable calculations for sufficiently large system sizes infer the behaviors at the thermodynamic limit that are largely done in specific cases such as systems at one dimension or at special filling in which the fermionic minus sign problem in the quantum Monte Carlo calculations can be subsided.

During the past two decades or so,several effective medium mean field methods have been developed as an alternative to direct numerical methods.   
For example, for systems with strong electron-electron interactions, over the past two decades or so, the  Dynamical Mean Field Theory (DMFT)~\cite{w_metzner_89a,e_mullerhartmann_89a,e_mullerhartmann_89b,a_georges_92a,m_jarrell_92a,t_pruschke_95,a_georges_96a}, constitutes a major development in the field of computational many body systems and materials science. The DMFT shares many similarities with the Coherent Potential approximation (CPA) for disordered systems~\cite{velicky_cpa, soven_cpa}. Conceptually,in both these methods, the lattice problem is approximated by a single site problem in a fluctuating local dynamical field (the effective medium). The fluctuating environment due to the lattice is replaced by the local energy fluctuation, and the dynamical field is determined by the condition that the local Green's function is equal to (in CPA, the disorder averaged) Green's function of the single site problem~\cite{Vollhardt_2010}.

DMFT has been extensively used on strongly correlated models, such as the Hubbard model\cite{m_jarrell_92a}, the periodic Anderson model\cite{m_jarrell_93b}, and the Holstein model~\cite{j_freericks_93b}. It provides a viable computational framework for strongly correlated systems in a wide range of parameters which were hitherto impossible to reach by Quantum Monte Carlo on lattice models.  Capturing the Mott-Hubbard transition in a non-perturbative fashion is a major triumph of the DMFT. A significant development of DMFT is its cluster extension, such as (momentum-space cluster extension of DMFT) Dynamical Cluster Approximation (DCA) and Cluster DMFT (real-space cluster extension of DMFT)~\cite{th_maier_05a,m_hettler_98a,g_kotliar_01,m_jarrell_93a}. 
Interesting physics which has non-trivial spatial structure, such as d-wave pairing in the cuprates can be studied by DCA \cite{th_maier_05b}. A very important feature of the DCA is that it is a controllable approximation with a small parameter of $1/N_c$ ($N_c$ is the cluster size), and its ability to provide systematic non-local corrections to the DMFT/CPA results.

For non-interacting but disordered systems, the first-principles analysis of defects in solids 
starts with the substitutional model of disorder. Here, the different atomic species occupy the lattice sites according to some probabilistic rules. The Coherent Potential Approximation (CPA) ~\cite{velicky_cpa,soven_cpa,yo.mo.73,r_elliott_74,ziman_79} proved to provide a scheme to obtain ensemble averaged quantities in terms of effective medium quantities satisfying analyticity and recovering exact results in appropriate limits. The effective medium (or coherent) ensemble averaged propagator is obtained from the condition of no extra scattering coming, on average, from any embedded impurities. Following the Anderson model Hamiltonian applications,~\cite{velicky_cpa,soven_cpa,d_taylor_67} the CPA was reformulated in the framework of the multiple scattering theory~\cite{Gyorffy72} and used to analyze real materials by combination with the Korringa-Kohn-Rostoker (KKR) basis~\cite{Johnson_1986a,vi.ab.01}  or linear muffin-tin orbital (LMTO) basis~\cite{si.go.93} sets.
It has been used to calculate thermodynamic bulk
properties~\cite{faul.82,jo.pi.93,ko.ru.95,ru.ab.95}, phase stability~\cite{gy.st.83,al.jo.95,ab.ru.93,vito.07}, magnetic properties~\cite{ak.de.93,tu.ku.94,ab.er.95}, surface electronic structures~\cite{ku.tu.92,ma.go.92,ab.sk.93,vito.07}, segregation~\cite{ru.ab.94,pa.dr.93} and other alloy characteristics with a considerable success. 
Recently, numerical studies of disordered interacting systems using the DFT+(CPA)DMFT method also become possible~\cite{j_minar_17}. As the CPA captures only the average presence of different atomic species, it cannot account for more subtle aspects connected to the actual distribution of atomic species, practically realized in materials. In a recent years, a considerable amount of theoretical effort has been directed towards the improvement of the original single-site CPA formulation, including the DCA ~\cite{m_jarrell_93a}. This is also the subject of the present review on a cluster development in the form of the typical medium DCA.

There are a number of excellent extensive research papers, reviews, and books covering different aspects of DMFT/CPA/DFT.  These include Ref.~\cite{t_pruschke_95,a_georges_96a} on DMFT aspects, Ref.~\cite{velicky_cpa,soven_cpa} concerning CPA, Wannier-function-based methods~\cite{marzari_1997,w_ku_02,Anisimov:2005ix} to extract a tight-binding Hamiltonian from the DFT calculation, multiple scattering theory~\cite{a_gonis_92}, and the combined LDA+DMFT approach\cite{kotliar_lda_dmft}, to enumerate just a few.  

Although these methods allow the study of various phenomena resulting from the interplay of disorder and interaction, they fail to capture the disorder-driven localization. As we will discuss in detail in the sections below, the fundamental obstacle in tackling the Anderson localization is the lack of a proper order parameter. Once the order parameter is identified as the typical density of states (Sec.\ref{sec:OPlocalization}), it can be incorporated into a self-consistency loop leading to the Typical Medium Theory~\cite{v_dobrosavljevic_03}. This was subsequently extended to clusters incorporating ideas of the DCA. This theory came to be known as the {\it Typical Medium Dynamical Cluster Approximation} (TMDCA) and is the major focus of current review. 

\begin{figure}[t]
\begin{center}
 \includegraphics[trim = 0mm 0mm 0mm 0mm,width=1\columnwidth,clip=true]{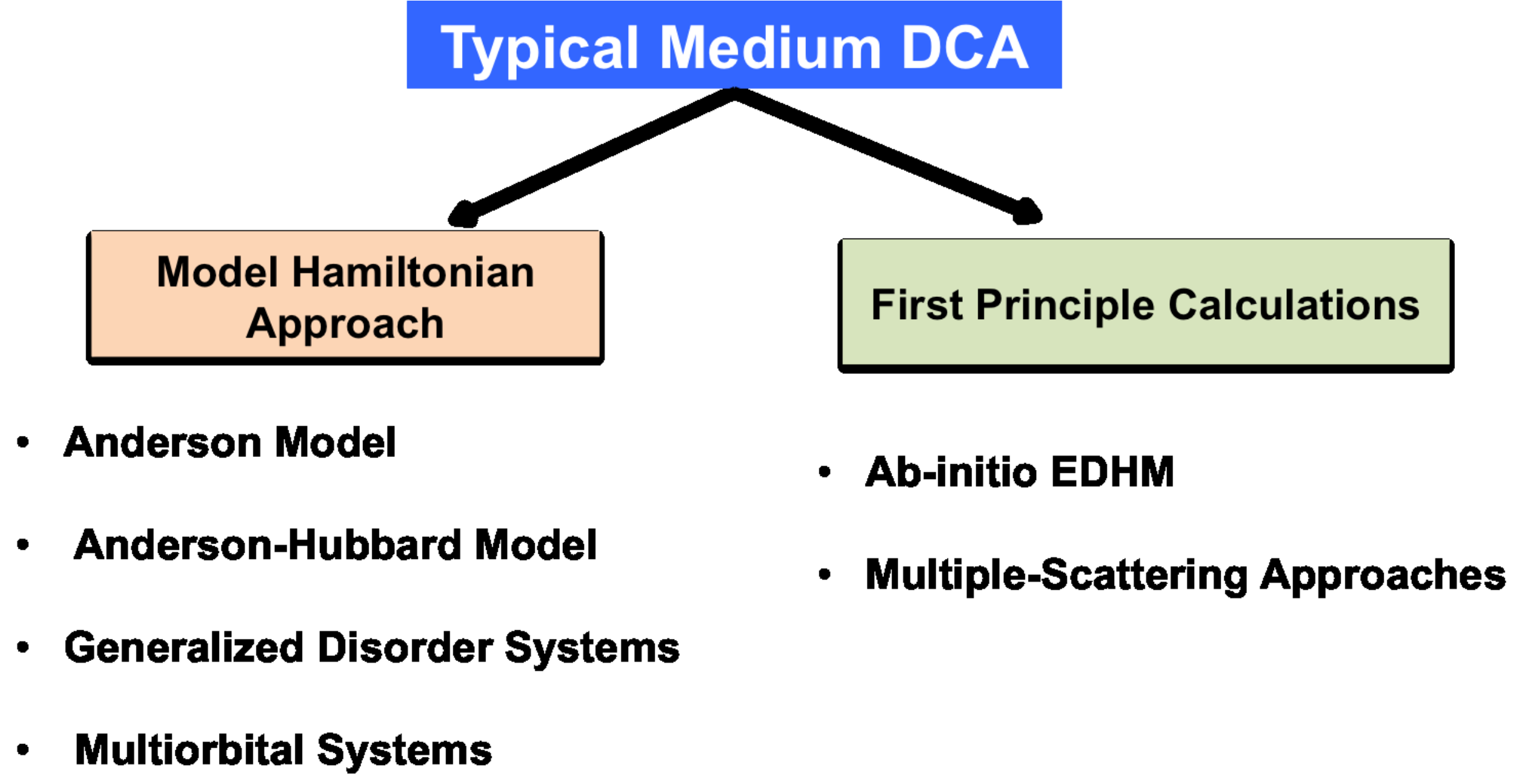}
\caption{The TMDCA may be used to study electron localization in both simple model Hamiltonians as well as those extracted from first principles calculations. }
\label{tmdca-chart-chart}
\end{center}
\end{figure}

In addition to being able to capture the Anderson localization properly, the TMDCA also allows the study of the interplay between disorder and interaction in both weak and strong coupling limits. Thus, it provides a new basis for studying the Mott and Anderson transitions on equal footing. As any cluster extension TMDCA inherits, so also the system size (i.e. the number of sites in the cluster $N_c$) dependence. In analogy with the DCA , the $1/N_{c}$ can be treated as a small parameter, therefore a systematic improvement of the approximation can be achieved by increasing the cluster size.  In addition, in contrast to direct numerical methods, the major strength of TMDCA lies in its flexibility to handle complex long range impurities and multi-orbitals systems which are unavoidable features of many realistic disordered system ~Figure \ref{tmdca-chart-chart}. This review collects the recent results of the TMDCA applied to the Anderson model and its extension, and to the real materials.

\subsection{Anderson localization} 
\label{sec:Anderson}
Strong disorder may have dramatic effects upon the metallic state~\cite{e_abrahams_10}:  the extended states that are spread over the entire system become exponentially localized, centered at one position in the material.  In the most extreme limit, this is obviously true. Consider for example a single orbital that is shifted in energy so that it falls below (or above) the continuum in the density of states (DOS).  Clearly, such a state cannot hybridize with other states since there are none at the same energy.  Thus, any electron on this orbital is localized, via this (deep) trapped states mechanism, and the electronic DOS at this energy will be a delta function.  
Of course this is an extreme limit. Even in the weak disorder limit, the resistivity of ideal metallic conductors decreases with lowering temperature. In reality, at very low temperatures, the resistivity saturates to a residual value. This is due to the imperfections in the formation of the crystal. If the disorder is not too strong, the perfect crystal still remains a good approximation. The imperfections can be considered as the scattering centers for the current-carrying electrons. Hence, the scattering processes between the electrons and defects lead to the reduction in the conduction of electrons. 

For low dimensional systems, the scattering can induce substantial change even for weak disorder. Within the weak localization theory, based on the Langer-Neal maximally crossed graphs, the correction to the conductivity can be rather large \cite{Bergmann_1984,Langer_1960,Langer_Neal_1966}. It can drive a metal into an insulator for dimension $D \leq 2$ (D is a dimensionality of the system) if the impurity does not break time reversal symmetry. 

Historically, it was first shown by Anderson that finite disorder strength can lead to the localization of electronic states in his seminal 1958 paper~\cite{p_anderson_58}. The technique involved can be considered as a locator expansion for the effective hopping element of Anderson model Hamiltonian around the limit of the localized state. He found a region of disorder strength in which the expansion is convergent and thus the localized state endures. Note that the probability distribution of the effective hopping element, instead of its average value, was discussed in the original paper by Anderson.  The importance of the distribution in disordered system is a critical insight in the development of the typical medium theory ~\cite{v_dobrosavljevic_10}.

Subsequently, Mott argued that the extended states would be separated from the localized states by a sharp mobility (localization) edge in energy \cite{Mott_1967,Cohen_Fritzsche_Ovshinsky_1969,Economou_Cohen_1972}.  His argument is that scattering from disorder is elastic, so that the incoming wave and the scattered wave have the same energy.  On the other hand, nearly all scattering potentials will scatter electrons from one wavevector to all others, since the strongest scattering potentials are local or nearly so.  If two states, corresponding to the same energy and different wavenumbers exist, then the scattering potential will cause them to mix, causing both to become extended.

An important development of the localization theory was the introduction of the concept of scaling. In 1972, Edwards and Thouless performed a numerical analysis on the dependence between the degree of localization and the boundary condition of the eigenstate of the Anderson model. They argued that the ratio of the energy shift from the change in the boundary conditions($\Delta E$) to the energy spacing ($\eta$) can be used as a measure for the degree of localization \cite{Edwards_Thouless_1972}. The ratio $\Delta E  / \eta$ now known as the Thouless energy is identified as a dimensionless conductance, $g(L)$, where $L$ is the liner dimension of a system \cite{Licciardello_Thouless_1974}. For a localized state, the Thouless energy decreases as the system size increases and tends to zero in the limit of a large system. For an extended state, the Thouless energy converges to a finite value as the system size increases. They further assume that $\Delta E/ \eta$ or the conductance $g(L)$ is the only relevant coupling parameter in the renormalization group sense. 

The assumption of a single coupling parameter leads to the development of the scaling theory for the conductance. It is based on the assumption that conductance at different length scales (say $L^{'}$ and $L$) are related by the scaling relation $g(L^{'})= f((L^{'}/L),g(L))$. In the continuum it can be written as $\frac{dlng(L)}{d lnL} = \beta(g(L))$. The $\beta$ function can be estimated from small and large $g$ limits. From these results, Abrahams, Anderson, Licciardello, and Ramakrishnan conclude that there is no true metallic behaviors in two dimensions, but a mobility edge exists in three dimensions \cite{e_abrahams_79}. The validity of the scaling theory gained further support after the discovery of the absence of $ln~L^2$ term from the perturbation theory.~\cite{Gorkov_Larkin_Khmelnitskii_1979}

The connection between the mobility edge and the critical properties of disorder spin models was realized in the 70's.~\cite{Aharony_Imry_1977} In a series of papers Wegner proposed that the Anderson transition can be described in terms of a non-linear sigma model.~\cite{Wegner_1979,Wegner_1980,Schafer_Wegner_1980}. Multifractality of the critical eigenstate was first proposed within the context of the sigma model \cite{Wegner_1980,Castellani_Peliti_1986}. All three Dyson symmetry classes were studied. Hikami, Larkin, and Nagaoka found that the symplectic class corresponds to the system with spin-orbit coupling that can induce delocalization in two dimensions.~\cite{Hikami_Larkin_Nagaoka_1980} In 1982, Efetov showed that tricks from super-symmetry can be employed to reformulate the mapping to a non-linear sigma model with both commuting and anti-commuting variables.~\cite{Efetov_1982}

Many of the recent efforts in studying Anderson localization, focus on the critical properties within an effective field theory--non-linear sigma model in different representations: fermionic, bosonic, and supersymmetric~\cite{f_evers_08}. While these works provide answers to important questions, such as the existence of mobility edges of different symmetry classes at different dimensions, they are not able to provide universal or off from criticality quantities, such as critical disorder strength, the correlation length and the correction to conductivity in the metallic phase. An important development to address these issues is the self consistent theory proposed by Vollhardt and W\"olffle.~\cite{Vollhardt_Wolfle_1980,Vollhardt_Wolfle_1992} It has also been shown that the results from this theory also obey the scaling hypothesis.~\cite{Vollhardt_Wolfle_1982}

More recent studies focus on classifying the criticality according to the local symmetry. Ten different symmetry classes based on classifying the local symmetry are identified generalizing the three Dyson classes including the Nambu space~\cite{Atland_Zirnbauer_1997}. The renormalization group study on the sigma model has been carried out on different classes and dimensions.~\cite{f_evers_08}. The importance of the topology of the sigma model target space is studied extensively in recent works~\cite{f_evers_08,Schnyder_etal_2009,Chiu_etal_2016}.

\subsection{Order parameter of Anderson localization}
\label{sec:OPlocalization}
As we discussed in the previous section, effective medium theories have been used to study Anderson localization, however progress has been
hampered partly due to ambiguity in identifying an appropriate
order parameter for Anderson localization, allowing for a clear distinction between localized and extended states ~\cite{v_dobrosavljevic_03}.

An order parameter function had been suggested about three decades ago, in the study of Anderson localization on the Bethe lattice.~\cite{Zirnbauer_1986,Efetov_1987} It has been shown that the parameter is closely related to the distribution of on-site Green's functions, in particular the local density of states.~\cite{Mirlin_Fyodorov_1994} Recently, following the work of Dobrosavljevic et. al ~\cite{v_dobrosavljevic_03}, there has been tremendous  progress along these ideas, with the local typical density of states identified as the order parameter.

\begin{figure}[h]
\includegraphics[width=0.5\textwidth,clip=true]{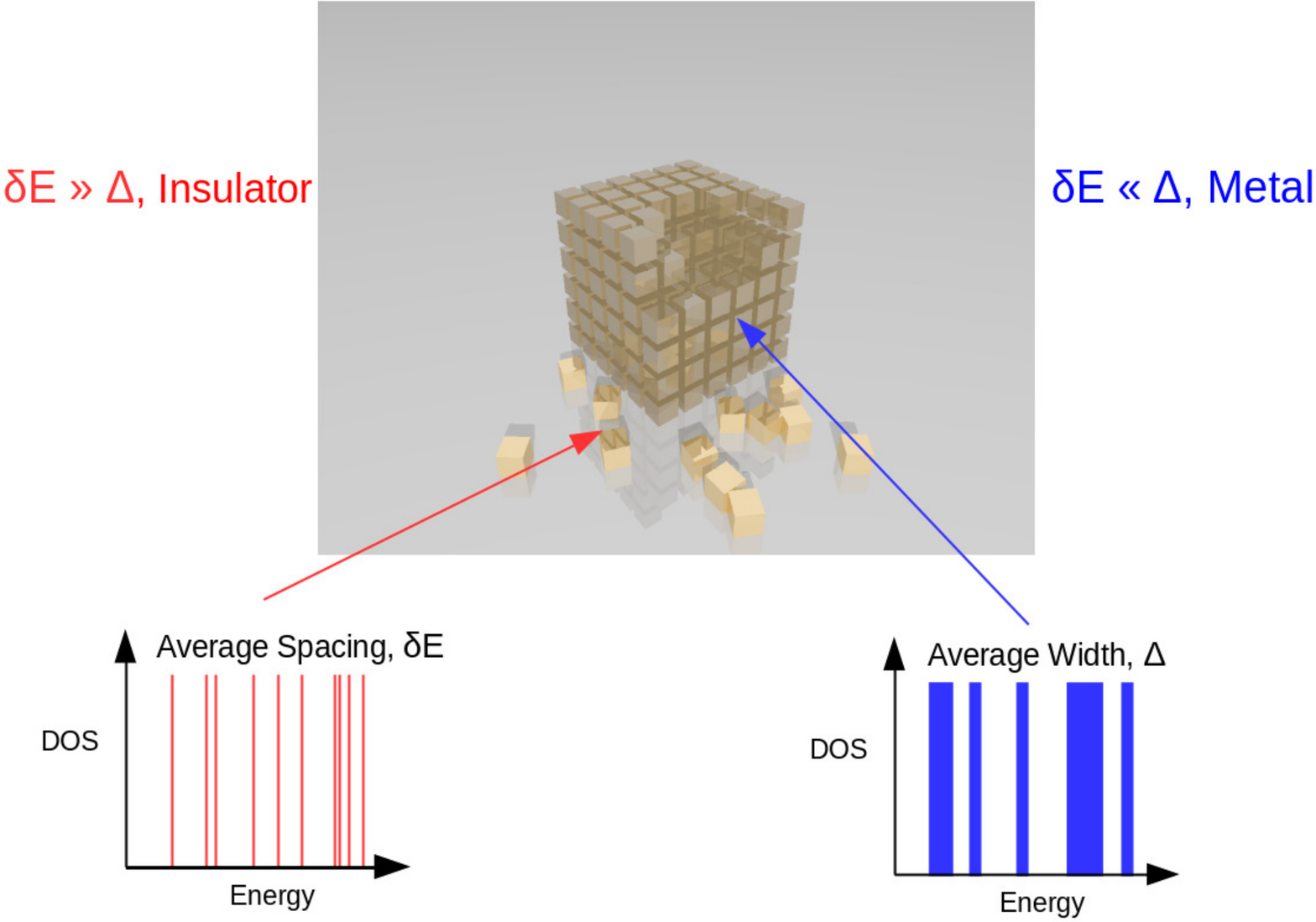}
\caption{To help understand localization, we divide the system into blocks. The average spacing of the energy levels of a block is $\delta E$ and the Fermi golden rule width of the levels is $\Delta$.  If $\Delta \gg \delta E$ then we have a metal and if $\Delta \ll \delta E$, an insulator.
}
\label{fig:blocks}
\end{figure}

To demonstrate how the local density of states and its typical (most probable value) can be utilized as an order parameter for Anderson localization, we consider a thought experiment. We imagine dividing the system up into blocks, as illustrated in Figure~\ref{fig:blocks}.  Later, when we construct our quantum cluster theory of localization, each of the blocks should be thought of as a cluster, and we construct the system by periodically stacking the blocks.  We make two controllable approximations.
\begin{enumerate}
\item We approximate the effect of coupling the block to the reminder of the lattice via Fermi's golden rule--coupling $\Delta$ which is proportional to the density of accessible states.  
\item Since on average each cluster is equivalent to all the others, this density will also be proportional to some appropriate block density of states.  
\end{enumerate}
Furthermore, imagine that the average level spacing of the states in a block is $\delta E$.  If $\Delta \gg \delta E$, then we have a metal since the states at this energy have a significant probability of escaping from this block, and the next one, etc. Alternatively if $\Delta \ll \delta E$ the escape probability of the electrons is low, so that an insulator forms.   

So what does this mean in terms of the local electronic density of states (LDOS) that is measured, i.e., via STM at one site in the system, and the average DOS (ADOS) measured, i.e., via tunneling (or just by averaging the LDOS)?  

\begin{figure}[h]
\includegraphics[width=0.45\textwidth,clip=true]{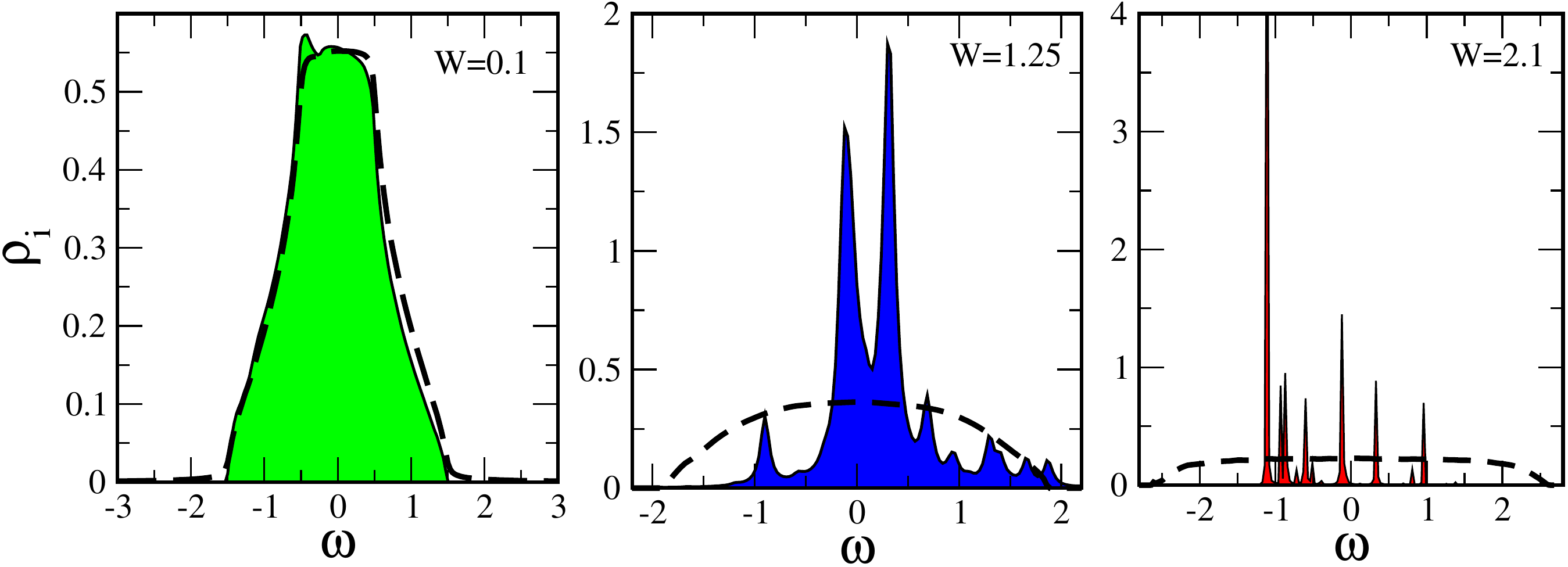}
\caption{
The global average (dashed lines) and the local (solid lines) DOS of the 3D Anderson model for small, moderate and large disorder strength $W$ with units $4t=1$ where $t$ is the near-neighbor hopping (see text for details).\label{fig:local}
}
\end{figure} 
In Figure~\ref{fig:local} we calculate the ADOS and TDOS for a simple (Anderson) single-band model on a cubic lattice with near-neighbor hopping $t$ (bare bandwidth $12t=3$ to establish an energy unit) and with a random site $i$ local potential $V_i$ drawn from a "box" distribution of width $2W$, with $P(V_i)=\frac{1}{2W}\Theta(W-|V_i|)$.  As can be seen from the Figure~\ref{fig:local}, as we increase the disorder strength $W$, the global average DOS (dashed lines) always favors the metallic state  (with a finite DOS at the Fermi level $\omega=0$) and it is a smooth (not critical) function even above the transition. In contrast to the global average DOS, the local density of states (solid lines), which measures the amplitude of the electron wave function at a given site,  undergoes significant qualitative changes as the disorder strength $W$ increases, and eventually becomes a set of the discrete delta-like functions as the transition is approached.

\begin{figure}[h]
\includegraphics[width=0.35\textwidth,clip=true]{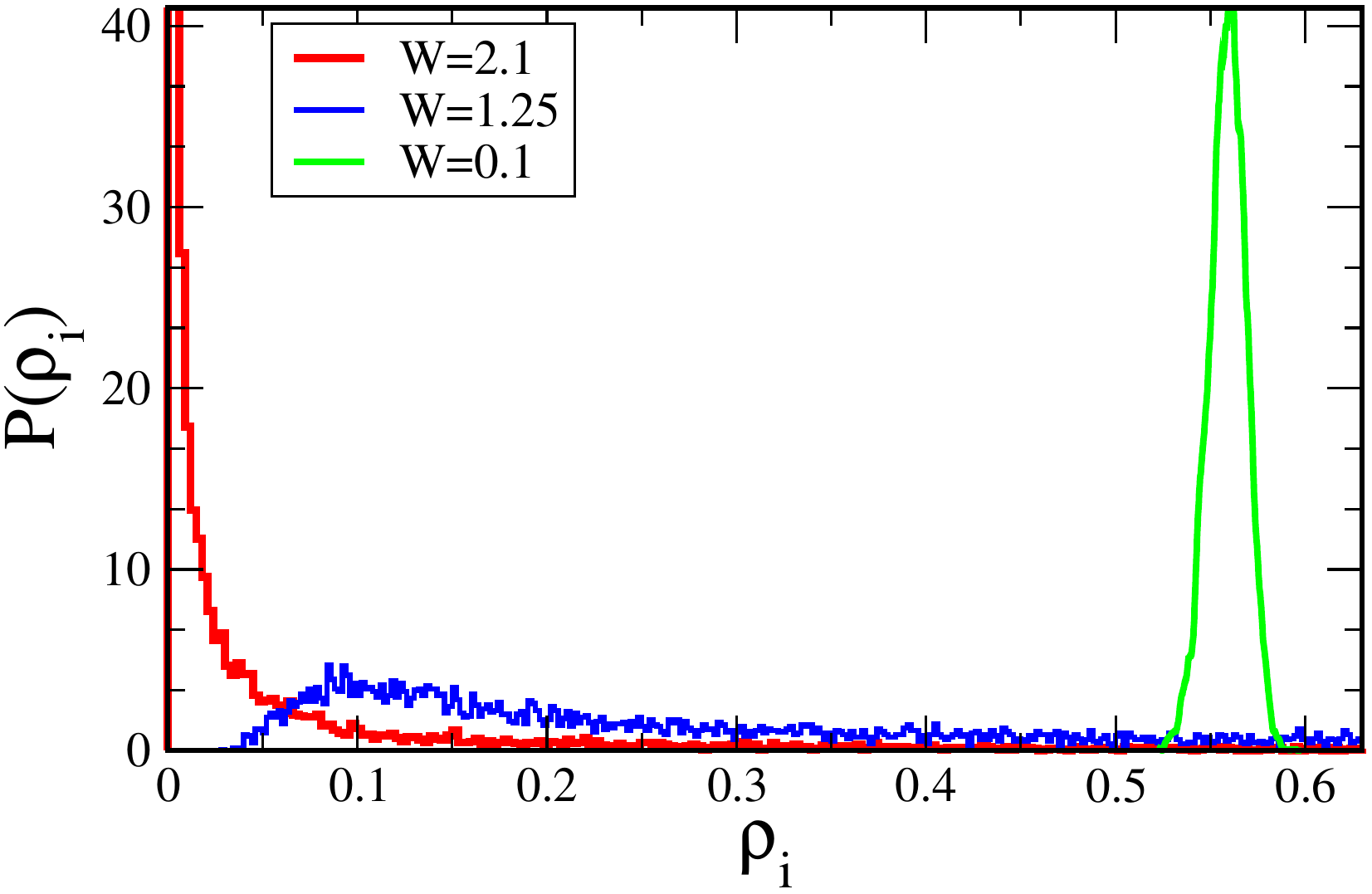}
\caption{
The evolution of the probability distribution function of the local DOS at the band center ($\omega=0$) with disorder strength $W$.  The data is the same as in Figure~\ref{fig:local}.
\label{fig:pdos_local}
}
\end{figure}

This must mean that the probability distributions of the local DOS for a metal and for an insulator is also very different.  This is illustrated in Figure~\ref{fig:pdos_local}. In particular, the most probable (typical) value of the local DOS in a metal is very different than the typical value in an insulator.  Consider again the local DOS in the metal and insulator.  In the metal, the probability distribution function is Gaussian-like form. The local DOS at any one energy the DOS at each site is a continuum.  It will change from site to site, but the most probable value and the average value, will be finite.  Now reconsider the local DOS in the insulator. It is composed of a finite number of delta functions.  For any energy in between the delta functions, the local DOS is zero.  Since the number of delta functions is finite, the typical value of the local DOS is zero, while the average value is still finite. Consequently, the probability distribution function of the local DOS is very much skewed towards zero and develops long tails. As a result, the order parameter for the Anderson metal-insulator transition is the typical local DOS, which is zero in the insulator and finite in the metal. This analysis also demonstrates one of the distinctive features of Anderson localization, i.e., the non-self-averaging nature of local quantities close to the transition. 

\begin{figure}[h]
\includegraphics[width=0.34\textwidth,clip=true]{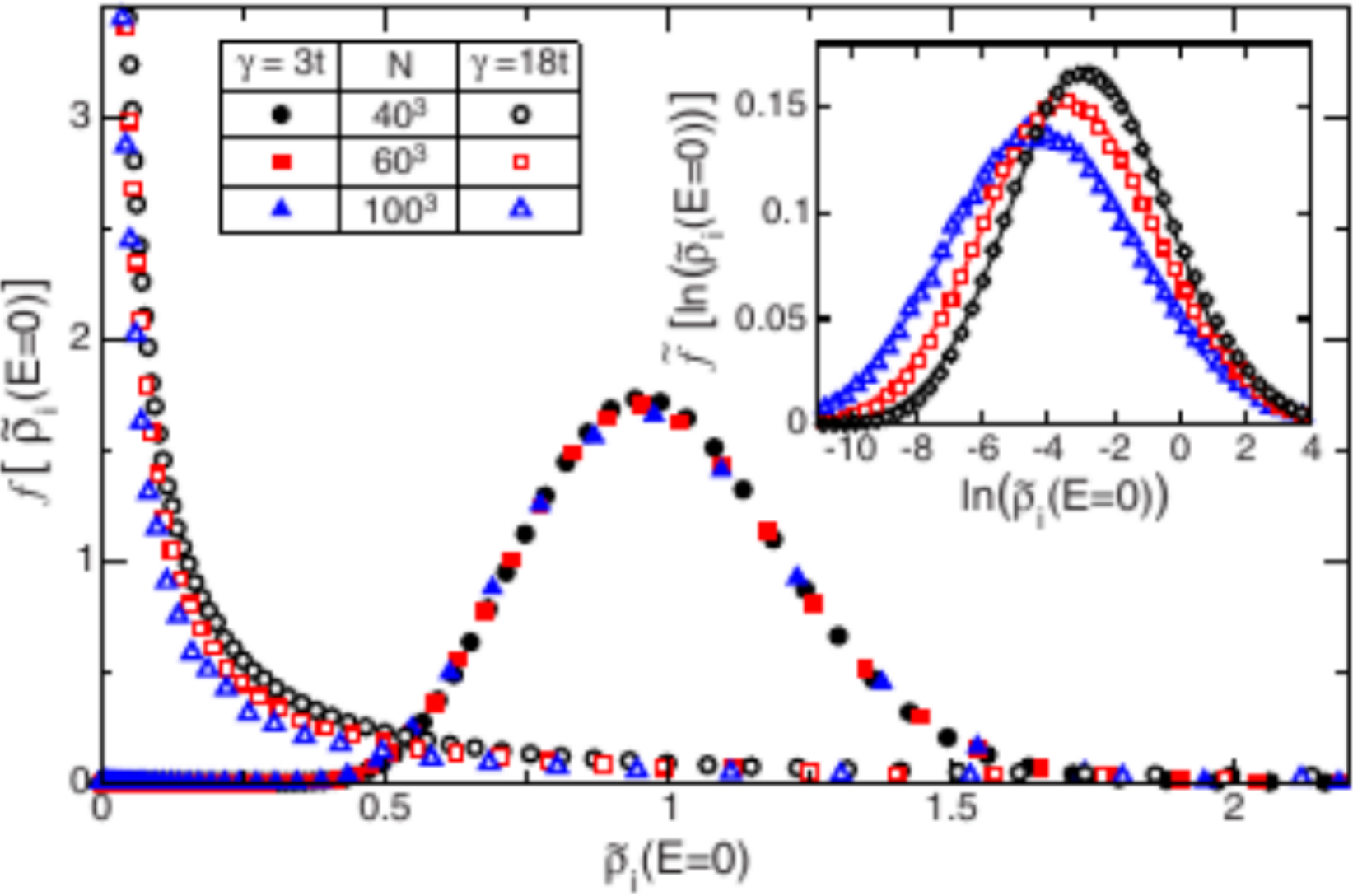}
\caption{The distribution of the local density of states at the band center (zero energy) in a single-band Anderson model with disorder strength $\gamma/t$ where $t=1$ is the near neighbor hopping. Near the localization transition, $\gamma/t=16.5$ the distribution becomes log-normal
(see also the inset) for over ten orders of magnitude, while for values well below the transition, $\gamma/3$ is shown, the distribution is normal~\cite{g_schubert_10}. 
}
\label{fig:lognormal}
\end{figure}

An alternative confirmation is also possible. Early on, Anderson realized that the distribution of the density of states in a strongly disordered metal would be strongly skewed towards smaller values. More recently, this distribution has been demonstrated to be log normal. Perhaps the strongest demonstration of this fact is that DOS near the transition has a log-normal distribution (Figure~\ref{fig:lognormal}) over 10 orders of magnitude~\cite{g_schubert_10}. Furthermore, one may also show that the typical value of a log-normal distribution can be approximated by  the geometric average which is particularly easy to calculate and can serve as an order parameter ~\cite{v_dobrosavljevic_03,g_schubert_10}.  

\subsection{On the role of interactions: Thomas-Fermi screening}
\label{sec:ThomasFermi}

Thus far, we have ignored the role of interactions in our discussion.  Surely the strongest such effect is screening.  In fact, its impact is so large that is often cited as the reason why a sea of electrons act as if they are non-interacting, or free, despite the fact that the average Coulomb interaction is as large or larger than the kinetic energy in many metals~\cite{Thomas_1927,Fermi_1928,Dirac_1930}.  

\begin{figure}[htb]
\centerline{\includegraphics[width=0.35\textwidth,clip=true]{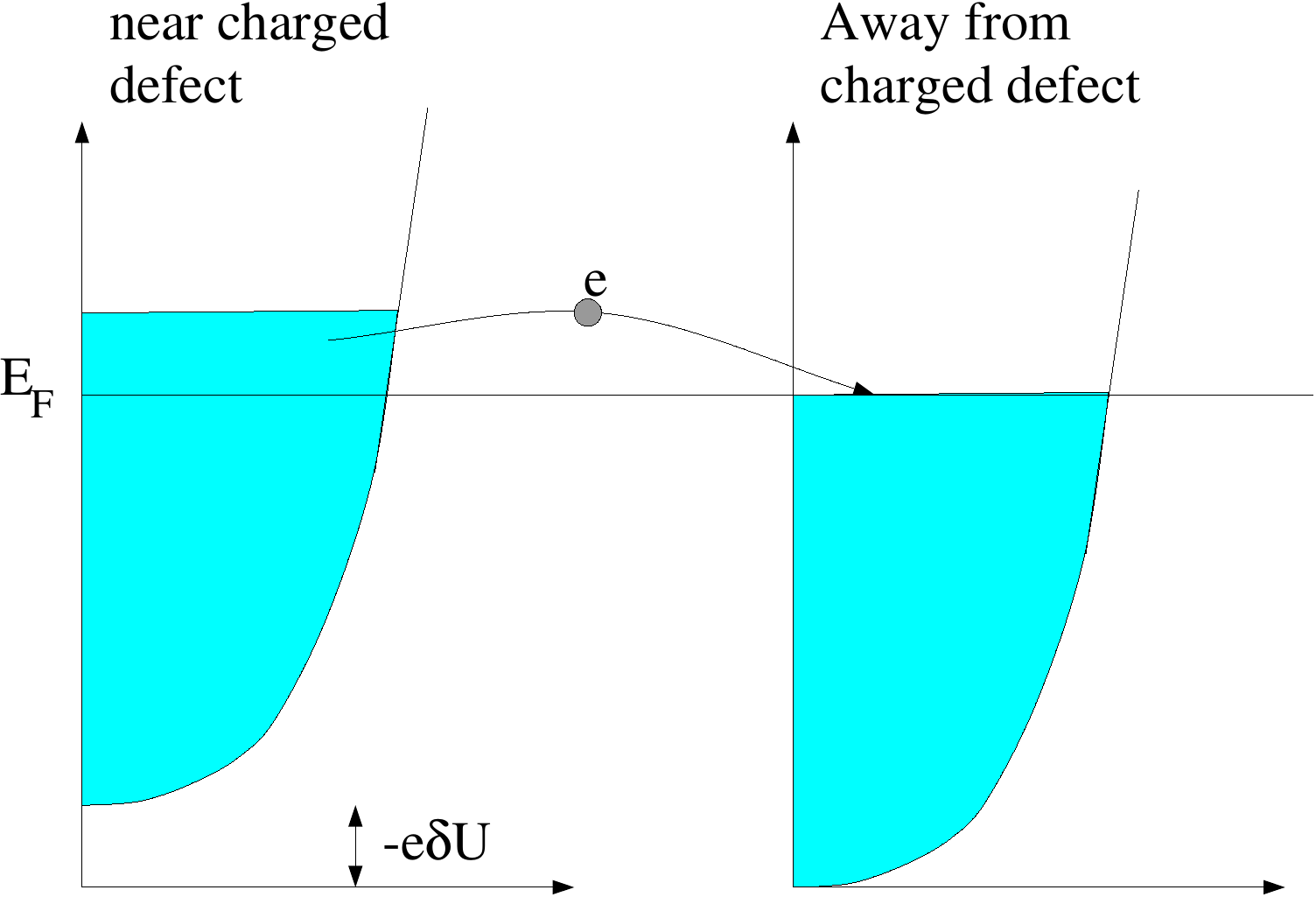}}
\caption[]{The shift in the DOS parabola near a charged defect causes electrons to move away from the defect.}
\label{fig:3DDOS}
\end{figure}

As an introduction to the effect of screening on electronic correlations, consider the effect of a charged defect in a conductor~\cite{h_ibach_09}. Assume that the defect is a cation, so that in the vicinity of the defect the electrostatic potential and the electronic charge density are reduced. We will model the electronic density of states in this material with the DOS of free electrons trapped in a box potential; we can think of this reduction in the local charge density in terms of raising the DOS parabola near the defect (cf.\ Figure~\ref{fig:3DDOS}).  

This will cause the free electronic charge to flow away from the defect. We will treat the screening as a perturbation to the free electron picture, so we assume that the electronic density is just given by an integral over the DOS which we will model with an infinite square well potential with a bare density of states:
\begin{equation}
\rho(E)=\frac{1}{2\pi^2} \lep\frac{2m}{\hbar^2}\rip^{3/2} E^{1/2}\,.
\end{equation}
with the Fermi energy $E_F=\frac{\hbar^2}{2m} \lep 3\pi^2 n\rip^{2/3}$.
If $|e\delta U| \ll E_F$, then we can find the electron density by integrating the bare DOS shifted by the change in potential $+e\delta U$ (c.f.\ Figure~\ref{fig:3DDOS}).
\beq
\delta n(\r) \approx  e\delta U \rho(E_F)\,.
\eeq
The change in the electrostatic potential is obtained by solving the Poisson equation.
\beq
\lap \delta U =4\pi e\delta n=4\pi e^2 \rho(E_F) \delta U\,.
\eeq
The solution is:
\beq
\delta U(\r)=\frac{q e^{-\lambda r}}{r}
\eeq
The length $1/\lambda=r_{TF}$ is known as the Thomas-Fermi screening length.
\beq
r_{TF}=\lep 4\pi e^2 \rho(E_F) \rip^{-1/2}
\eeq
Within this simplified square-well model, $r_{TF}$ in Cu can be estimated to be about 0.5$\AA$.  Thus, if we add a charge defect to Cu metal, its ionic potential is screened away for distances $r>\frac12\AA$.  

\subsection{The Mott transition}
\label{sec:mott}

\begin{figure}[htb]
\centerline{\includegraphics[width=0.45\textwidth,clip=true]{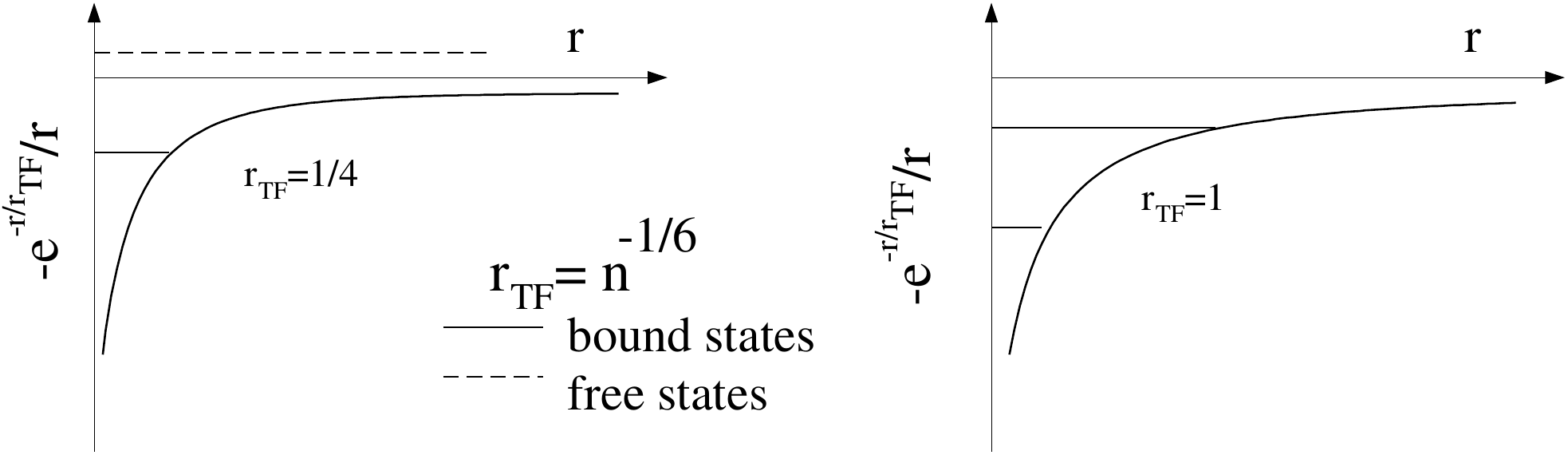}}
\caption[]{Screened defect potentials. The screening length increases with decreasing electron density $n$, causing states that were free to become bound.}
\label{fig:screenpot}
\end{figure}

Consider further, an electron bound to an ion in Cu or some other metal.  As shown in Figure~\ref{fig:screenpot}, as the screening length decreases, the bound states rise up in energy. In a weak metal, in which the valence state is barely free, a reduction in the number of carriers (electrons) will increase the screening length, since
\beq
r_{TF}\sim n^{-1/6}\,.
\eeq
This will extend the range of the potential, causing it to trap or bind more states--making the one free valance state bound.  

Now imagine that instead of a single defect, we have a concentrated system of such ions, and suppose that we decrease the density of carriers (i.e., in Si-based semiconductors, this is done by doping certain compensating dopants, or even by modulating the pressure). This will in turn, increase the screening length, causing some states that were free to become bound, leading to an abrupt transition from a metal to an insulator, and is believed to explain the metal-insulator transition in some transition-metal oxides, glasses, amorphous semiconductors, etc. This metal-insulator transition was first proposed by N.\ Mott, and is called the Mott transition. More significantly Mott proposed a criterion based on the relevant electronic density such that this transition should occur~\cite{n_mott_49,n_mott_68}. 
In Mott's criterion, a metal-insulator transition occurs when the potential generated by the addition of an ionic impurity binds an electronic state. If the state is bound, the impurity band is localized. If the state is not bound, then the impurity band is extended. The critical value of $\lambda=\lambda_c$ may be determined numerically~\cite{y_li_06} with $\lambda_c/a_0 \approx 1.19$, which yields the Mott criterion of
\beq
2.8 a_0 \approx n_c^{-1/3} \, ,
\eeq 
where $a_0$ is the Bohr radius. Despite the fact that electronic interactions are only incorporated in the extremely weak coupling limit, Thomas-Fermi Screening, Mott's criterion still works for moderately and strongly interacting systems~\cite{a_pergament_14}.

While the Mott and Anderson localization mechanisms are quite different, the TDOS can be used as an order parameter in both cases. In the Anderson metal-insulator transition, the transition is entirely due to disorder, with no interaction effects.  In the Mott metal-insulator transition, although the described system is surely strongly disordered, these effects do not contribute to the mechanism of localization.  Nevertheless, both transitions share the same order parameter. On the insulating side of the transition the localized states are discrete so that the typical DOS is zero, while on the extended side of the transition, these states mix and broaden into a band with a finite typical and average DOS. So, both transitions are characterized by the vanishing typical DOS, thus it may serve as an order parameter in both cases.  

Finally, note that while the Mott transition is quite often associated with strong electronic correlations (in clean systems), for impurities in metals with screened Coulomb interactions, such transition occurs already in the weak coupling regime. Thus, any cluster solver which captures interaction effects, at least at the Thomas-Fermi level, (including DFT), with the additional condition to self-consist the impurity potentials, should be able to capture the physics of this transition.

\subsection{Interacting disordered systems: beyond the single particle description}
\label{sec:interacting}

The interplay of strong electronic interactions and disorder and its relevance to the metal-insulator transition, remains an open and challenging question in condensed matter physics. There was an exciting revival of the field after the pioneering experiments by Kravchenko~\etal in low-density high mobility MOSFETs~\cite{Kravchenko_etal_1995,Kravchenko_etal_1994,dobrosavljevic_mosfet,wigner_mott}. These experiments 
provided a clear evidence for a metal-insulator transition in such 2D systems, which 
contradicted the paradigmatic scaling theory of localization according to which the absence of metallic behavior is expected in non-interacting disordered electron systems in $D\leq 2$. 

Incorporating electron-electron interactions into the theory has been problematic mainly due to the fact that when both disorder and interactions are strong, the perturbative approaches break down. Perturbative renormalization group calculations found indications of metallic behavior, but in the case without a magnetic field or magnetic impurities, the runaway flow was towards a strong coupling region outside of the controlled perturbative regime and hence the results were not conclusive~\cite{Finkelstein_1984a,Finkelstein_1984b,Finkelstein_1984c,Finkelstein_1983,p_lee_85,Belitz_Kirkpatrick_1993,Castellani_etal_1984a}. 

Numerical methods for the study of systems with both interactions and disorder are rather limited. Accurate results are largely based on some variants of exact diagonalization on small clusters.
Given this difficulty, the effective medium DMFT-like approaches for localization would be particularly helpful. In particular, the approaches which employ the typical density of states in the dynamical mean field theory present a new opportunity for the study of interacting disordered systems. Consequently, interesting questions which are controversial in the effective field theory approach, can be studied from an entirely different perspective. These include the density of states of the disordered Fermi liquid at low dimensions, the existence of a direct metal to Anderson insulator transition, and the criticality in the transition between the metallic phase and the Anderson phase. 

In refs.~\cite{Vlad_tmt_critical,k_byczuk_05,k_byczuk_09} the generalized DMFT, using the numerical renormalization group as the impurity solver, was used to study the Anderson-Hubbard model.  Here, a typical medium calculated from the geometric averaged density of states instead of the usual linear averaged density of states as that in the CPA~\cite{k_byczuk_05}, was used to determine the effective medium. The effect of disorder and interactions on the Mott and Anderson transitions is investigated, and it is shown that the typical density of states can be treated as an order parameter even for the interacting system. However, all these calculations were performed with a local single-site approximation. In Sec.~\ref{sec:TMDCAinteracting} we show that the cluster extension, within the TMDCA framework can treat the effects of disorder and interaction on an equal footing. It thus provides a new framework for the study of interplay between Mott-Hubbard and Anderson localization.

\section{Direct numerical methods for strongly disordered systems} 
\label{sec:DirectMethods}

Here we provide a brief overview of some of the popular numerical methods proposed for the study of disordered lattice models, including the transfer matrix, kernel polynomial, and exact diagonalization methods. These methods will be used to benchmark and verify our quantum cluster method. We will outline the main steps of these methods, highlighting their advantages and limitations, particularly for applying to materials with disorder.

\subsection{Transfer matrix method}
\label{sec:TMM}
The transfer matrix method (TMM) is used extensively on various disorder problems \cite{Kramer_etal_2010,Markos_2006,Kramer_MacKinnon_1993}. Unlike brute force diagonalization methods, the TMM can handle rather large system sizes. When combined with finite-size scaling, this method is very robust for detecting the localization transition and its corresponding exponents. Most of the accurate estimates of critical disorder and correlation length exponents for disorder models in the literature are based on this method \cite{Kramer_MacKinnon_1993,Markos_2006}.

The simplifying assumption of the TMM is that the system can be decomposed into many slices, and each slice only connects to its adjacent slice. Precisely for this reason, the TMM is not ideal for models with long range hopping, or long range disorder potentials or interactions.

\begin{figure}[h!]
 \includegraphics[trim = 45mm 60mm 45mm 60mm,width=1\columnwidth,clip=true]{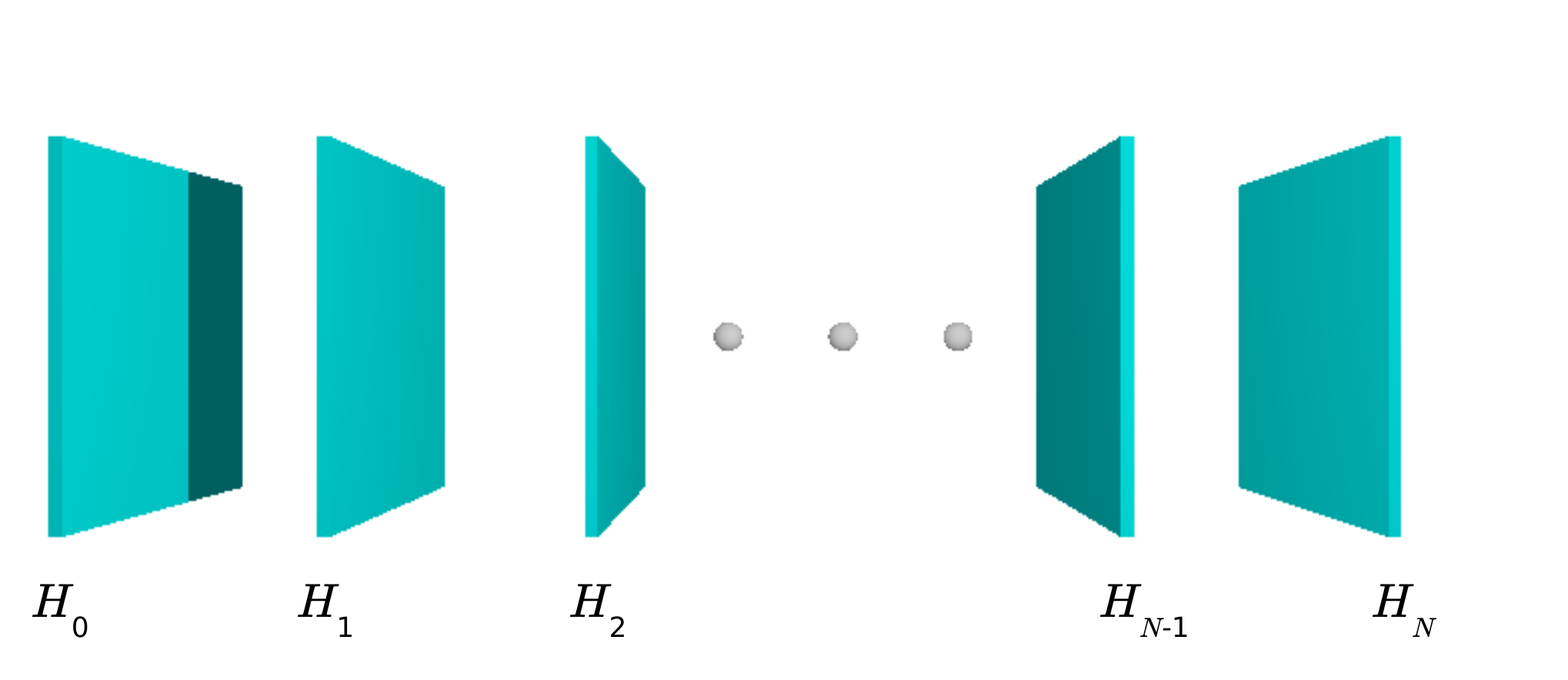}
 \caption{Schematic of a transfer matrix method (TMM) calculation. Assuming the system has a width and height equal to $M$ for each slice of a $N$-slice cuboid, forming a ``bar'' of length $N$, the amplitude of the wavefunction in the 0-th slice can be related to that in the N-th slice via the transfer matrix, Eq.~\ref{tmm_eq}.  }
 \label{fig:KMscaling}
\end{figure}

We can understand the computational scaling of the TMM by a simple 3D example without an explicit interaction. We assume the system has a width and height equal to $M$ for each slice of a $N$-slice cuboid, forming a ``bar'' of length $N$.  The Hamiltonian can be decomposed into the form 
\begin{equation}
H = \sum_{i} H_{i} + \sum_{i} (H_{i,i+1}+H.c.),
\end{equation}
where $H_{i}$ describes the Hamiltonian for slice $i$ and $H_{i,i+1}$ contains the coupling terms between the $i$ and $i+1$ slices. The Schr\"odinger equation can be written as 
\begin{equation}
{H}_{n,n+1} \psi_{n+1} = (E - { H}_n) \psi_n - { H}_{n, n-1} \psi_{n-1} \,, 
\label{andSchro} 
\end{equation}
where $\psi_i$ is a vector with $M^{2}$ components which represent the wavefunction of the slice $i$. This may be reinterpreted as an iterative equation 
\begin{equation}
\left[  \begin{array}{c} \psi_{i+1}\\ \psi_{i} \end{array} \right] =  T_i \times \left[ \begin{array}{c} \psi_{i} \\ \psi_{i-1} \end{array} \right].
\label{tmm_eq}
\end{equation}
where the transfer matrix
\begin{equation}
T_i = \begin{bmatrix} H_{i,i+1}^{-1} (E-H_{i}) & -H_{i,i+1}^{-1} H_{i,i-1} \\ 1 & 0 
\end{bmatrix}\,.
\label{transferMatrix}
\end{equation}

The goal of the transfer matrix method is to calculate the localization length, $\lambda_{M}(E)$ for a system with linear size $M$ at energy $E$, from the product of $N$ transfer matrices
\begin{equation}
\tau_{N} \equiv \prod_{i=1}^{N} T_{i}.
\label{Qn}
\end{equation}
The Lyapunov exponents, $\alpha$, of the matrix $\tau_{N}$ is given by the logarithm of its eigenvalues, $Y$, at the limit of $N \rightarrow \infty$, $ \alpha = lim_{N\to\infty} \frac{ ln(Y)}{N}$. The smallest exponent corresponds to the slowest exponential decay of the wavefunction and thus can be identified as corresponding to the localization length, $\lambda_{M}(E) = 1 / \alpha_{min}$ \cite{Derrida_Gardner_1984,Oseledets_1968,Pichard_1986,Furstenberg_1963,MacKinnon_Kramer_1983,Furstenberg_Kesten_1960,Pichard_Sarma_1981}.

Since the repeated multiplication of $T_{i}$ is numerically unstable, periodic reorthogonalization is needed in the numerical implementation \cite{Kramer_etal_2010,Markos_2006,Kramer_MacKinnon_1993}. For the 3D Anderson model, the reorthogonalization is done for about every 10 multiplications. This is the major bottleneck for the TMM method, as reorthogonalization scales as the third power of the matrix size. Therefore, the method in general scales as $M^3$.

\subsection{Kernel polynomial method}
\label{sec:KPM}
The kernel polynomial method (KPM) is a procedure for fitting a function onto an orthogonal set of polynomials of finite order. For the study of disordered systems, the functions which are routinely calculated by the KPM include the density of states and the conductance \cite{Wang_1994,Silver_Roder_1994,Silver_roeder_Voter_etal_1996,Silver_Roeder_1997,a_weisse_06}. These quantities are not representable by smooth functions, indeed they are often the sum of a set of delta functions. Two outstanding characteristics of fitting such functions to orthogonal polynomials are that the delta functions are smoothed out, and that the fitted function is usually accompanied with undesirable Gibbs oscillations. Different kernels for reweighing the coefficients of the polynomial are devised to lessen such oscillations.

Here we highlight the main steps for calculating the density of states by the KPM. For such a polynomial expansion it is more convenient to rescale the Hamiltonian so that the eigenvalues fall in the range of $[-1,1]$. We assume that the eigenvalues of the Hamiltonian are properly scaled and shifted to be within this range. The density of states is given as a sum of delta functions, 
\begin{equation}
\rho(E) = \sum_{i} \delta(E-E_{i})
\approx \sum_{n=0}^{n_{max}} g_{n} \mu_{n} T_{n}(E), 
\end{equation}
where $g_{n}$ is the kernel function, $\mu_{n}$ is the expansion coefficient, and $T_{n}$ is the Chebyshev polynomial. Jackson's kernel is usually used for the $g_{n}$ \cite{Jackson_1930}. The expansion coefficient is given as $\mu_{n}=\int_{-1}^{1}\rho(E)T_{n}(E)dE = \frac{1}{D}\sum_{k=0}^{D-1} \langle k|T_{n}(H)| k \rangle$, where $D$ is the size of the Hilbert space. The efficiency of the KPM is based on a simple sampling of a small number of basis functions instead of the full summation. The $T_{n}(H)|k \rangle$ for different values of $n$ can be calculated with the recursion relation of the Chebyshev polynomial. The dominant part in using the recursion relation is the matrix vector multiplication.

The Hamiltonian matrix is usually very sparse. For example, the number of non-zero matrix elements for a 3D Anderson model on a simple cubic lattice is seven for each row.   This number does not change with system size. The method is rather versatile and can be adapted for almost any Hamiltonian. Unlike the TMM, the KPM can handle long-range hopping and long-range disorder potentials. It can also be used for interacting systems; however, the matrix size grows exponentially \cite{a_weisse_06}, limiting practical calculations to a few tens of orbitals.

\subsection{Diagonalization methods }
\label{sec:diagmethods}

Diagonalization methods are designed to solve the matrix problem, $H\psi = E\psi$, directly. A full matrix diagonalization scales with the third power of the matrix size. So, practical calculations are often limited to matrix sizes of the order of ten thousand. For the study of the localization transition, we are usually interested in the states close to the Fermi level. Indeed, most of the numerical studies of the Anderson model are focused on the energy at the band center \cite{Kramer_MacKinnon_1993}.
Methods have been proposed for calculating the eigenvalues and eigenvectors for sparse matrices in the vicinity of a target eigenvalue, $\sigma$. Particularly, the Lanczos~\cite{Lanczos_1950} and Arnoldi~\cite{Arnoldi_1951} methods have been widely used for strongly correlated systems~\cite{Lin_etal_1993,Weisse_Fehske_2008,Noack_Manmana_2005}. The feature common to these methods is the Krylov subspace, $K$, generated by repeatedly multiplying a matrix, $H$, on an initial trial vector, $\psi_{t}$, 
\begin{equation}
K^{j} = \{\psi_{t}, H\psi_{t}, H^{2}\psi_{t}, H^{3}\psi_{t}, \cdots\, H^{j-1}\psi_{t}\}.
\end{equation}
As all the vectors generated converge towards the eigenvector with the lowest eigenvalue, the basis set that is generated is ill-conditioned for large $j$. 

The solution is to orthogonalize the basis at each step of the iteration via the Gram-Schmidt process. In essence, the difference between the Lanczos and Arnoldi methods is in the number of vectors in the Gram-Schmidt process. The Arnoldi method uses all the vectors and the Lanczos method only uses the two most recently generated vectors. The original matrix can then be projected into the Krylov subspace of much smaller size, where it may be fully diagonalized~\cite{Ericsson_Ruhe_1980}.

The dominant component of the computation is the matrix-vector multiplication described above.  This scales only linearly with the matrix size. For the ground state calculation, matrix sizes of over one billion are routinely done \cite{Kawamura_etal_2017}; however, calculating the inner spectrum is somewhat more difficult. The matrix has to be shifted and then inverted to transform the target eigenvalue to the extremal eigenvalue. 
\begin{equation}
(H-\sigma \I)^{-1} \psi = \frac{1}{E-\sigma} \psi, 
\end{equation}
The inverse of the Hamiltonian with a shifted spectrum is generally not known. Then, instead of expanding the basis in the Krylov subspace, the Jacobi-Davidson method (JDM) is often employed~\cite{Davidson_1975}.  It expands the basis ($\mathbf{u_{0}, u_{1}, u_{2}, \cdots}$) using the Jacobi orthogonal component correction which may be written as

\begin{equation}
H ( \mathbf{u_{j}}+\boldsymbol{\delta}) = (\theta_{j}+\epsilon) (\mathbf{u_{j}}+\boldsymbol{\delta}) \;\;\;\; \forall \;\;\;\; \mathbf{u_{j}} \perp \boldsymbol{\delta},
\end {equation}
where $( \mathbf{u_{j}},\theta_{j})$ and $( \mathbf{u_{j}} +  \boldsymbol{\delta}$,$\theta_{j}+\epsilon)$ are the approximate and the exact eigenvector and eigenvalue pairs, respectively. Upon solving the equation for the vector $\boldsymbol{\delta}$, a new basis vector $\mathbf{u_{j+1}}=\mathbf{u_{j}}+\boldsymbol{\delta}$ is included in the subspace. Matrix inversion is again involved in solving the equation. Various pre-conditioner are proposed for a quick approximation of the  matrix inverse~\cite{Davidson_1975}. JADAMILU is a popular package which implements the JDM with an incomplete LU factorization \cite{Dupont_Kendall_Rachford_1968,Meijerink_Vorst_1977} as a pre-conditioner~\cite{Bollhofer_Notay_2007}.

The scaling of this method seems to be strongly dependent on the Hamiltonian. It tends to be more efficient for matrices which are diagonally dominant, but much less so when off-diagonal matrix elements are large. This is probably due to the difficulty of obtaining a good approximation of the inverse based on the incomplete LU factorization used as a pre-conditioner.

Exact diagonalization methods provide an accurate variational approximation for the eigenvalues and eigenvectors of the Hamiltonian, thus allowing the calculation of quantities such as multifractal spectrum and entanglement spectrum which are difficult to obtain from other approaches \cite{Rodriguez_etal_2010,Ujfalusi_Varga_2015}. On the other hand, Krylov subspace methods are not a good option for calculating the density of states as only one, or a few, eigenstates are targeted at each calculation. A self-consistent treatment of the interaction, even at a single particle level, would also be rather challenging. Clearly, the major obstacle for applying it to systems with an explicit interaction is again the exponential growth of the matrix size with respect to the system size. 

While these numerical methods can provide very accurate results for the models which are non-interacting, single band, and with local or short-ranged disorder, applying them to chemically specific calculations is a major challenge. None of these conditions is satisfied for realistic models of materials with disorder. In this case, the complexity of these methods increases drastically and obtaining accurate results for sufficiently large system sizes to perform a finite size scaling analysis is often impossible. This highlights the importance, or perhaps necessity, of the coarse grained methods described below.

\section{Coarse grained methods}
\label{sec:CGmethods}
In this section and corresponding subsections, we discuss coarse-graining as a unifying concept behind quantum cluster theories such as the CPA and DMFT as well as their cluster extension, the DCA, which preserve the translational invariance of the original lattice problem.  All quantum cluster theories are defined by their mapping of the lattice to a self-consistency embedded cluster problem, and the mapping from the cluster back to the lattice.  The map from the lattice to the cluster in these quantum cluster methods may be obtained when the coarse-graining approximation is used to simplify the momentum sums implicit in the irreducible Feynman diagrams of the lattice problem (see subsection~\ref{sec:fundamentals}).  As discussed in Secs.~\ref{sec:DMFT} and \ref{sec:DCA} this approximation is equivalent to the neglect of momentum conservation at the internal vertices, which is exact in the limit of infinite dimensions, and systematically restored in the DCA.  The resulting diagrams are identical to those of a finite-sized cluster embedded in a self-consistently determined dynamical host.  The cluster problem is then defined by the coarse-grained interaction and bare Green's function of the cluster.  The mapping from the cluster back to the lattice is motivated in Sec.~\ref{sec:DCAderiv} by the observation that irreducible or compact diagrammatic quantities are much better approximated on the cluster than their reducible counterparts. This mapping may also be obtained by optimizing the lattice free energy, as discussed in Sec~\ref{sec:DCAfromPhi}.

\subsection{A few fundamentals sec:fundamentals}
\label{sec:fundamentals}

In this section, we will introduce two central paradigms in the physics of many-body systems: the Anderson and Hubbard models of disordered and interacting electrons on a lattice, respectively.  We will then use perturbation theory to prove and demonstrate some fundamental ideas.

	Consider an Anderson model with diagonal disorder, described by the Hamiltonian
\begin{equation}
H =
\displaystyle - \sum_{<ij>,\sigma} 
t \left( 
c^{\dagger}_{i,\sigma} c_{j,\sigma} + 
c^{\dagger}_{j,\sigma} c_{i,\sigma}
\right)  \\[5mm]
\displaystyle+
 \sum_{i \sigma}(V_i- \mu) n_{i,\sigma}
 \label{eq:AMHamiltonian}
\end{equation}
where $c^{\dagger}_{i,\sigma}$ creates a quasiparticle on site $i$ with spin $\sigma$, and $n_{i,\sigma}=c^{\dagger}_{i,\sigma} c_{i,\sigma}$.  The disorder occurs in the local orbital energies $V_i$, which we assume are independent quenched random variables distributed  according to some specified probability distribution  $P(V)$. 

\begin{figure}[htb]
\includegraphics*[width=3.5in]{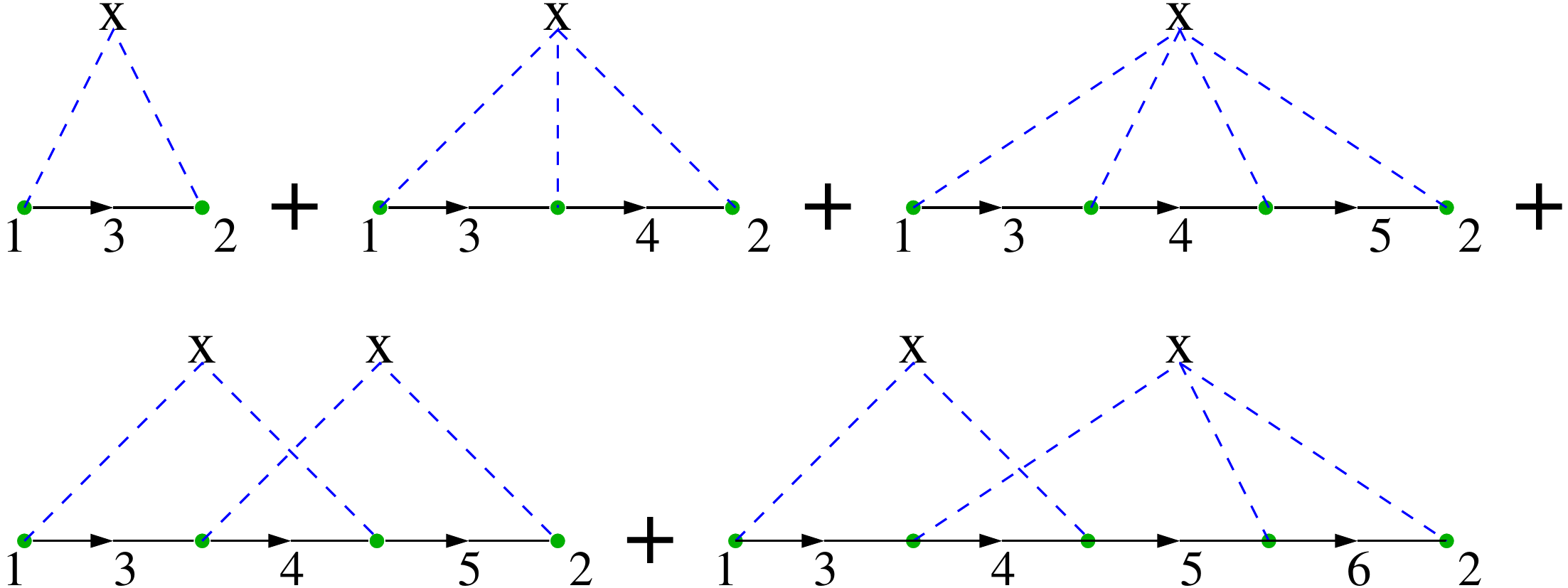}
\caption{The first few graphs in the irreducible self energy of a diagonally disordered system.  Each $\circ$ represents the scattering of a state $\k$ from sites (marked $X$) with a local disorder potential distributed  according to some specified probability distribution  $P(V)$.  The numbers label the $\k$ states of the fully-dressed Green's functions, represented by solid lines with arrows.}
\label{fig:dis_dia}
\end{figure}

The effect of the disorder potential $\sum_{i \sigma}V_i n_{i,\sigma}$ can be described using standard diagrammatic perturbation theory (although we will eventually sum to {\em{all}} orders).  It may be re-written in reciprocal space as
\begin{equation}
H_{dis} = \frac{1}{N} \sum_{i,\k,\k',\sigma} V_i
c^{\dagger}_{\k,\sigma} c_{\k',\sigma} e^{i\r_i(\k-\k')}
\end{equation}
The corresponding irreducible (skeleton) contributions to the self energy may be represented diagrammatically\cite{a_gonis_92} and the first few are displayed in Figure~\ref{fig:dis_dia}.  Here each $\circ$ represents the scattering of an electronic Bloch state from a local disorder potential at some site $X$. The dashed lines connect scattering events that involve the same local potential.  In each graph, the sums over the sites are restricted so that the different $X$ 's  represent scattering from {\em{different}} sites. No graphs representing a single scattering event are included since these may simply be absorbed  as a renormalization of the chemical potential $\mu$ (for single band models).

	Translational invariance and momentum conservation are restored by averaging over all possible values of the disorder potentials $V_i$.  For example\cite{m_jarrell_01a}, consider the second diagram in Figure~\ref{fig:dis_dia}, given by 
\begin{equation}
\frac{1}{N^3}\sum_{i,\k_3, \k_4} \langle V_i^3 \rangle
G(\k_3)G(\k_4) e^{i\r_i\cdot(\k_1-\k_3+\k_3-\k_4+\k_4-\k_2)}\,,
\end{equation}
where $G(\k)$ is the disorder-averaged single-particle Green's function for state $\k$.  The average over the distribution of scattering potentials $\langle V_i^3 \rangle = \langle V^3 \rangle$ is independent of the position $i$ in the lattice. After summation over the remaining labels, this becomes
\begin{equation}
\langle V^3 \rangle G(\r=0)^2 \delta_{\k_1,\k_2}\,,
\end{equation}
where $G(\r=0)$ is the local Green's function.  Thus the second diagram's contribution to the self energy involves only local correlations. Since the internal momentum labels always cancel in the exponential, the same is true for all non-crossing diagrams shown in the top half of Figure~\ref{fig:dis_dia}.
  
Only the diagrams with crossing dashed lines have non-local contributions.  Consider the fourth-order diagrams such as those shown on the bottom left and upper right of Figure~\ref{fig:dis_dia}. During the disorder averaging, we generate potential terms $\langle V^4 \rangle$ when the scattering occurs from the same local potential (i.e.\ the third diagram) or $\langle V^2 \rangle^2$ when the scattering occurs from different sites, as in the fourth diagram.  When the latter diagram is evaluated, to avoid overcounting, we need to subtract a term proportional to $\langle V^2 \rangle^2$ but corresponding to scattering from the same site.  This term is needed to account for the fact that the fourth diagram should really only be evaluated for sites $i \neq j$. For example, the fourth diagram yields
\begin{eqnarray*}
\large\langle \frac{1}{N^4} \sum_{i\neq j \k_3 \k_4 \k_5} 
V_i^2 V_j^2
e^{i\r_i\cdot(\k_1+\k_4-\k_5-\k_3)}
e^{i\r_j\cdot(\k_5+\k_3-\k_4-\k_2)}\nonumber\\
G(\k_5)G(\k_4)G(\k_3)
\large\rangle
\end{eqnarray*}
Evaluating the disorder average $\langle\rangle$, we get the following two terms:
\begin{eqnarray}
\frac{1}{N^4} \sum_{i j \k_3 \k_4 \k_5} 
\langle V^2 \rangle^2
e^{i\r_i\cdot(\k_1+\k_4-\k_5-\k_3)}e^{i\r_j\cdot(\k_5+\k_3-\k_4-\k_2)}\nonumber\\
G(\k_5)G(\k_4)G(\k_3) \nonumber\\
-
\frac{1}{N^4} \sum_{i \k_3 \k_4 \k_5} 
\langle V^2 \rangle^2 e^{i\r_i\cdot(\k_1-\k_2)}
G(\k_5)G(\k_4)G(\k_3)
\end{eqnarray}
Momentum conservation is restored by the sum over $i$ and $j$; i.e.~over all possible locations of the two scatterers.  It is reflected by the Laue functions, $\Lambda=N\delta_{\k+\cdots}$, within the sums
\begin{eqnarray}
\frac{\delta_{\k_2,\k_1}}{N^3} \sum_{\k_3 \k_4 \k_5} 
\langle V^2 \rangle^2
N\delta_{\k_2+\k_4,\k_5+\k_3}\nonumber\\
G(\k_5)G(\k_4)G(\k_3)\nonumber \\
-
\frac{\delta_{\k_2,\k_1}}{N^3} \sum_{\k_3 \k_4 \k_5} 
\langle V^2 \rangle^2
G(\k_5)G(\k_4)G(\k_3) 
\label{eval_dia4}
\end{eqnarray}
Since the first term in Eq.~\ref{eval_dia4} involves convolutions of $G(\k)$ it 
reflects non-local  correlations. Local contributions such as the second term in Eq.~\ref{eval_dia4} can be combined together with the contributions from the corresponding local diagrams such as the third diagram in Figure~\ref{fig:dis_dia}  by replacing $\langle V^4 \rangle$ in the latter by the cumulant $\langle V^4 \rangle - \langle V^2 \rangle^2$ .  Given the fact that different $X$'s must correspond to different sites, it is easy to see that all crossing diagrams must involve non-local correlations.

\begin{figure}[htb]
\includegraphics*[width=3.5in]{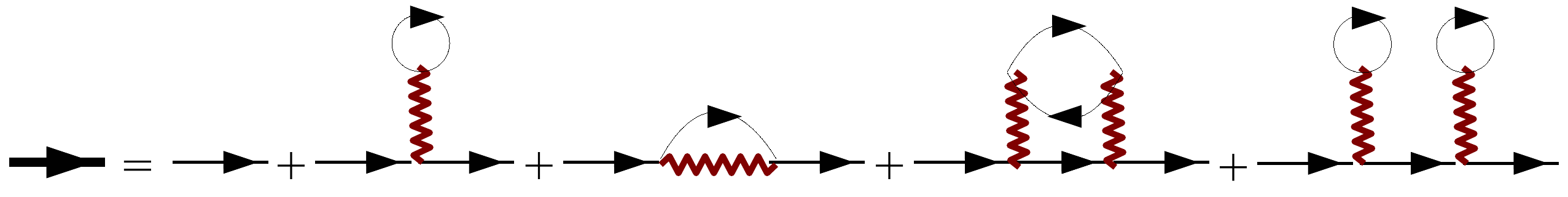}
\caption{The first few diagrams for the Hubbard model single-particle Green's function.  Here, the solid black line with an arrow represents the single-particle Green's function and the wavy line the Hubbard $U$ interaction.}
\label{fig:GPT}
\end{figure}
The developed formalism also works for interacting systems.  Again we will use perturbation theory to illustrate some of these ideas. Consider the Hubbard model ~\cite{j_hubbard_63} which is the simplest model of a correlated electronic lattice system.  Both it and the $t-J$ model are thought to at least qualitatively describe some of the properties of transition metal oxides, and high temperature superconductors\cite{p_anderson_06}. The Hubbard model Hamiltonian is given as 
\begin{equation}
H = - t \sum_{\langle j,k\rangle \sigma} ( c_{j\sigma}^{\dag }
c_{k\sigma} + c_{k\sigma}^{\dag  } c_{j\sigma} ) 
+
U\sum_in_{i\uparrow}n_{i\downarrow}
\label{eq:HMHamiltonian}
\end{equation}
where $c_{j\sigma}^{\dag}$ ($c_{j\sigma}$) creates (destroys) an electron at site $j$ with spin $\sigma$, $n_{i\sigma}=c_{i\sigma}^{\dag  } c_{i\sigma}$ stands for the particle number at a given site $i$. The first term describes the hopping of electrons between nearest-neighboring sites $i$ and $j$, and the $U$ term describes the interaction between two electrons once they meet at a given site $i$. 

As for the disordered case described above, the effect of the local Hubbard $U$ potential can be described using standard diagrammatic perturbation theory. The first few diagrams for the single-particle Green's function are shown in Figure~\ref{fig:GPT}.  Very similar arguments to those employed above may be used to show that the first self energy correction to the Green's function is local whereas some of the higher order graphs reflect non-local contributions.

\subsection{The Laue function and the limit of infinite dimension}
\label{sec:DMFT}

The local approximation for the self energy was used by various authors in 
perturbative calculations as a simplification of the k-summations which render the problem intractable. 
It was only after the work of Metzner and Vollhardt~\cite{w_metzner_89a,w_metzner_89b} and M\"uller-Hartmann~\cite{e_mullerhartmann_89a,e_mullerhartmann_89b} who showed that this approximation becomes exact in the limit of infinite dimension that it received extensive attention. Precisely in this limit, the spatial dependence of the self energy disappears, retaining only its variation with time. Please see the reviews by Pruschke {\it et al}~\cite{t_pruschke_95} and Georges {\it et al}~\cite{a_georges_96a} for a more extensive treatment. 

In this section, we will show that the DMFT and CPA share a common interpretation as coarse graining approximations in which the propagators used to calculate the self energy $\Sigma$ and its functional derivatives are coarse-grained over the entire Brillouin zone.  M\"uller-Hartmann~\cite{e_mullerhartmann_89a,e_mullerhartmann_89b} showed that it is possible to completely neglect momentum conservation so that this coarse-graining becomes exact in the limit of infinite-dimensions.  For simple models like the Hubbard and Anderson models, the properties of the bare vertex are completely characterized by the Laue function $\Lambda$ which expresses the momentum conservation at each vertex.  In a conventional diagrammatic approach
\begin{eqnarray}
\Lambda(\k_1,\k_2,\k_3,\k_4) &=&
\sum_{\r} \exp{\left[ \nonumber
i\r\cdot(\k_1+\k_2-\k_3-\k_4)\right] }\nonumber \\
&=& N \delta_{\k_1+\k_2,\k_3+\k_4} \,,
\label{Laue_3D}
\end{eqnarray}
where $\k_1$ and $\k_2$ ($\k_3$ and $\k_4$) are the momenta entering (leaving) each vertex through its legs of Green's function $G$.  However as the dimensionality $D\to\infty$, M\"uller-Hartmann showed that the Laue function reduces to\cite{e_mullerhartmann_89a}
\begin{eqnarray}
\label{Laue_infD}
\Lambda_{D\rightarrow\infty}({\bf k}_1,{\bf k}_2,{\bf k}_3,{\bf k}_4)=
1+{\cal O}(1/D)\quad\mbox{.}
\end{eqnarray}
\begin{figure}[htb]
\includegraphics*[width=3.5in]{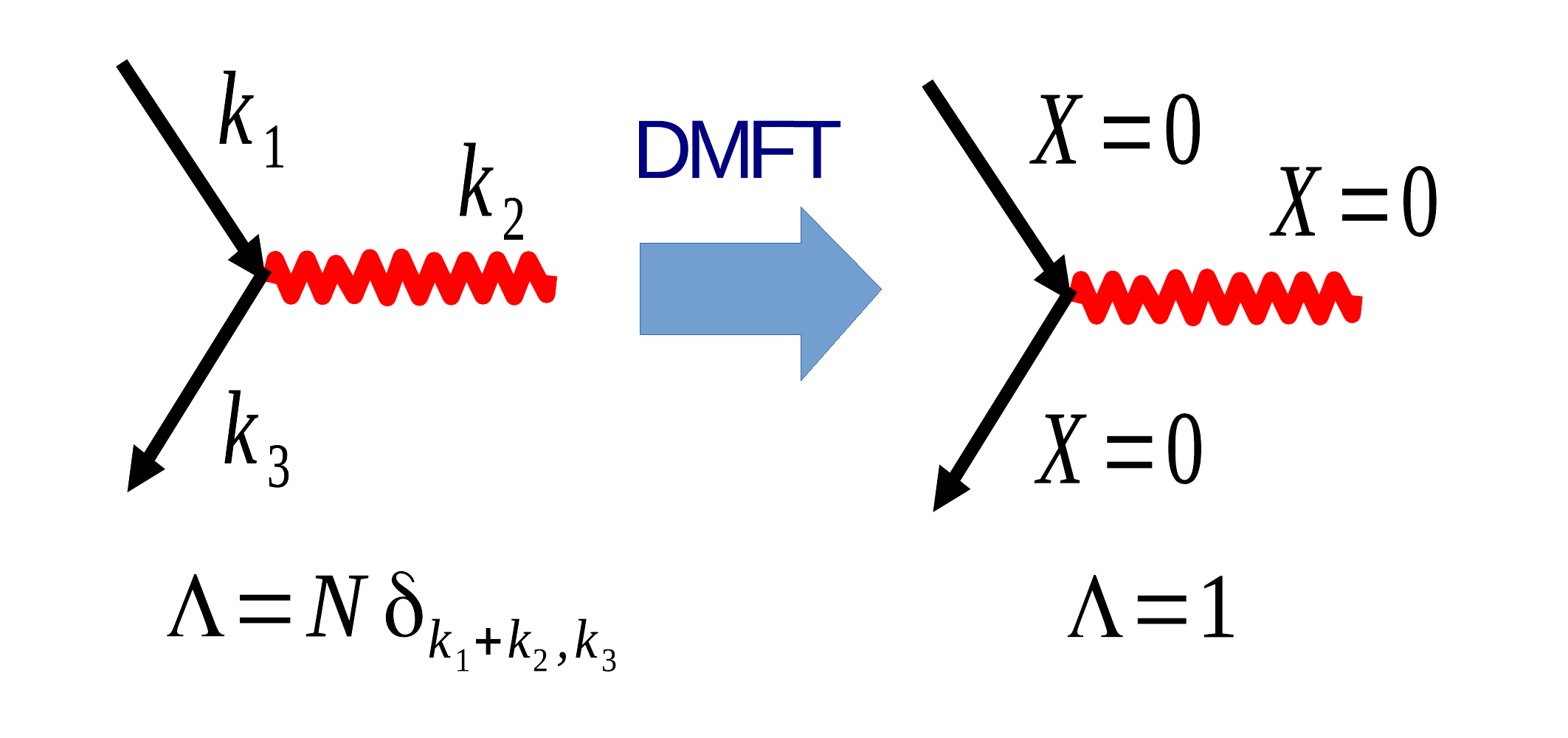}
\caption{The Laue function $\Lambda$, which described momentum conservation at a vertex (left) with two Green's function solid lines and a wiggly line denoting an interaction (perhaps mediated by a Boson). In the DMFT/CPA we take $\Lambda=1$, so momentum conservation is neglected for irreducible graphs (right) so that we may freely sum over the momentum labels $\kt,\kt '\cdots$ leaving only local ($\X=0$) propagators and interactions.}
\label{fig:DMFT_Laue}
\end{figure}
The DMFT/CPA assumes the same Laue function, $\Lambda_{DMFT}(\k_1,\k_2,\k_3,\k_4)=1$, even in the context of finite dimensions.  More generally, for an electron scattering from an interaction (boson) pictured in Figure~\ref{fig:DMFT_Laue}, $\Lambda_{DMFT}(\k_1,\k_2,\k_3)=1$.  Thus, the conservation of momentum at internal vertices is neglected.  We may freely sum over the internal momentum labels of each Green's function leg and interaction leading to a collapse of the momentum dependent contributions leaving only local terms.

\begin{figure}
\includegraphics*[width=3.5in]{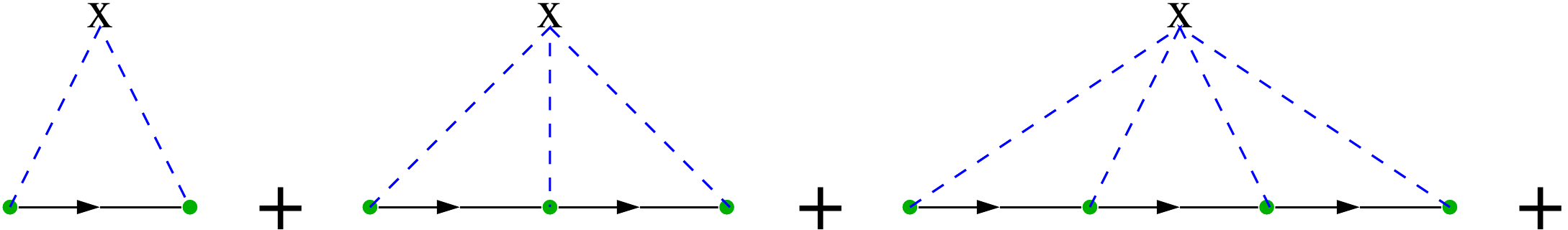}
\caption[a]{The first few graphs of the CPA local self energy of the Anderson model.  Here the solid Green's function line represents the average local propagator and the dashed lines the impurity scattering.  These graphs may be obtained from the full set of graphs shown in Figure~\ref{fig:dis_dia} by replacing each graphical element (Green's function and impurity scattering lines) with its local analog coarse-grained through the entire first Brillouin zone.}
\label{fig:cpa_dia}
\end{figure}

These arguments may then be applied to the self energy $\Sigma$, which becomes a local (momentum-independent) function. For example, in the CPA for the Anderson model, nonlocal correlations involving different scatterers are ignored.  Thus, in the calculation of the self energy, we ignore all of the crossing diagrams shown on the bottom of Figure~\ref{fig:dis_dia}; and retain only the class of diagrams such as those shown on the top representing scattering from a single local disorder potential.  These diagrams are shown in Figure~\ref{fig:cpa_dia}.  

It is easy to show this reduction in the number and complexity of the graphs is fully equivalent to the neglect of momentum conservation at each internal vertex.  This is accomplished by setting each Laue function within the sum (eg., in Eq.~\ref{eval_dia4}) to 1.  We may then freely sum over the internal momenta, leaving only local propagators.  All non-local self energy contributions (crossing diagrams) must then vanish.  For example, consider again the fourth graph at the bottom of Figure~\ref{fig:dis_dia}. If we replace the Laue function $N\delta_{\k_1+\k_4,\k_5+\k_3}\to 1$ in Eq.~\ref{eval_dia4}, then the two contributions cancel and this diagram vanishes.  

Thus an alternate definition of the CPA, in terms of the Laue functions $\Lambda$, is
\begin{equation}
\Lambda = \Lambda_{CPA} = 1
\end{equation}
I.e., the CPA is equivalent to the neglect of momentum conservation at all internal vertices of the disorder-averaged irreducible graphs.  It is easy to see that this same definition applies to the DMFT for the Hubbard model.  This will be done below in the context of a generating functional based derivation.


Now it is easy to see that both DMFT and CPA employ the locality of the self energy $\Sigma(\omega)$ in their construction. As a result, the two algorithms are very similar, they both employ the mapping of the lattice problem onto an impurity embedded in an effective medium, described by a local self energy $\Sigma(\omega)$ which is determined self-consistently.  The perturbative series for the self energy $\Sigma$ in the DMFT/CPA are identical to those of the corresponding impurity model, so that conventional impurity solvers may be used.   However, since most impurity solvers can be viewed as methods that sum all the graphs, not just the skeleton ones,  it is necessary to exclude $\Sigma(\omega)$ from the bare local propagator ${\cal G (\omega)}$ input to the impurity solver in order to avoid overcounting the local self energy $\Sigma(\omega)$~\cite{m_jarrell_92a} corrections. This is typically done via the Dyson's equation, $ {\cal G}(\omega)^{-1}=G(\omega)^{-1} + \Sigma(\omega)$ where $G(\omega)$ is the full local Green's function. Hence, in the local approximation, the Hubbard model has the same diagrammatic expansion as an Anderson impurity with a bare local propagator ${\cal G}(\omega;\Sigma)$  which is determined self-consistently.

\begin{figure}[htb]
\includegraphics*[width=3.5in]{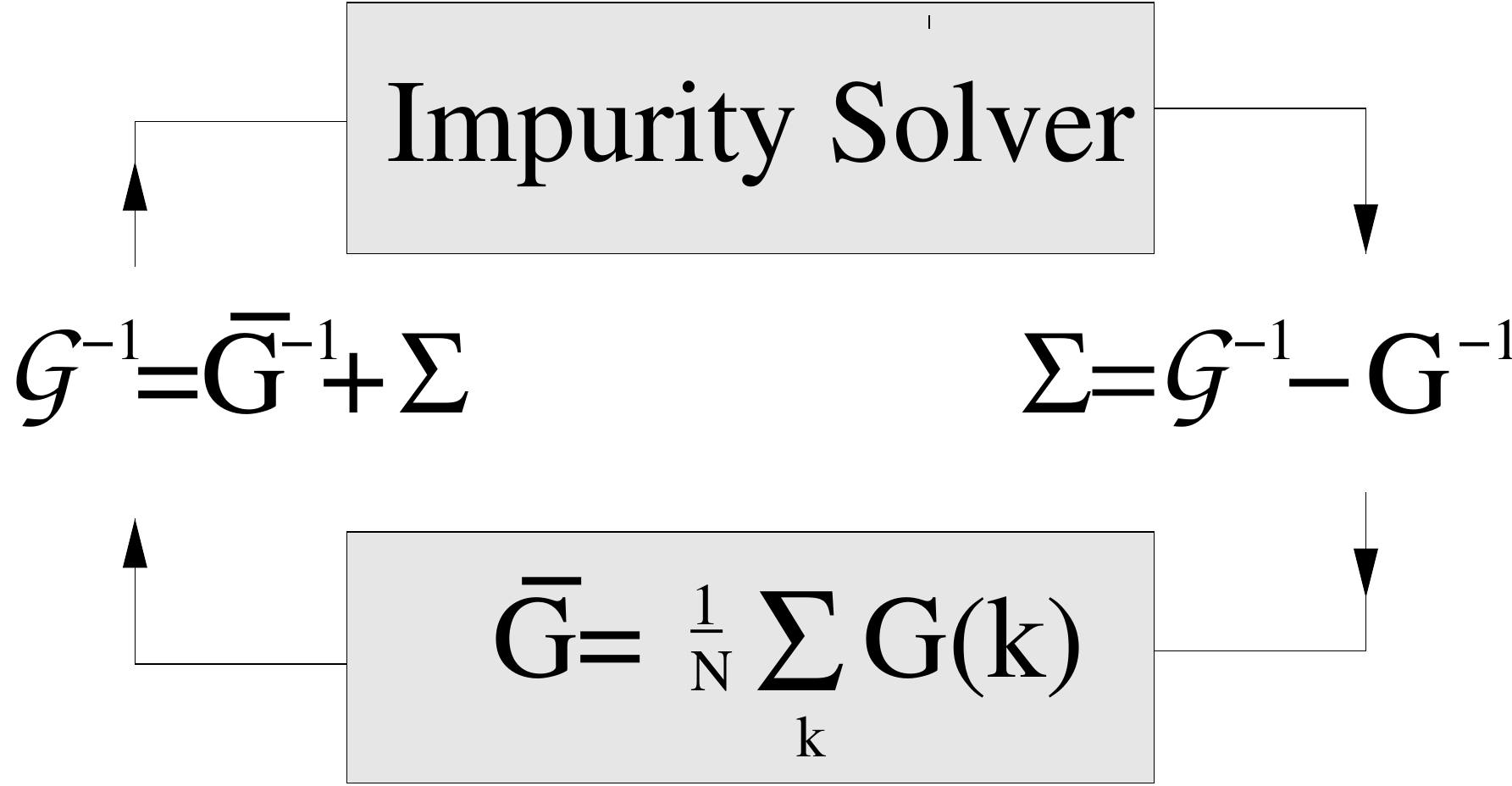}
\caption{The DMFT/CPA self-consistency algorithm}
\label{fig:algorithm_DMFT}
\end{figure}

A generalized algorithm constructed for such local approximations is the following (see Figure ~\ref{fig:algorithm_DMFT}): 
(i) An initial guess for $\Sigma(\omega)$ is chosen (usually from perturbation theory).
(ii) $\Sigma(\omega)$ is used to calculate the corresponding coarse-grained local Green's function
\begin{equation}
\label{eq:gloc}
\bar{G}(\omega)=\frac{1}{N}  \sum_{\k} G(\k,\omega) \, .
\end{equation}
(iii) Starting from $\bar{G}(\omega)$ and $\Sigma(\omega)$ used in the second step, the host Green's function ${\cal G}(\omega)^{-1}=\bar{G}(\omega)^{-1} + \Sigma(\omega)$ is calculated.  It serves as the bare Green's function of the impurity model. (iv) starting with ${\cal G}(\omega)$ as an input, the impurity problem is solved for the local Green's function $G(\omega)$ (various impurity solvers are available, including  QMC, enumeration of disorder, NRG, etc..). (v) Using the  impurity solver output for the impurity Green's function $G(\omega)$ and the host Green's function ${\cal G}(\omega)$ from the third step, a new $\Sigma(\omega)={\cal G}(\omega)^{-1}-G(\omega)^{-1}$ is calculated, which is then used in step (ii) to reinitialize the process. Steps (ii) - (v) are repeated until convergence is reached. 

\subsection{The Dynamical cluster approximation }
\label{sec:DCA}

        In this section, we will review the dynamical cluster approximation (DCA) formalism\cite{m_hettler_98a,m_hettler_00a,m_jarrell_96b,th_maier_05a}.  We motivate the fundamental idea of the DCA which is coarse-graining and then use it to define the relationship between the cluster and lattice at the one and two-particle level.

\subsubsection{Coarse-graining }
\label{sec:CG}

\begin{figure}[htb]
\parbox[b]{3.5in}{\includegraphics*[width=3.4in]{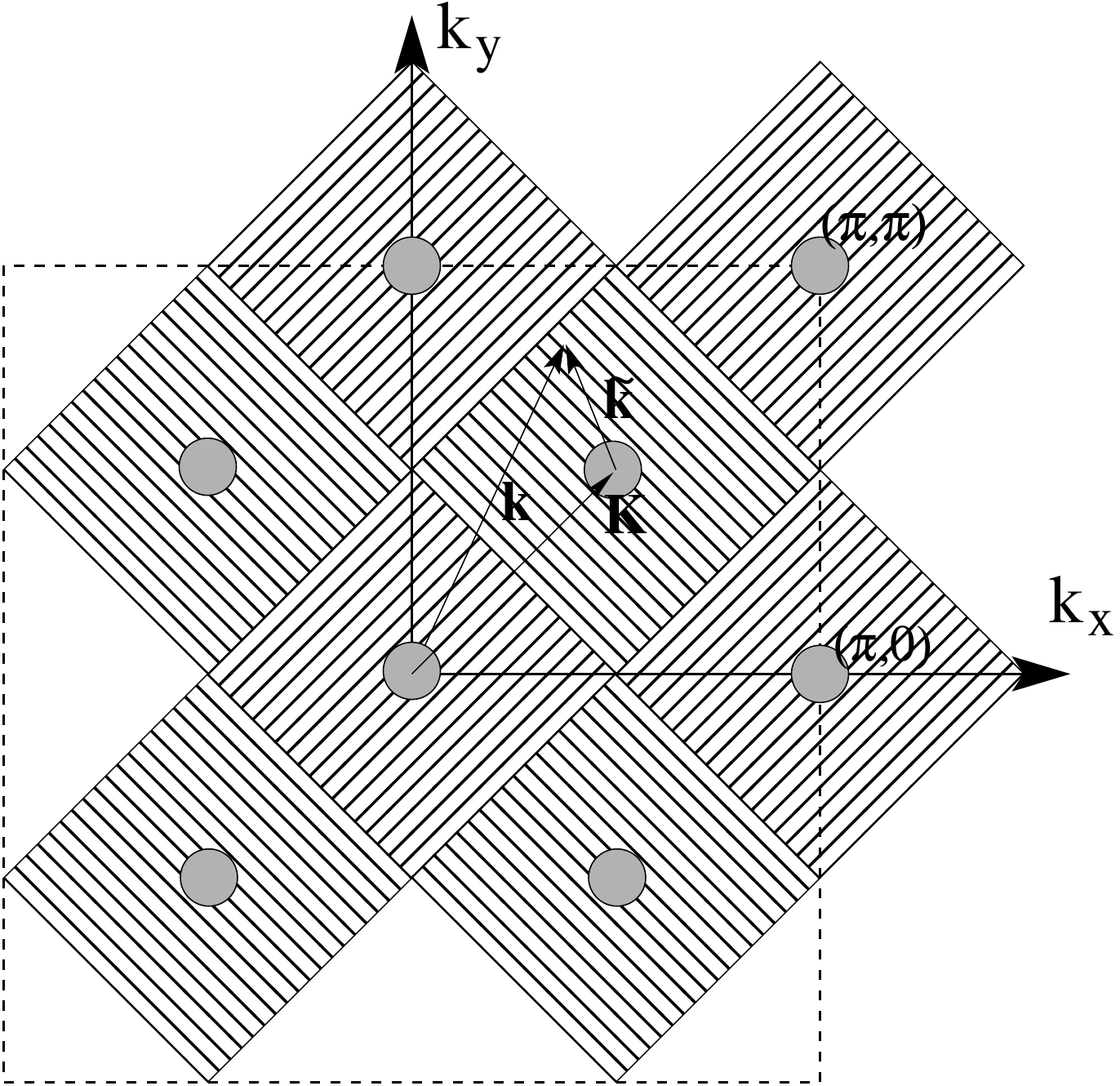}}
\caption{Coarse-graining cells for $N_c=8$ (differentiated by alternating fill patterns) that partition the first Brillouin Zone (dashed line).  Each cell is centered on a cluster momentum $\K$ (filled circles). To construct the DCA cluster (e.g.\ for $N_c=8$) we map a generic $\k$ to the nearest cluster point $\K=\M(\k)$ (c.f.\ \ref{fig:CG_Nc8}) so that $\kt=\k-\K$ remains in the cell around $\K$. }
\label{fig:BZ_Nc8}
\end{figure}
Like the DMFT/CPA, in the DCA the mapping from the lattice to the cluster diagrams is accomplished via a coarse-graining transformation. In the DMFT/CPA, the propagators used to calculate $\Sigma$ and its functional derivatives are coarse-grained over the entire Brillouin zone, leading to local (momentum independent) irreducible quantities.  In the DCA, we wish to relax this condition, and systematically restore momentum conservation and non-local corrections.  

Thus, in the DCA, the reciprocal space of the lattice (Figure~\ref{fig:BZ_Nc8}) which contains $N$ points is divided into $N_c$ cells of identical linear size $\Delta k$.  The geometry and point groups of these clusters may be determined by considering real-space finite size clusters of size $N_c$ that are able to tile the lattice of size $N$.  The tiling momenta $\K$ are conjugate to the location of the sites in the cell labeled by $\X$, while the coarse-graining wavenumbers $\kt$ label the wavenumbers within each cell surrounding $\K$ and are conjugate to the real-space labels of the cell centers $\tilde{x}$.

\begin{figure}[htb]
\parbox[b]{3.5in}{\includegraphics*[width=3.4in]{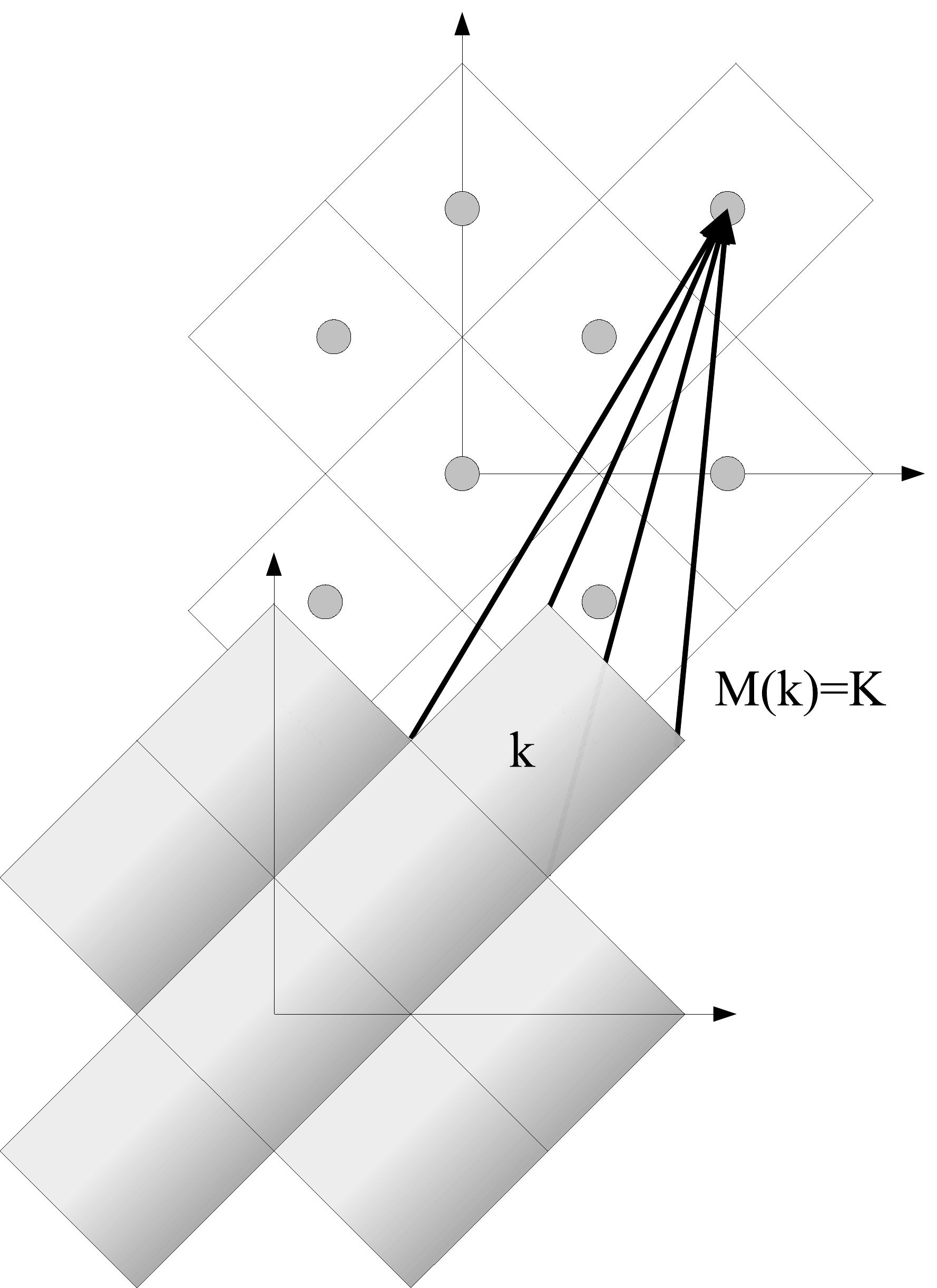}}
\caption{The DCA many-to-few mapping of an arbitrary point in the first Brillioun zone to one of $N_c=8$ cluster momenta $\K$. }
\label{fig:CG_Nc8}
\end{figure}

The coarse-graining transformation is set by averaging the function within each cell as illustrated in Figure~\ref{fig:CG_Nc8}.  For an arbitrary function $f(\k)$ (with $\k=\K+\kt$), this corresponds to 
\begin{equation}
{\bar {f}} (\K) = \frac{N_c}{N} \sum_{{\kt}} f(\K+\kt)
\label{eq:CG}
\end{equation}
where $\kt$ label the wavenumbers within the coarse-graining cell adjacent to $\K$.  According to Nyquist's sampling theorem\cite{m_weik_01}, to reproduce the function $f$ at lengths $\alt L/2$ in Eq.~\ref{eq:CG}, we only need to sample the reciprocal space at intervals of $\Delta k\approx 2\pi/L$.  Eq.~\ref{eq:CG} may be interpreted as the sum of $N/N_c$ such samplings. 

Knowledge of $f$ on a finer scale in momentum than $\Delta k$ is unnecessary, and may be discarded to reduce the complexity of the problem.  For example, convolutions of periodic functions $f$ may be approximated as
\begin{equation}
g(\q) = \frac{1}{N}\sum_\k f(\k+\q) f(\k) \approx \frac{1}{N_c}\sum_\K {\bar {f}} (\K+\Q) {\bar {f}} (\K) + {\cal{O}} (\Delta k^2)\, ,
\label{eq:CGconvolution}
\end{equation}
where $\Q=M(\q)$. Eq.~\ref{eq:CGconvolution} is an approximation where we first average the function over a set of D dimensional cells and then perform a sum over the cells.  Thus, reducing the numerical complexity from order $N$ to order $N_c$ floating point operations.  

\subsubsection{DCA: a diagrammatic derivation}
\label{sec:DCAderiv}
        This coarse graining procedure and the relationship of the DCA to the local approximations (DMFT/CPA) is illustrated by a microscopic diagrammatic derivation ~\cite{m_jarrell_01a} of the DCA. We chose disorder case for the demonstration.  Quantum cluster theories are defined by two mappings:  one from the lattice to the cluster and the other from the cluster back to the lattice.  
        
\paragraph{Map from the lattice to the cluster} To define the first mapping, we start from the diagrams in the irreducible self energy $\Sigma(V,G)$ of the Anderson model illustrated in Figure~\ref{fig:dis_dia}.  We saw above, that when we completely neglect momentum conservation by first coarse graining the interactions and Green's functions over the entire first Brillioun zone, the diagrams corresponding to non-local corrections vanish, leaving the reduced set of local diagrams which constitute the CPA illustrated in Figure~\ref{fig:cpa_dia}.  The resulting approximation shares the limitations of a local approximation, described above, including the neglect of non-local correlations. 
        
The DCA systematically incorporates such neglected non-local correlations by systematically restoring the momentum conservation at the internal vertices of the self energy $\Sigma$.  To this end, the Brillouin-zone is divided into $N_c=L_c^D$ cells of size $\Delta k=2\pi/L_c$ (c.f.~Figure~\ref{fig:BZ_Nc8} for $N_c=8$).  Each cell is represented by a cluster momentum $\bf K$ in the center of the cell. We require that momentum conservation is (partially) observed for momentum transfers between cells, i.e., for momentum transfers larger than $\Delta k$, but neglected for momentum transfers within a cell, i.e., less than $\Delta k$. This requirement can be established by using the Laue function \cite{m_hettler_00a}
\begin{equation}
\label{Laue_DCA}
\Lambda_{DCA}(\k_1,\k_2,\k_3,\k_4)=
N_c \delta_{\M(\k_1)+\M(\k_2),\M(\k_3)+\M(\k_4)} \quad\mbox{,}
\end{equation}
where $\M(\k)$ is a function which maps $\k$ onto the momentum label $\K$ of the cell containing $\k$ (see, Figure~\ref{fig:BZ_Nc8}).  This choice for the Laue function systematically interpolates between the exact result, Eq.~\ref{Laue_3D}, which it recovers when $N_c\to N$ and the DMFT result, Eq.~\ref{Laue_infD}, which it recovers when $N_c=1$.  With this choice of the Laue function the momenta of each internal leg may be freely summed over the cell.  

This procedure accurately reproduces the physics on short length scales and provides a cutoff of longer length scales where the physics is approximated with the mean field.  For short distances $r\alt L_c/2$, where $L_c$ is now the linear size of the cluster, the Fourier transform of the Green's function $\bar{G}(r) \approx G(r) +{\cal{O}}((r\Delta k)^2)$, so that short ranged correlations are reflected in the irreducible quantities constructed from $\bar{G}$; whereas, longer ranged correlations $r>L_c/2$ are cut off by the finite size of the cluster \cite{m_hettler_00a}.  Longer ranged interactions are also cut off when the transformation is applied to the interaction.  To see this, consider an extended Hubbard model on a (hyper)cubic lattice with the addition of a near-neighbor interaction $V\sum_{\left\langle ij \right \rangle} n_i n_j$ where $ \left\langle ij \right \rangle $ denotes near-neighbor pairs.  When the point group of the cluster is the same as the lattice the coarse-grained interaction takes the form $V \sin(\Delta k/2)/(\Delta k/2) \sum_{\left\langle ij \right \rangle} n_i n_j$.  It vanishes when $N_c=1$ so that $\Delta k = 2\pi$. If  $N_c$ is larger than one, then non-local corrections of length $\approx \pi/\Delta k$ to the DMFT/CPA are introduced. 

\begin{figure}[htb]
\includegraphics*[width=3.5in]{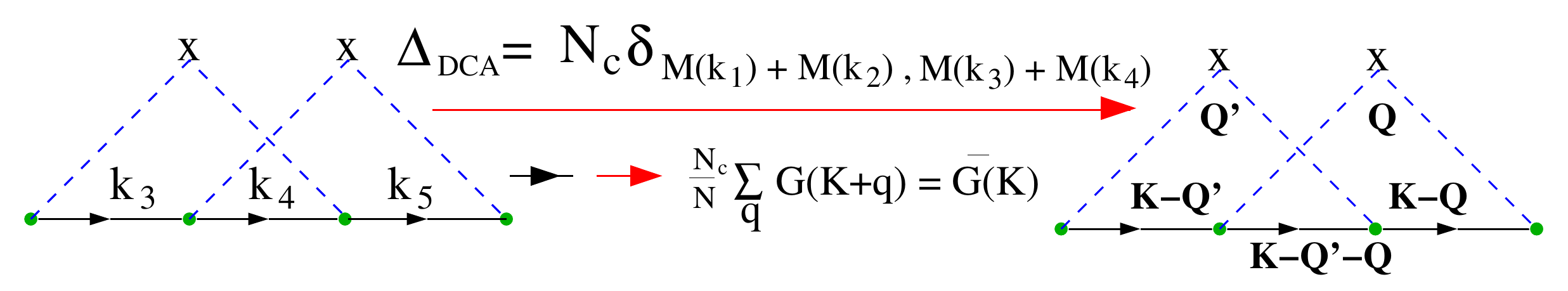}
\caption{ Use of the DCA Laue function $\Lambda_{DCA}$ leads to the replacement of the lattice propagators $G({\bf k}_1)$, $G({\bf k}_2)$, ... by coarse grained propagators $\bar{G}({\bf K})$, $\bar{G}({\bf K}^\prime)$, ... The impurity scattering dashed lines and unchanged by coarse-graining since the scatterings are local.}
\label{fig:collapse_dis}
\end{figure}
When applied to the DCA, the cluster self energy will be constructed from the {\em{coarse-grained average}} of the single-particle Green's function within the cell centered on the cluster momenta.  This is illustrated for a fourth-order term in the self energy shown in Figure~\ref{fig:collapse_dis}.  Each internal leg $G(\k)$ in a diagram is replaced by the coarse--grained Green's function ${\bar G}(\M(\k))$, defined by
\begin{equation}
\label{eq:Gbar}
\bar{G}(\K) \equiv \frac{N_c}{N}\sum_{\kt}G(\K+\kt) \quad\mbox{,} 
\end{equation}
and each interaction in the diagram is replaced by the coarse-grained interaction
\begin{equation}
\label{eq:Vbar}
\bar{V}(\K) \equiv \frac{N_c}{N}\sum_{\kt}V(\K+\kt) \quad\mbox{,} 
\end{equation}
where $N$ is the number of points of the lattice, $N_c$ is the number of cluster $\K$ points, and the $\kt$ summation runs over the momenta of the cell about the cluster momentum  $\K$ (see, Figure~\ref{fig:BZ_Nc8}). For the Anderson model, where the scattering potential is local, the interaction is unchanged by coarse-graining. The diagrammatic sequences for the self energy and its functional derivatives are unchanged; however, the complexity of the problem is greatly reduced since $N_c\ll N$.  

Provided that the propagators are sufficiently weakly momentum dependent, this is a good approximation. If $N_c$ is chosen to be small, the cluster problem can be solved using conventional techniques such as QMC. This averaging process also establishes a relationship between the systems of size $N$ and $N_c$.  When $N_c=N$ a finite size simulation is recovered.  So, there are no mean-field embedding effects, etc.\  

\paragraph{Map from the cluster back to the lattice}  Once the cluster problem is solved, we use the solution of the cluster problem to approximate the lattice problem.  This may be done in a number of ways, and its not \emph{a priori} clear which way is optimal.  At the single-particle particle level, we could, e.g., calculate the cluster single particle Green's function and use it to approximate the lattice result, $G^l(\k,\omega) \approx G^c(M(\k),\omega) $.  Or, at the other extreme, we could calculate the self energy on the cluster, and use it to first approximate the lattice result $\Sigma^l(\k,\omega)\approx \Sigma^c(M(\k),\omega) $, and then use the Dyson equation $G^l(\k,\omega) = \left( 1- \Sigma^c(M(\k),\omega) G^{l,0}(\k,\omega) \right)^{-1} $ to calculate the lattice Green's function ($G^{l,0}(\k,\omega)$ is the bare lattice Green's function).  The second way is far better. We will motivate this mapping with more rigor in the next part, where we calculate and minimize the free energy, but here we offer a physically intuitive motivation.

\begin{figure}[htb]
\includegraphics[trim = 0mm 0mm 0mm 0mm,width=1\columnwidth,clip=true]{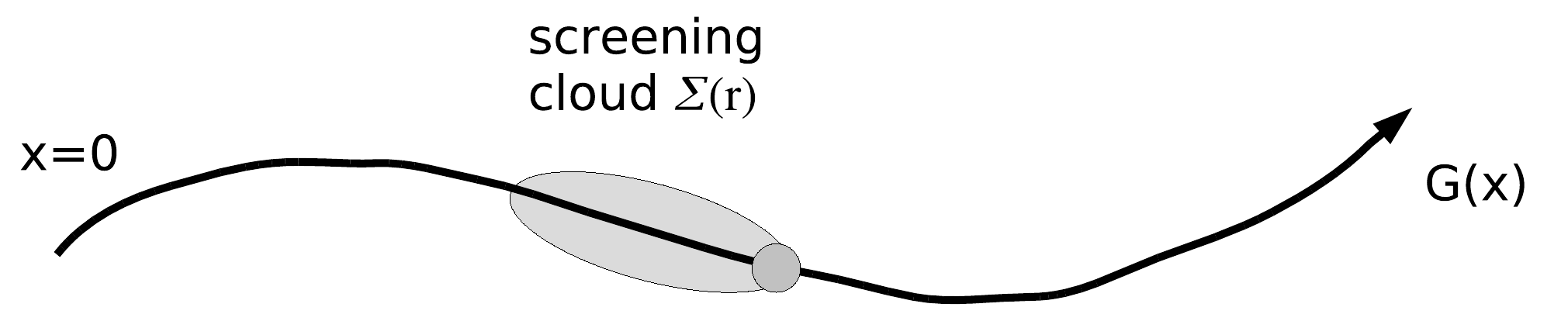}
\caption{Path-integral interpretation of the screening of a propagating particle.  The single particle lattice Green's function, $G^l$, describes the quantum phase and amplitude the particle accumulates along its path as it propagates from space-time location $0$ to $x$.  It is poorly approximated by the cluster Green's function from a small cluster calculation, $G^l\approx G^c$, especially when $x, r \leq  L_c$, the linear cluster size.  Its self energy, which describes generally short ranged $r$ screening processes, is well approximated $\Sigma^l\approx\Sigma^c$, by a small cluster calculation, especially when the cluster size $L_c$ is greater than the screening length.  As discussed in Sec.~\ref{sec:intro} this screening length $f_{TF}\approx r$ which may be less than an Angstrom for a good metal. So, rather than directly approximating the lattice Green's function by the cluster Green's function, the cluster self energy is used to approximate the lattice self energy in a Dyson equation for the lattice Green's function $G^l =G^l + G^{l0} + G^{l0} \Sigma^l G^l $, where $G^{l0}$ is the bare lattice Green's function.}
\label{fig:screening}
\end{figure}

Physically, this is justified by the fact that irreducible terms like the self energy are short ranged, while reducible quantities the $G$ must be able to reflect the long length and time scale physics.  This is motivated in Figure~\ref{fig:screening}.  As the particle propagates from the origin to space-time location $x$, the quantum phase and amplitude it accumulates is described by the single-particle Green's function $G(x)$.  Consequently if $x$ is larger than the size of the DCA cluster, then $G(x)$ is poorly approximated by the cluster Green's function.  However, the Self energy $\Sigma$ describes the many-body processes that produce the screening cloud surrounding the particle.  As we saw in Sec.~\ref{sec:ThomasFermi} these distances are typically very short, on the order of an Angstrom or less, so the lattice self energy is often well approximated by the cluster quantity.

\subsubsection{DCA: a generating functional derivation }
\label{sec:DCAfromPhi}

Finally, in this section, we will derive the DCA for the Hubbard model using the Baym generating functional formalism.  The generating functional $\Phi$ is the collection of all compact closed graphs that may be constructed from the fully dressed single-particle Green's function and the bare interaction.  Starting from the generating functional, it is quite easy to generate the diagrams in the fully irreducible self energy and the irreducible vertex function needed in the calculation of the phase diagram.  Note that in terms of Feynman graphs, each functional derivative $\delta / \delta G_\sigma$ is equivalent to breaking a single Green's function line.  So, the self energy $\Sigma_\sigma$ is obtained from a functional derivative of $\Phi$, $\Sigma_\sigma= \delta \Phi/\delta G_\sigma$, and the irreducible vertices $\Gamma_{\sigma\sigma'} = \delta \Sigma_\sigma/\delta G_{\sigma'}$.  Since we obtain the free energy, Baym's formalism is also quite useful for proving a few essentials. 

\paragraph{Map from the lattice to the cluster}

\begin{figure}[htb]
\includegraphics*[width=3.5in]{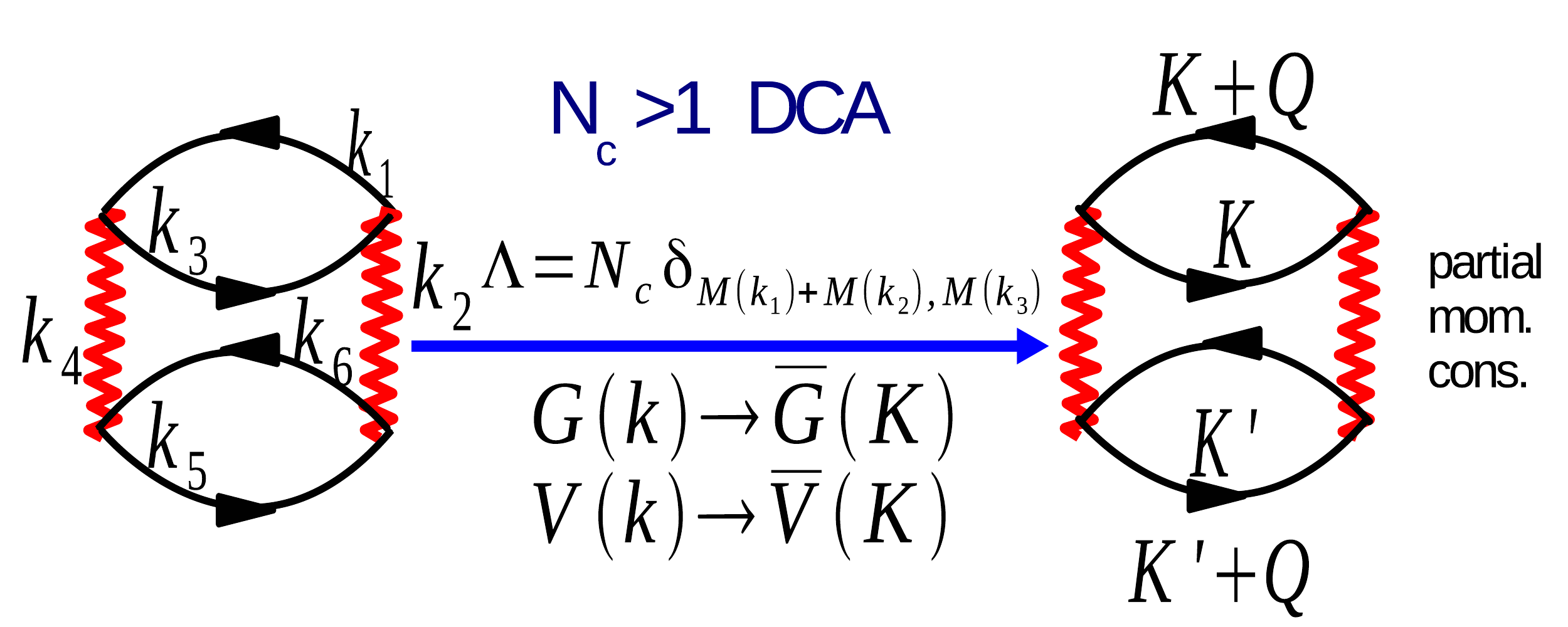}
\caption{A second-order term in the generating functional of the Hubbard model.  Here the undulating line represents the interaction $U$, and on the LHS (RHS) the solid line the lattice (coarse-grained) single-particle Green's functions.  When the DCA Laue function is used to describe momentum conservation at the internal vertices, the momenta collapse onto the cluster momenta and each lattice Green's function and interaction is replaced by the corresponding coarse-grained result. }
\label{fig:collapse_DCA}
\end{figure}

\begin{figure}[htb]
\includegraphics*[width=3.5in]{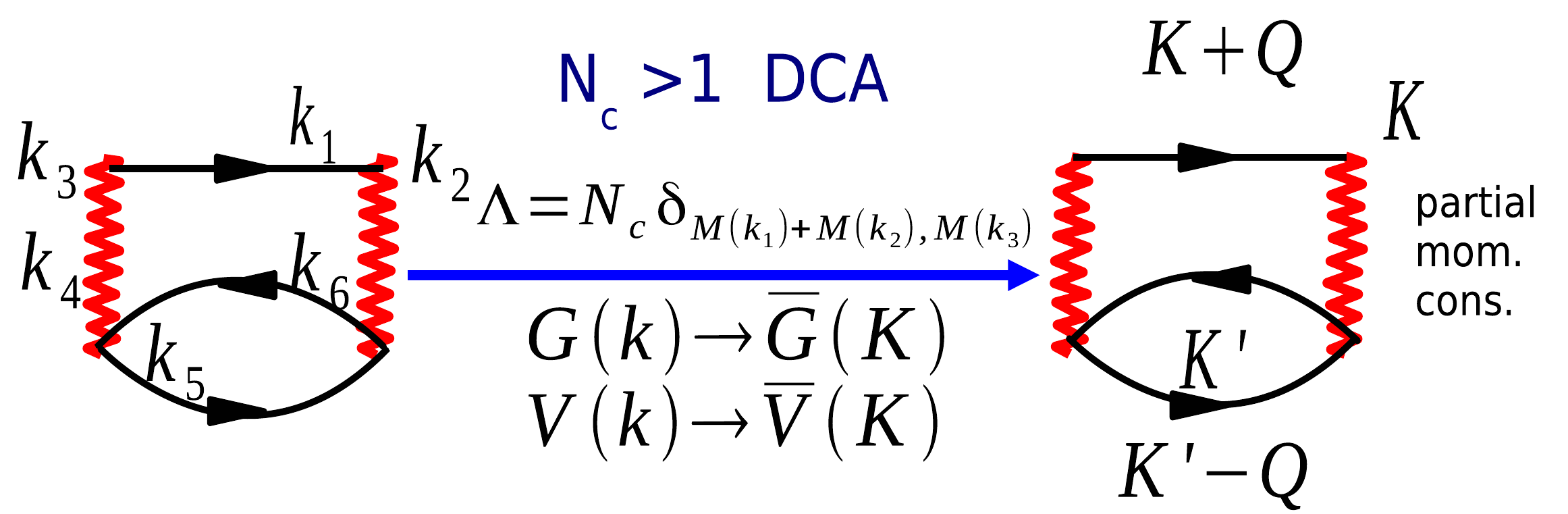}
\caption{A second-order term in the self energy of the Hubbard model obtained from the first functional derivative of the corresponding term in the generating functional $\Phi$ (Figure~\ref{fig:collapse_DCA}).  When the DCA Laue function is used to describe momentum conservation at the internal vertices, the momenta collapse onto the cluster momenta and each lattice Green's function and interaction is replaced by the corresponding coarse-grained result. }
\label{fig:SE_CG}
\end{figure}

To derive the DCA, we first apply the DCA coarse-graining procedure to the diagrams in the generating functional $\Phi(G,U)$. In the DCA, we obtain an approximate $\Phi^c$ by applying the DCA Laue function to the internal vertices of the lattice $\Phi^l$. This is illustrated for the second order term in Figure~\ref{fig:collapse_DCA}
It is easy to see that the corresponding term in the self energy $\Sigma^{(2)}$ is obtained from a functional derivative of $\Phi^{(2)}$, $\Sigma^{(2)}_\sigma= \delta \Phi^{(2)}/\delta G_\sigma$, and the irreducible vertices $\Gamma^{(2)}_{\sigma\sigma'} = \delta \Sigma^{(2)}_\sigma/\delta G_{\sigma'}$.  This is illustrated for the second order self energy in Figure~\ref{fig:SE_CG}.

Above, we justified these approximations in wavenumber space; however, one may also make a real-space argument.  In high spatial dimensions $D$, one may show~\cite{w_metzner_89a,e_mullerhartmann_89a} that $G(r,\tau)$ falls of exponentially quickly with increasing $r$ $G(r,\tau) \sim t^r \propto d^{-r/2}$ while the interaction remains local.  Thus, when $D=\infty$ all non-local graphs vanish.  In finite $D$, due to causality, we may expect the Green's functions to fall exponentially for large time displacements; whereas, the decay of the quaisparticle ensures that it also fall exponentially with large spacial displacements.  So, one may safely assume that longer range graphs are "smaller" in magnitude.  

Now, consider a non-local correction to the local approximation where only graphs constructed from $G(r=0,\tau)$ enter.  The first such graph would be when all vertices are at $r=0$ apart from one which is on a near neighbor to $r=0$, which we will label as $r=1$.  We allow $G(r=1)/G(r=0)$ to be the "small" parameter.  It is easy to see that the first non-local correction to $\Phi$ is fourth order in $G(r=1)/G(r=0)$. 

Likewise, the first such corrections to the self energy are third order while those for the Green's function itself are first order in $G(r=1)/G(r=0)$.  Thus, the approximation where lattice quantities are approximated by cluster quantities, is much better for the self energy than for the Green's function.  Thus, the most accurate approximation is to replace the lattice generating functional with the cluster result, $\Phi^l \approx \Phi^c$ and the lattice self energy as the cluster result $\Sigma^l(\k) \approx \Sigma^c(\K)$ and use it in the lattice Dyson's equation to form the lattice single particle Green's function. 

Summarizing, the map from the lattice to the cluster is accomplished by replacing $G(\k)$ by $\bar{G}(\K)$ and the interaction $V(\k)$ by $\bar{V}(\K)$ in the diagrams for the generating functional.  These are precisely the generating functional, self energy and vertex diagrams of a finite size cluster with a bare Hamiltonian defined by $\cal{G}$, and an interaction determined by the bare coarse-grained $\bar{V}(\K)$. In this mapping from the lattice to the cluster, the complexity of the problem has been greatly reduced since this cluster problem may often be solved exactly and with multiple methods including quantum Monte Carlo~\cite{m_jarrell_01c} 

\paragraph{Map from the cluster back to the lattice}  We may accomplish the mapping from the cluster back to the lattice problem by minimizing the lattice estimate for the self energy.  The corresponding DCA estimate for the free energy is 
\begin{equation}
F_{DCA}= -k_B T\left(
\Phi^c-\mbox{Tr}\left[{\Sigma}^l_\sigma {G}_\sigma\right] 
+\mbox{Tr}\ln\left[-{G}_\sigma\right]\right)
\end{equation}
where $\Phi^c$ is the cluster generating functional.  The trace indicates summation over frequency, momentum and spin.  

We may prove that the corresponding optimal estimates of the lattice self energy and irreducible lattice vertices are the corresponding cluster quantities. $F_{DCA}$ is stationary with respect to ${\bf{G}}_\sigma$,\(\textit{}\)
\begin{equation}
\frac{-1}{k_B T}\frac{\delta F_{DCA}}{\delta G_\sigma(\k)}=
\Sigma^c_{\sigma}(M(\k))-\Sigma^l_{\sigma}(\k)=0,
\end{equation}
which means that $\Sigma^l(\k)=\Sigma^c(M(\k))$ is the proper approximation for the lattice self energy corresponding to $\Phi^c$. The corresponding lattice single-particle propagator is then given by 
\begin{equation}
G^l(\k,z) =\frac{1}{z-\ep_\k-\Sigma^c(M(\k),z) } \,.
\label{G_DCA}
\end{equation}
\textbf{}
A similar procedure is used to construct the two-particle quantities needed to determine the phase diagram or the nature of the dominant fluctuations that can eventually destroy the quasi-particle.  This procedure is a generalization of the method of calculating response functions in the DMFT~\cite{v_zlatic_90,m_jarrell_92a}. In the DCA, the introduction of the momentum dependence in the self energy will allow one to detect some precursor to transitions which are absent in the DMFT; but for the actual determination of the nature of the instability, one needs to compute the response functions. These susceptibilities are thermodynamically defined as second derivatives of the free energy with respect to external fields.  $\Phi^c(G)$ and $\Sigma^c_{\sigma}$, and hence $F_{DCA}$ depend on these fields only through $G_\sigma$ and $G_\sigma^0$.  Following Baym\cite{g_baym_61,g_baym_62} it is easy to verify that, the 
approximation 
\begin{equation}
\Gamma_{\sigma,\sigma'}\approx \Gamma^c_{\sigma,\sigma'} 
\equiv \delta\Sigma^c_{\sigma}/\delta G_{\sigma'}
\end{equation}
yields the same estimate that would be obtained from the second derivative of $F_{DCA}$ with respect to the applied field.  For example, the first derivative of the free energy with respect to a spatially homogeneous external magnetic field $h$ is the magnetization,
\begin{equation}
m=\mbox{Tr}\left[\sigma {G}_\sigma\right].
\end{equation}
The susceptibility is given by the second derivative, 
\begin{equation}
\frac{\delta m}{\delta h}=
\mbox{Tr}\left[\sigma\frac{\delta {G}_\sigma}{\delta h}\right].
\end{equation}
We substitute ${G}_\sigma =
\left( {G}_\sigma^{0-1}- {\Sigma}^c_{\sigma} \right)^{-1}$, and
evaluate the derivative, 
\begin{equation}
\frac{\delta m}{\delta h}=
\mbox{Tr}\left[\sigma\frac{\delta {G}_\sigma}{\delta h}\right]
=\mbox{Tr}\left[
{G}_\sigma^2
\left( 1 +
\sigma\frac{\delta {\Sigma}^c_{\sigma}}{\delta {G}_{\sigma'}}
  \frac{\delta {G}_{\sigma'}}{\delta h}
\right)
\right].
\end{equation}
If we identify $\chi_{\sigma,\sigma'}=\sigma \frac{\delta{G}_{\sigma'}}{\delta h}$, and $\chi_{\sigma}^0= {G}_\sigma^2$, collect all of the terms within both traces, and sum over the cell momenta $\tk$, we obtain the two--particle Dyson's equation \(\)
\begin{eqnarray} 
2\big({\bar{\chi}}_{\sigma,\sigma}&-&{\bar{\chi}}_{\sigma,-\sigma}\big)\\
&=&
2{\bar{\chi}}_{\sigma}^0 +
2{\bar{\chi}}_{\sigma}^0 \left( {\Gamma}^c_{\sigma,\sigma} 
-{\Gamma}^c_{\sigma,-\sigma}\right)
\left({\bar{\chi}}_{\sigma,\sigma}-{\bar{\chi}}_{\sigma,-\sigma} \right)\,.
\nonumber
\end{eqnarray}
We see again it is the irreducible quantity, this time the irreducible vertex function $\Gamma$, for which cluster and lattice correspond.

\begin{figure}[htb]
\includegraphics*[width=3.5in]{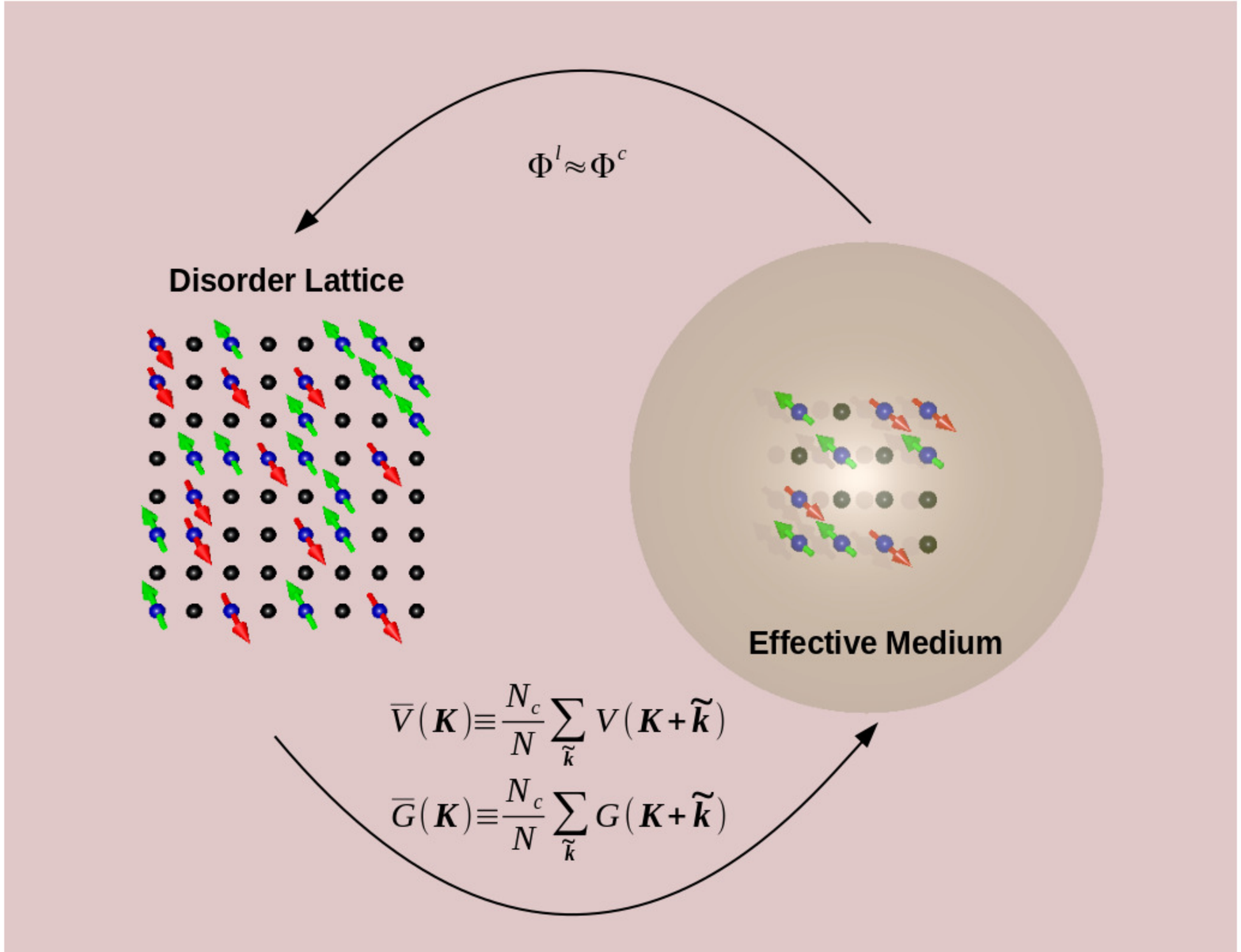}
\caption{The mapping from the cluster to the lattice is accomplished by replacing the Green's function and interaction by their coarse-grained analogs in the diagrams for the generating functional, self energy and irreducible vertices. In the map back to the cluster, this self energy is used to calculate a new cluster host Green's function.
}
\label{cluster_lattice_map}
\end{figure}
Summarizing, the mapping from the cluster back to the lattice problem is accomplished by approximating the lattice generating functional by the cluster result $\Phi^c$ 
\begin{equation}
\Phi^l \approx \Phi^c 
%
\end{equation}
and then optimizing the resulting free energy for its functional derivatives
yields
\begin{equation}
\Sigma^l(\k,\omega) \approx \Sigma^c(M(\k),\omega) \,\,\,\,
\Gamma^l(\k,\k') \approx \Gamma^c (M(\k),M(\k'))  \,\,\,\,\,
\end{equation}

\paragraph{The DCA algorithm.}
%
Thus the algorithm for the DCA is the same as that of the CPA/DMFT, but with coarse-grained propagators and interactions which are now functions of $\K$: 
(i) An initial guess for $\Sigma(\K,z)$ is chosen (usually from perturbation theory). 
(ii) $\Sigma(\K,z)$ is used to calculate the corresponding cluster Green's function
\begin{eqnarray}
\label{eq:gloc2}
{\bar{G}}(\K,\omega)=\frac{N_c}{N}\sum_{\tilde{\k}}G(\K+\tilde{\k},\omega)
\end{eqnarray}  
(iii) Starting from ${\bar{G}}(\K,z)$ and $\Sigma(\K,z)$ used in the second step, the host Green's function ${\cal G}(\K,z)^{-1}=G(\K,z)^{-1} + \Sigma(\K,z)$ is calculated which serves as bare Green's function of the cluster model. 
(iv) starting with ${\cal G}(\K,z)$, the cluster Green's function $G^c(\K,z)$ is obtained using the Quantum Monte Carlo method (or another technique). 
(v) Using the QMC output for the cluster Green's function $G^c(\K,z)$ and the host Green's function ${\cal G}(\K,z)$ from the third step, a new $\Sigma(\K,z) ={\cal G}(\K,z)^{-1}-G^c(\K,z)^{-1}$ is calculated, which is then used in step (ii) to reinitialize the process. Steps (ii) - (v) are 
repeated until convergence is reached. In step (iv) various QMC algorithms, exact enumeration of disorder, etc.\ may be used to compute the cluster Green's function $G^c(\K,z)$ or other physical quantities in imaginary Matsubara frequency $z=i\omega_n$.  Local dynamical quantities are then calculated by analytically continuing the corresponding imaginary-time quantities using the Maximum-Entropy Method (MEM)~\cite{m_jarrell_96a}.

This generating-functional based derivation of the DCA is appealing, since it requires the least initial assumptions.  Quantum cluster theories are defined by the maps between the lattice and cluster.  The map from the lattice to the cluster is obtained from a coarse-graining approximation for the generating functional $\Phi^l\approx\Phi^c$.  The map from the cluster back to the lattice is obtained by optimizing the free energy.  One may derive the same algorithm for a disordered system following the same prescription as described above\cite{h_terletska_13}. However, the treatment of a system with both disorder and interactions requires Keldysh \cite{j_rammer_86,Keldysh_1965}, or Wagner formalism \cite{m_wagner_91} via the replica trick \cite{Edwards_Anderson_1975,m_jarrell_01a,h_terletska_df} which is beyond the scope of this review.

\section{Typical medium theories of Anderson localization: model studies 
}\textit{\textbf{}} 
\label{sec:tmdca}

In this section via a series of subsections, we develop a formalism which incorporates the typical medium analysis into the DCA.  The resulting formalism enables the study of electron localization in models derived from first principles DFT calculations.  As summarized in Table~\ref{table:ansatze}, a progression of quantum cluster theories are proposed, each incorporating more chemical details of the model, including both diagonal and off diagonal disorder, multiple bands, and electronic interactions.  This culminates in a formalism able to deal with the full chemical details provided by modern electronic structure calculations. 

These developments are hampered by the lack of a limit where these mean field theories are exact.  Typically, we develop mean field theories which are exact in some physically meaningful limit, such as the limit of infinite dimensions.  The resulting theory then inherits some features due to this exactness even when applied in finite dimensions, such as thermodynamic consistency, translational invariance, etc.  

However, in order to be most useful, the mean field theory must yield results that are reasonably consistent with the real solution in finite dimensions.  Magnetism is a good example.  Here, the Weiss mean field theory becomes exact in infinite dimensions.  With a proper scaling of the model parameters with the dimensionality D, the phase diagram of the 3D model can be qualitatively reproduced by the mean field formalism.  However, the details of the transition, such as the universality class, may change with D, even becoming mean-field like above the upper critical dimension.  Despite this, since the transition persists, the mean field theory may be used to study it.  

For localization, the problem is complicated by the fact that the phenomena does not persist into infinite dimensions.  As we have seen, the CPA/DMFT becomes exact in the infinite dimensional limit.  However, as discussed in Sec.~\ref{sec:Anderson}, they fail to capture localization due to the self-averaging nature of the average DOS used to define their effective medium.  As a cluster extension of these formalisms, the DCA also fails to capture Anderson localization phenomena~\cite{m_jarrell_01a} and so fails to provide an adequate mean field theory for localization.  

A significant step towards this goal was developed by Dobrosavljevic et.al.~\cite{v_dobrosavljevic_03}.  They demonstrated that the typical density of states (TDOS) vanishes as the disorder strength increases, and hence can serve as a proper order parameter for Anderson Localization.   The authors constructed the typical medium theory (TMT), where they incorporated the geometric averaging over disorder in the CPA self-consistency loop. 

The TMT is the first successful mean-field theory for Anderson Localization.  Nevertheless, because of its local single-site nature, it suffers several drawbacks. It underestimates the critical disorder strength by about twenty percent, and does not capture the re-entrance features in the mobility edge (see Sec.~\ref{sec:Applications}), The lack of a non-trivial limit where it becomes exact, can make the results difficult to interpret.  For example, the TMT predicts a transition in any dimension, but it is not clear {\emph{a priori}} whether this is more likely true in high or low finite dimensions.  The CPA, which is exact in high dimensions, inherits a number of features from this exact limit.  For example, without {\emph{a priori}} knowledge of the upper critical dimension, we might be more inclined to believe its predictions for a 3D model over those for a 1D model.  This lack of an exact limit makes the imposition of any other {\emph{a priori}} known constraints significant.

\subsection{Building quantum cluster theories for the study of localization sec:criteria}
\label{sec:criteria}

In this section, we address these difficulties associated with the construction of a mean field theory with no known non-trivial exact limiting solution.  Our approach will be to construct a theory which inherits the desirable properties of the DMFT/CPA and DCA in the weak disorder limit, while also incorporating the TDOS order parameter into the mean field host ensuring that the method is also able to capture localization phenomena. The natural way to improve upon the local TMT is to construct a cluster extension which satisfies the constraints mentioned in Sec.~\ref{sec:Anderson} which when rephrased in terms of clusters are:
\begin{enumerate}
\item We approximate the coupling of the clusters to their lattice environment at the single-particle level (akin to the Fermi golden rule) neglecting two-particle and higher processes. This coupling is proportional to the square of a matrix element between the cluster and its host, times an appropriate DOS which describes the states available on the surrounding clusters.  
\item Since on average each cluster is equivalent to all the others, this DOS will also be proportional to some appropriate cluster density of states.  And, since the distribution of the DOS is highly skewed, the typical DOS is quite different than the average DOS.  The typical cluster DOS, which is clearly more representative of the local environment, will be used to define the effective medium.
\end{enumerate}
In addition, there are several additional desirable properties of a cluster theory, some of which appear in Ref.~\onlinecite{a_gonis_92} which should also be satisfied if possible:
\begin{enumerate}
\label{table:DP}
\setcounter{enumi}{2}
\item Maintain the translational invariance of the impurity averaged cluster.  I.e., there should be no distinction between, e.g., sites in the center and those at the boundary of the cluster.
\item The clusters should maintain the point group symmetries of the lattice.
\item The method should be fully causal, with positive definite spectra $A(\K,\omega) = -1/\pi \Im G(\K,\omega) > 0$
\item It should recover the DCA when the disorder is weak.
\item it should recover the TMT when $N_c=1$
\item In lieu of interactions, the scatterings at different energies are completely independent of each other.
\item For large $N_c\to \infty$ it should become exact while avoiding self averaging effects.
\item It should be extensible to multiple bands, and realistic models with longer ranged diagonal and off-diagonal disorder 
\end{enumerate}

Based on these criteria, we have constructed a set of TMDCA algorithms, listed in Table~\ref{table:ansatze}. By construction,  all of the algorithms listed in the table satisfy the first two criteria.  Furthermore, since they each map the periodic lattice problem onto a self-consistently embedded periodic cluster, they all maintain translational invariance.

The point group symmetry of the cluster is a matter of choice.  By allowing the cluster to have a lower symmetry than the lattice, there are far more clusters that can be used, e.g., in cluster size scaling calculations.  The quality and the selection criteria for the clusters have been addressed by D.D.\ Betts ~\cite{d_betts_97,d_betts_99a,p_kent_05}.

All proposed algorithms are fully causal.  The first two algorithms discussed below may be shown to be causal with a proof involving two conformal maps~ \cite{m_hettler_00a,m_jarrell_01a}.  This proof is not applicable to the multiband methods; however, we have not observed any causality violations in the iteration of the resulting equations.  

All of the algorithms recover the DCA in the weak disorder limit, whereas they do not all recover the TMT when $N_c=1$. There appears to be a trade-off between this and maintaining the independence of the scatterings at different energies.  The algorithms which use a Hilbert transform to calculate the imaginary part of the cluster Green's function, including the original TMT, violate this rule. The ones that calculate the cluster typical Green's function directly (and not the typical DOS), both imaginary and real parts,  satisfy the rule.  The algorithms which avoid the Hilbert transform are far more numerically stable, and both are equivalent for large clusters, so we tend to strongly favor the algorithms which directly calculate the cluster typical Green's function, avoiding the Hilbert transform.

Each of the algorithms become equivalent to a finite size simulation when $N=N_c$, so they all recover the exact result in this limit, and the thermodynamic limit for large $N$. On the other hand, the injunction against self averaging in item 9 is a bit subtle, which can be illustrated by an example.  Consider another apparently good Ansatz $\rho_{typ}(\k,\omega) = \exp\langle\left ( \ln \rho^c(\K,\omega)\right\rangle$ where $\rho^c(\K,\omega)=\frac{-1}{\pi} \Im G^c(\K,\K,\omega)$. $\Im G^c(\K,\K,\omega) $ is the diagonal part of the Fourier transform of the cluster Green's function.  The sum over sites in this transform involves an average of $G^c(\X,\X,\omega)$ over all cluster sites $\X$.  Thus the local part of this transform contains an average of the DOS over all cluster sites.  For large clusters, this is an average quantity, which as we argue above, is not critical at the transition.  Thus, an effective medium of this type fails to describe the localization transition, especially in three spatial dimensions~\cite{c_ekuma_14a}.

\subsection{Typical Medium Dynamical Cluster Approximation (TMDCA)}
\label{sec:TMDCA}
In this section we develop a cluster extension of the TMT, the typical medium DCA formalism (TMDCA) for the single-band Anderson model in 3D with diagonal disorder (the Hamiltonian was given in Sec.~\ref{sec:fundamentals}).  Due to the lack of a limit where the formalism becomes exact, the defining Ansatz for this formalism is not uniquely defined. In consideration of this, we will be guided by the desirable properties listed above.  We found two Ansatze which satisfy most of these desirable properties.

\begin{itemize}

\item 
Ansatz 1
\begin{equation}
\rho_{typ}^c(\K,\omega)=exp{\frac{1}{N_c}\sum_{I}^{N_c}
\langle \ln \rho_I^c(\omega, V) \rangle
\left\langle \frac{\rho^c(K,\omega,V)}{\frac{1}{N_c}\sum_I \rho_I^c(\omega,V)}\right\rangle}\,.
\label{eq:ansatz1}
\end{equation}
When the cluster size N$_c$ = 1, this Ansatz ~\cite{c_ekuma_15b} recovers the local TMT with $\rho_{typ}(\omega)=e^{\langle\ln \rho(w,V)\rangle}$. For weak disorder, the TMDCA recovers the average DCA results, with $\rho_{typ}(\K,\omega)\approx \langle\rho(\K,w,V) \rangle$. And in the limit of N$_c$ $\rightarrow$ $\infty$, the TMDCA becomes exact. Hence, between these limits, this Ansatz 1 of the TMDCA systematically incorporates non-local correlations into the local TMT.  Since, this Ansatz uses the TDOS, to get typical cluster Green's function $G_{typ}^c(\K,\omega)$, we use a Hilbert transformation, with
\begin{equation} 
G_{typ}^c(\K,\omega) =
\int d \omega' \frac{\rho_{typ}^c(\K,\omega')}{\omega - \omega'}\, .
\label{eq:Hilbert}
\end{equation}

\item
Ansatz 2 \newline
While Ansatz 1 works rather well for simple single-band models with local and non-local disorder, we find that it can suffer from numerical instabilities when applied to complex first-principle effective Hamiltonians with many orbitals and non-local disorder potentials. Such numerical instabilities arise due to the Hilbert transformation which is used to calculate the Green's function from the typical density of states $\rho_{typ}^c(\K,\omega)$. To avoid such numerical instabilities, we constructed the following Ansatz~2~\cite{y_zhang_16} where we calculate $G_{typ}^c(\K,\omega)$ directly as
\begin{equation}
G^c_{typ}(\K,\omega)=exp{\frac{1}{N_c}\sum_{I}^{N_c} 
\langle \ln \rho_I^c(\omega, V)\rangle 
\left\langle \frac{G^c(\K,\omega,V)}{\frac{1}{N_c}\sum_I \rho_I^c(\omega,V)}\right\rangle}\,.
\label{eq:ansatz2}
\end{equation}

This Ansatz 2 again incorporates the typical value of the local density of states, the resulting formalism again becomes exact in the limit of $N_c \rightarrow \infty$, promotes numerical stability of the algorithm, and converges quickly with cluster size. As noted in Table~\ref{table:ansatze} it does not reproduce the TMT when $N_c=1$.  This is due to the lack of a limit where the formalism is exact so that the Ansatz may be uniquely defined.
\end{itemize}

\begin{table*}[ht!]
\label{table:ansatze}
\begin{center}
\begin{tabular}{|c|c|c|c|}
\hline 
System/Ansatz &  Characteristics  & ODP &  VDP 
 \\
\hline \hline
Single Band &
Recovers TMT at $N_c=1$. &
  & 
 \\
Local (diagonal) Disorder  & 
 Recovers DCA for $W<<W_c$ &
 8 &   
 7  \\
Ansatz Eq.~\ref{eq:ansatz1} &
Calculate $\rho_{typ}$   &
&
\\
&
Hilbert trans.\ for $G_{typ}^c$ &
&
\\ \hline
Single Band &
Not TMT when $Nc=1$. &
  & 
 \\
Local (diagonal) Disorder  & 
 Recovers DCA for $W<<W_c$ &
 7 &   
 8  \\
Ansatz Eq.~\ref{eq:ansatz2} &
Calculate $G_{typ}^c$ directly   &
&
\\ \hline
Single Band &
$2\times 2$ matrix & 
&  
\\
Off-Diagonal Disorder &
Calculate $\rho_{typ}$ matrix & 
 8 &   
 7 \\
Ansatz Eq.~\ref{rhotyp_BEB} &
HT to get $G_{typ}^c$ matrix &
 &
 \\ \hline
Multi-band Systems &
Matrix in orbital space &
 &
 \\
Local Disorder  & 
Calculate $\rho_{typ}$ matrix &
 8 &   
 7 \\
Ansatz Eq. ~\ref{eqn:ansatz_mo} &
HT to get $G_{typ}^c$ matrix &
&
 \\
&
Recovers DCA for $W<<W_c$ &
   &
   \\ \hline
Realistic Material Systems &
Matrix in orbital space $G_{typ}^c$ &
&
\\
Complex Disorder Potentials &
Recovers DCA for $W<<W_c$ &
 7 &   
 8 \\  
with full DFT detail & 
   & 
   &  
   \\
Ansatz Eq. ~\ref{eq:ansatz2} &
&
&
 \\ \hline 
 
\end{tabular}
\end{center}
\caption{A progression of TMDCA algorithms, with each one able to incorporate greater chemical detail as we go down the list.  The first column lists systems that may be studied together with the label of the defining Ansatze.  The second column lists some additional characteristics including a brief discussion of the desirable properties. The columns labeled VDP and ODP identify the desirable properties, discussed above, which are notably violated and observed.  
}
\end{table*}

These two Ansatze will be used below as paradigms for the development of Ansatze for more realistic systems and will be referred to as Ansatz 1 and 2, respectively.

The main modification of the DCA self-consistency loop for the TMDCA involves the calculation of the cluster typical Green's function $G_{typ}^c(\K,\omega)$ using Eq.\ref{eq:ansatz1} and Eq.\ref{eq:Hilbert} or Eq.~\ref{eq:ansatz2}.  The typical Green's function is then used to complete the self-consistency loop. A schematic diagram of the TMDCA self-consistency loop is shown in Figure~\ref{algorithm-self_consistency}. 
The TMDCA iterative procedure is described as follows: 
\begin{enumerate}
 \item 
We start with a guess for the cluster self energy $\Sigma (\K, \omega)$, usually set to zero.

\item
Then we calculate the coarse-grained cluster Green's function $\bar{G}(\K,\omega)$ as
\begin{equation}
\bar{G}(\K,\omega)=\frac{N_c}{N}\sum_{\tilde{k}}\frac{1}{\omega+\mu-\varepsilon(\tilde{k}+\K)-\Sigma(\K,\omega)}\,.
\end{equation}

\item 
The cluster problem is now set up by calculating the cluster-excluded Green's function ${\cal G}(\K,\omega)$ as
\begin{equation} \label{eqn:clusterexG}
{\cal G}(\K,\omega)
=\frac{1}{\frac{1}{\bar{G}(\K,\omega)}+\Sigma(\K,\omega)}\,.
\end{equation}

\item
Since the cluster problem is solved in real space, we then Fourier transform ${\cal G}$(\K,$\omega$) to real space:  ${\cal G}_{I,J} = \sum_{\K} {\cal G}(\K)\exp(i \K\cdot(\R_I-\R_J))$.

\begin{figure}[b]
\begin{center} \includegraphics[trim = 0mm 0mm 0mm 0mm,width=1\columnwidth,clip=true]{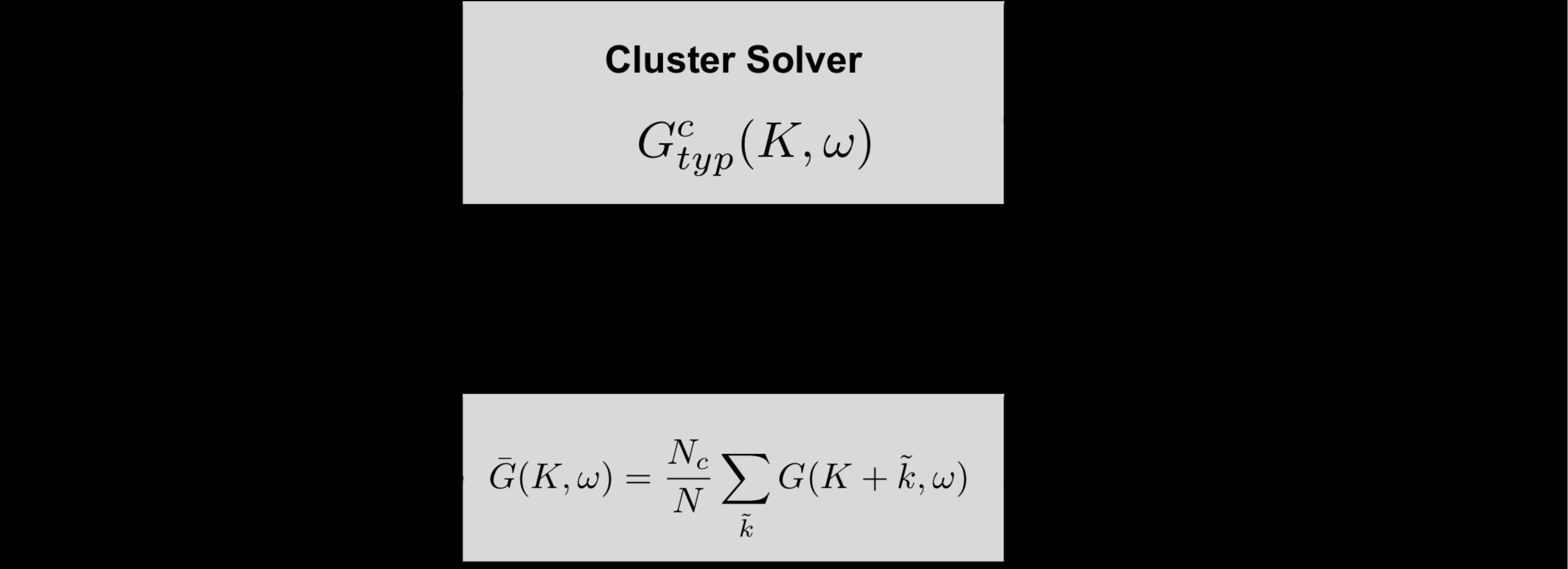}
\caption{The TMDCA self-consistent loop.}
\label{algorithm-self_consistency}
\end{center}
\end{figure}

\item
We solve the cluster problem using, e.g., a random sampling simulation.  Here, we stochastically generate random configurations of the disorder potential $V$.  For each disordered configuration, we 
construct the new fully dressed cluster Green's function as 
\begin{equation}
G^c(V) = ({\cal G}^{-1} - V)^{-1}.
\label{eq:dyson_nint}
\end{equation}

We then calculate the disorder-averaged, typical cluster Green's function $G_{typ}^c(\K,\omega)$ via the Hilbert transform using Eq.~\ref{eq:Hilbert} for Ansatz 1, or we can directly calculate the $G_{typ}^c(\K,\omega)$ from Eq.~\ref{eq:ansatz2} if we use Ansatz 2.

\item
With the cluster problem solved, we use the obtained typical cluster Green's function $G_{typ}^c(\K,\omega)$ to obtain a new estimate for the cluster self energy 
\begin{equation}
\Sigma(\K,\omega)={\cal G}^{-1}(\K, \omega)-(G_{typ}^c(\K,\omega))^{-1}
\end{equation}

\item We repeat this procedure starting from 2, until $\Sigma(\K,\omega)$ converges to the desired accuracy.
\end{enumerate}

We note that instead of using the self energy  
in the self-consistency, one can also use the hybridization function $\Delta(\K,\omega)$. Both procedures are observed to converge to the same solution.  

\subsection{Off-diagonal disorder}
\label{sec:DCAodd}

In this section, we extend the DCA and TMDCA formalisms to enable the study of off-diagonal disorder.  The simplest model used to study the effects of disorder in materials is a single-band tight binding model with a random on-site disorder potential. Such a model is justified when the disorder is introduced by substitutional impurities, as in a binary alloy where the substitution of host atoms by impurities only leads to changes of the local potential on the substitutional site and, on average, does not affect the neighbors. Then, the disorder appears only in the diagonal terms of the Hamiltonian coupling to the electronic density and hence is referred to as diagonal disorder.  However, when the bandwidth of the dopant is very different from that of the pure host, such substitution results not only in the change of the local potential but may also affect the neighboring sites. A simple model to capture such effects should include both random local potentials and random hopping amplitudes which depend on the occupancy of the sites. The dependence of the hopping amplitude on the disorder configuration is usually referred to as off-diagonal disorder \cite{Blackman_1917}. Of course, a proper theoretical description of realistic disordered materials requires the inclusion of both diagonal and off-diagonal randomness.

To illustrate these ideas, we will employ a simple binary alloy model with random nearest-neighbor hoppings.  Each site may be one of two types, A and B, with random diagonal potential depending on the type, $V_A$ and $V_B$, and hoppings between nearest neighbors $i$ and $j$, $t_{ij}$, are introduced as
\begin{eqnarray}
t_{ij}&=&  \nonumber
t^{AA}  ,~{\rm if} \quad i\in A,\quad j\in A \\ \nonumber
& &t^{BB}  ,~{\rm if} \quad i\in B,\quad j\in B \\ \nonumber
& &t^{AB}  ,~{\rm if} \quad i\in A,\quad j\in B \\ 
& &t^{BA}  ,~{\rm if} \quad i\in B,\quad j\in A,
\label{eq:BEBtmatrix}
\end{eqnarray}
with all others being zero. The hopping depends on the type of ion occupying sites $i$ and $j$.  We will assume that the alloy is completely random without clustering, with the concentration of $A$ sites, $c_A=1-c_B$.  

\begin{figure}[t!]
 \includegraphics[trim = 30mm 50mm 30mm 10mm,width=0.9\columnwidth,clip=true]{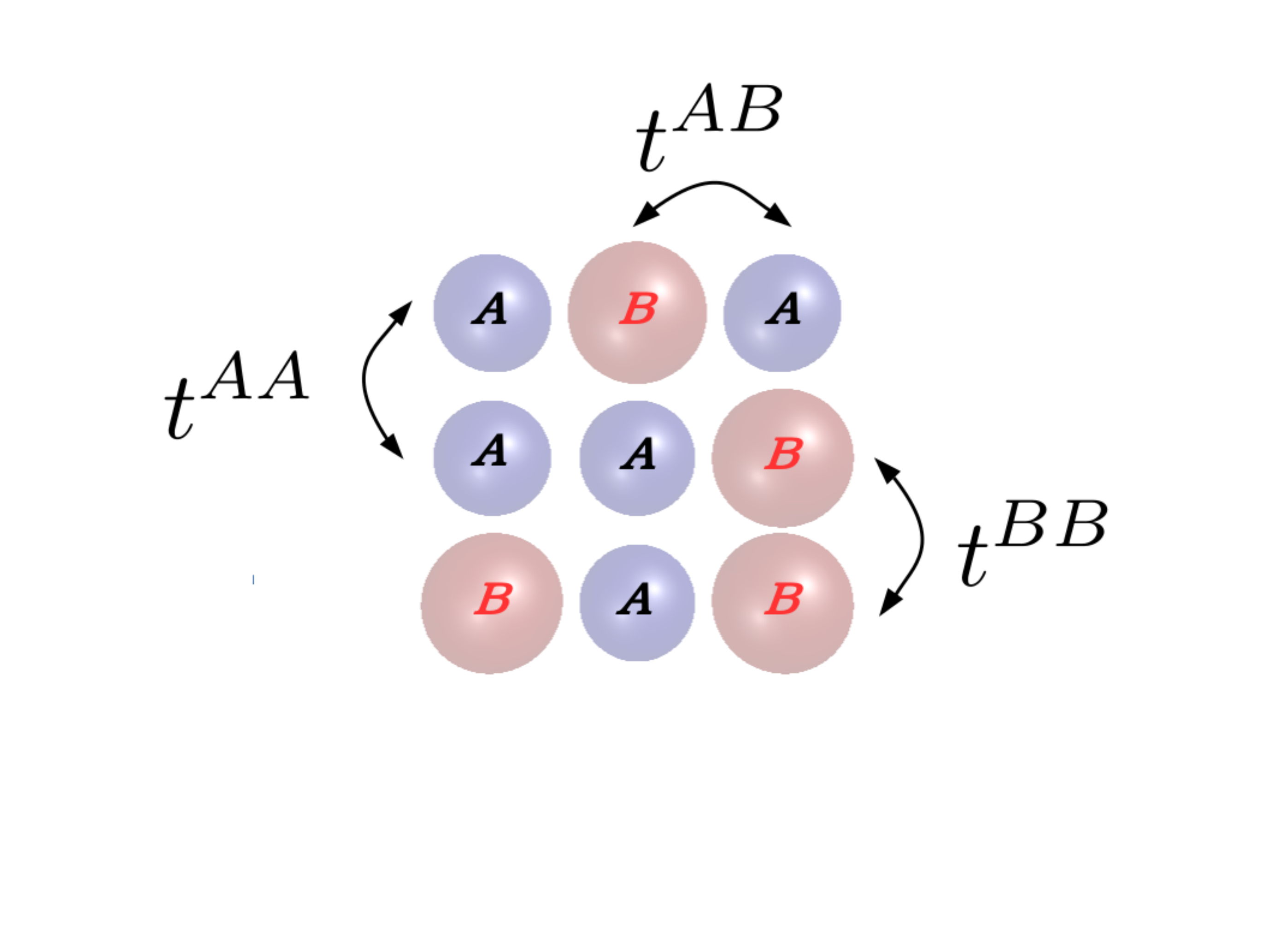}
 \caption{For off-diagonal disorder the hopping amplitude depends on the occupancy of the neighboring sites. 
 }
 \label{fig:sopt_diagram}
\end{figure}

We may immediately see the difficulty that the off-diagonal disorder poses: the mean field, contained within ${\cal{G}}$, depends upon the configuration of a site. Physically, the reason for this is clear. Consider the CPA ($N_c=1$) in our binary disorder model. Since the cluster/impurity site couples to the host only through the near-neighbor hoppings, it will depend on the occupancy of the impurity and neighboring sites. If we approximate the mean field coupling with the Fermi's golden rule, then we might expect the coupling to depend on the square of the relevant near-neighbor hoppings multiplied by the local density of states. In the CPA with nearest-neighbor hoppings, this matrix element is just the nearest neighbor hoppings. Since it depends on the occupancies, A or B, of the neighboring sites involved, we expect the mean-field coupling to depend strongly upon the type of impurity and its neighbors.

\subsubsection{DCA with off-diagonal disorder }
\label{sec:DCAODD}

This poses problems when formulating a Green's function formalism. Even after averaging over the disorder, the Green's functions depend on the type, $A$ or $B$, of the sites involved.  Blackman, Esterling and Berk~\cite{Blackman_1917} (BEB) extended the CPA to systems with off-diagonal disorder. They developed an elegant formalism to address the problem in multicomponent alloys. BEB showed the scalar CPA equation becomes a $2\times2$ matrix equation. For example, for our binary alloy model, the BEB single-particle Green's function is a $2\times 2$ matrix 
\begin{eqnarray}
\underline{G(\k,\omega)} & = \left(\begin{array}{cc} 
G^{AA}(\k,\omega) & G^{AB}(\k,\omega) \\ [1.0em]
G^{BA}(\k,\omega) & G^{BB} (\k,\omega) \end{array}\right) \,.
\label{eq:GBEB}
\end{eqnarray}
Since physically the Green's function describes the amplitude and phase the particle accumulates as it propagates, we can expect, i.e., $\int d\omega \frac{-1}{\pi} \Im G^{AA}(\k,\omega) = c_A$, $\int d\omega \frac{-1}{\pi} \Im G^{BB}(\k,\omega) = c_B$, etc.  

In momentum space, if there is only nearest-neighbor hopping between all ions as in our simple example, the bare dispersion can be written as (the under-bar denotes matrices)
\begin{eqnarray}
\underline{\varepsilon_\k} & = \left(\begin{array}{cc} 
t^{AA} & t^{AB} \\ [1.0em]
t^{BA} & t^{BB} \end{array}\right) \varepsilon_\k
\label{eq:dispersion}
\end{eqnarray}
where in three dimensions for our simple model $\varepsilon_k=-2t(\cos(k_x) +  \cos(k_y) +  \cos(k_z))$ with $4t=1$ which sets our unit of energy, and $t^{AA}$, $t^{BB}$, $t^{AB}$, and $t^{BA}$ are unitless prefactors. Using this, we may define a bare lattice propagator, and a corresponding diagrammatic perturbation theory for the lattice single-particle propagator $G(\k,\omega)$.  


As done in previous sections, the CPA or BEB formalism may be derived by replacing the Laue function by one at each internal vertex of the irreducible quantities, including the generating functional, and its functional derivatives the self energy and the vertex functions.  However, being single-site approximations, the CPA and the BEB theories neglect all disorder induced non-local correlations.

The DCA systematically incorporates such missing non-local corrections by mapping the lattice problem onto a self-consistently embedded cluster problem.  The mapping is accomplished by replacing the Laue function in the internal vertices of the irreducible quantities by the DCA Laue function. This causes all the Green's functions and vertices to be replaced by their coarse-grained counterparts.  The remaining details of the DCA formalism for off-diagonal disorder may then be defined by following the same procedures discussed in Sec.~\ref{sec:DCA}.  

To define the mean-field coupling between the cluster and its host, we introduce a DCA hybridization matrix $\Delta$. 
\begin{equation}
\begin{split}
\underline{\Delta(\K,\omega)}=
\left(\begin{array}{cc}
\Delta^{AA}(\K,\omega) & \Delta^{AB}(\K,\omega)\\
\Delta^{BA}(\K,\omega) & \Delta^{BB}(\K,\omega)
\end{array}\right)
\label{eq:-2-1}
\end{split}
\end{equation} 
which is related to the cluster Green's function, through the $2\times 2$ matrix equation
\begin{equation}
\underline{G^c(\K,\omega)} = \left (\omega- \underline{\bar{\varepsilon}_{\k}}-\underline{\Delta(\K,\omega)}-\underline{\Sigma(\K,\omega)} \right )^{-1}
\end{equation}

With this result, the mapping between the lattice and the cluster is established, and the cluster problem may be solved with a variety of methods.  We choose to average over the disorder configurations stochastically.  It is possible to enumerate all  configurations of the cluster.  For a binary alloy, there are $2^{N_c}$ such configurations, and an algorithm which enumerates all of them would scale exponentially in $N_c$. To avoid the exponential scaling that would come from enumeration, we randomly sample the configurations.  We draw the configurations purely at random, and calculate the corresponding components of the cluster $G^c(\X,\X')$, an $N_c \times N_c$ matrix.  We then average over the translations and point group operations of the cluster to restore the expected symmetries of a disorder-averaged system.  Our goal is to calculate the average $G^c(\X-\X')$ for each link $\X-\X'$.  This may be done by assigning the components according to the occupancy of the sites in the cluster $I$ and $J$ 
\begin{eqnarray}
(G^{c,AA})_{IJ} &=&(G^{c})_{IJ}\: ~{\rm if} \quad I\in A,\quad J\in A\nonumber \\
(G^{c,BB})_{IJ} &=&(G^{c})_{IJ}\: ~{\rm if} \quad I\in B,\quad J\in B\nonumber \\
(G^{c,AB})_{IJ} &=&(G^{c})_{IJ}\: ~{\rm if} \quad I\in A,\quad J\in B\nonumber \\
(G^{c,BA})_{IJ} &=&(G^{c})_{IJ}\: ~{\rm if} \quad I\in B,\quad J\in A\,
\label{eq:11}
\end{eqnarray}
with the other components being zero (for any disorder configuration, only 1/4 of the $G^{c,\alpha \beta}(\X-\X')$ are non-zero).  

Once the average cluster $G^c$ Green's function is obtained, we can get the cluster self-energy $\Sigma(\K,\omega)$ or the hybridization function matrix $\Delta(\K,\omega)$ using the Dyson's equation.  

We then close the loop on the DCA algorithm by calculating the coarse-grained lattice Green's function as
\begin{eqnarray}
\underline{\bar{G}(\K,\omega)} & = & \left(\begin{array}{cc}   
\bar{G}^{AA}(\K,\omega) & \bar{G}^{AB}(\K,\omega) \nonumber \\ 
\bar{G}^{BA}(\K,\omega) & \bar{G}^{BB}(\K,\omega)\end{array}\right) \nonumber\\
& = & \frac{N_{c}}{N}\sum_{\tilde{\k}}\Big(\underline{ G^{c}(\K,\omega)}^{-1}+\underline{\Delta_{}(\K,\omega)} \nonumber  \\
& - & \underline{\varepsilon_{\k}}+\underline{\overline{\epsilon}(\K+\tilde{\k})}\Big)^{-1}.
\label{coarsegraining}
\end{eqnarray}
A new estimate of the  hybridization function is then formed from $ \underline{\Delta}_{new} =\underline{\Delta}_{old}  + \underline{G^c(\K,\omega)}^{-1} - \underline{\bar{G}(\K,\omega)}^{-1}$. This may be used to define a new cluster problem, etc.  This procedure continues until $\Delta$ converges.

\subsubsection{TMDCA with off-diagonal disorder }
\label{sec:TMDCAodd}
In this section, we will discuss the modifications needed for the above DCA off-diagonal disorder formalism in order to incorporate the typical medium analysis ~\cite{c_ekuma_14b} 

In the presence of off-diagonal disorder, following BEB, the typical density of states becomes a $2 \times 2$ matrix, which we define as
\begin{widetext}
\begin{eqnarray} 
\addtolength{\jot}{1em}
\underline{\rho^c_{typ}(\K,\omega)} & = \exp\left(\dfrac{1}{N_c} \sum_{I=1}^{N_c} \left\langle \ln \rho_{II} (\omega) \right\rangle\right) 
\times & \left(\begin{array}{cc}
\left\langle \dfrac{-\dfrac{1}{\pi}\Im G^{c,AA}(\K,\omega)}{\frac{1}{N_c} \sum_{I=1}^{N_c}(-\dfrac{1}{\pi}\Im G_{II}(\omega))}\right\rangle  & \left\langle \dfrac{-\frac{1}{\pi}\Im G^{c,AB}(\K,\omega)}{\frac{1}{N_c} \sum_{I=1}^{N_c}(-\frac{1}{\pi}\Im G_{II}(\omega))}\right\rangle \\ [1.8em]
\left\langle \dfrac{-\dfrac{1}{\pi}\Im G^{c,BA}(\K,w)}{\frac{1}{N_c} \sum_{I=1}^{N_c}(-\dfrac{1}{\pi}\Im G_{II}(\omega))}\right\rangle  & \left\langle \dfrac{-\frac{1}{\pi}\Im G^{c,BB}(\K,\omega)}{\frac{1}{N_c} \sum_{I=1}^{N_c}(-\frac{1}{\pi}\Im G_{II}(\omega))}\right\rangle \end{array}\right).
\label{rhotyp_BEB}
\end{eqnarray}
\end{widetext}
Here the scalar prefactor depicts the local typical (geometrically averaged) density of states, while
the matrix elements are linearly averaged over the disorder. Also notice that 
the cluster Green's function $(\underline{G^c})_{IJ}$ and its components $G^{c,AA}$, $G^{c,BB}$ 
and $G^{c,AB}$ are defined in the same way as in Eqs.~\ref{eq:GBEB}-\ref{eq:11} above.

For $N_c=1$ with only diagonal disorder ($t^{AA}=t^{BB}=t^{AB}=t^{BA}$)  the above procedure reduces to the local TMT scheme. In this case, the diagonal elements of the matrix in Eq.~\ref{rhotyp_BEB} will contribute $c_A$ and $c_B$, respectively, with the off-diagonal elements being zero (for $N_c=1$ the off-diagonal terms vanish because a given site can only be either $A$ or $B$). Hence, the typical density reduces to the local scalar prefactor only, which has exactly the same form as in the local TMT scheme.

Another limit of the proposed Ansatz for the typical density of states of Eq.~\ref{rhotyp_BEB} is obtained at small disorder. In this case, the TMDCA reduces to the DCA for off-diagonal disorder, as the geometrically averaged local prefactor term  cancels by the contribution from the linearly averaged local term in the denominator of Eq.~\ref{rhotyp_BEB}. 

Once the first Ansatz is used to calculate the typical spectra, $\rho_{typ}^{\alpha\beta}$, the typical Green's function $G^c_{typ}(\K,\omega)$ is then obtained by performing Hilbert transform 
for each component
\begin{eqnarray} 
\addtolength{\jot}{1em}
\underline{ G^c_{typ}(\K,\omega)} & =  & \left(\begin{array}{cc}
\int d\omega'\frac{\rho_{typ}^{AA}(\K,\omega')}{\omega-\omega'}  
& \int d\omega'\frac{\rho_{typ}^{AB}(\K,\omega')}{\omega-\omega'} \\ [1.8em]
\int d\omega'\frac{\rho_{typ}^{BA}(\K,\omega')}{\omega-\omega'}
& \int d\omega'\frac{\rho_{typ}^{BB}(\K,\omega')}{\omega-\omega'}
\end{array}\right).
\label{Gtyp_BEB}
\end{eqnarray}

Once the disorder averaged cluster Green's function $G^c_{typ}(\K,\omega)$  is obtained from Eq.~\ref{Gtyp_BEB}, the self-consistency steps are the same as in the procedure for the off-diagonal disorder DCA.  I.e., we calculate the coarse-grained lattice Green's function $\bar{G}(\K,\omega)$
using Eq.~\ref{coarsegraining}.  Then, we use the obtained coarse-grained lattice  Green's function $\bar{G}(\K,\omega)$ to update the hybridization function with the effective medium as 
$ \underline{\Delta_{new}} =\underline{\Delta_{old}}  + \underline{G^c_{typ}(\K,\omega)}^{-1} - \underline{\bar{G}(\K,\omega)}^{-1}$, which is used to construct a new input to the cluster problem. The procedure is repeated, until numerical convergence is reached.

\subsection{TMDCA for multi-orbital systems} 
\label{sec:TMDCAmo}
Since realistic materials also have multiple orbitals, the TMDCA formalism has been generalized to multi-orbital system at the simple model level~\cite{y_zhang_15a} as well as for realistic materials~\cite{y_zhang_16}.  For the standard DCA, where the Green's function is averaged over disorder algebraically, the multi-orbital generalization is as simple as replacing all the quantities in the single orbital system with their matrix form.  This is due to the fact that all the linear operations performed in the single orbital system are also valid in the matrix system. However, in the TMDCA, the order parameter is constructed from the typical values of the LDOS \ie the TDOS, approximated as the geometric average of the LDOS.  So, we need to construct a multi-orbital generalization of the typical Green's function with an imaginary part that can properly reflect the TDOS so that it captures the localization of electrons. Since the off-diagonal elements of the LDOS are not positive definite, an extension of single band TMDCA to multi-orbital systems is not straightforward.  Despite the difficulty described above, it has been shown that\cite{y_zhang_15a} the critical behavior of the TDOS is independent of the local basis and the vanishing of the TDOS is equivalent to the vanishing of the typical value of the LDOS for all the orbitals, leaving some freedom to construct the appropriate typical Green's function. 

For the simple multi-orbital Anderson model with local diagonal disorder and guided by the selection criteria discussed in Sec.~\ref{sec:criteria}, we construct the following Ansatz for the typical DOS for the multi-orbital case ~\cite{y_zhang_15a}:
\begin{widetext}
\begin{equation}
\rho^{c,nn'}_{typ}(\K,\omega)=\begin{cases}
\begin{array}{c}
e^{\frac{1}{Nc}\sum_{I}\left\langle ln\rho_{II}^{nn}(\omega)\right\rangle}\left\langle\frac{\rho^{nn}(\K,\omega)}{\frac{1}{Nc}\sum_{i}\rho_{II}^{nn}(\omega)}\right\rangle,\ \ \ \ if\ n = n'\\
\\
e^{\frac{1}{Nc}\sum_{I}\left\langle ln|\rho_{II}^{nn'}(\omega)|\right\rangle}\left\langle\frac{\rho^{nn'}(\K,\omega)}{\frac{1}{Nc}\sum_{i}|\rho_{II}^{nn'}(\omega)|}\right\rangle,\ \ \ \ if\  n\neq n'
\end{array}\end{cases}
\label{eqn:ansatz_mo}
\end{equation}
with
\begin{equation}\label{eq:rho}
\rho_{II}^{nn'}(\omega)=-\frac{1}{\pi}\mathrm{Im}[G_{II}^{c,nn'}(\omega)]\,.
\end{equation}
\end{widetext}
Here, $n$ and $n'$ are orbital indices. As one can see, the orbital diagonal part ($n=n'$) takes the same form as the single-band TMDCA Ansatz 1, while the orbital off-diagonal part (with $n\neq n'$) is of a similar form, but involves the absolute value of the off-diagonal `local' density of states.  The typical cluster Green's function is then constructed through a Hilbert transformation 
\begin{equation}
G_{typ}^{c,nn'}(\K,\omega)= 
\int d\omega'\frac{\rho_{typ}^{c,nn'}(\K,\omega')}{\omega-\omega'}   
\label{eqn:Hilbert_mo}
\end{equation}
This Ansatz has been tested in the two-band Anderson model and it was shown that it successfully captures the localization of electrons with relatively fast convergence with the cluster size (more details are described in Sec.~\ref{sec:moresults}).

However, for more complicated materials such as (Ga,Mn)N, where the disorder potential contains both diagonal and off-diagonal parts, if a direct generalization of the Blackman off-diagonal disorder Ansatz above is applied, severe numerical instabilities arise when solving the self-consistent TMDCA equations. The main source of the instability comes from the Hilbert transformation used to calculate the full typical Green's function from the TDOS $\rho_{typ}^{c,nn'}$ of Eq.~\ref{eqn:Hilbert_mo}. Since the Hilbert transformation connects the typical Green's function at all the frequencies and makes the real component of the typical Green's function a functional of its imaginary part, this means a small error at certain frequency can spread to its neighbor frequencies, which makes the calculation numerically unstable, especially for systems with multiple bands and complicated disorder potentials. This frequency mixing is also somewhat unphysical, since the scattering processes are purely elastic, and processes at different energy are independent.  

To overcome such numerical instability, an alternative Ansatz for the multi-orbital typical Green's function is proposed in Ref.~\cite{y_zhang_16}.  It has the form:
\begin{widetext}
\begin{equation}\label{eq:ansatz2mb}
G_{typ}^{nn'}(\K,\omega)=e^{\frac{1}{N_{c}}\sum_{I}\left\langle \ln\left(\sum_{m}\rho_{II}^{nn}(\omega)\right)\right\rangle }
\left(\begin{array}{cc}
\left\langle \frac{G_{AA}^{c,nn'}(\K,\omega)}{{\displaystyle {\textstyle {\scriptstyle \frac{1}{N_c}\sum_{I,m}\rho_{II}^{nn}(\omega)}}}}\right\rangle  & \left\langle \frac{G_{AB}^{c,nn'}(\K,\omega)}{{\displaystyle {\textstyle {\scriptstyle \frac{1}{N_c}\sum_{I,m}\rho_{II}^{nn}(\omega)}}}}\right\rangle \\
\left\langle \frac{G_{BA}^{c,nn'}(\K,\omega)}{{\displaystyle {\textstyle {\scriptstyle \frac{1}{N_c}\sum_{I,m}\rho_{II}^{nn}(\omega)}}}}\right\rangle  & \left\langle \frac{G_{BB}^{c,nn'}(\K,\omega)}{{\displaystyle {\textstyle {\scriptstyle \frac{1}{N_c}\sum_{I,m}\rho_{II}^{nn}(\omega)}}}}\right\rangle 
\end{array}\right)
\end{equation}
with 
\begin{equation}\label{eq:rho2}
\rho_{II}^{nn'}(\omega)=-\frac{1}{\pi}\mathrm{Im}[G_{II}^{c,nn'}(\omega)]
\end{equation}
\end{widetext}

This Ansatz is an extension of Ansatz 2 (Eq.~\ref{eq:ansatz2}) for a single band model to the multi-orbital system. It incorporates the Blackman formalism so that off-diagonal disorder can also be included. For the diagonal disorder case, all four elements in Eq.~\ref{eq:ansatz2mb} are identical, so that it reduces to the multi-orbital version of Ansatz 2.

Since in this Ansatz we directly calculate the typical Green's function without invoking a Hilbert transformation, the calculated TDOS for each frequency is completely independent of the others.  This is consistent with the elastic scattering in the disordered system and greatly improves the numerical stability of the calculation. Note, that this Ansatz does not recover the TMT in the limit of $N_c$=1, but as shown in ~\cite{y_zhang_16}, for large cluster sizes, it converges quickly and approaches the exact results.

This Ansatz is one of many tried;  and it proved to be the most usable of the different Ansatze that we could formulate, and most importantly, it is able to treat the complex potentials extracted from a supercell DFT calculation.  It converges quickly with cluster size and yields a stable numerical iteration scheme. 

\subsection{Disorder in interacting systems. }
\label{sec:TMDCAinteracting}
In this section, we review the modifications of the TMDCA that are required for the study of interacting disordered systems.  As an example, to model the interplay between disorder and electron-electron interactions, we consider the Anderson-Hubbard model given by the Hamiltonian,
\begin{equation}
H=-\sum_{<ij>,\sigma} t_{ij}\left ( c_{i\sigma}^{\dagger}c_{j\sigma} +h.c.  \right ) 
+\sum_{i\sigma}\left (V_i -\mu \right )n_{i\sigma}+U\sum_i n_{i\uparrow} n_{i\downarrow},
\label{eq:AHM}
\end{equation}
here as before, $V_i$ describes the random disorder potential, and $U$ is the strength of electron-electron interactions between electrons at site $i$.

Electron-electron interactions are unavoidable in any realistic situation, and might have a dramatic effect on the MIT ~\cite{Belitz_Kirkpatrick_1993,Altshuler_Aronov_1979, Efros_Shklovskii_1975, v_dobrosavljevic_97}.  The important question is, to what extent do they change the nature of the localization transition.  In fact, as we have seen, near the transition, the hybridization between the cluster and its host vanishes, so that $U/\Delta$  becomes large suggesting that interaction effects become more important near the transition.  

  Great care must be taken while calculating disorder averaged quantities in the presence of interactions.  This is especially true when there is a need to mix linear and non-linear operations.  Examples include the calculation of typical (as opposed to arithmetically averaged) spectra, or when performing measurements in a QMC simulation when there is a minus sign problem. 

This problem arises since disorder averaging is inherently different than the thermodynamic averaging used in the calculation of the partition function $Z$.  The latter is always linear but only applied to the arguments of $Z$.  The situation is somewhat less clear when we must also perform averaging over disorder.  However, we may be guided by our desire to formulate a theory which properly describes experiments.  Nearly all experimental measurements are described by response functions, which may be expressed as derivatives of the free energy.  Furthermore, in order to obtain a large signal, most experiments, such as light scattering, are done on relatively large samples.  If the sample is disordered, then this means that the response function, $A(\k,\omega)$ in our example, is averaged over the sample which has many local disorder configurations.  The same is true for most experiments, including bolometry, nearly any scattering experiment including ARPES, neutrons, etc.  Therefore, to describe these experiments, we disorder average not the partition function, but the logarithm \cite{Edwards_Anderson_1975,atland} of the partition function and its functional derivatives which include all of the observable response functions.  

 This rule may easily be applied to quantum cluster calculations.   We start by generating disorder configurations of the cluster potential $V$ stochastically.  For a given interaction strength $U$ and randomly chosen disorder  configuration $V$, we solve the interacting cluster problem, obtaining a set of response functions, e.g., $G^c(\K,\omega, V)$.  When we have the final response functions for each disorder configuration $V$, we then take the average over the disorder. 

One of the prominent advantages of the TMDCA is that electron-electron interactions can be included in a very straightforward way while respecting these rules for disorder-averaging.  Within the TMDCA, the only modification to the algorithm for the inclusion of interactions is through the calculation of the cluster Green's function for each disorder configuration
\begin{equation}
G^c(V,U) = \left({\cal G}^{-1} - V - \Sigma^{Int}(U)+U/2 \right )^{-1}\,,
\label{eq:dyson_int}
\end{equation}
where $\Sigma^{Int}(U)$ is a thermodynamically averaged self energy matrix that may be derived through a real-space, real-frequency cluster solution of the electron-electron interaction term $U$ in the Hamiltonian of Eq.~\ref{eq:AHM}. Note that the adoption of this form involves no further approximation, despite the fact that when viewed in terms of Feynman diagrams, the self energy $\Sigma^{Int}$ contains only electron-electron interaction graphs and $V$ only disorder potentials.  The crossing diagrams (where interaction and disorder diagrams cross each other) are introduced by disorder averaging. The inclusion of these diagrams is essential for a proper description of the interplay between interactions and disorder.

Below, we review in some detail, two perturbation-theory-based cluster solvers for the interacting problem: a second order perturbation theory (SOPT) ~\cite{c_ekuma_15c}, and a statistical DMFT (stat-DMFT) ~\cite{e_miranda_12} based solver which needs to be supplemented with a local impurity solver such as local moment approach (LMA)\cite{m_galpin_09}, or the  numerical renormalization group ~\cite{r_bulla_08} etc. The SOPT based solver, albeit perturbative, incorporates dynamical non-local corrections properly; while the stat-DMFT based solver, despite employing non-perturbative impurity solvers does not capture true dynamical non-local corrections (that arise through interactions).

\subsubsection{Second order perturbation theory}
\label{sec:SOPT}

In order to understand the effect of weak interaction effects on the critical disorder concentration, as well as to investigate the effect on the mobility edge, we have incorporated a straight second order perturbation theory in the cluster momentum space into TMDCA formalism ~\cite{c_ekuma_15c}. In the constructed SOPT formalism, the interacting self energy $\Sigma^{Int}$ is obtained using the first and the second order perturbation theory contributions (shown in ~Figure~\ref{fig:sopt_diagram2})
\begin{equation}
\Sigma^{Int}=\Sigma^H+\Sigma^{(2)}\, .
\label{eq: sopt}
\end{equation}
Here the first term is the static Hartree correction $\Sigma^{H}=U\tilde{n}_I/2$.   The second term is the non-local second order contribution, defined as

\begin{figure}
\begin{fmffile}{first}
\begin{center}
    
\begin{fmfgraph*}(90,45)
\fmftop{t}
\fmfstraight
\fmfbottom{b0,b1,b2}
\fmffreeze
\fmf{photon,label=$I$}{b1,v2}
\fmf{fermion,right,tension=0.45}{v2,t}
\fmf{fermion,right,tension=0.45}{t,v2}
\end{fmfgraph*}\qquad {\Huge +}
\begin{fmfgraph*}(90,100)
\fmfstraight
\fmftop{t0,t1,t2,t3}
\fmfbottom{b0,b1,b2,b3}
\fmf{phantom}{t1,v1,b1}
\fmf{phantom}{t2,v2,b2}

\fmf{fermion,right=0.5,tension=0}{v2,v1,v2}

\fmf{fermion}{b1,b2}
\fmf{photon,label=$I$}{b1,v1}
\fmf{photon,label=$J$}{b2,v2}
\end{fmfgraph*}\qquad
\end{center}
\end{fmffile}

\smallskip

\caption{The diagrams for the first and second-order self energy labeled in real space. The indices $I,J$ indicate sites in the real-space cluster, while the lines are Hartree-corrected propagators $\tilde{\cal{G}}$. }
\label{fig:sopt_diagram2}
\end{figure}

\begin{align}
\Sigma_{I,J}^{(2)}(i\omega_n) = \frac{U^2}{\beta^2} 
\sum_{mp}\tilde{\cal{G}}_{IJ}(i\omega_n+i\nu_p)
\tilde{\cal{G}}_{IJ}(i\omega_m)
\tilde{\cal{G}}_{JI}(i\omega_m+i\nu_p),
\label{eq:sigma2}
\end{align}
where ${\tilde{{\cal G}}}(i\omega_n,V,U)$ is the Hartree-corrected host Green's function, $ {\tilde{{\cal G}}}^{-1}(i\omega_n)={\cal{G}}^{-1}-V-\epsilon_d(U)$, with $\epsilon_d(U)=\mu+U/2-U\tilde{n}_I/2$ and the cluster Green's function is finally given by $G^c(V)=({\cal{G}}^{-1}- \epsilon_d(U) - V - \Sigma^{Int})^{-1}$.

Although the above expression (equation~\ref{eq:sigma2}) appears to imply that we evaluate the self energy on the Matsubara frequency axis, it is not really so. We use the spectral representation of the propagators within a Hilbert transform to get a real-frequency expression for the imaginary part of the self energy (for more details, see Appendix of ~\cite{c_ekuma_15c}). Further, the real part of the self-energy is obtained through a Kramers-Kr\"{o}nig transform.

Once the cluster self energy due electron-electron interaction  $\Sigma^{Int}$  is obtained via Eq.~\ref{eq: sopt}, we then use Eq.~\ref{eq:dyson_int} to get the interaction-corrected cluster Green's function for the given disorder configuration $V$. This is then used to calculate the typical density of states Ansatz 1 of Eq.~\ref{eq:ansatz1}, with $\rho^c(K,\omega,V,U)=-\frac{1}{\pi}ImG^c(K,\omega,V,U)$.

The other parts of the TMDCA algorithm, namely the disorder averaging, coarse graining etc.\ remain exactly the same as in the non-interacting case described above in section~\ref{sec:tmdca}. A second order (in $U$) self energy evaluated on the full cluster, either in real or momentum space, is capable of incorporating non-local dynamical effects. However, by construction, such a cluster solver would only be valid for weakly interacting systems. If the system is strongly renormalized close to a metal-insulator transition, due to the reduction in $\Delta$ then this method might break down, since the assumption of weak coupling  is not valid for large $U/\Delta$. 

\subsubsection{Stat DMFT approach }
\label{sec:LMA}

The SOPT method described above is applicable only in the weakly interacting regime. Unfortunately for the strong coupling regime, there are very few cluster solvers available for disordered interacting electron systems.  The two most extensively used solvers capable of treating a wide range of energy and length scales, and are numerically exact, are quantum Monte Carlo methods~\cite{e_gull_11,Assaad_2014} and exact diagonalization \cite{a_georges_96a,Weisse_Fehske_2008,Noack_Manmana_2005,Lin_etal_1993}.

Quantum Monte Carlo methods have been extended to clusters~\cite{m_jarrell_01c,th_maier_05a}. However, since the typical averaging has to be performed on the real-frequency spectral function, the ill-posed step of analytic continuation is required for every disorder configuration and in every TMDCA iteration, rendering them unusable.  Alternatively, exact diagonalization may be used, but as is well-known, the cluster sizes that can be treated are very modest, and the associated computational expense is quite substantial. At present, the only fully non-local cluster solver available that is computationally feasible, and yields a real frequency self energy is a straight perturbation theory.

Thus, one has to resort to approximate cluster solvers, especially for investigating the strong coupling regime. Such a solver may be constructed by combining a non-perturbative real frequency single-site solver and statistical DMFT~\cite{e_miranda_12}. The former must be capable of treating the moment formation and Kondo physics characteristic of the strong coupling regime. It must also properly incorporate the eventual many body screening of the local moment leading to a singlet ground state. The resulting formalism is then able to capture these local dynamical correlations due to $U$, while treating the corresponding non-local correlations at a static level.  On the other hand, the correlations due to the disorder are captured exactly up to a length scale given by the linear cluster size.

There are several excellent real-frequency solvers available to treat the strong coupling regime of the single impurity Anderson model. Amongst them are the numerical renormalization group, non-crossing approximation and the local moment approach (LMA).  Since we have used the LMA for our investigations, we provide a brief introduction to this method here. The LMA~\cite{m_galpin_09} is a diagrammatic perturbation theory based impurity solver, starting at the unrestricted Hartree-Fock static mean field solution. The symmetry, broken at the mean field level, is restored through the inclusion of transverse spin flip dynamics. This symmetry restoration step, equivalent to restoring adiabatic continuity to the non-interacting limit, leads to the emergence of a low energy Kondo scale, $T_K$. The latter is an exponentially small scale in strong coupling, proportional to $\exp\left(-\alpha U/\Gamma\right)$, where $\alpha$ is a number $\sim{\cal{O}}(1)$, $U$ is the local Hubbard repulsion, and $\Gamma$ is the hybridization of the impurity with the local reservoir at the chemical potential. Since, within stat-DMFT, the hybridization is site-dependent. Rather than a single Kondo scale for the entire system, a distribution of Kondo scales, $P(T_K)$ is obtained. The form of such a distribution and its consequences on the properties of the disordered system have been extensively investigated using  slave-boson methods and phenomenological arguments~\cite{Miranda_Kondo,Miranda_Kondo_prb,Fisk_Kondo}

It has been seen in the above mentioned studies that typical medium theory based calculations yield a Kondo scale distribution $P(T_K)$ exhibiting a long tail at higher Kondo scales, while diverging at a specific, lower bound scale. This is determined by the solution of the impurity problem in the particle-hole symmetric limit \cite{s_sen_16a}. Extensions to statistical DMFT combined with the slave-boson solver yields a $P(T_K)$ that also has a long tail at larger $T_K$, but is not divergent at lower scales~\cite{Vlad_tmt_critical}.

Instead, it is highly skewed, has a maximum at a specific scale, and has either a vanishing or a finite intercept depending on whether the disorder is below or above a critical disorder value. Such a distribution with a finite intercept has been shown to be a sufficient condition for the system to exhibit non-Fermi liquid (nFL) behavior in transport and thermodynamics. Thus, these theories have provided a route to explain the crossover from conventional metallic behavior at low disorder to singular, non-Fermi liquid behavior at strong disorder~\cite{v_dobrosavljevic_97,Vlad_scaling_tmt}.

Nevertheless, since slave-boson methods are just a renormalized version of the non-interacting limit, and hence fail to capture dynamics at all energy scales, the above theories do not provide an insight into the role of dynamics in the Fermi liquid to non-Fermi liquid crossover. Additionally, since the stat-DMFT does not incorporate an embedding of the disordered cluster into a translationally invariant medium, it does not allow access to Anderson localization unless the cluster is prohibitively large. The TMDCA combined with a cluster solver based on stat-DMFT and the LMA does not suffer from the two shortcomings of the previous work. A rapid convergence with increasing cluster size, ensured by the embedding of the cluster in  a medium, ensures the feasibility of the solver, thus allowing the replacement of the slave-boson solver by a non-perturbative, albeit more expensive method such as the LMA. Additionally, the TMDCA captures Anderson localization almost exactly in the non-interacting case, as discussed in the previous sections. 

\begin{figure}[h!]
 \includegraphics[trim = 0mm 0mm 0mm 0mm,width=1\columnwidth,clip=true]{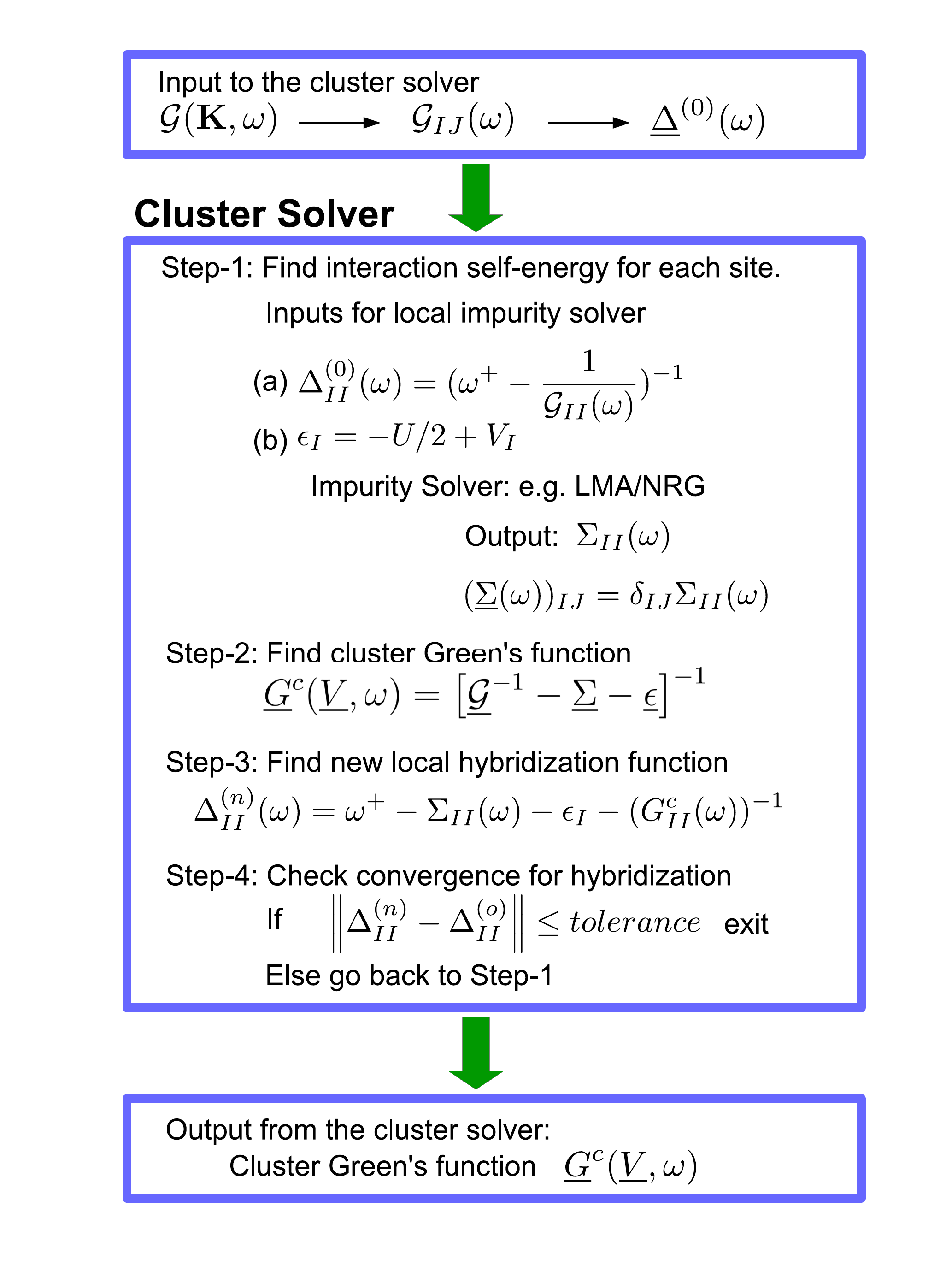}
\caption{The detailed algorithm implemented to solve the interacting disordered problem with a cluster solver built by combining statistical DMFT and a local impurity solver which could be, for example  LMA or NRG. Note the self-consistency loop within the stat-DMFT cluster algorithm.}
 \label{fig:algo_int}
\end{figure}

The stat-DMFT based TMDCA algorithm is illustrated in Figure~\ref{fig:algo_int}~\cite{s_sen_16a}. 
The input to the cluster solver is the real space hybridization matrix, derived through the real space host Green's function, which in turn can be obtained through a Fourier transform of the ${\mathbf K}$-space host Green's function, ${\cal G}({\bf{K}},\omega)$. The cluster solver begins with a solution of $N_c$ impurity problems, for which the two required inputs are the local orbital energy, $\epsilon_I=-U/2+V_I$, and the local hybridization function, $\Delta^{\rm (o)}_{II}(\omega)$. The output of this step is a diagonal self energy matrix,
$\underline{\Sigma}(\omega)$. The second step uses the modified Dyson's equation, namely
\begin{equation}
\underline{G}^c(V,\omega) =
\left[\underline{{\cal G}}^{-1} - \underline{\Sigma} -
\underline{\epsilon} \right]^{-1}\,
\end{equation}
which incorporates the effects of interactions and disorder on an equal footing, and yields the real space cluster Green's function.
This can now be inverted as shown in Step-3 of Figure~\ref{fig:algo_int}, to get a new local hybridization function, $\Delta^{\rm (n)}_{II}(\omega)$. The final step involves a stat-DMFT self-consistency check of the local hybridization function, as shown in Step-4. If $\Delta_{II}(\omega)$ is found to have been converged (within a numerical tolerance), the cluster solver is exited, with the output being the cluster Green's function found at Step-2, else the new local hybridization function is plugged back into the Step-1 of the cluster solver, and these steps are repeated until the convergence is reached \cite{Sen_etal_2018}. The last box in Figure~\ref{fig:algo_int} shows that the output of the cluster solver is the converged (within stat-DMFT) cluster Green's function for a single disorder configuration (as obtained in step-2). Subsequently this is then Fourier transformed to cluster momentum space, and the disorder average is carried out, as in the standard TMDCA algorithm (see section~\ref{sec:tmdca}).

In practice, since the number of disorder realizations is very large ($\sim$3000) and the largest cluster size used was $N_c=38$, a very large set of impurities ($\sim 10^5$) need to be solved. Each such solution yields a Kondo scale, expected to be statistically different from the others due to the unique local hybridization function `generated within the cluster solver. The histogram of all the Kondo scales yields a very reliable Kondo scale distribution, as well as a physical self energy which encompasses disorder and interaction effects on an equal footing. Some of these results are reviewed in Sec.~\ref{sec:ResultsInteracting}.

\subsection{Two-particle calculations }
\label{sec:2particle}
\begin{figure}[t]
    \centerline{\includegraphics[scale=0.65]{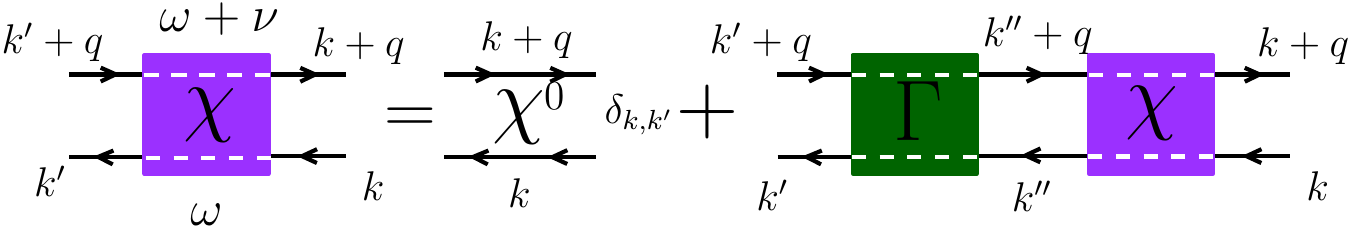}}
\caption{
Bethe-Salpeter equation relating the two-particle Green's function $\chi$ and the irreducible vertex $\Gamma$. While $\k$, $\k'$ and $\q$ represent momentum indices, $\omega$ and $\nu$ represent frequency indices (for fermionic and bosonic frequencies respectively) and the spin indices are suppressed. Note that for the disordered systems considered here, the scatterings are elastic and thus the energy is conserved following any fermionic Green's function line. Therefore, we only need two frequency indices to represent the frequency degree of freedom of the system.
}
\label{fig:bs}
\end{figure}

Up to this point, the theory has  focused on the calculation of single particle quantities, \ie the TDOS to capture the localization transition. However, most experimental measurements are described by two-particle Green's functions, including transport, most X-ray and neutron scattering, NMR, \etc. Therefore, the TMDCA has also been extended to include the description of two-particle quantities including vertex corrections~\cite{y_zhang_17} in a similar fashion as that in the CPA and DCA~\cite{m_jarrell_01c,th_maier_05a}. In conventional mean-field theories such as the CPA and DCA, the order parameters are constructed from the lattice Green's function defined as
\begin{equation}
G^l_{\sigma}(\k,\omega) = \frac{1}{\omega - h\sigma - \epsilon_\k - \Sigma_\sigma(M(\k),\omega)}\,,
\label{eq:Glattice}
\end{equation}
where $M(\k)=\K$ maps an arbitrary wave number $\k$ to the closest DCA cluster $\K$ and $\Sigma(M(\k),\omega)$ is the self energy calculated on the cluster.  If the order parameter is local, the order parameters may also be constructed from the cluster single-particle Green's function
\begin{equation}
G^c_{\sigma}(\K,\omega) = \frac{1}{\omega - h\sigma - \bar \epsilon_\K - \Delta_\sigma(\K,\omega) - \Sigma_\sigma(\K,\omega)}\,.
\label{eq:Gcluster}
\end{equation} 
For example, for the magnetization $m$
\begin{equation}
m=\sum_{\k,\omega,\sigma}\sigma  G^l_{\sigma}(\k,\omega) = \sum_{\K,\omega,\sigma}\sigma G^c_{\sigma}(\K,\omega).
\label{eq:m}
\end{equation}
Since these equations depend on $h$ through the Green's function and through the dependence of $\Sigma$ and $\Delta$ on $G$, in order to calculate the susceptibility $\left. dm/dh\right|_{h=0}$ using the cluster Green's function, we need to know both $\delta G/\delta \Sigma$ and $\delta \Delta/\delta G$.  The former is the irreducible vertex function
\begin{equation}
\Gamma_{\sigma,\sigma'}(\K,\omega;\K',\omega') = \frac{\delta  G_\sigma(\K,\omega)}{\delta \Sigma_{\sigma'}(\K',\omega')}\,.
\end{equation}
but the lack of information on $\delta \Delta/\delta G$ prevent us from using this representation for the extended states. However, for the localized states, $\Delta$ vanishes, so that $\delta \Delta/\delta G$ is not needed and we can use the cluster Green's function for the localized states.  Since the scattering events at different $\omega$ are completely independent, to avoid using $\delta \Delta/\delta G$ for the extended states, we introduce a mixed representation with
\begin{equation}
m=\sum_{\k,\omega,\sigma}\sigma G^p_{\sigma}(\k,\omega) 
\end{equation}where 
\begin{equation}
G^p_{\sigma}(\k,\omega) = \left\{ \begin{array}{ll}
         G^l_{\sigma}(\k,\omega) & \mbox{if $ |\omega| < \omega_e$};\\
         G^c_{\sigma}(M(\k),\omega) & \mbox{if $|\omega| > \omega_e$}.\end{array} \right.
         \label{eq:Gp}
\end{equation}
and $\omega_e$ is the mobility edge energy. Physically, this is more meaningful than the use of one of the formulas in Eq.~\ref{eq:Glattice},\ref{eq:Gcluster} alone. Below the mobility edge, $\omega<\omega_e$, all of the states are extended, and they may be described as states with a dispersion $ \epsilon_\k$ renormalized by $\Sigma$. However, for localized states $\omega>\omega_e$, above the mobility edge, the electrons are localized to the cluster with $\Delta_\sigma(\K,\omega)=0$ so that $\frac{\delta\Delta\sigma(\K,\omega)}{\delta h}=0$. These states may not be described as extended states with a renormalized dispersion. So the usual interpretation fails, and it is much better to think in terms of states localized to the cluster described by the cluster Green's function for frequencies above the localization edge. This leads to the main difference between the typical analysis of the two-particle quantities and the conventional CPA and DCA, where for the states above the mobility edge, the TMDCA average cluster Green's function $G^c_{\sigma}(\K,\omega)$ is used to construct the two-particle susceptibility matrix
\begin{equation}
\left. \frac{\delta G^p_\sigma(\k,\omega)}{\delta h}\right|_{h=0} = \sum_{\k',\omega',\sigma'}\chi_{\sigma,\sigma'}(\k,\omega;\k',\omega') \sigma'
\end{equation}
Based on this, and the observation that at convergence, $G^c=\bar{G}$ so that for the $\frac{\delta }{\delta G^c_\sigma}=\frac{\delta }{\delta \bar{G}_\sigma}=\frac{\delta }{\delta G^l_\sigma}$ the Bethe-Salpeter equation can be derived with $G^p$ Green's function
\begin{equation}
\sigma \chi_{\sigma,\sigma'}\sigma'=
\sigma \chi^{p0}_\sigma \sigma +
\sigma \chi^{p0}_{\sigma \sigma} \Gamma_{\sigma,\sigma''}\sigma''
\sigma'' \chi_{\sigma'',\sigma'} \sigma'
\label{eq:BSEspin}
\end{equation}
where $\chi^{p0}_{\sigma \sigma}= \left(G^p_{\sigma}(\k,\omega)\right)^2$.  This equation may be described diagrammatically as in Figure~\ref{fig:bs}.  Again, the lattice momentum sums on $\kt$, where $\k=M(\k)+\kt$, render the direct solution to Eq.~\ref{eq:BSEspin} intractable.  Fortunately, since the irreducible vertex function above depends only on the momentum cell centers $\K$, this equation may be coarse-grained, by summing over the $\kt,\kt ',\cdots$ labels.  The corresponding coarse-grained Bethe-Salpeter equation becomes 
\begin{equation}
\sigma \bar{\chi}_{\sigma,\sigma'}\sigma'=
\sigma \bar{\chi}^{p0}_\sigma \sigma +
\sigma \bar{\chi}^{p0}_{\sigma \sigma} \Gamma_{\sigma,\sigma''}\sigma''
\sigma'' \bar{\chi}_{\sigma'',\sigma'} \sigma'
\label{eq:BSEspinCG}
\end{equation}
where $\bar{\chi}^{p0}_{\sigma \sigma} =\sum_{\kt}  \left(G^p_{\sigma}(\K+\kt,\omega)\right)^2 $

The susceptibility corresponding to different physical quantities can be constructed through the two-particle Green's function. For instance, the charge susceptibility can be constructed as
\begin{equation}
\chi_c= \sum_{\k,w,\sigma;\k',\omega',\sigma'} \chi_{\sigma,\sigma'}(\k,\omega;\k',\omega')
\end{equation}, 
which is also used to calculate the DC conductivity at zero temperature for a single band Anderson model with results shown in Sec.~\ref{sec:2presults}.  In this typical analysis, the inclusion of the vertex corrections follows the same procedure as that described in ~\cite{m_jarrell_01c,th_maier_05a}.

\section{Methodology for first-principles studies of localization}
\label{sec:edhm} 
 
There are two general methods which may be used to study localization from first principles. The first is a  \emph{component-based approach} wherein the calculation is split into three basic components, as depicted in Figure~\ref{fig:EDHM} and described in Secs.,~\ref{sec:DFT2edhm} and~\ref{sec:EDHM2TMDCA} below.  Here, the DFT and TMDCA calculations are performed separately, connected by the second step where a tight-binding model is extracted from the DFT to be solved in the third, TMDCA step. The first two steps of this process are quite mature, allowing researchers to focus on the third step, as we have done thus far in this review.

Alternatively, in the \emph{integrated approach}, the coarse-graining ideas behind the DCA, the typical medium analysis, and multiple scattering theory based DFT are integrated together to form a fully self consistent treatment of the problem.  This multiple-scattering formalism has been developed~\cite{h_terletska_17}, but as it has not yet been implemented in a real materials calculation, it is beyond the scope of this review.

In this section we focus on the component-based approach based approach illustrated in Figure~\ref{fig:EDHM}). Specifically the first sub-section will describe how to extract low energy effective models of disordered materials using the Effective Disordered Hamiltonian Method (EDHM)~\cite{naxcoo2}. The second sub-section will describe how these models with real material parameters are inserted into the Effective Medium Solver, in this case the TMDCA framework.

\subsection{From Density Functional Theory to the EDHM }
\label{sec:DFT2edhm}
To describe the effect of disorder within realistic first-principles simulations, we utilize our recently developed Effective Disordered Hamiltonian Method (EDHM)~\cite{naxcoo2}． The EDHM maps Density Functional Theory (DFT) calculations of ordered materials onto low-energy effective tight binding Hamiltonians. These, then in turn, can be used as input for the TMDCA calculations. 

The EDHM is a Wannier-function based method~\cite{marzari_1997,w_ku_02,Anisimov:2005ix}. It makes the TMDCA more tractable by significantly reducing the number of basis functions (i.e., from hundreds of plane-waves to a few Wannier functions per atom). Besides the EDHM, there are other electronic structure methods that aim at reducing the number of basis functions such as Numerical Atomic Orbitals~\cite{soler_2002, nao_2001,blum_2009} and Density Functional Tight-Binding theory~\cite{dftb_2009}. 

\begin{figure}[h]
\includegraphics[width=0.5\textwidth,clip=true]{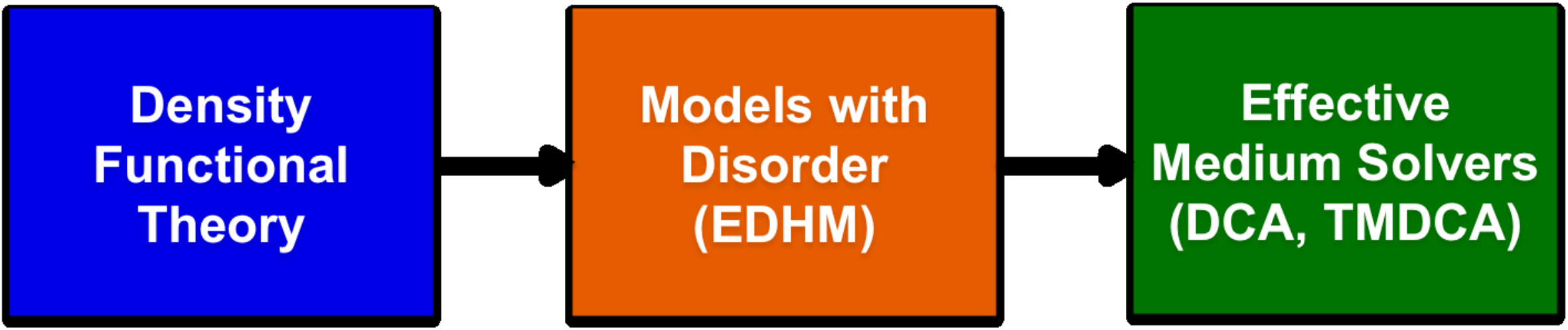}
\caption{Organization of the modular approach to first-principles calculations of localization.  A DFT of the pure system and a DFT supercell calculation of a single impurity are performed as the first step.  In the second step, the EDHM converts the DFT output into model parameters of the disordered system. In the third step, the TMDCA is used to study the materials-specific localization properties.}
\label{fig:EDHM}
\end{figure}

Conceptually the EDHM is based on a cluster expansion approximation~\cite{ceder_1998} (not to be confused with the clusters embedded in the effective medium theories discussed in the previous sections).
In this approximation a physical quantity, the low energy effective Hamiltonian in this case, is expanded in impurity clusters of increasing size. Specifically, the effective Hamiltonian of an arbitrary configuration of $N$ impurities, positioned at $(x_{1},...,x_{N})$, can be exactly rewritten as
\begin{eqnarray}\label{eq:tom_ce}
H^{(x_{1},...,x_{N})}=H^{0}+\sum_{i=1}^{N}V^{(x_{i})}+\sum_{i> j=1}^{N}V^{(x_{i},x_{j})}+...
\end{eqnarray}
where $H^{0}$ denotes the Hamiltonian of the system with no impurities,  $V^{(x_{i})}=H^{(x_{i})}-H^{0}$, denotes the potential of an impurity at $x_{i}$ and $V^{(x_{i},x_{j})}=H^{(x_{i},x_{j})}-V^{(x_{i})}-V^{(x_{j})}-H^{0}$ denotes the two-impurity correction of a pair of impurities at ($x_{i}$,$x_{j}$), etc. 
We have found that for many materials it is already highly accurate to retain only the single impurity potentials and neglect the higher order corrections~\cite{naxcoo2,tm122,kfe2se2,ru122,11sto}.  Furthermore, we are typically interested in  very dilute impurity concentrations for which Anderson and Mott localization take place. In this limit it is unlikely that multi-impurity corrections to the Hamiltonian need to be taken into account. Here we emphasize keeping in Eq.~\ref{eq:tom_ce} only the single impurity potentials does not mean that multi-impurity scattering is not taken into account. At this point we are deriving the low-energy Hamiltonian which can, in principle, be solved by exact diagonalization that takes into account multi-impurity scattering exactly to all orders.  

In practice, the EDHM consists of three steps. 
\begin{enumerate}
\item In the first step two DFT calculations are performed: a normal cell calculation of the pure host material and a supercell calculation of the host material with a single impurity in it. For example for KFe$_{2-y}$Se$_{2}$, an iron based superconductor that contains Fe vacancies, the normal cell of the host will be KFe$_2$Se$_2$. To capture the impurity potential of an Fe vacancy one can run a DFT calculation for a K$_8$Fe$_{15}$Se$_{16}$ supercell containing a single Fe vacancy~\cite{kfe2se2}. 
\item The second step is to derive the low-energy Hamiltonians using a projected Wannier function transformation in which a set of atomic orbitals is projected on the bands close to the Fermi level~\cite{w_ku_02,Anisimov:2005ix,thesistberlijn}． For the case of KFe$_{2-y}$Se$_{2}$, one can project Fe-$d$ and Se-$p$  orbitals on the bands within [-6,2]eV ~\cite{kfe2se2}. This results in two ordered tight-binding Hamiltonians. One for the normal cell $H^0$, and one for the single-impurity supercell $H^{(x_{j})}$.  
\item Finally, a superposition of these ordered Hamiltonians is used to build Hamiltonians of arbitrary impurity configurations. Specifically, the difference between the single impurity and pure Hamiltonian is taken to derive the single impurity potential: $V^{(x_{j})}=H^{(x_{j})}-H^{0}$. To remove the influence of the periodically repeated impurities in the single-impurity supercell calculation a partitioning procedure is necessary. A detailed account of this procedure is given in ~\cite{thesistberlijn}. From single impurity potential the effective Hamiltonian of a disordered impurity configuration with $N$ impurities can be assembled as follows: $H_{\rm{eff}}^{(x_{1},...,x_{N})}=H^{0}+\sum_{j=1}^{N}V^{(x_{j})}$. 
\end{enumerate}

\begin{figure}[h]
\includegraphics[width=0.5\textwidth,clip=true]{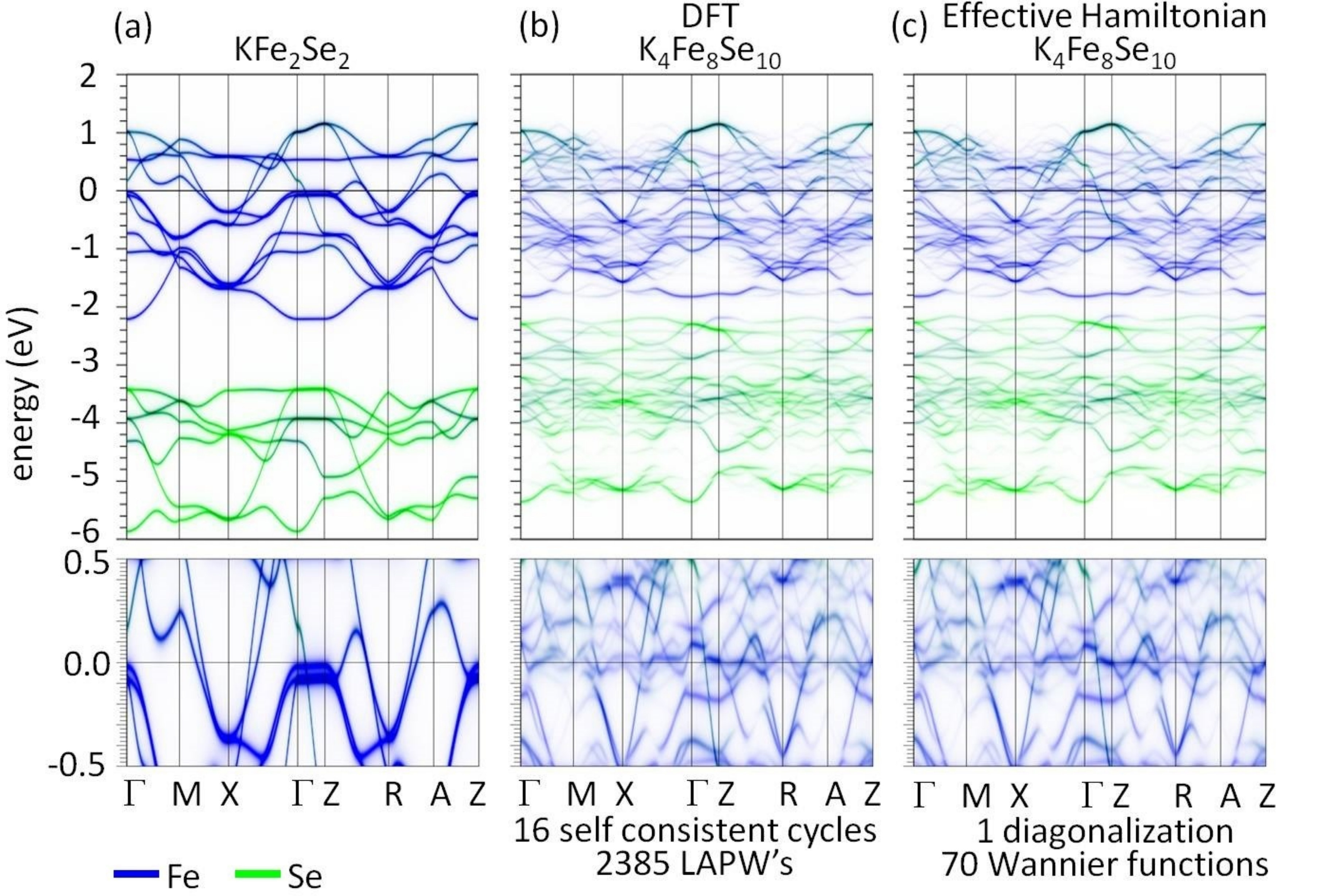}
\caption{Spectral functions of the clean reference system KFe$_2$Se$_2$ (a) and K$_4$Fe$_{8}$Se$_{10}$ with 
one K vacancy and two Fe vacancies obtained from DFT (b) and the effective Hamiltonian method (c). Reprinted from ~\cite{kfe2se2}.
}
\label{fig:figtom1}
\end{figure}

To illustrate the accuracy and efficiency of the EDHM we present in Figure~\ref{fig:figtom1} a comparison of spectral functions for a K$_4$Fe$_{8}$Se$_{10}$ supercell calculated from the full DFT and the effective Hamiltonian. The size of the deviations between the spectral functions obtained from the full DFT and the EDHM should be compared with the size of the impurity-induced changes. For this purpose the spectral function of the undoped KFe$_2$Se$_2$ is also plotted as a reference.  As can be seen from Figure~\ref{fig:figtom1}, the effective Hamiltonian describes the influence of the Fe and K vacancies with high accuracy. All the detailed gap openings and shadow bands induced by the vacancies are captured. However, the basis set of Linear Augmented Plane Waves (LAPW's) used in the full DFT is $\sim30$ times larger then the basis set of Wannier functions used in the EDHM.  This reduction in the size of the basis set dramatically improves the efficiency of model-based calculations, especially when combined with model solvers such as the TMDCA.  Many more benchmarks can be found in the supplementary materials of Ref.~\cite{naxcoo2,tm122,kfe2se2,ru122,11sto} demonstrating the high accuracy and efficiency of the method.

In addition to chemical disorder it is also possible to take into account the influence of magnetic disorder by mapping the DFT onto a generalized spin-fermion model as we describe below. This is relevant for dilute magnetic semiconductors in which a strongly interacting impurity is embedded into a weakly interacting host.  

In practice, the generalized spin-fermion model is derived as follows. First we perform spin-density functional theory (using for example a LDA+U~\cite{tbanisimov,tbcococcioni} exchange correlation functional). Then we perform a Wannier transformation of the low energy bands by projecting only the host orbitals and not the impurity orbitals. 

This effectively integrates out the charge degrees of freedom corresponding to the impurity.  For example in the case of Ga$_{1-x}$Mn$_{x}$N ~\cite{rykynelson} we project only on the N$-sp3$ host orbitals thereby integrating out the charge degrees of freedom of the strongly interacting Mn-$d$ impurity orbitals.  Next, one derives the impurity potential in each of the two spin-channels resulting in $V^{x_{j}}_\uparrow$ and $V^{x_{j}}_\downarrow$ corresponding to the impurity at site $x_{j}$. 
In the generalized spin-fermion model the impurity potential is given by:
\begin{eqnarray}
\label{eq:sf_model}
V^{x_{j}}=\sum_{\i\i' nn'}\big(T_{\j\i\i'}^{nn'}c^\dagger_{\i n\sigma}c_{\i' n'\sigma}
+J_{\j\i\i'}^{nn'}c^\dagger_{\i n\sigma}\pmb{\tau_{\sigma\sigma'} } c_{\i' n'\sigma'} \cdot \pmb{S_{\j}}\big)
\nonumber \\
\end{eqnarray}
which incorporates the effect of the strong Coulomb repulsion at the impurity site. 
As usual, $ c_{\i n\sigma}$ ($ c^{\dagger}_{\i n\sigma}$) annihilates (creates) an electron with spin $\sigma$ in unit-cell $r_{i}$ in the $n$-th host orbital. $\pmb{\tau_{\sigma\sigma'} }$ and $\pmb{S_{\j}}$ are the Pauli matrices and the spin-vector operator.  The non-magnetic and magnetic coefficients are determined  
$T_{\j\i\i'}^{nn'}=\langle r_{i}n|V^{x_{j}}_\uparrow+V^{x_{j}}_\downarrow|r_{i'}n'\rangle$
and
$J_{\j\i\i'}^{nn'}=\langle r_{i}n|V^{x_{j}}_\uparrow-V^{x_{j}}_\downarrow|r_{i'}n'\rangle$ respectively.
Here we note that the impurity potential involves three spatial points labelled by $i$, $i'$ and $j$, meaning that if we place an impurity at site $j$ the processes from site $i$ to $i'$ will be modified.
We have recently performed such a derivation for Ga$_{1-x}$Mn$_{x}$N to resolve a long standing debate on the valence state of Mn~\cite{rykynelson}.  The main advantage of this approach compared to deriving a multi-orbital Hubbard model~\cite{crpa} is that by treating the impurity spins classically one can avoid the fermion sign problem ~\cite{tbchandrasekharan} and thus greatly reduce the computational expense of including interactions in the typical medium dynamical cluster approximation.

Recently, we also generalized the EDHM to include the treatment of phonons~\cite{fesi}. 
Rather than making a cluster expansion of the Wannier function based Hamiltonian of the electrons, a cluster expansion can be made in the force constant matrices of the phonons. This opens the way for studying disorder induced localization of phonons from first principles.

\subsection{From the EDHM to TMDCA}
\label{sec:EDHM2TMDCA}
In order to incorporate the EDHM into the TMDCA, we first need to convert the parameters derived from the EDHM into the form of the multi-orbital Anderson model used in the TMDCA. Moreover, since the impurity potentials derived are usually quite long ranged, an appropriate coarse-graining procedure is needed to map the effective impurity  potential from the lattice to the DCA cluster (c.f.~\ref{sec:DCA}).  In the following, we outline the procedure of these two steps.
\paragraph{Extraction of the impurity potential\\}

We start from the effective EDHM Hamiltonian: $H_{eff}=H_0+V$, where
\begin{equation}\label{eq:DFT1}
 H_{0}=\sum_{\mathbf{i,i'} n,n',\sigma}t_{\mathbf{ii'}}^{nn'}c_{in\sigma}^{\dagger}c_{\mathbf{i'} n'\sigma}+h.c.
\end{equation}
is the Hamiltonian of the pure host material with $\i$, $\i'$ corresponding to the site indices and $n$, $n'$ corresponding to the orbital indices. $V$ is defined in Eq.~\ref{eq:sf_model} which contains the impurity potential induced by the impurity located at site $\j$. Since for each impurity, the induced impurity potential on neighboring sites
has the same form, we can rewrite the parameters in Eq.~(\ref{eq:sf_model}) as:
\begin{equation}
T_{\j\i\i'}^{nn'}= T_{\i-\j,\i'-\j}^{nn'}
\end{equation}
\begin{equation}
J_{\j\i\i'}^{nn'}= J_{\i-\j,\i'-\j}^{nn'}\, .  
\end{equation}
Here, since the spin-independent and spin-dependent parameters have similar structures, we only show the transformation for the spin-independent parameter. The spin-dependent component can be inferred by analogy.

To investigate the structure of the impurity potential, we first look at the terms induced by a single impurity located at the origin $V_{0}$ by letting $\j=0$ in Eq.~\ref{eq:sf_model}, and further split it into three parts:
\begin{equation}
\begin{split}
V_{0}&=\sum_{\i,\i',n,n',\sigma}T_{\i\i'}^{nn'}c_{\i n\sigma}^{+}c_{\i' n'\sigma}\\&
=\sum_{\i,n,n',\sigma}T_{\i\i}^{nn'}c_{\i n\sigma}^{+}c_{\i n'\sigma}
+\sum_{\i\ne0,n,n',\sigma}T_{0\i}^{nn'}c_{0n\sigma}^{+}c_{\i n'\sigma}\\&
+\sum_{\i,\i'\ne0,\i\ne \i',n,n',\sigma}T_{\i\i'}^{nn'}c_{\i n\sigma}^{+}c_{\i' n'\sigma}+h.c.\,.
\end{split}
\label{eq:vzero}
\end{equation}
The first term is diagonal disorder which in general, extends to a finite region from the origin. The second term is the off-diagonal disorder associated with hopping between the impurity site and a host site. The disorder induced by this term can be properly described in the Blackman formalism~\cite{Blackman_1917}. The last term is the off-diagonal disorder associated with the hopping between two host sites that are induced by the impurity located on the sites other than these two host sites.  Due to this feature, the disorder caused by this term can not be described properly in the original Blackman formalism so a slight
modification is made to include these terms in our calculation. 

To extend the Blackman formalism we first write $H_{eff}$ for a specific disorder configuration, with impurities labeled by $\j$,
\begin{equation}
\begin{split}
H_{eff}&=H_{0}+\sum_{\j}V_{\j}=\sum_{\i,n,n',\sigma}\epsilon_{\i\sigma}^{nn'}c_{\i n\sigma}^{+}c_{\i n'\sigma}\\&
+\sum_{\i \ne \i',n,n',\sigma}W_{\i,\i',\sigma}^{nn'}c_{\i n\sigma}^{+}c_{\i' n'\sigma}
\end{split}
\end{equation}
where, 
\begin{equation}\label{eq:dd}
\epsilon_{\i\sigma}^{nn'}=t_{\i\i}^{nn'}+\sum_{\j}T_{\j\i\i}^{nn'},
\end{equation}
\begin{equation}\label{eq:odd}
W_{\i,\i',\sigma}^{nn'}
=t_{\i\i'}^{nn'}+\sum_{\j=\i,or,\i'}T_{\j\i\i'}^{nn'}+\sum_{\j\ne \i,\j\ne \i'}T_{\j\i\i'}^{nn'}.
\end{equation}
Here, in Eq.~\ref{eq:odd} the first term is independent of the disorder configuration.  The third term depends on the disorder configuration but is independent of the chemical occupation of sites $\i$ and $\i'$.  The second term only depends on the chemical occupation of sites $\i$ and $\i'$.  If we denote the site as A if it is occupied by the host atom and B if it is occupied by the impurity atom, then we can see there are only four possible values for the second term:
\begin{equation}
\sum_{\j=\i,or,\i'}T_{\j\i\i'}^{nn'}=\begin{cases}
\begin{array}{c}
0,\ \ \ \ if\ \i \in A,\ \i' \in A\\
T_{\i'\i\i'}^{nn'},\ \ \ \ if\ \i \in A,\ \i' \in B\\
T_{\i\i\i'}^{nn'},\ \ \ \ if\ \i \in B,\ \i' \in A\\
T_{\i'\i\i'}^{nn'}+T_{\i\i\i'}^{nn'},\ \ \ \ if\ \i \in B,\ \i' \in B,
\end{array}\end{cases}
\end{equation}
so in the Blackman formalism, the hopping term $W_{\i,\i',\sigma}^{nn'}$can
be written as a 2 by 2 block matrix:
\begin{equation}
\begin{split}
\underline{W}_{\i,\i',\sigma}^{nn'}&=t_{\i\i'}^{nn'}\left[\begin{array}{cc}
1 & 1\\
1 & 1
\end{array}\right]+\left[\begin{array}{cc}
0 & T_{\i'\i\i'}^{nn'}\\
T_{\i\i\i'}^{nn'} & T_{\i'\i\i'}^{nn'}+T_{\i\i\i'}^{nn'}
\end{array}\right]\\&+\sum_{\j\ne \i,\j\ne \i'}T_{\j\i\i'}^{nn'}\left[\begin{array}{cc}
1 & 1\\
1 & 1
\end{array}\right].
\end{split}
\end{equation}
Here, we use underscore to denote the 2 by 2 matrix in Blackman formalism and we use overbar to denote the quantities that are coarse-grained.  We can see that the first two terms are configuration independent and translationally invariant in the Blackman formalism, because
\begin{equation}
T_{\i'\i\i'}^{nn'}=T_{\i-\i',0}^{nn'}
\end{equation}
\begin{equation}
T_{\i\i\i'}^{nn'}=T_{0,\i'-\i}^{nn'},
\end{equation}
so we can combine the first two terms as
\begin{equation}
\underline{W}_{\i,\i',\sigma}^{1,nn'}=\left[\begin{array}{cc}
t_{\i\i'}^{nn'} & t_{\i\i'}^{nn'}+T_{\i-\i',0}^{nn'}\\
t_{\i\i'}^{nn'}+T_{0,\i'-\i}^{nn'} & t_{\i\i'}^{nn'}+T_{\i-\i',0}^{nn'}+T_{0,\i'-\i}^{nn'}
\end{array}\right],
\end{equation}
and we identify the remaining term as
\begin{equation}
\underline{W}_{\i,\i',\sigma}^{2,nn'}=\sum_{\j\ne \i,\j\ne \i'}T_{\j\i\i'}^{nn'}\left[\begin{array}{cc}
1 & 1\\
1 & 1
\end{array}\right]=\sum_{\j\ne \i,\j\ne \i'}T_{\i-\j,\i'-\j}^{nn'}\left[\begin{array}{cc}
1 & 1\\
1 & 1
\end{array}\right],
\end{equation}
so that 
\begin{equation}
\underline{W}_{\i,\i',\sigma}^{nn'}=\underline{W}_{\i,\i',\sigma}^{1,nn'}+\underline{W}_{\i,\i',\sigma}^{2,nn'}.
\end{equation}
Note, $\underline{W}_{\i,\i',\sigma}^{2,nn'}$ which is related to the last term of Eq.~\ref{eq:vzero}, is not translational invariant even in the Blackman formalism, and cannot be described in the original Blackman method, so a slight modification is made to account for these terms in DCA/TMDCA calculations.

\paragraph{Coarse-graining the impurity potential\\}
Then, $\underline{W}_{\i,\i',\sigma}^{nn'}$ is coarse-grained in the DCA cluster with periodic boundary conditions to obtain the cluster parameters $\underline{\overline{W}}_{\I,\I',\sigma}^{nn'}$ used for the DCA and TMDCA calculations in the Blackman formalism, where the capital indices correspond to the lattice sites in the periodic TMDCA cluster.

Here, since $\underline{W}_{\i,\i',\sigma}^{1,nn'}$ is translationally invariant, it can be coarse-grained easily in the same manner as the regular kinetic energy terms:
\begin{eqnarray}
 \underline{W}_{\k,\sigma}^{1,nn'}=\sum_{\i}\underline{W}_{\i,\i',\sigma}^{1,nn'}
 e^{i\k\cdot(\mathbf{r}_{i}-\mathbf{r}_{i'})},
 \\
 \underline{\overline{W}}_{\K,\sigma}^{1,nn'}=\frac{N_c}{N}\sum_{\k}\underline{W}_{\K+\k,\sigma}^{1,nn'},
 \\
 \underline{\overline{W}}_{\I,\I',\sigma}^{1,nn'}=\frac{1}{N_{c}}\sum_{\K}\underline{\overline{W}}_{\K,\sigma}^{1,nn'}
 e^{-i\K\cdot(\mathbf{R}_{I}-\mathbf{R}_{I'})}.
\end{eqnarray}
But $\underline{W}_{i,i',\sigma}^{2,nn'}$ still depends on the disorder configuration, and is not translationally invariant, so it needs to be coarse-grained differently. We carry out the the coarse-graining according to the following procedure: 
\begin{equation}
\underline{W}_{\k,\k',\sigma}^{2,nn'}=\sum_{\i,\i'}\underline{W}_{\i,\i',\sigma}^{2,nn'}
e^{i(\k\cdot\mathbf{r}_{i}-\k'\cdot\mathbf{r}_{i'})},
\end{equation}
\begin{equation}
\underline{\overline{W}}_{\K,\K',\sigma}^{2,nn'}=(\frac{N_{c}}{N})^{2}\sum_{\k,\k'}
\underline{W}_{\K+\k,\K'+\k',\sigma}^{2,nn'},
\end{equation}
\begin{equation}
\underline{\overline{W}}_{\I,\I',\sigma}^{2,nn'}=(\frac{1}{N_{c}})^{2}\sum_{\K,\K'}
\underline{W}_{\K,\K',\sigma}^{2,nn'}e^{-i(\K\cdot\mathbf{R}_{I}-\K'\cdot\mathbf{R}_{I'})}.
\end{equation}
The diagonal disorder component from Eq.~(\ref{eq:dd}) includes also an extended contribution, 
$T_{\j\i\i}^{nn'}=T_{\i-\j,\i-\j}^{nn'}$, which needs to be coarsed grained. We implement
the following procedure:
\begin{equation}
T_{\k}^{nn'}=\sum_{\i}T_{\i\i}^{nn'}e^{i\k\cdot\mathbf{r}_{i}},
\end{equation}
\begin{equation}
\overline{T}_{\K}^{nn'}=\frac{N_{c}}{N}\sum_{\k}T_{\K+\k}^{nn'},
\end{equation}
\begin{equation}
\overline{T}_{\I\I}^{nn'}=\frac{1}{N_{c}}\sum_{\K}T_{\K}^{nn'}e^{-i\K\cdot\mathbf{R}_{I}} \,.
\end{equation}
Then the coarse-grained version of Eq.~(\ref{eq:dd}) is just
\begin{equation}
\begin{split}
\overline{\epsilon}_{\I\sigma}^{nn'}&=t_{\I\I\sigma}^{nn'}+\sum_{\J}\overline{T}_{\I-\J,\I-\J}^{nn'}
\\&
=\epsilon_{0\sigma}^{nn'}+\overline{V}_{\I}^{nn'},
\end{split}
\end{equation}
where
\begin{equation}
\overline{V}_{\I}^{nn'}=\sum_{\J}\overline{T}_{\I-\J,\I-\J}^{nn'}
\end{equation}
is the diagonal disorder potential in the cluster. Since 
$t_{\I\I\sigma}^{nn'}$ is local and translationally invariant, 
it is not modified by coarse graining, so we set it to $\epsilon_{0\sigma}^{nn'}$.
For the spin-dependent part, the same procedure can be carried out completely by analogy.

From the procedure above, we get the parameters needed for the DCA/TMDCA calculation. These are $\overline{\epsilon}_{\I\sigma}^{nn'}=\epsilon_{0\sigma}^{nn'}+\overline{V}_{\I,\sigma}^{nn'}$
for the diagonal component and $\underline{W}_{I,J,\sigma}^{1,nn'}$ and $\underline{W}_{I,J,\sigma}^{2,nn'}$
for the off-diagonal component of the disorder potential. The self-consistent loop is similar to the multi-orbital TMDCA and more details are described in the Appendix of ~\citep{y_zhang_16}

\section{ Applications of the Typical Medium DCA to Systems with Disorder (sec:Applications) }
\label{sec:Applications}

In this section we review the applications of the typical medium formalism to a selection of systems with disorder. We start our discussion with the application of TMDCA to single-band 3D Anderson model. Then we show how the TMDCA can be used with complex systems, including those with more generalized types of disorder, multiple orbitals, and electron-electron interactions.

\subsection{Results for the Anderson model.} 
\label{sec:AM}

\subsubsection{Typical DOS as an order parameter for Anderson localization} 
\label{sec:TDOSasOP}
\begin{figure}[t]
\begin{center}
\includegraphics[trim = 0mm 0mm 0mm 0mm,width=1.0\columnwidth,clip=true]
{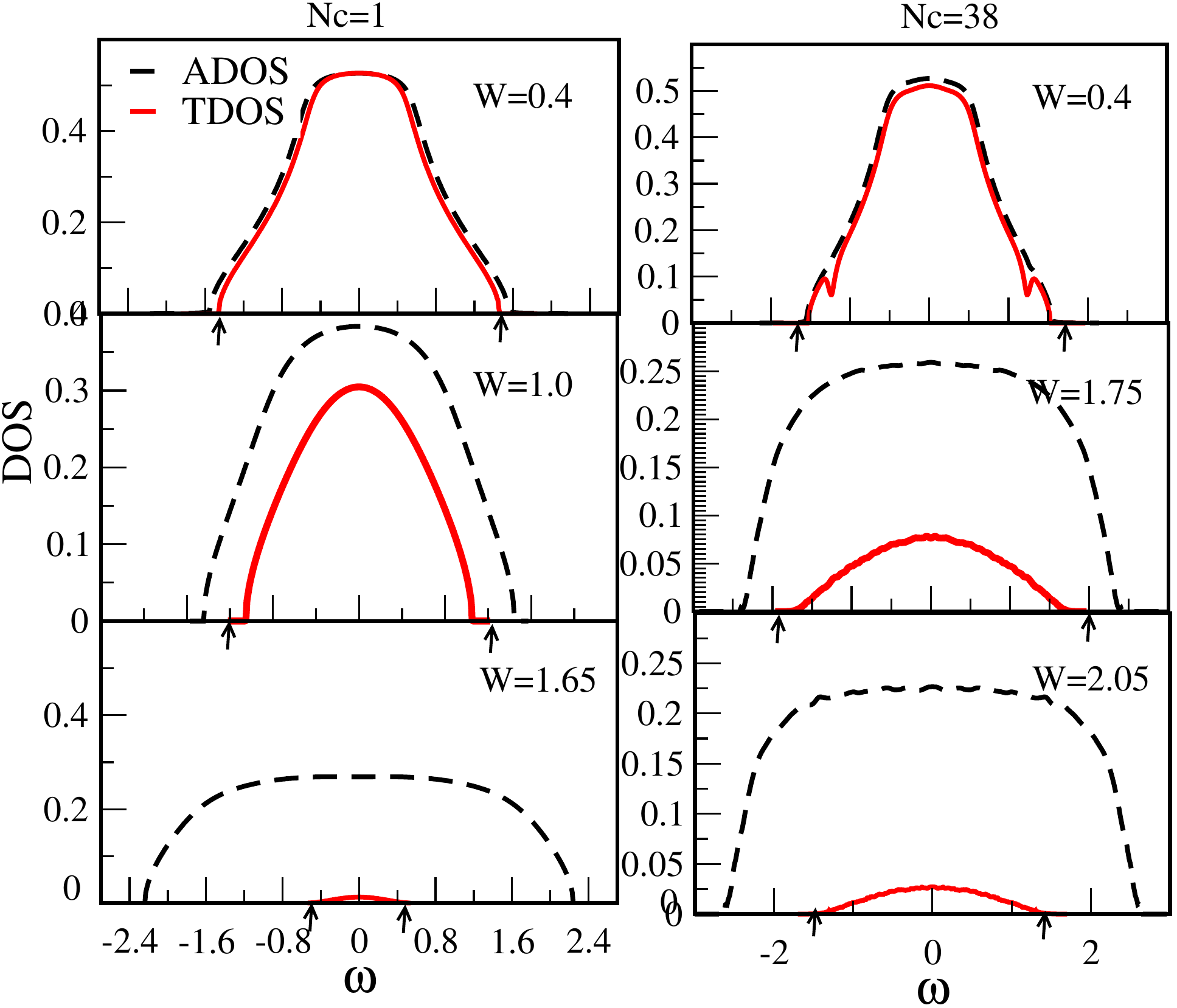}
\caption{TMT (left) and TMDCA (right with N${}_c = 38$) DOS of the 3D Anderson model for different disorder strengths $W$ in units where $4t=1$ .  The ADOS and TDOS coincide for weak disorder. While as $W$ increases the ADOS becomes suppressed.  In the TMDCA the mobility edge, indicated by the arrows, first moves to higher energy.  For roughly $W > 1.75$ (in units where $4t=1$) it starts moving towards the band center, indicating that TMDCA can successfully capture the re-entrance behavior missing in the TMT scheme. Reprint from ~\cite{c_ekuma_14b}.}
\label{fig:tdos}
\end{center}
\end{figure}

We start our discussion of the results by presenting the application of the TMDCA to a single site Anderson Model in 3D. First we demonstrate that the typical and not the average DOS can serve as a proper order parameter for defining the Anderson localization transition.  In Figure~\ref{fig:tdos}, we compare the algebraically averaged DOS (ADOS) calculated using the conventional DCA scheme (dashed lines) and the TDOS (solid lines) obtained from both a single site TMT (left panel, $N_c = 1$) and finite clusters obtained from the TMDCA (right panel, $N_c = 38$).  The TMDCA employed Ansatz 1 for various disorder strengths $W$ for the box disorder distribution with $ P(V)=\frac{1}{2W}\theta(W-|V|)$.  

As seen from Figure~\ref{fig:tdos},  as the disorder strength increases, the ADOS broadens but remains finite while the TDOS obtained from both the TMT ($N_c=1$) and the TMDCA ($N_c=38$) continuously decreases.  It eventually vanishes even at the band center at the critical disorder strength with $W_c(N_c=1)\approx 1.65$ and $W_c(N_c=38)\approx 2.25$ (in units $4t=1$).  Below the transition, for $W<W_c$, the part of the spectrum with vanishing TDOS corresponds to localized states, while the part of spectrum with a finite TDOS corresponds to the extended states. As one can see the band tail localize first. Also, notice that at small disorder with $W<<W_c$, e.g. $W=0.4$ the ADOS and the TDOS are almost the same. This indicates that at small disorder the TMDCA reduces to the standard DCA scheme, which is consistent with the analysis used to construct Ansatz 1 in Sec.~\ref{sec:TMDCA}.  

Comparing the local TMT ($N_c=1$) and the non-local TMDCA ($N_c>1$) results, one observes a crucial difference between them. For the local TMT, the mobility edge (indicated by arrows) delineating the  region with extended states where the TDOS is finite, always becomes narrower with increasing disorder strength $W$. For a finite cluster TMDCA, the mobility edge first expands and then
decreases, hence giving rise to the re-entrance behavior, missing in the single-site TMT. 

\begin{figure}[t]
\begin{center}
\includegraphics[trim = 0mm 0mm 0mm 0mm,width=0.9\columnwidth,clip=true]
{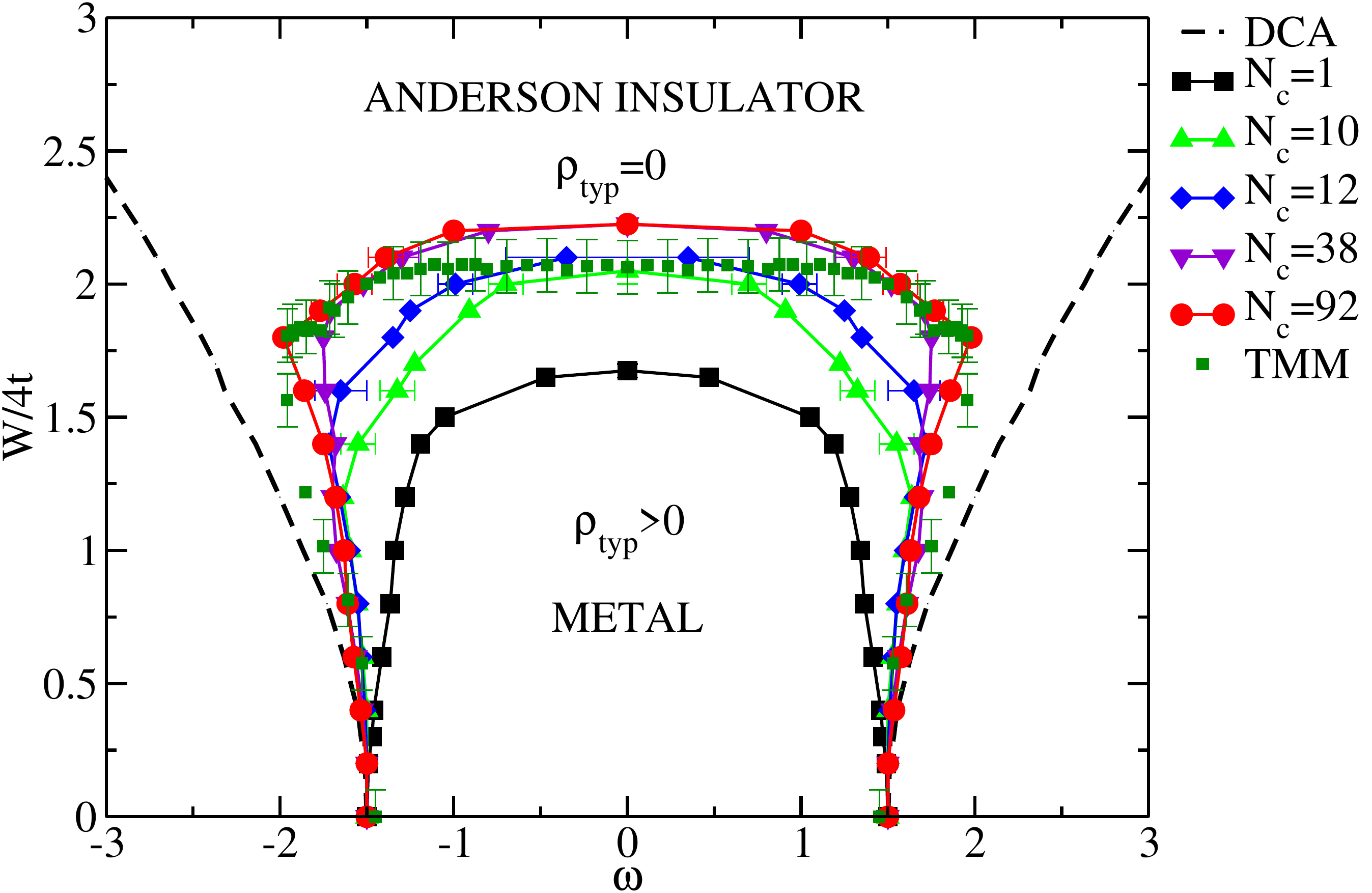}
\caption{Phase diagram of the Anderson localization transition in 3D obtained 
from TMDCA simulations. As N$_c$ increases, a systematic improvement of the trajectory of the mobility edge is achieved. At large enough N$_c$ and within computation error, our results converge to those determined by the TMM~\cite{b_bulka_85}. }
\label{fig:phasediagram}
\end{center}
\end{figure}

The resulting $W-\omega$ (disoder-energy) phase diagram is shown in Figure~\ref{fig:phasediagram}. Here, we show the mobility edge trajectories, (obtained by the frequencies $\omega$ where the TDOS vanishes at a given disorder strength $W$), and the band edge trajectories, (where the ADOS calculated within the DCA scheme vanishes).  To benchmark our results, we also present the mobility edge trajectories obtained from the transfer matrix method.  The finite cluster TMDCA trajectories gradually approach the TMM results with the re-entrance behavior, (missing in $N_c=1$ case)  recovered with increasing cluster size.  For a large clusters $N \geq 92$ our TMDCA results converge to TMM trajectories  within the errors of both approaches. 

\subsubsection{Cluster size convergence}
\label{sec:CSConv}
\begin{figure}[h!]
\begin{center}
\includegraphics[trim = 0mm 0mm 0mm 0mm,width=1.0\columnwidth,clip=true]
{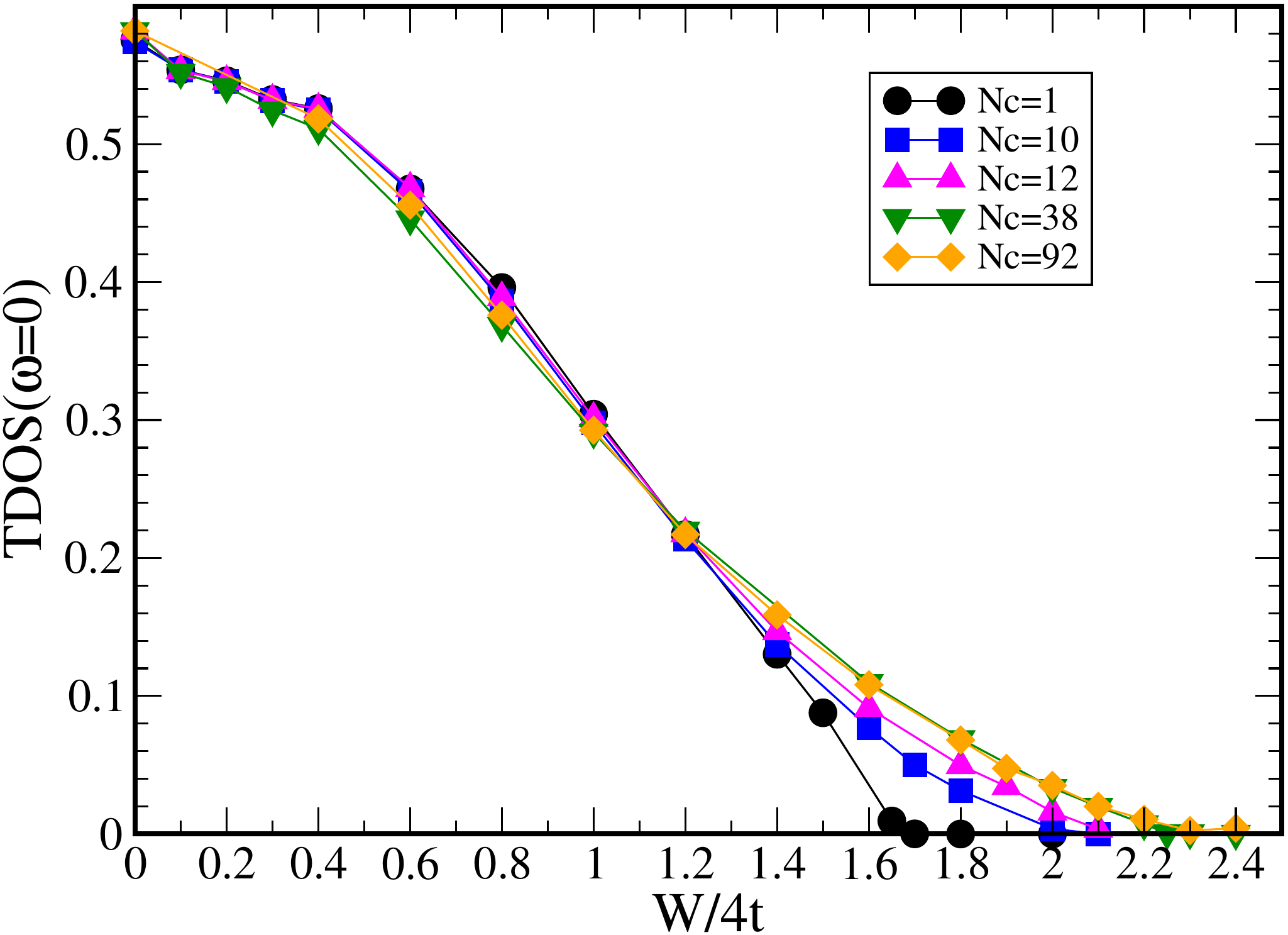}
\caption{The TDOS at the band center $TDOS(\omega = 0)$ vs. disorder strength \ $W$ for the 3D Anderson model calculated with the TMDCA using Ansatz 1 for different cluster sizes $N_c =1, 10,12, 38,92$ with units where $4t=1$.  The TDOS ($\omega$ = 0) vanishes at the critical disorder strength $W_c$ when all states become localized. For $N_c = 1$, which corresponds to the TMT method, the critical disorder strength $W_c(N_c=1)\approx 1.65$. As cluster size $N_c$ increases, the critical disorder strength $W_c$ increases quickly to $\approx 2.25$, which is in very good agreement with the results from the transfer matrix method  $W_c\approx2.1$ \cite{k_slevin_14}.}
\label{fig:Wc}
\end{center}
\end{figure}

We now consider how the critical disorder strength $W_c$  converges with the cluster size $N_c$. Since $W_c$ is defined by the vanishing $TDOS(\omega=0)=0$, in Figure~\ref{fig:Wc} we plot the local $TDOS(\omega=0)$ at the band center as a function of disorder strength $W$ for several clusters $N_c$.
The presented results are obtained using Ansatz 1. We also did calculations with Anzats 2 (data not shown) and obtained very similar results.  
Our results show that as cluster size $N_c$ increases, the $W_c$ systematically increases until it converges to $W_c \approx 2.25 $ which is in good agreement with the $W_c\approx 2.1$ values reported in the literature ~\cite{b_bulka_85}.
The data presented in Ref.~\cite{c_ekuma_14b} for large cluster sizes does not attain full self-consistency.  We pay extra attention to the convergence of the self-energy and redo the calculations for the data as shown in Figs.~\ref{fig:phasediagram} and \ref{fig:Wc}.

\subsection{ Results for models with more realistic parameters}
\label{sec:Real}

In this section we apply the typical medium analysis to more complex disordered systems, including those with off-diagonal disorder, multiple orbitals, and interactions. We continue this section by showing application of TMDCA to calculate two-particle quantities and explore the effect of interactions.  Finally, we discuss the simulation of some select high temperature superconductors and dilute magnetic semiconductors.
\subsubsection{Off-diagonal disorder}
\label{sec:ODD}
So far, we have presented the TMDCA results for systems with local disorder having potentials coupling only the density operators.  As they are diagonal in the creation and annihilation operators, this is called diagonal disorder.  However, in many materials, the disorder not only affects the strength of the local potential, but it also impacts the strength of the hopping of electrons between different sites.  Since this involves the creation of an electron on one site and the annihilation on another site, the associated disorder is called non-local or off-diagonal disorder. To demonstrate that our TMDCA scheme can properly treat such generalized cases of disorder and to understand how the off-diagonal disorder affects the electron localization, we first present the results for the 3D single band Anderson model with disorder and hopping defined by the Hamiltonian Eq.~\ref{eq:AMHamiltonian} 

\begin{figure}[tbh]
\centering{} \includegraphics[trim = 0mm 0mm 0mm 0mm,width=1\columnwidth,clip=true]{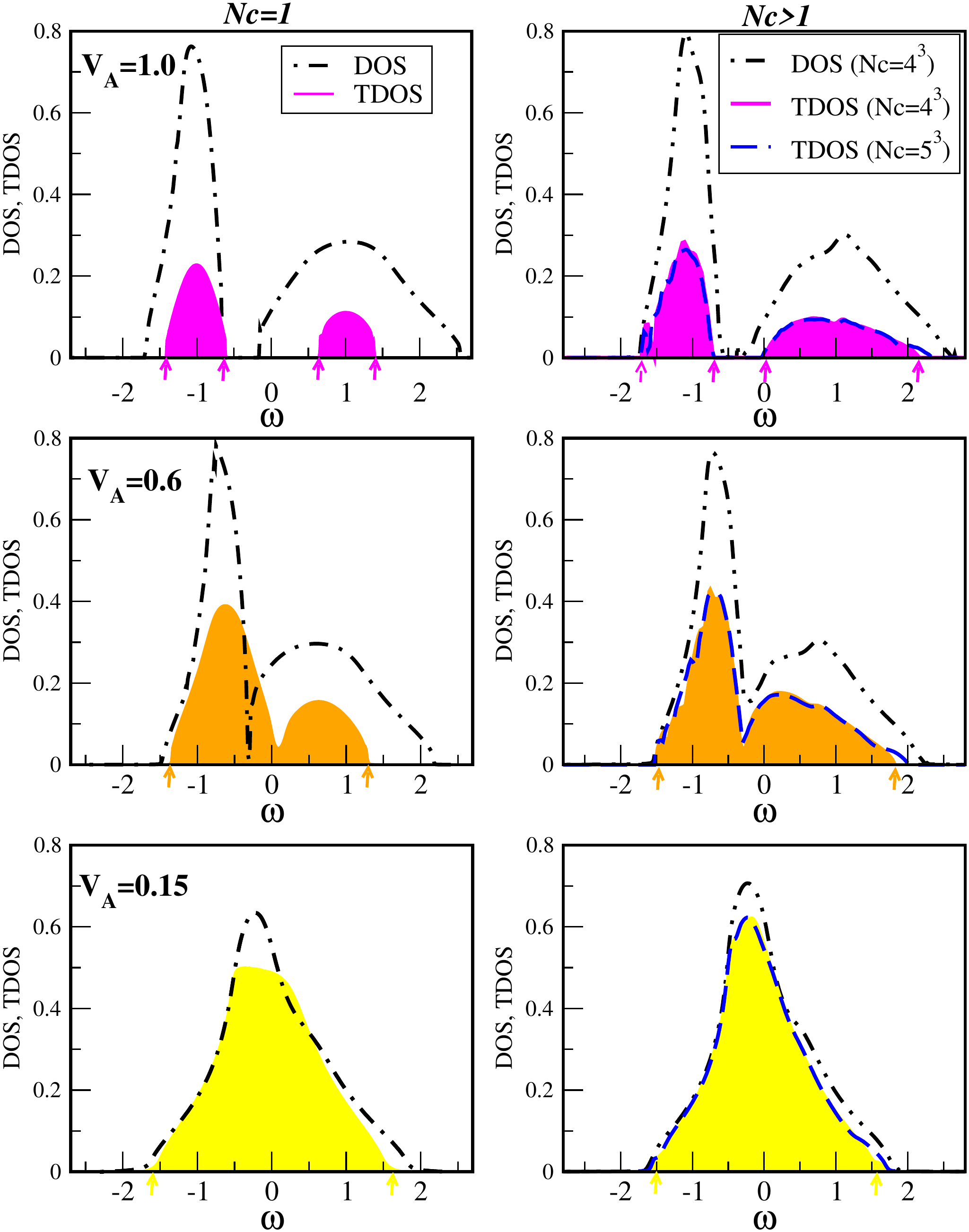}
\caption{The average DOS and the typical TDOS of the A-B binary alloy model with off-diagonal disorder.  The left panel displays results for $N_c=1$ (corresponding to TMT local method) and the right panel for $N_c>1$ (TMDCA results). The data show the average DOS (dash-dotted line) and the typical density of states (shaded regions) for $N_c=1$ (left panel), $N_c=4^3$ (right panel) and blue dash lines for $N_c=5^3$ (left panel) for various values of the local potential $V_A$ with off-diagonal disorder parameters: $t^{AA}=1.5$, $t^{BB}=0.5$, $t^{AB}=0.5(t^{AA}+t^{BB})$, and $c_{A}=0.5$. We show the TDOS for several cluster sizes N$_c=1$, $4^3$, and $=5^3$ in order to demonstrate its systematic convergence with increasing cluster size N$_c$. The average DOS converges within our numerical precision for cluster sizes beyond N$_c=4^3$. As in the diagonal disorder case, the TDOS is finite for the extended states and zero for localized states. Reprint from ~\cite{h_terletska_14a}. 
\label{fig:Fig4}}

\end{figure}

To illustrate the method, we return to our simple model of an AB binary alloy.  
In Figure~\ref{fig:Fig4}, we present the results for the TDOS obtained from the generalized TMDCA and the ADOS obtained from the DCA schemes for several values of the diagonal disorder strength $V_A=0.15, 0.6, 1.0$ at fixed off-diagonal disorder amplitudes $t^{AA}=1.5$, $t^{BB}=0.5$, $t^{AB}=1.0$.

We also present data for the local $N_c=1$ case, in order to demonstrate the effect of non-local correlations captured within the finite cluster  $N_{c}=4^3$ and $5^3$  DCA and TMDCA algorithms. The ADOS data for $N_c>1$ shows that non-local multisite effects lead to the development of finite detailed structures in the density of states and the partial filling of the gap at larger values of disorder strength.    

Comparing TDOS and ADOS, we observe that for small disorder $V_{A}$, both are practically the same. This is consistent with our analytical construction of the Ansatz (Eq.~\ref{rhotyp_BEB}), where for small disorder strength, the TMDCA should converge to the DCA scheme. As the disorder strength $V_{A}$ increases, significant differences start to emerge. Increasing $V_A$ leads to the gradual opening of a gap which is more pronounced in the $N_c=1$,  For weaker disorder, $V_{A}=0.6$, it is partially filled for the $N_c>1$ clusters. As compared to the diagonal disorder case (\cite{c_ekuma_14b}), the average DOS and TDOS become asymmetric with respect to zero frequency due to the off-diagonal randomness.  We again observe that the local TMT ($N_c=1$) underestimates the extended states regime by having a narrower TDOS as compared to the case when $N_c>1$.

\begin{figure}[t!]
\includegraphics[trim = 0mm 0mm 0mm 0mm,width=.45\textwidth,clip=true]{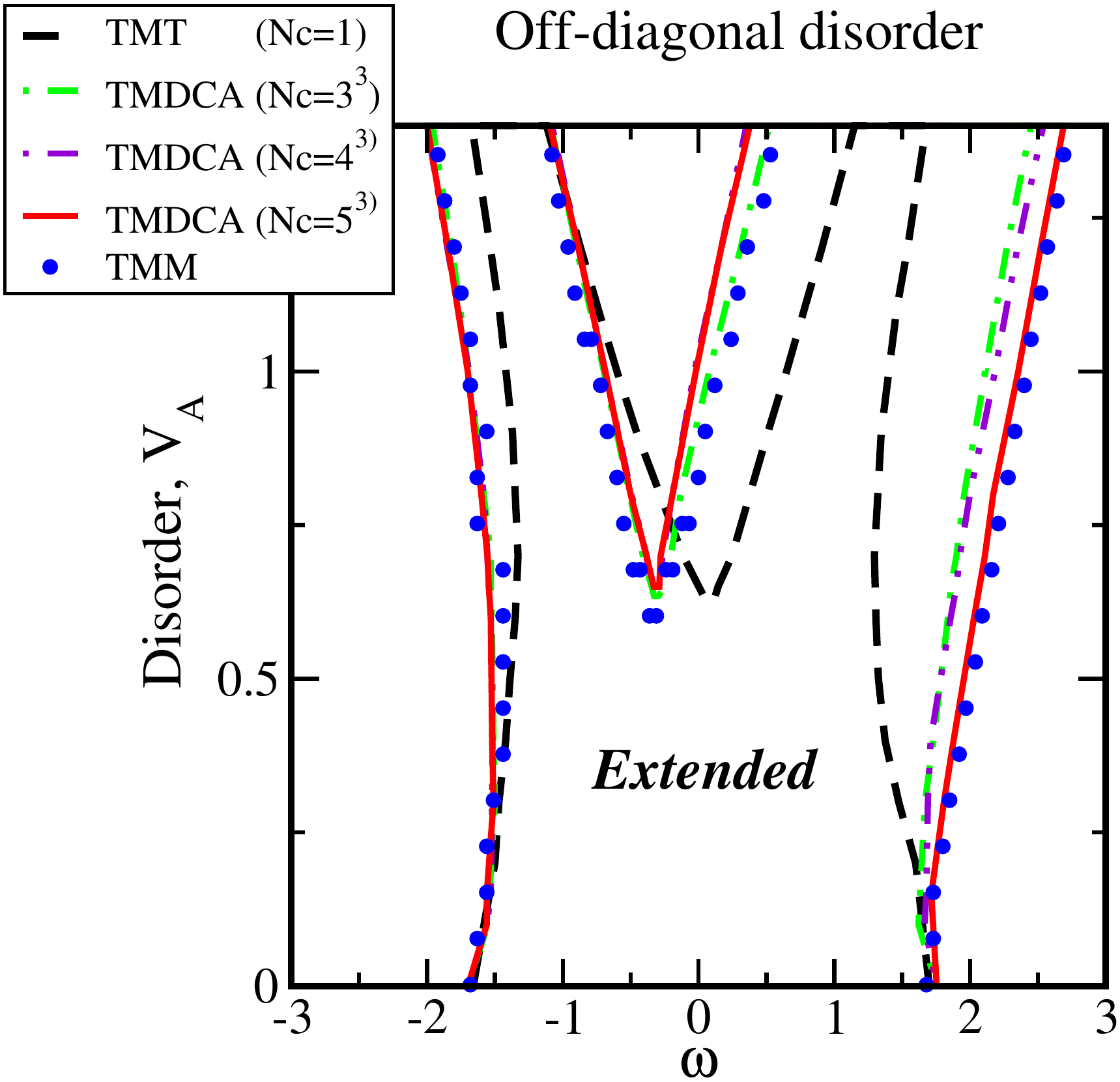}
\caption{Disorder $(V_A)$-energy $(\omega)$ phase diagram of the A-B binary alloy model with off-diagonal disorder.  Parameters used are $t^{AA}=1.5$, $t^{BB}=0.5$, $t^{AB}=1.0$, and $c_A=0.5$.  The mobility edges obtained from the TMT $N_c=1$ (black dashed line), TMDCA $N_c=3^3$ (green dot-dashed line), $N_c=4^3$ (purple double-dot-dashed line) and $N_c=5^3$ (red solid line), and the transfer-matrix method (TMM) (blue dotted line).  The single site TMT method $(N_c=1)$ strongly underestimates the extended states region, with the finite TDOS, especially for higher values of disorder potential $V_A$. The mobility edges obtained from the finite cluster TMDCA ($N_c>1$) converge gradually with increasing cluster size $N_c$ and show good agreement with those obtained from the TMM, in contrast to the single site TMT. Reprint from ~\cite{h_terletska_14a}. 
\label{fig:mobility}}
\end{figure}

We performed a similar analysis for a range of $V_A$ values, and our final result for the $V_A-\omega$ parameter space is shown in  Figure~\ref{fig:mobility}. Here for comparison we present the mobility edge boundaries (extracted from boundaries where the TDOS vanishes) from the single TMT ($N_c=1$) and the non-local TMDCA ($N_c>1$) results, and benchmark with the TMM results.  The mobility edges shown in Figure~\ref{fig:mobility} were extracted from the TDOS, with boundaries being defined by zero TDOS.  As can be seen from Figure~\ref{fig:mobility}, while the single-site TMT does not change much under the effect of off-diagonal disorder, the TMDCA results are significantly modified.  The bands for a larger cluster become highly asymmetric with significant widening of the A sub-band. The local $N_c=1$ boundaries are narrower than those obtained for $N_c>1$ indicating that the TMT strongly underestimates the extended states regime in both diagonal and off-diagonal disorder.  On the other hand, comparing  the mobility edge boundaries for $N_c>1$ with those obtained using TMM, we find very good agreement. This again confirms the validity of our generalized TMDCA.

\subsubsection{Multiple orbitals }
\label{sec:moresults}
\begin{figure}[h!]
 \includegraphics[trim = 0mm 0mm 0mm 0mm,width=.45\textwidth,clip=true]{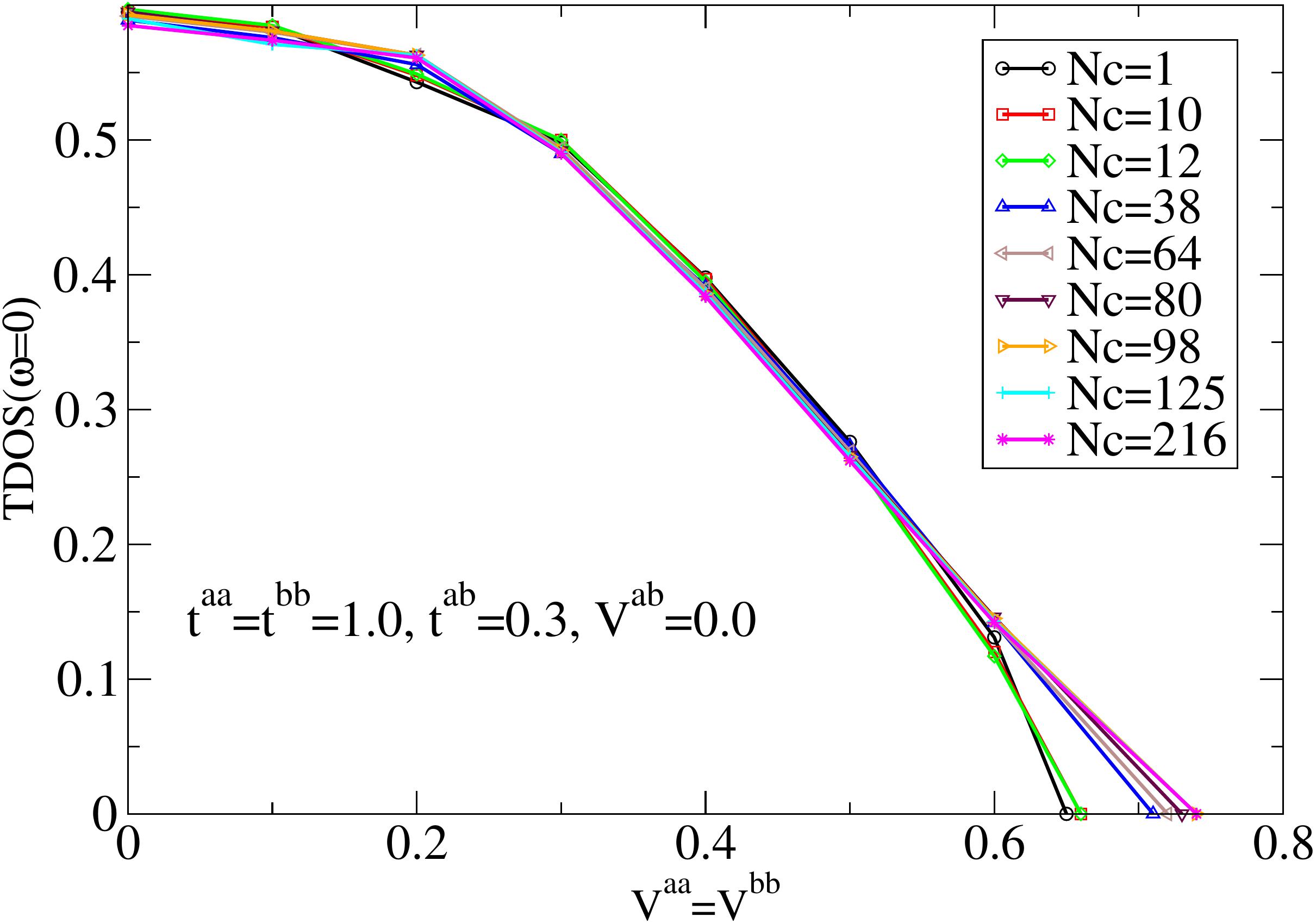}
 \caption{The TDOS at the band center ($\omega=0$) vs.\ $V^{aa}=V^{bb}$ in the a-b two-orbital model with increasing cluster size, for $t^{aa}=t^{bb}=1.0$, $t^{ab}=0.3$, $V^{ab}=0.0$. For $N_c=1$, the critical disorder strength is 0.65 and as $N_c$ increases, it increases and converges to 0.74 for $N_c=98$. Reprint from~\cite{y_zhang_15a}.}
 \label{fig:convergence}
\end{figure}

The multi-orbital TMDCA with the Ansatz defined in Eq.~\ref{eqn:ansatz_mo} and Eq.~\ref{eqn:Hilbert_mo} has been tested for a 3D Anderson model with two degenerate bands (denoted by a and b), so that both nearest neighbor hopping and disorder potential in this case are 2 $\times$ 2 matrices in the band basis given by
\begin{equation}
\underline{t_{ij}}=\underline{t}=\left(\begin{array}{cc}
t^{aa} & t^{ab}\\
t^{ba} & t^{bb}
\end{array}\right),
\end{equation}
and
\begin{equation}
\underline{V_i}=\left(\begin{array}{cc}
V_i^{aa} & V_i^{ab}\\
V_i^{ba} & V_i^{bb}
\end{array}\right),
\end{equation}
respectively.  The intra-band hopping is set as $t^{aa}=t^{bb}=1$, with
finite inter-band hopping $t^{ab}$. The local inter-band disorder $V_i^{ab}$ is set to to be zero considering the two bands orthogonal to each other so that the randomness only comes from the local intra-band disorder potential $V_i^{aa(bb)}$ that follow independent binary probability distribution functions with equal strength, $V^{aa}=V^{bb}$ and impurity concentration $x=0.5$.  As shown in Figure~\ref{fig:convergence}, in this two-band system the TMDCA again captures localization, where the TDOS at the band center gradually decreases as the disorder strength increases, and eventually vanishes at the critical point. The critical disorder strength reaches convergence within our numerical precision for a cluster size of roughly $N_c$=98.

\begin{figure}[h!]
 \includegraphics[trim = 0mm 0mm 0mm 0mm,width=.45\textwidth,clip=true]{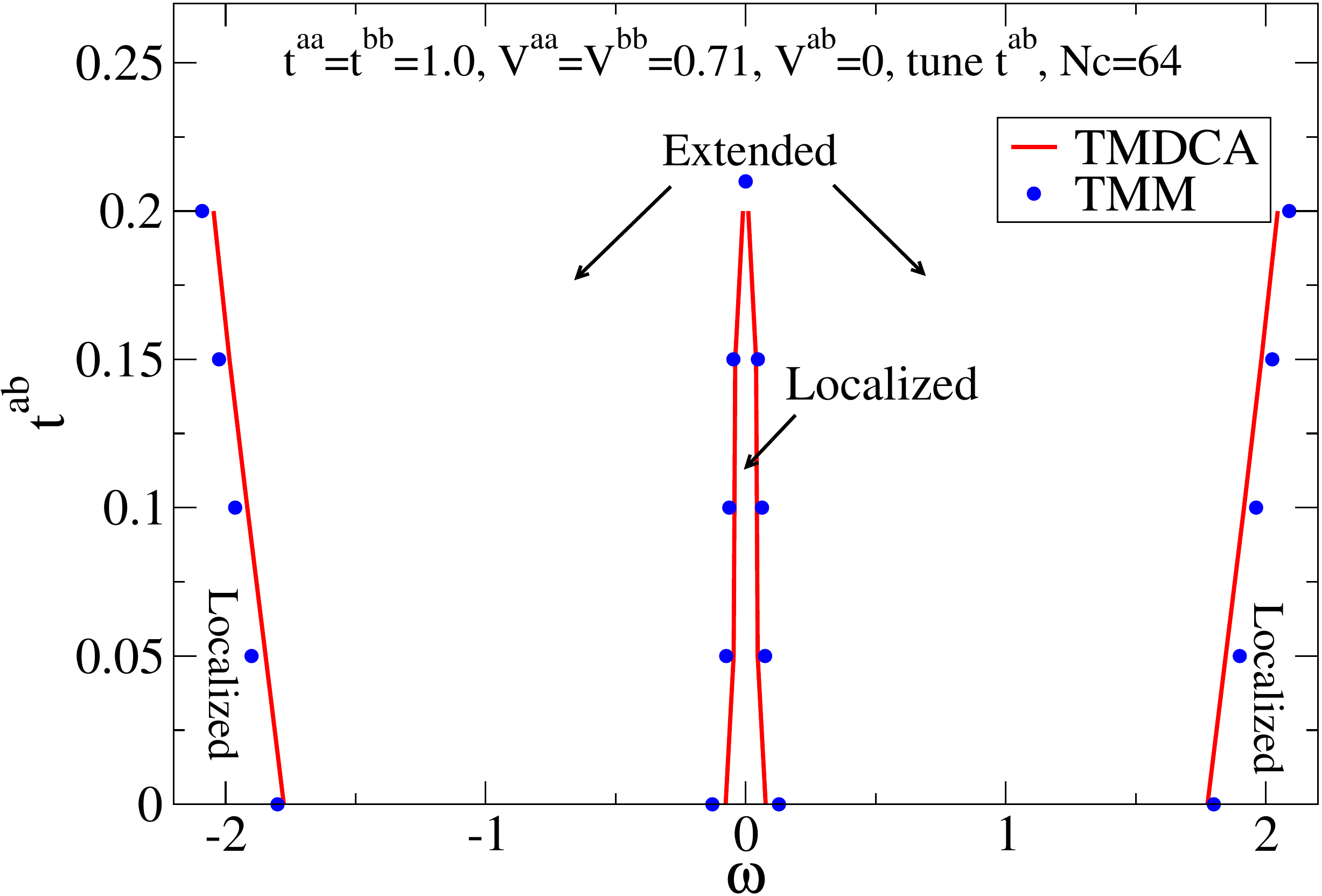}
 \caption{Evolution of the mobility edge of the a-b two-orbital model as $t^{ab}$ increases, while $V^{aa}$ and $V^{bb}$ are fixed. The results are calculated for $N_c=64$. A dome-like shape shows up around the band center, signaling the closing of the TDOS gap. Reprint from~\cite{y_zhang_15a}.}
 \label{fig:pd_tune_tab}
\end{figure}

In order to demonstrate the effect of inter-band hopping in this two-band model, the evolution of the mobility edge as a function of t$^{ab}$ with a fixed disorder strength is also studied and shown in Figure~\ref{fig:pd_tune_tab}. The dome-like shape around the band center reflects the delocalization effect of the inter-band hopping which is again in excellent agreement with results from the TMM method.

\begin{figure}[h!]
 \includegraphics[trim = 0mm 0mm 0mm 0mm,width=1\columnwidth,clip=true]{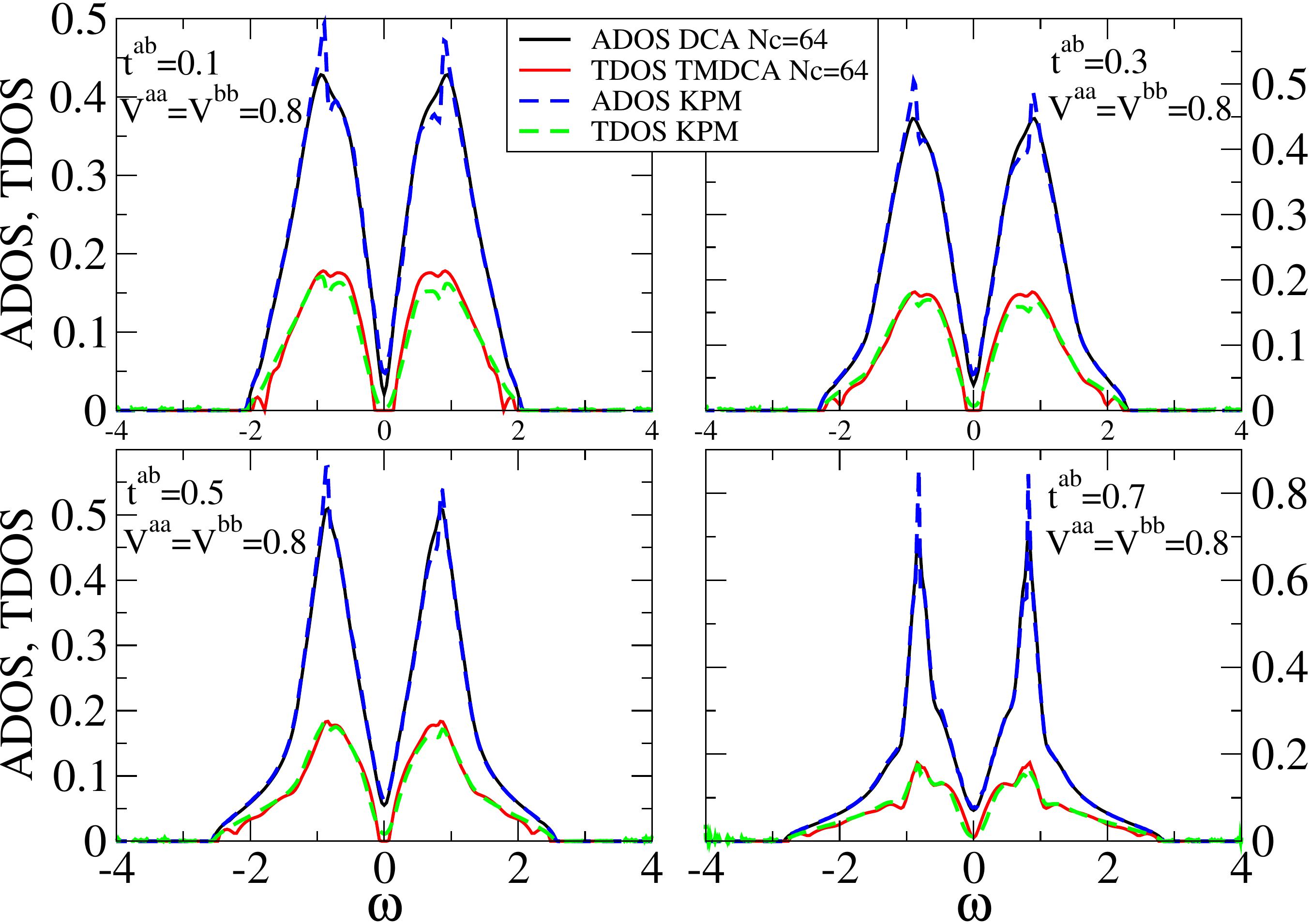}
 \caption{Comparison of the ADOS and TDOS of the a-b two-orbital model calculated with the DCA, TMDCA and KPM with fixed disorder strength $V^{aa}=V^{bb}=0.8$ with impurity concentration $x=0.5$ and various values of the inter-band hopping $t^{ab}$. The KPM uses $2048$ moments on a cubic lattice of size $48^3$ and $200$ independent realizations generated with $32$ sites randomly sampled from each realization. Reprint from~\cite{y_zhang_15a}.}
 \label{fig:kpm2band}
\end{figure}

To further benchmark the method, the calculated ADOS and TDOS using the DCA and TMDCA are also compared with those calculated using the KPM which is shown in Figure~\ref{fig:kpm2band}. As shown in the plot, a nice agreement between the (TM)DCA and KPM are achieved.

\subsection{Results for two-particle calculations }
\label{sec:2presults}
\begin{figure}[t]
    \centerline{\includegraphics[trim = 0mm 0mm 0mm 0mm,width=1\columnwidth,clip=true]{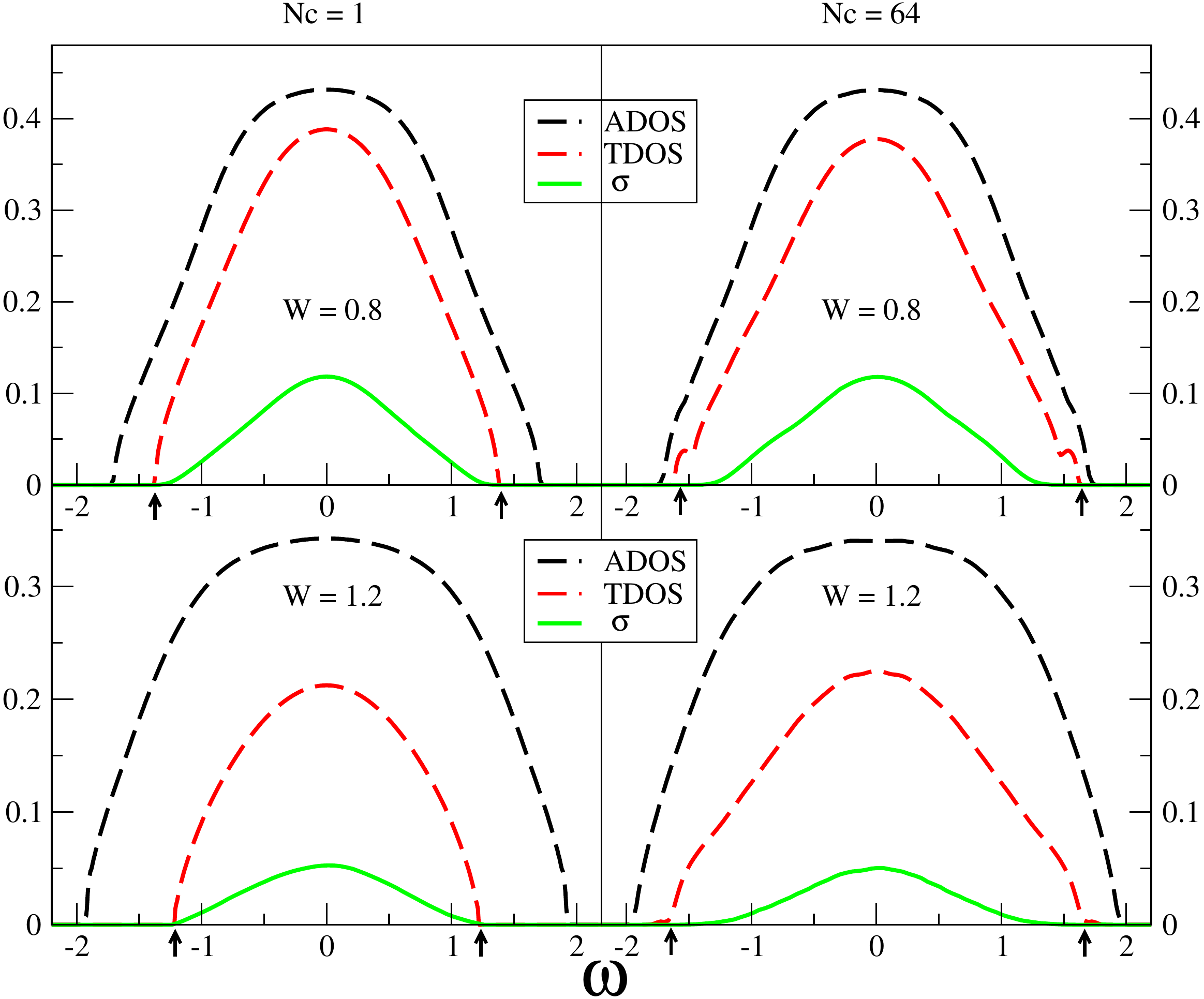}}
\caption{
    The evolution of the ADOS, TDOS and DC conductivity of the single-band 3D Anderson model at various disorder strengths $W$ for the single-site TMT and the TMDCA with cluster size $N_c=64$.   Here, for the DC conductivity, $\omega$ corresponds to the chemical potential used in the calculation. Arrows indicate the position of the mobility edge, which separates the extended electronic states from the localized ones. Reprint from~\cite{y_zhang_17}.}
\label{fig:energy}
\end{figure}

The typical analysis has been applied to the single band Anderson model to calculate the DC conductivity~\cite{y_zhang_17}. As shown in Figure~\ref{fig:energy}, the DC conductivity vanishes in the region where the TDOS is zero. This is expected since when the TDOS is zero, meaning all states are localized on the cluster, the hybridization function also becomes zero and all clusters are isolated.

\begin{figure}[t]
    \centerline{\includegraphics[scale=0.35]{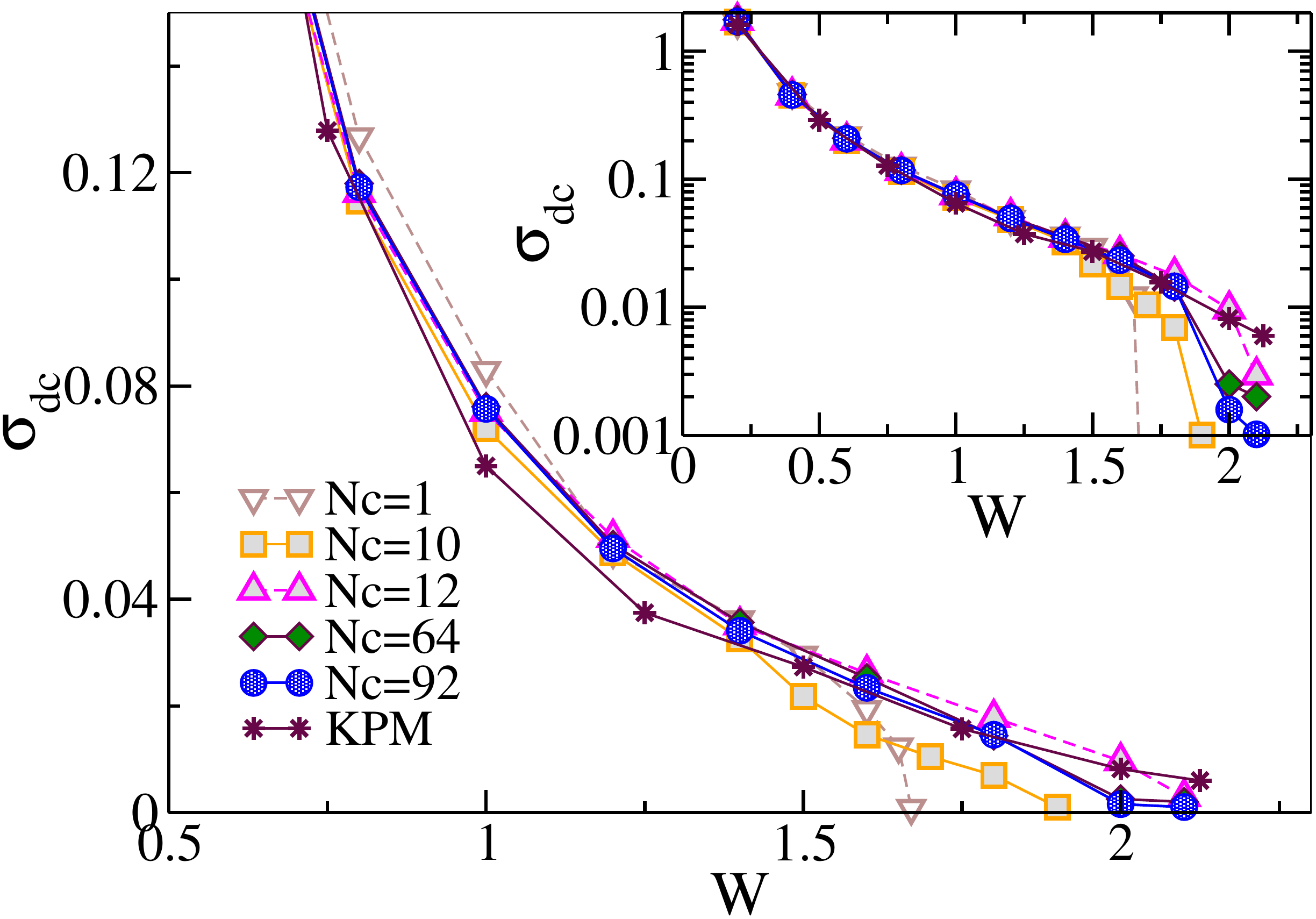}}
\caption{
DC conductivity of the 3D Anderson model at $T=0$ and $\mu=0$ (band center) vs.\ disorder $W$ for different cluster size $N_c=1,10,12,64,92$. The DC conductivity vanishes at $W_{c}$ where all states become localized.  For $N_c=1$ (TMT), the critical disorder  strength $W_c^{N_c=1} \approx 1.65$ (units $4t=1$). As the cluster size increases, $W_c$ systematically increases with $W_c^{N_c\gg12}\approx 2.10\pm 0.10$ (in units of $4t=1$), showing a quick convergence with cluster size to the KPM result. Reprint from~\cite{y_zhang_17}.  }
\label{fig:disorder}
\end{figure}

The convergence of the critical disorder strength $W_c$ with the cluster size $N_c$ is also studied.  Figure~\ref{fig:disorder} shows the DC conductivity at zero chemical potential as a function of disorder strength $W$ for several $N_c$. $W_c$ is defined by the vanishing of the DC conductivity. The results show that as cluster size $N_c$ increases for $N_c\ge12$, the $W_c$ systematically increases until it converges to $W_c\approx 2.1$.  This is consistent with the values reported in the literature\cite{b_bulka_85}. From this cluster onward, $W_c$ converges to $\approx$ 2.1.  The TMDCA results are also compared with the KPM~\cite{a_weisse_06,Weisse_2004,Ferreira_2016,Garcia_2015} which leads to excellent agreement for most values of the disorder strength. The results get noisy near the transition (Figure~\ref{fig:disorder}), but the deviation from the KPM calculations is in the correct direction given that the KPM is a finite-sized approximation and the conductivity vanishes near the critical disorder strength.

\subsection{Results for interacting models}
\label{sec:ResultsInteracting}
\subsubsection{Results from SOPT}
\label{sec:ResultsSOPT}
As discussed in the introduction, the interplay between disorder and interactions can be quite subtle and counterintuitive. Using the TMDCA, we explored the effect of weak interactions in a strongly disordered Anderson-Hubbard model through second order perturbation theory, described in Sec.~\ref{sec:SOPT}. A thorough benchmarking study reveals excellent agreement of the perturbation theory results until $U \lesssim 1.0$ (in units of $4t=1$) with results from the DCA-CTQMC results\cite{c_ekuma_15c}. Beyond $U\sim 1.0$, deviations begin to appear, and the SOPT does not remain reliable. 

\begin{figure}[h]
\includegraphics[width=0.48\textwidth,clip=true]{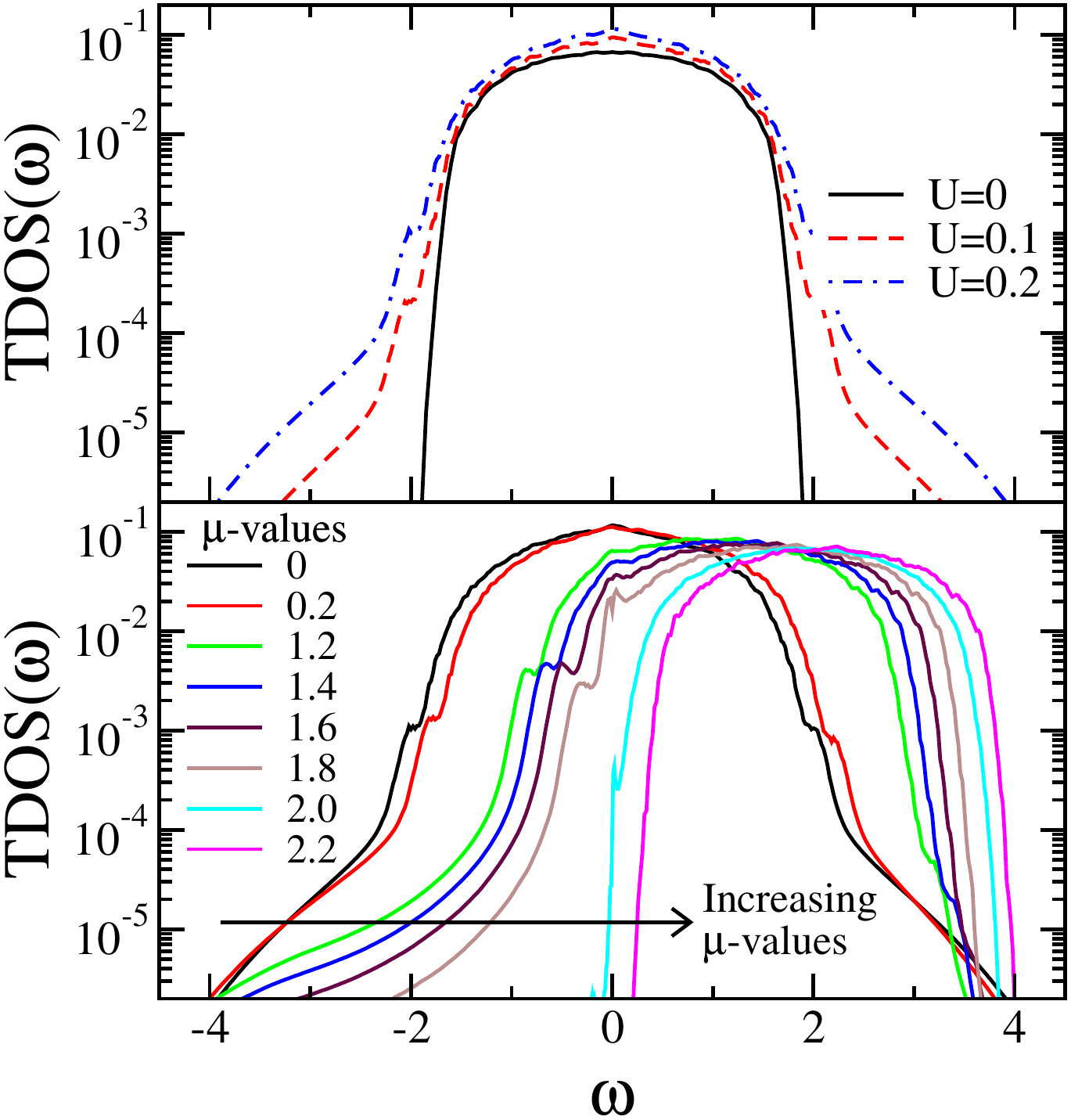}
\caption{Top: The typical DOS as a function of frequency, for the non-interacting case ($N_c=38, U=0.0$, units $4t=1$) and two weakly interacting cases ($U=0.1, 0.2$) are shown for a disorder value $W$ that is close to the critical disorder, i.e $W/W_c(U)=0.86$\cite{c_ekuma_15c} of the 3D Anderson-Hubbard model. The $U=0$ TDOS shows a sharp band edge, while for $U>0$, exponential tails are seen, indicating the broadening of the mobility edge. Bottom: The typical DOS as a function of frequency, for the interacting case ($N_c=38, U=0.2$, units $4t=1$) at various chemical potentials ($\mu$). As the $\mu$ approaches the non-interacting mobility edge, the exponential tail seen in the top panel is replaced by a sharp edge.  }
\label{fig:mob_edge}
\end{figure}  

 One of the main results of this study was the absence of a sharp mobility edge separating the localized from the delocalized spectrum if the chemical potential is at or beyond the mobility edge of the corresponding non-interacting system. We show the result for both p-h symmetric and away from p-h symmetry cases in Figure~\ref{fig:mob_edge}.
  In Figure~\ref{fig:no_mob} the typical density of states on a logarithmic scale {\it vs.} $\omega$ on a linear scale, for a fixed cluster size of $N_c=38$, various $U$ values and a fixed disorder ratio $W/W_c(U)=0.86$ is displayed. The non-interacting case shows a sharp drop of the TDOS at the band edges, thus exhibiting a sharp mobility edge. However,  for $U>0$, the TDOS is seen to have exponential tails at the band edges.
  

  We also found that the width of the mobility edge depends on the location of the chemical potential\cite{c_ekuma_15b} (not shown here), and goes continuously to zero as the energy approaches the chemical potential.  Here, the decay of the states via interactions is suppressed by the lack of phase space for which energy is conserved and the Pauli principle satisfied.  This is similar to the situation in a Fermi liquid.  However, here, the Pauli principle, together with energy and momentum conservation means that the scattering rate vanishes quadratically with the energy measured relative to the Fermi energy.   As a result, the Fermi liquid has a resistivity which is quadratic in temperature, a linear in temperature electronic specific heat, etc.  In our case, the momentum conservation is lost since the impurities break translational invariance.   So, we might expect a different power law; perhaps, a lower power reflecting the fact that the phase space will open more quickly than in a Fermi liquid, due to the reduced number of constraints.   The absence of a sharp mobility edge may also be understood through a perturbation theory argument (which should be valid in weak coupling), where the starting point is the non-interacting disordered system having a clear mobility edge. A perturbation theory in $U$ involves convolutions which mix the localized states below and extended states above the mobility edge, thus leading to a smearing of the TDOS band edge, and hence to a complete absence of a sharp division between the extended and localized states.  
  
  Since only these states very close to the Fermi surface are probed by most experiments,  this phenomena may be difficult to distinguish from the non-interacting case.  The difficulty is that since the width goes to zero as the chemical potential approaches the remnant of the mobility edge.  So, that experiments (most of them) that probe only the states near the Fermi energy will see a sharp mobility edge.  However, the low energy excitations may exhibit non-Fermi liquid behavior.  To our knowledge, this phenomena has not yet been explored.

\begin{figure}[h]
\includegraphics[width=0.45\textwidth,clip=true]{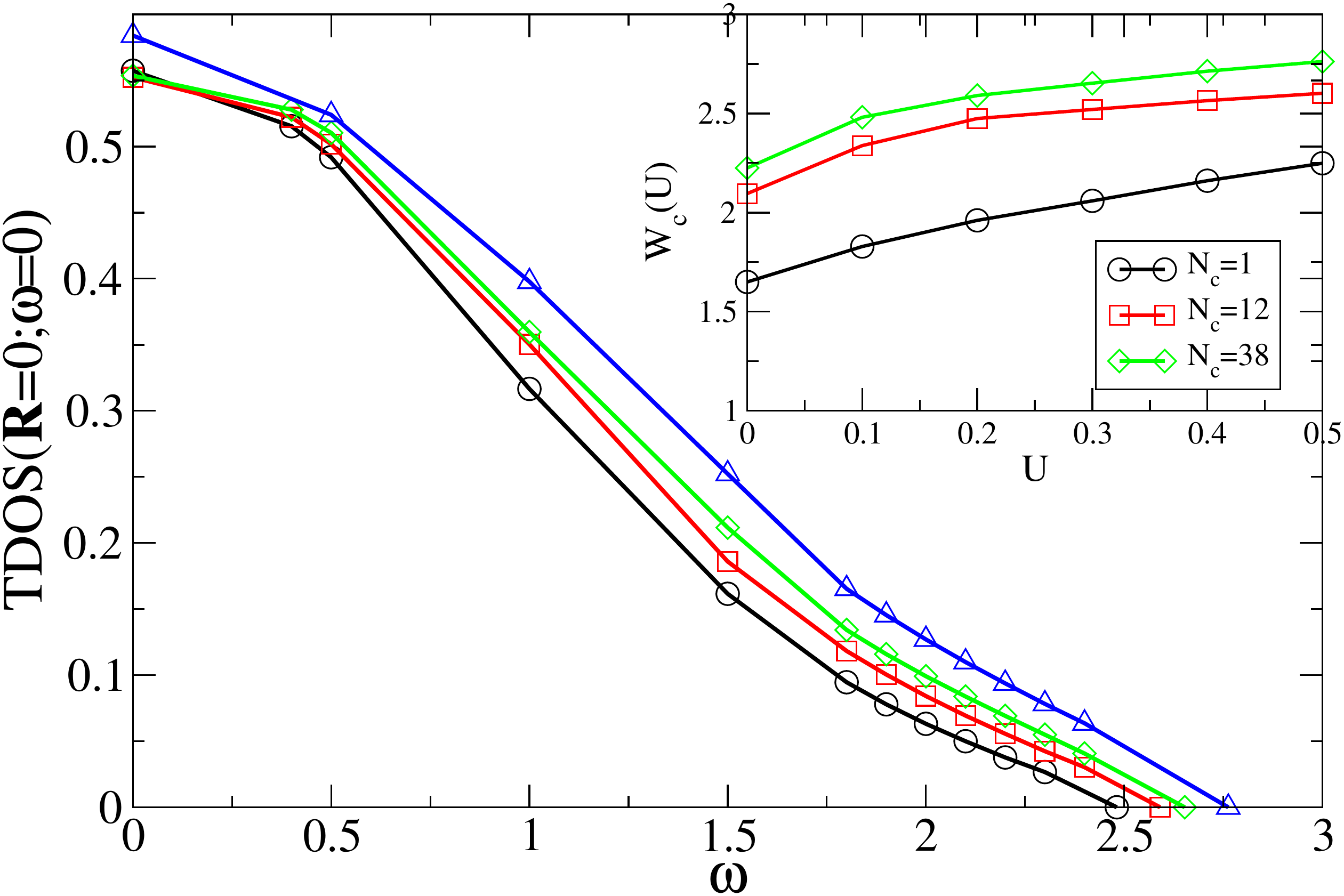}
\caption{
Screening of disorder effects by weak interactions in the 3D Anderson-Hubbard model: The main panel shows the momentum integrated typical DOS, TDOS({\bf R}=0; $\omega=0$) for $N_c=38$ as a function of disorder, $W$ for various $U$ values (units $4t=1$). The inset shows that the critical disorder value, $W_c(U)$ increases with increasing $U$ for three cluster sizes. }
\label{fig:wcrit}
\end{figure} 
  The lack of a sharp mobility edge due to interactions may also be interpreted as a delocalization of states that would have otherwise been localized by disorder. Further support for such a role of interactions is also found in the increase of the critical disorder, $W_c(U)$ with increasing $U$. In figure~\ref{fig:wcrit}, the integrated typical DOS for $N_c=38$ as a function of disorder for various interaction strengths is seen to decrease sharply and vanish at a critical disorder strength, $W_c$, whose value depends on $U$. The inset shows that the $W_c(U)$ increases with increasing $U$.  Using the TMT with an NRG impurity solver, Byczuk \etal had also found the same result\cite{k_byczuk_05}; however, since the TMT is a local theory, and hence corresponds to $N_c=1$, it was not clear if their result was robust against inclusion of non-local dynamical correlations due to disorder and interactions. The TMDCA results for $N_c=38$, which fully incorporate these correlations,  shown in Fig.~\ref{fig:wcrit} confirm that, indeed interactions can screen disorder effects, and hence a larger disorder value is needed to localize the system in the presence of interactions. 

Interestingly, we also found a dip in the density of states at the chemical potential, akin to a pseudogap, at disorder values that were very close to the critical disorder. Since this is the weak coupling regime, this pseudogap could be a precursor of the Efros-Shklovskii  Coulomb gap\cite{Efros_Shklovskii_1975}, however the present model has purely local interactions, while the Coulomb gap is found for long-range interactions, which have not been explored yet.

\subsubsection{Results from Stat-DMFT }
\label{sec:ResultsStatDMFT}

The role of strong interactions is also of great interest. Unfortunately, the second order perturbation theory based cluster solver is, naturally, restricted to the weakly interacting regime. Hence, to investigate the interplay of disorder and interactions in the strong coupling regime, we developed a real-space cluster solver based on statistical DMFT coupled with an impurity solver, namely the local moment approach, that is capable of capturing local Kondo physics in a non-perturbative way. 

\begin{figure}[h]
\includegraphics[width=0.45\textwidth,clip=true]{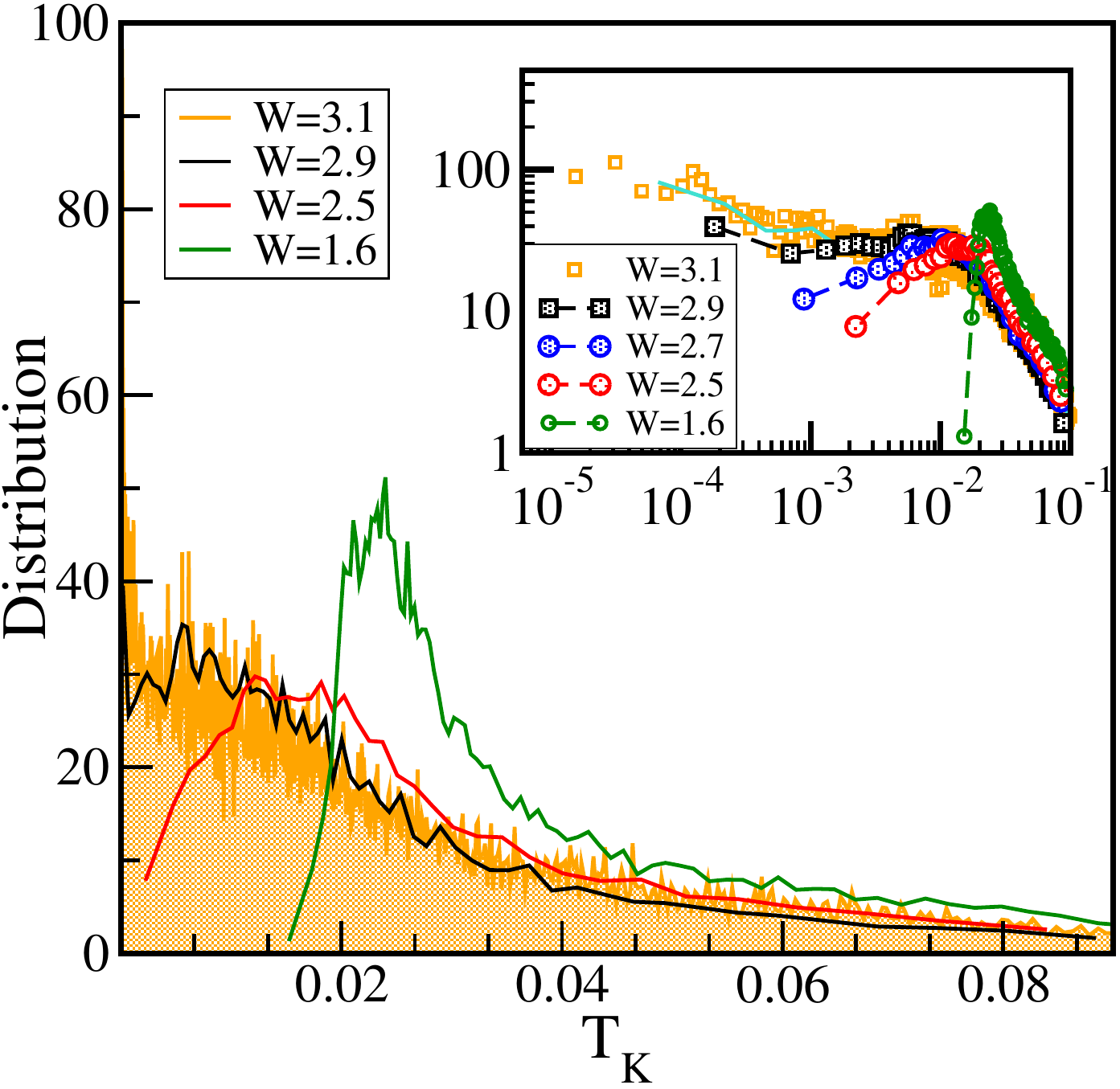}
\caption{Distribution of Kondo scales {\it vs.} $T_K$ for various disorder values in the 3D Anderson-Hubbard model (units $4t=1$) with $U=1.6$.  For larger $W$ values, the distribution develops a finite intercept. The inset shows the same data on a log-linear scale. Reprint from ~\cite{Sen_etal_2018}. }
\label{fig:tk_distrib}
\end{figure}

\begin{figure}[h]
\includegraphics[width=0.47\textwidth,clip=true]{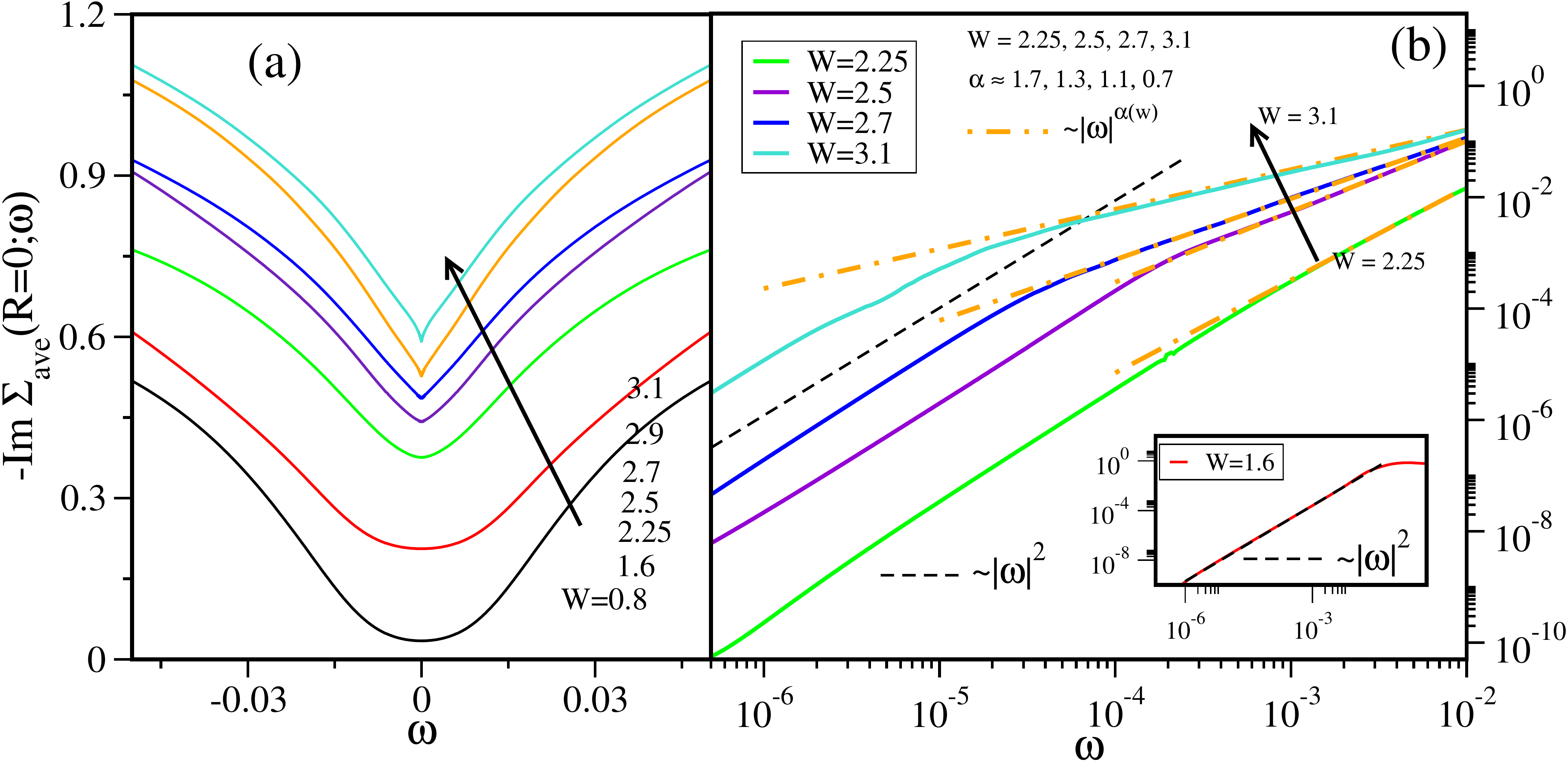}
\caption{The negative of the imaginary part of the low-frequency ($\omega$) self energy as a function of $\omega$ for various disorder values (legends), $N_c=38$ and $U=1.6$. The left panel shows that the $-{\rm Im}\Sigma(\omega)$ is quadratic close to the Fermi level and crosses over to  a power law form (see more clearly in the right panel) with an exponent $\alpha(W) < 2$, that is
disorder-dependent. Reprint from ~\cite{Sen_etal_2018}.}
\label{fig:nfl_se}
\end{figure}

\begin{figure}[h]
\includegraphics[width=0.47\textwidth,clip=true]{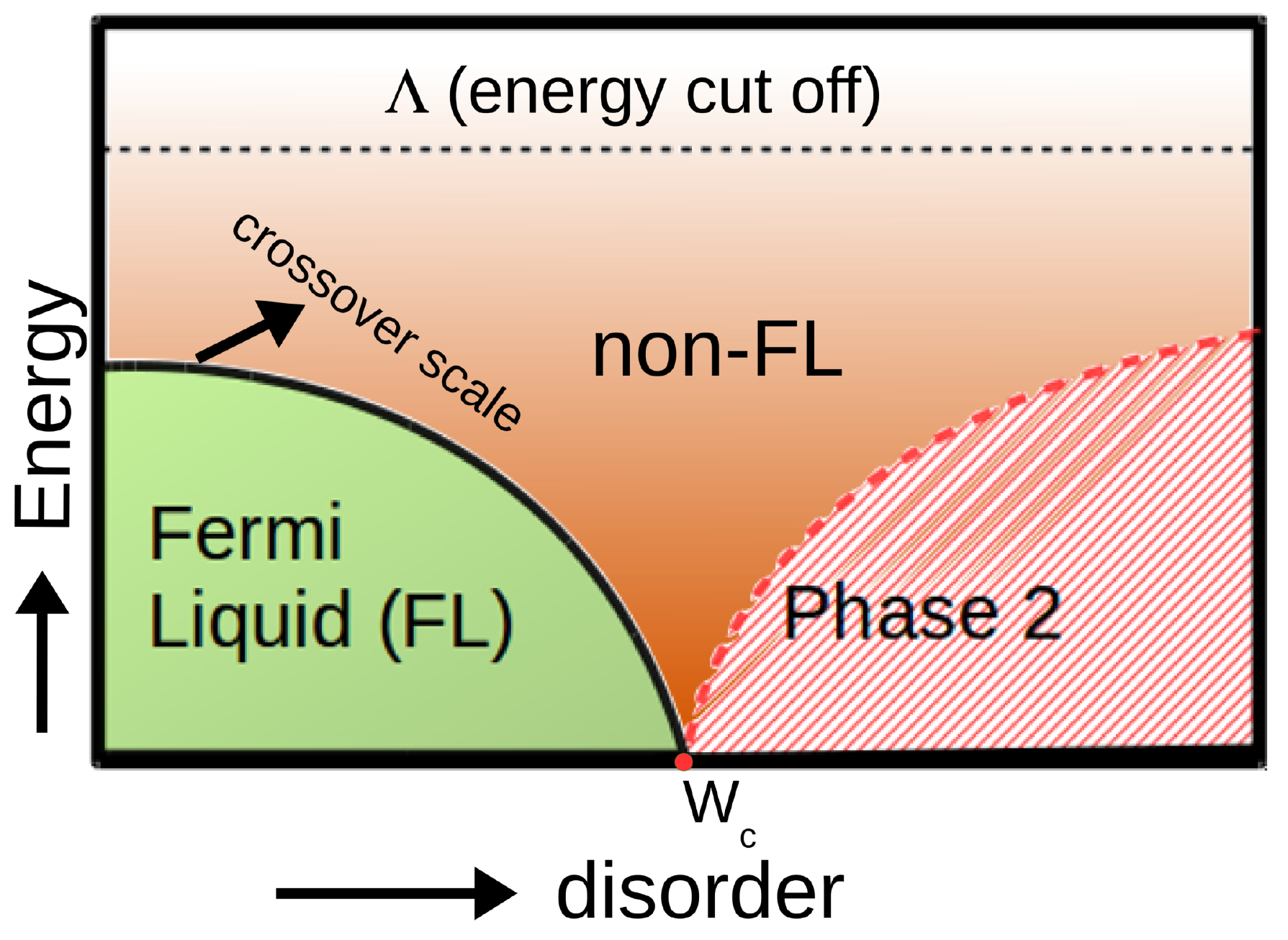}
\caption{A schematic phase diagram in the disorder-energy plane of the Anderson-Hubbard model showing a disorder-driven QCP separating a Fermi liquid from an as yet unidentified Phase-2. Reprint from ~\cite{Sen_etal_2018}. }
\label{fig:sch_ph}
\end{figure}

Since, within stat-DMFT, the hybridization is different for each site, the Kondo scale, $T_K$, acquires a highly non-trivial and skewed distribution, $P(T_K)$,  as shown in Figure~\ref{fig:tk_distrib}. For a fixed $U=1.6$, the distribution of Kondo scales as a function of $T_K$\cite{Sen_etal_2018} 
is shown for increasing disorder values and a cluster size, $N_c=38$. 
The figure shows that the distribution of $T_K$s develops a finite intercept at larger disorder values, indicating the formation of local moments. Many studies have shown that a sufficient condition for non-Fermi liquid behavior is a non-zero value of $P(T_K=0)$\cite{Miranda_Kondo_prb,d_tanaskovic_03}. Indeed, the corresponding self energy shows a crossover from low frequency Fermi liquid  to high frequency non-Fermi liquid behavior at a crossover scale $\omega_c$.  This is shown in Figure~\ref{fig:nfl_se}, where the negative of the imaginary part of the self energy, $-{\rm Im}\Sigma(\omega)$ is shown on a linear and log-log scale in the left and right panels respectively. The right panel shows clearly that the frequency dependence is Fermi liquid like ($\omega^2$) at low frequencies, and crosses over to $|\omega|^\alpha$, with a disorder-dependent $\alpha < 2$ at higher frequencies. The crossover scale, $\omega_c(W)$ decreases with increasing $W$, leading us to speculate the existence of a disorder-driven quantum critical point where $\omega_c(W)=0$. Our results for the crossover scale along with inferences from previous works may be combined to get a schematic phase diagram (shown in figure~\ref{fig:sch_ph}) of the quantum-critical region of the Anderson-Hubbard model.  

As the schematic suggests, a quantum critical point at $W_c$, identified by the vanishing of the crossover scale, separates a Fermi liquid phase from a second phase which we simply call Phase-2.This second phase could not be identified within the TMDCA calculations, but can be speculated to be some kind of a quantum spin liquid. It was also argued in the work that the quantum criticality cannot be of a local type or a Hertz-Millis-Moriya type, and hence has to be of a new type. 

\subsection{Results of the first-principles studies of localization }
\label{sec:edhmresults}
The combined method EDHM+TMDCA (described in Sec.~\ref{sec:edhm}) has so far been applied to study localization from first principles in two types of functional materials: superconductors~\cite{y_zhang_15a} and diluted magnetic semiconductors~\cite{y_zhang_16}. Due to its ability to access systems with multiple orbitals and complicated disorder potentials, it provides a powerful approach to study localization caused by the impurities in these functional materials in an unbiased and material-specific way.

\subsubsection{Application to K$_{\lowercase{y}}$F\lowercase{e}$_{2-\lowercase{x}}$S\lowercase{e}$_2$ sec:KFe2Se2}
\label{sec:KFe2Se2}

\begin{figure}[h!]
 \includegraphics[trim = 0mm 0mm 0mm 0mm,width=.45\textwidth,clip=true]{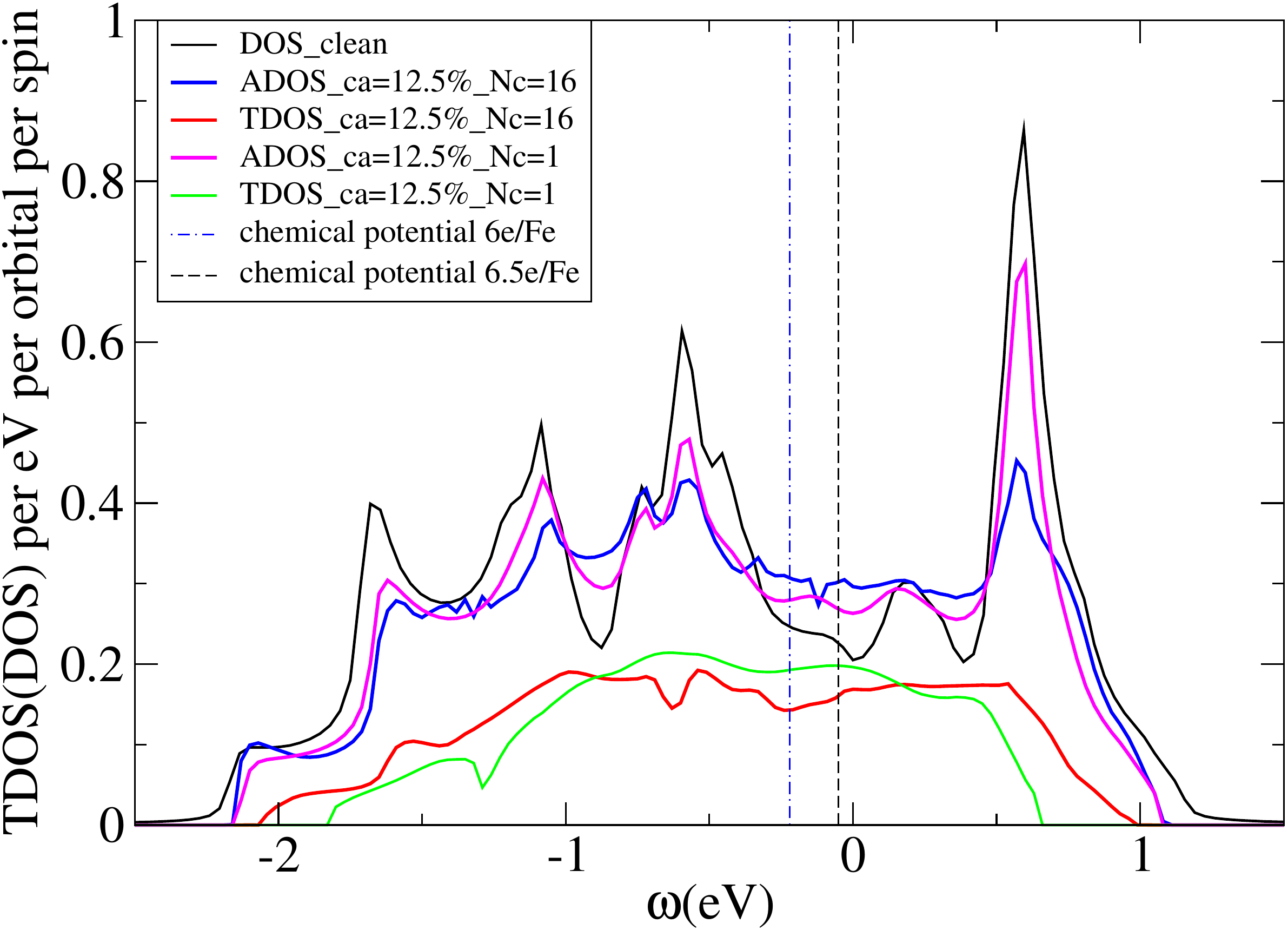}
 \caption{The average and typical density of states of KFe$_2$Se$_2$ with 12.5\% Fe vacancy concentration calculated by 
 multiband DCA and TMDCA with cluster size $N_c=1$ and $N_c=16$, compared with the average density of states of the clean (no vacancy) KFe$_2$Se$_2$. Reprinted from ~\cite{y_zhang_15a}. 
 }
 \label{fig:KFeSe_tdos}
\end{figure}

For example, among the iron based superconductors, K$_x$Fe$_{2-y}$Se$_2$ has been studied intensely because of its unique properties. It has a relatively high $T_c$ of 31 Kelvin~\cite{j_guo_10} and an exotic type of antiferromagnetic order. It was the first iron based superconductor that only has electron pockets and no hole pockets. Moreover, K$_x$Fe$_{2-y}$Se$_2$ is strongly disordered due to a significant amount of Fe vacancies and it is the only iron based superconductor whose parent compound is an anti-ferromagnetic insulator instead of a anti-ferromagnetic metal~\cite{Wei_2011}. Like other iron based superconductors,  it is quasi two dimensional which makes it more sensitive to the disorder. This leads to the question whether it can be an Anderson insulator. Due to the presence of the strong disorder, the precise number of electrons in K$_x$Fe$_{2-y}$Se$_2$ is difficult to quantify, we consider two extreme cases with fillings of 6.0 and 6.5 electrons per Fe. The true electron concentration should fall in between these cases. As shown in Figure~\ref{fig:KFeSe_tdos}, the calculated DCA and TDOS indicate that despite the strong Fe vacancy disorder and the low dimensionality, for both fillings, there are very few states that are Anderson localized in the Fe bands. Since those states reside far away from the Fermi level it can be concluded that K$_x$Fe$_{2-y}$Se$_2$ is not an Anderson insulator.

\subsubsection{Application to (Ga,Mn)N }
\label{sec:GaMnN}
Another class of functional materials in which disorder plays an important role are diluted magnetic semiconductors (DMS). Magnetic impurities give rise to magnetic order in these systems via the creation of a magnetic impurity band. To study localization of the impurity band is not only important for the transport properties, but is also essential to understand the magnetic exchange mechanism these materials. When the carriers in the impurity band are localized, itinerant mechanisms of magnetism, such as double exchange, are ruled out, in favor of other mechanisms such as superexchange.~\cite{jungwirth_rmp_2006} 

Among the DMS materials, (Ga,Mn)N is of particular interest since Dietl ~\cite{t_dietl_01a} predicted its Curie temperature to be above room temperature. However, until now, this prediction remains far from being fulfilled as various experiments lead to controversial conclusions concerning the ferromagnetism. ~\cite{m_zajac_01,s_dhar_03,m_overberg_01,s_stefanowicz_13,t_sasaki_02} 

\begin{figure}[h!]
 \includegraphics[trim = 0mm 0mm 0mm 0mm,width=1\columnwidth,clip=true]{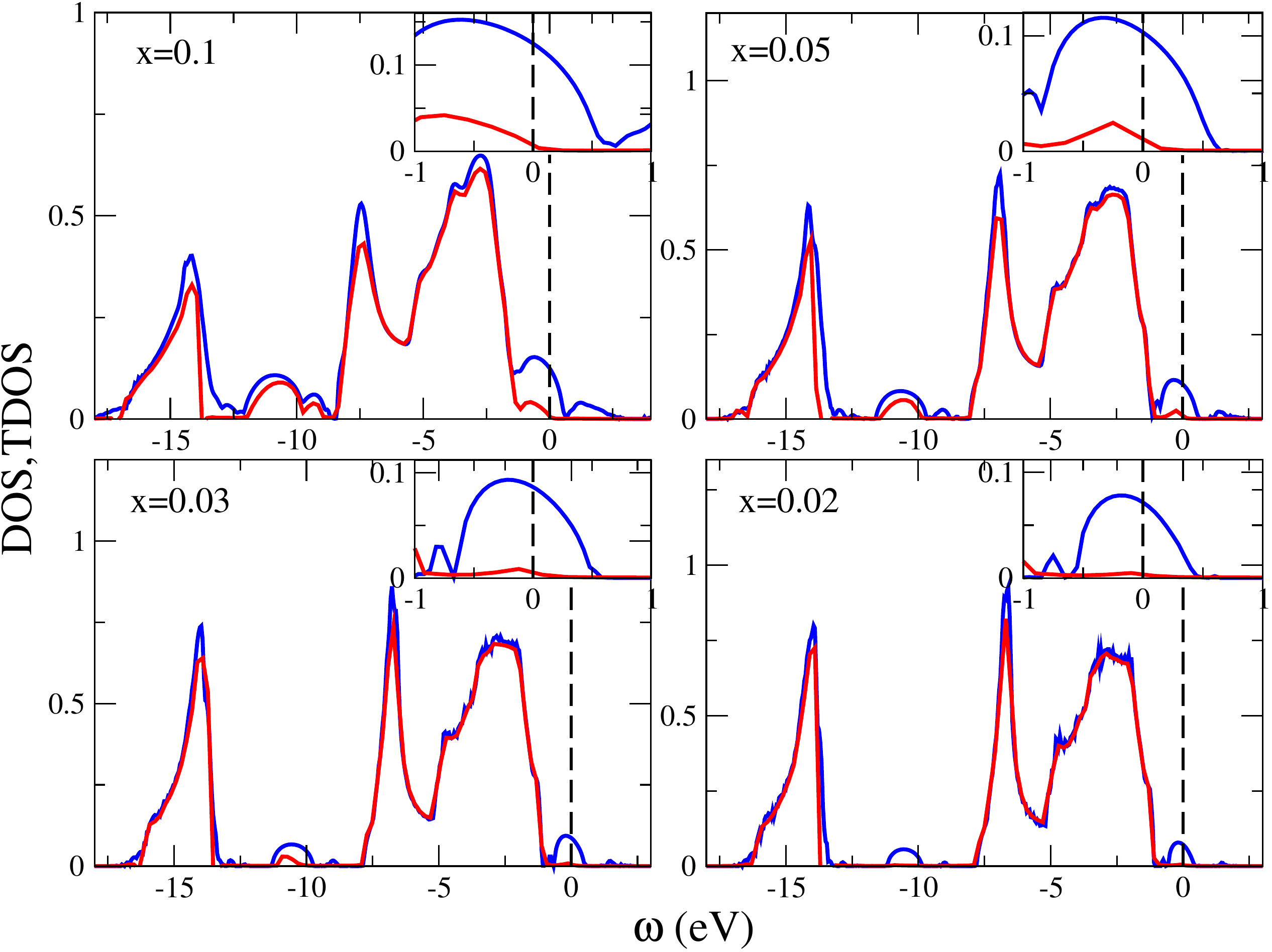}
 \caption{DOS (blue) and typical DOS (red) of Ga$_{1-x}$Mn$_x$N for various Mn concentrations: x=0.02, 0.03, 0.05, 0.1, with N$_c$=32, showing that the impurity band is completely localized for $x\le 0.03$.
 The chemical potential is set to be zero and denoted as the dash line. Inset: Zoom in of the DOS and TDOS
 around the chemical potential. Reprinted from ~\cite{y_zhang_16}. }
 \label{fig:dos_tdos_mu_x}
\end{figure}

To enhance the understanding of magnetism in (Ga,Mn)N we have studied localization in this material from first principles. Figure~\ref{fig:dos_tdos_mu_x} shows the calculated ADOS and TDOS of the minority band for various Mn concentrations. We can see that for Mn impurity concentrations less than 10\% (the compositional limit of (Ga,Mn)N), the chemical potential always sits above the mobility edge, indicating that it is insulating due to localization. Moreover, when the Mn concentration is below 3\%, the TDOS of the impurity band vanishes completely, leading to the complete localization of the impurity band supporting the dominance of the ferromagnetic superexchange mechanism over the double exchange mechanism for the low concentration. 

\section{Conclusions }
\label{sec:Conclusion}

Over the past couple of decades, dynamical mean field theory and its generalization, the DCA have become a major paradigm in the field of computational strongly correlated systems. They provide a new framework for the study of strong interaction. Interesting phenomena such as the metal-Mott insulator transition can be studied in a controllable fashion.

A glaring shortcoming of the CPA (a DMFT analog for disordered system) is its limitation for treating strong disorder. The Anderson insulator due to disorder is completely absent not only due to the local nature of the method but also because the average DOS used in the CPA does not serve as an order parameter for Anderson localized states. There have been cluster extensions of the CPA, including the DCA and MCPA. The DCA is the momentum-space quantum cluster theory, which is based on a mapping from the lattice models onto the quantum cluster embedded in self-consistently determined effective medium. Such mapping involves the concept of coarse-graining, and has been used in the CPA, DMFT and their cluster extensions. A very important feature of the DCA is that it is a controllable approximation with a small parameter of $1/L_c$ ($L_c$ is the linear cluster size), and its ability to provide systematic non-local corrections to the CPA and DMFT.  This is significant, since while the CPA and DMFT are exact in the infinite dimensional $D$ limit, a physically meaningful systematic expansion in $1/D$ has yet to be formulated.  Thus, when viewed as an extension of the DMFT/CPA, the DCA is significant in that it adds a control or small parameter to these quantum cluster approaches.  

When applied to disordered systems, the DCA incorporates the non-local correlations missed in the CPA, and as a result it provides a better qualitative description of the average spectra, it still can not capture the large disorder effects, including Anderson localization. This limitation from the fact that the average DOS used in the DCA is not critical at the transition, and hence can not serve as an order parameter.

The proposal to identify the typical density of states (with the geometrical not algebraic averaging over disorder) as the order parameter of Anderson localization has inspired the development of the TMT which incorporates the typical density of states within the CPA formalism. The TMT is an important development in generalizing the CPA for capturing the Anderson metal to insulator transition. But a single site approximation cannot provide a quantitatively accurate calculation at finite dimensions. And thus, a cluster extension along the lines of the DCA which can handle both strong interactions and disorder is desired.

The TMDCA, which is a main focus of this review, is such cluster extensions, for disordered and interacting systems. Inheriting some properties from the DCA, the TMDCA is a controlled approximation with a small parameter of $1/L_c$, and it systematically includes the non-local corrections to the TMT results. We discuss various benchmarks of the accuracy of the TMDCA against other conventional methods for the Anderson model, including KPM and TMM methods. The versatility of the TMDCA makes it a superior choice when dealing with more complicated models and systems. We survey a series of extensions of the TMDCA to include more chemical details of the model, including off-diagonal disorder, multiple orbitals, long ranged disorder potential and electronic interactions. These extensions make it possible to incorporate the TMDCA with first principles calculations to study the localization in a material-specific way. We also discuss the calculation of two-particle response functions, such as the conductivity, which can be directly measured in experiments.

A prominent advantage of the TMDCA is that it can include electronic interactions and treat the disorder and interaction on equal footing. Since in the TMDCA a geometric average of the local DOS is used for the self consistency, it requires a real-frequency cluster solver to provide reliable spectra for each disorder configuration. A general real frequency cluster solver that can cover the whole range of electronic interaction will greatly improve the TMDCA results to study the interplay between disorder and correlation effect.

We presented two calculations for the Anderson Hubbard model using two perturbation based cluster solvers each of which is suitable for weak or strong interaction respectively.  Most significantly, we show that in the limits of strong disorder and weak interactions treated perturbatively, that the phenomena of 3D localization, including a mobility edge, remains intact.  However, the metal-insulator transition is pushed to larger disorder values by the local interactions.  We also study the limits of strong disorder and strong interactions capable of producing moment formation and screening, with a non-perturbative local approximation.  Here, we find that the Anderson localization quantum phase transition is accompanied by a quantum-critical fan in the energy-disorder phase diagram.

The TMDCA has been successfully combined with the Density Functional Framework to study functional materials including the iron based superconductors and diluted magnetic semiconductors. This opens a broad venue of various applications of the developed method to realistic systems with disorder.  In the future it can be applied to systems where disorder plays an important role, such as intermediate band semiconductors, topological Anderson insulators~\cite{Jian_Li_2009,H_Guo_2010}. Combinations of this method with other first-principle methods, including multiple-scattering theory for disordered systems is underway. 





\section*{Acknowledgements}
We thank Vladimir Dobrosavljevic, Dieter Vollhardt, Juana Moreno, Wei Ku, Vaclav Janis, Richard Scalletar for useful conversations. This work (H.T., Y.Z., and M.J.) was supported by the National Science Foundation under the NSF EPSCoR Cooperative Agreement No.\ EPS-1003897 with additional support from the Louisiana Board of Regents.  M.J.\ gratefully acknowledges support from NSF Materials Theory grant DMR1728457 for the work involving quantum criticality in interacting systems, and the US DOE  DE-SC0017861 for the work involving the first-principles study of strongly disordered energy materials.
A portion of the work (T.B.) was conducted at the Center for Nanophase Materials Sciences, which is a DOE Office of Science User Facility. 
L.C.\ acknowledges the financial support by the Deutsche Forschungsgesellschaft (DFG) through the Transregional Collaborative Research center TRR80/F6.
We thank D.\ Vollhardt and J.\ Jarrell for careful read of the manuscript.

\section{Author Contributions}
The paper was written with input from all authors. 
\section{Conflicts of Interest} The authors declare no conflict of interest.

\bibliographystyle{apsrev4-1}
\bibliography{main}

\pagebreak

\section{Appendix}
This appendix contains the tables with the acronyms and their descriptions used in this manuscript.

\begin{widetext}

\begin{table}[h]
\caption{Table of Acronyms (left), Table of Symbols (right).}
\label{tab:acronyms}
\begin{tabular}{lr}
\begin{minipage}{0.5\textwidth}
\begin{tabular}{|c|c|}
\hline 
Acronym & Description \\
\hline \hline
ADOS & Average Density of States \\
AL & Anderson Localization \\
BEB & Blackman Esterling Berk\\
CPA & Coherent Potential Approximation \\
DCA & Dynamical Cluster Approximation\\
DFT & Density Functional Theory\\
CDMFT & Cluster Dynamical Mean Field Theory\\
EDHM & Effective Disorder Hamiltonian Method\\
JDM & Jacobi-Davidson Method \\
KKR & Korringa-Kohn-Rostoker method\\
KPM & Kernel Polynomial Method\\
LAPW & Linear Augmented Plane Wave\\
LDOS & Local Density of States \\
LMA & Local Moment Approach \\
MCPA & Molecular Coherent Potential Approximation\\
MS & Multiple-Scattering\\
NLCPA & Non-Local Coherent Potential Approximation\\
ODD & Off-Diagonal Disorder\\
QC & Quantum Critical \\
QMC  & Quantum Monte Carlo \\
SOPT & Second Order Perturbation Theory\\
TDOS & Typical Density of States \\
TMDCA & Typical Medium Dynamical Cluster Approximation \\
TMM & Transfer Matrix Method \\
TMT & Typical Medium Theory\\
\hline
\end{tabular}
\end{minipage}  
  & 
\begin{minipage}{0.4\textwidth}
\begin{tabular}{|c|c|}
\hline 
Symbol & Description \\
\hline \hline
$ \k$    & wavenumber \\
$ \K$    & Cluster wavenumber \\
$ \x $   & lattice site coordinate \\
$\X$     & Cluster site coordinate \\
$N$      & Number of lattice sites \\
$N_c$    & Number of cluster sites \\
$\omega, \omega_n,  z$    & Complex and real frequencies \\
$M(\k)$  & DCA coarse-graining many to one map \\
$\rho $  & Density of states \\
$V$      & Electronic potential \\
$\epsilon$      &  Electronic energy \\
$\mu$      &  Electronic chemical potential \\
$\sigma$  & spin index \\
$t$      & Electronic Hopping matrix element (energy)  \\
$m$      & Magnetization \\
$h$      & Magnetic Field \\
$\chi $  & Two-particle Green's function (tensor) \\
$F $  & Full vertex function (tensor) \\ 
$G $       & Single-particle Green's function \\  
$A $       & Single-particle spectral function \\  
$\Delta $  & Mean field hybridization between cluster and host \\
${\cal{G}} $  & Host or cluster excluded Green's function \\
$\Sigma $  & Single-particle self energy \\ 
$\Gamma$   & Irreducible vertex function \\ 
$\Lambda$  & Laue function \\
\hline
\end{tabular}
\end{minipage}  
\end{tabular}
\end{table}

\end{widetext}

\begin{table*}[!ht]
\caption{Table of Usage.  To be consistent with other papers, and not to introduce new notations, we employ subscripts to label type or characteristic, i.e., $\rho_{typ}$ to indicate the result of a geometric average.  So, we will subscripts for indices, i.e., for real space $I,J$ labels }
\label{tab:usage}
\begin{tabular}{|c|c|}
\hline 
Usage & Description \\
\hline \hline
$O^c$          & A superscript ``c'' designates a cluster quantity\\
$O^l$          & A superscript ``l'' designates a lattice quantity\\
$O_{typ}$      & A subscript ``typ'' designates a cluster quantity\\
${\bar{O}}$    & denotes a coarse-grained quantity \\  
$O_{I,J,\cdots} $  & uppercase subscripts indicate indices in cluster space\\
$O_{i,j,\cdots} $  & lowercase subscripts indicate indices in lattice space\\
${\underline{O}}$    & denotes a matrix in the Blackman formalism or in the multi-orbital system \\
\hline
%
\end{tabular}
\end{table*}

 
\end{document}

%% file: defs.tex
\def\AA{\stackrel{\circ}{A}}
\def\barr{\begin{array}}
\def\btab{\begin{tabular}}
\def\etab{\end{tabular}}
\def\bbi{\begin{thebibliography}{99}}
\def\ebi{\end{thebibliography}}
\def\bce{\begin{center}}
\def\ece{\end{center}}
\def\bab{\begin{abstract}}
\def\eab{\end{abstract}}
\def\ear{\end{array}}
\def\earr{\end{array}}
\def\beq{\begin{equation}}
\def\beqa{\begin{eqnarray}}
\def\eeqa{\end{eqnarray}}
\def\eeq{\end{equation}}
\def\ben{\begin{enumerate}}
\def\een{\end{enumerate}}
\def\bdo{\begin{document}}
\def\edo{\end{document}}
\def\bit{\begin{itemize}}
\def\eit{\end{itemize}}
\def\ble{\begin{letter}}
\def\ele{\end{\letter}}
\def\ol#1{\overline{#1}}
\def\fig#1{\newpage\vspace*{7.4in}\bce Figure #1\ece}
\def\rta{\rightarrow}
\def\lra{\Leftrightarrow}
\def\eg{{\it e.g., }}
\def\ap{{\it a priori }}
\def\cf{{\it cf}.~}
\def\ie{{\it i.e., }}
\def\etal{{\it et al. }}
\def\etc{{\it etc}}
\def\vs{{\it vs.}~}
\def\via{{\it via} }
\def\hsph{\hspace*{0.5in}}
\def\hspq{\hspace*{0.25in}}
\def\andh{\hspace*{0.25in}\mbox{and}\hspace*{0.25in}}
\def\orh{\hspace*{0.25in}\mbox{or}\hspace*{0.25in}}
\def\wheh{\hspace*{0.25in}\mbox{where}\hspace*{0.25in}}
\def\vr{\vec{r}}
\def\vR{\vec{R}}
\def\vq{\vec{q}}
\def\vG{\vec{G}}
\def\sd{d\mbox{\put(-3.5,6){\line(1,0){5}}}}
\def\soli{\mbox{(\put(0,3){\line(1,0){49}}\hsp{0.7})}}
\def\dali{\mbox{(\put(0,3){\line(1,0){7}\,\,\line(1,0){7}\,\,\line(1,0){7}
          \,\,\line(1,0){7}\,\,\line(1,0){7}}\hsp{0.7})}}
\def\sdli{\mbox{(\put(0,3){\line(1,0){3}\,\line(1,0){3}\,\line(1,0){3}
          \,\line(1,0){3}\,\line(1,0){3}\,\line(1,0){3}\,\line(1,0){3}
          \,\line(1,0){3}\,\line(1,0){3}\,\line(1,0){3}\,\line(1,0){3}}
          \hsp{0.7})}}
\def\udli{\mbox{(\put(0,3){\line(1,0){5}\,\line(1,0){2}\,\line(1,0){5}
          \,\line(1,0){2}\,\line(1,0){5}\,\line(1,0){2}\,\line(1,0){5}
          \,\line(1,0){2}\,\line(1,0){5}\,\line(1,0){2}}
          \hsp{0.7})}}
\def\doli{\mbox{(\put(0,2.5){.\,.\,.\,.\,.\,.\,.\,.\,.\,.\,.}\hsp{0.7})}}
\def\ddli{\mbox{(\put(0,2.5){\_\,.\,\_\,.\,\_\,.\,\_\,.\,\_\,.}\hsp{0.7})}}
\def\pll{_\parallel}
\def\per{_\perp}
\def\al{\alpha}
\def\all{\;\;\al}
\def\bel{\;\;\be}
\def\gal{\;\;\ga}
\def\del{\;\;\de}
\def\allu{^{\;\;\al}}
\def\alld{_{\;\;\al}}
\def\belu{^{\;\;\be}}
\def\beld{_{\;\;\be}}
\def\galu{^{\;\;\ga}}
\def\gald{_{\;\;\ga}}
\def\delu{^{\;\;\de}}
\def\deld{_{\;\;\de}}
\def\umi{^{(-)}}
\def\upl{^{(+)}}
\def\upm{^{(\pm)}}
\def\ump{^{(\mp)}}
\def\uze{^{(0)}}
\def\be{{\beta}}
\def\bed{\dot{\beta}}
\def\ga{{\gamma}}
\def\de{{\delta}}
\def\ep{\epsilon}
\def\vep{\varepsilon}
\def\ze{\zeta}
\def\et{\eta}
\def\th{\theta}
\def\io{\iota}
\def\ka{{\kappa}}
\def\la{\lambda}
\def\om{omicron}
\def\rh{\rho}
\def\si{{\sigma}}
\def\sib{\bar{\sigma}}
\def\ta{\tau}
\def\up{\upsilon}
\def\ph{\phi}
\def\ch{\chi}
\def\ps{\psi}
\def\om{{\omega}}
\def\wo{{\omega}_o}
\def\prl#1#2#3{, Phys.\ Rev.\ Lett.\ {\bf #1}, #2 (#3)}
\def\prb#1#2#3{, Phys.\ Rev.\ B {\bf #1}, #2 (#3)}
\def\pra#1#2#3{, Phys.\ Rev.\ A {\bf #1}, #2 (#3)}
\def\jcp#1#2#3{, J.\ Chem.\ Phys.\ {\bf #1}, #2 (#3)}
\def\rmp#1#2#3{, Rev.\ Mod.\ Phys.\ {\bf #1}, #2 (#3)}
\def\ssc#1#2#3{, Solid State Commun.~{\bf #1}, #2 (#3)}
\def\jpc#1#2#3{, J.~Phys.~Chem.~{\bf #1}, #2 (#3)}
\def\cpl#1#2#3{, Chem.~Phys.~Lett.~{\bf #1}, #2 (#3)}
\def\AM#1{Ashcroft and Mermin, p.\ #1}
\def\ab{\bar{a}}
\def\bb{\bar{b}}
\def\Tb{\bar{T}}
\def\Jb{\bar{J}}
\def\Kb{\bar{K}}
\def\Sb{\bar{S}}
\def\Ab{\bar{A}}
\def\At{\tilde{A}}
\def\Et{\tilde{E}}
\def\yt{\tilde{y}}
\def\kt{\bf \tilde{k}}
\def\tk{\tilde{k}}
\def\Fb{\bar{F}}
\def\gb{\bar{g}}
\def\xb{\bar{x}}
\def\pb{\bar{p}}
\def\vSb{{\bf \bar{S}}}
\def\vKb{{\bf \bar{K}}}
\def\f{{\bf f}}
\def\ftp{{\bf f}(\th,\ph)}
\def\aone{{\bf a}_1}
\def\atwo{{\bf a}_2}
\def\athree{{\bf a}_3}
\def\a{{\bf a}}
\def\at{{\bf a}(t)}
\def\ao{{\bf a}(\om)}
\def\F{{\bf F}}
\def\E{{\bf E}}
\def\B{{\bf B}}
\def\C{{\bf C}}
\def\x{{\bf x}}
\def\z{{\bf z}}
\def\xd{{\dot{\bf x}}}
\def\vd{{\dot{\bf v}}}
\def\xt{{\bf x}(t)}
\def\tx{(t,\x)}
\def\tpxp{(t',\xp)}
\def\txu{(t_1,\x_1)}
\def\txd{(t_2,\x_2)}
\def\k{{\bf k}}
\def\q{{\bf q}}
\def\kp{{\bf k'}}
\def\xp{{\bf x'}}
\def\A{{\bf A}}
\def\H{{\bf H}}
\def\G{{\bf G}}
\def\C{{\bf C}}
\def\D{{\bf D}}
\def\I{{\bf I}}
\def\K{{\bf K}}
\def\L{{\bf L}}
\def\S{{\bf S}}
\def\J{{\bf J}}
\def\M{{\bf M}}
\def\P{{\bf P}}
\def\V{{\bf V}}
\def\U{{\bf U}}
\def\X{{\bf X}}
\def\R{{\bf R}}
\def\Q{{\bf Q}}
\def\Om{\Omega}
\def\Ph{\Phi}
\def\mn{\mu\nu}
\def\nm{\nu\mu}
\def\CC{{\cal C}}
\def\CE{{\cal E}}
\def\CF{{\cal F}}
\def\CH{{\cal H}}
\def\CO{{\cal O}}
\def\CL{{\cal L}}
\def\CN{{\cal N}}
\def\CP{{\cal P}}
\def\CS{{\cal S}}
\def\CW{{\cal W}}
\def\CV{{\cal V}}
\def\CZ{{\cal Z}}
\def\ul{\underline}
\def\aboverel#1\above#2{\mathrel{\mathop{#2}\limits^{#1}}}
\def\beneathrel#1\under#2{\mathrel{\mathop{#2}\limits_{#1}}}
\def\vev#1{{\left\langle{#1}\right\rangle}}
\def\state#1{|#1\rangle}
\def\vac{\state0}
\def\sol{\state s}
\def\div{\nabla\cdot}
\def\divp{\nabla'\cdot}
\def\grad{\nabla}
\def\curl{\nabla\times}
\def\curlp{\nabla'\times}
\def\lap{\nabla^2}
\def\dal{\Box^2}
\def\pde#1#2{\frac{\partial #1}{\partial #2}}
\def\pdes#1#2#3{\lep\frac{\partial #1}{\partial #2}\rip_{#3}}
\def\pop{\partial}
\def\der#1#2{\frac{d #1}{d #2}}
\def\dpdo{\der{{\cal P}}\Om}
\def\spd#1#2{\frac{\partial^2 #1}{\partial #2^2}}
\def\spdm#1#2#3{\frac{\partial^2 #1}{\partial #2\partial #3}}
\def\spds#1#2#3{\lep\frac{\partial^2 #1}{\partial #2^2}\rip_{#3}}
\def\sq{{\vbox {\hrule height 0.6pt\hbox{\vrule width 0.6pt\hskip 3pt
        \vbox{\vskip 6pt}\hskip 3pt \vrule width 0.6pt}\hrule height 0.6pt}}}
\def\square{\kern1pt\vbox{\hrule height 1.2pt\hbox{\vrule width 1.2pt\hskip 3pt
   \vbox{\vskip 6pt}\hskip 3pt\vrule width 0.6pt}\hrule height 0.6pt}\kern1pt}
\def\nn{{\bf n}}
\def\nnz{{\bf n}_0}
\def\np{{\bf n'}}
\def\epb{\bar{\ep}}
\def\thb{\bar{\th}}
\def\scl{{\cal L}}
\def\thu{\mbox{{\boldmath $\th$}}}
\def\phu{\mbox{{\boldmath $\ph$}}}
\def\epi{\mbox{{\boldmath $\ep_i$}}}
\def\epv{\mbox{{\boldmath $\ep$}}}
\def\rhv{\mbox{{\boldmath $\rh$}}}
\def\epz{\mbox{{\boldmath $\ep$}}_0}
\def\epzs{\mbox{{\boldmath $\ep$}}^*_0}
\def\epj{\mbox{{\boldmath $\ep_j$}}}
\def\epk{\mbox{{\boldmath $\ep_k$}}}
\def\epu{\mbox{{\boldmath $\ep_1$}}}
\def\epd{\mbox{{\boldmath $\ep_2$}}}
\def\ept{\mbox{{\boldmath $\ep_3$}}}
\def\epr{\mbox{{\boldmath $\ep_r$}}}
\def\epph{\mbox{{\boldmath $\ep_{\phi}$}}}
\def\epth{\mbox{{\boldmath $\ep_{\theta}$}}}
\def\eppm{\mbox{{\boldmath $\ep_\pm$}}}
\def\epmp{\mbox{{\boldmath $\ep_\mp$}}}
\def\epp{\mbox{{\boldmath $\ep_+$}}}
\def\epm{\mbox{{\boldmath $\ep_-$}}}
\def\vbed{\dot{\mbox{{\boldmath $\be$}}}}
\def\vbe{\mbox{{\boldmath $\be$}}}
\def\vom{\mbox{{\boldmath $\om$}}}
\def\vze{\mbox{{\boldmath $\ze$}}}
\def\vpi{\mbox{{\boldmath $\pi$}}}
\def\bsi{\mbox{{\boldmath $\sigma$}}}
\def\epti{\tilde{\ep}}
\def\epo{\ep(\om)}
\def\epop{\ep(\om')}
\def\gone{{\bf {g}}_1}
\def\gtwo{{\bf g}_2}
\def\gthree{{\bf g}_3}
\def\g{{\bf g}}
\def\dV{{\bf\bar{V}}}
\def\dT{{\bf\bar{T}}}
\def\iniv{\int d^3x\,}
\def\inia{\int d^2x\,}
\def\inivp{\int d^3x'\,}
\def\iniap{\int d^2x'\,}
\def\inik{\int d^3k\,}
\def\inxt{\int d^3x\,dt\,}
\def\inxtp{\int d^3x'dt'}
\def\dvp{d^3x'\,}
\def\dap{d^2x'\,}
\def\inv{\int_V d^3x\,}
\def\invp{\int_V d^3x'\,}
\def\invy{\int_V d^3y\,}
\def\ina{\int_S d^2x\,}
\def\inac{\oint_S d^2x\,}
\def\inacp{\oint_Sd^2x'\,}
\def\inap{\int_S d^2x'\,}
\def\inay{\int_S d^2y\,}
\def\invl{\int_C d{\bf l}}
\def\invlc{\oint_C d{\bf l}}
\def\inlc{\oint_C dl}
\def\inopc{\oint_C d\om'\,}
\def\invv{\int d^3x\,d^3x'\,}
\def\inko{\int d^3k\,d\om}
\def\intmp{\int_{-\infty}^\infty}
\def\intzp{\int_0^\infty}
\def\ftk{\frac1{\sqrt{2\pi}}\int_{-\infty}^\infty dk\,}
\def\ftz{\frac1{\sqrt{2\pi}}\int_{-\infty}^\infty dz\,}
\def\fto{\frac1{\sqrt{2\pi}}\int_{-\infty}^\infty d\om\,}
\def\ftt{\frac1{\sqrt{2\pi}}\int_{-\infty}^\infty dt\,}
\def\stp{\sqrt{2\pi}}
\def\fpi{{4\pi}}
\def\tpi{{2\pi}}
\def\embe{e^{-\be E}}
\def\emot{e^{-i\om t}}
\def\eot{e^{i\om t}}
\def\eotp{e^{i\om t'}}
\def\etot{e^{-2i\om t}}
\def\ekxot{e^{i(\k\cdot\x-\om t)}}
\def\emkxot{e^{-i(\k\cdot\x-\om t)}}
\def\ekzot{e^{i(kz-\om t)}}
\def\ekzokt{e^{i(kz-\om(k)t)}}
\def\ekdod{e^{i[\k\cdot(\x-\xp)-\om(t-t')]}}
\def\ekxxpa{e^{ik\xxpa}}
\def\qko{q(\k,\om)_}
\def\ekx{e^{i\k\cdot\x}}
\def\emkx{e^{-i\k\cdot\x}}
\def\ekz{e^{ikz}}
\def\emkz{e^{-ikz}}
\def\ekr{e^{ikr}}
\def\ekrp{e^{ikr'}}
\def\ekR{e^{ikR}}
\def\ekRp{e^{ikR'}}
\def\emknxp{e^{-ik(\nn\cdot\xp)}}
\def\emkxp{e^{-i\k\cdot\xp}}
\def\emkzot{e^{-i(kz+\om t)}}
\def\ekrot{e^{i(kr-\om t)}}
\def\uvk{u_\k}
\def\uk{u_k}
\def\lep{\left(}
\def\rip{\right)}
\def\Phx{\Ph(\x)}
\def\Phxt{\Ph(\x,t)}
\def\Phko{\Ph(\k,\om)}
\def\Phmx{\Ph_M(\x)}
\def\Phxp{\Ph(\xp)}
\def\phx{\ph(\x)}
\def\psx{\ps(\x)}
\def\fxt{f(\x,t)}
\def\Fxt{F(\x,t)}
\def\Fx{\F(\x)}
\def\Fxo{F(\x,\om)}
\def\Fko{F(\k,\om)}
\def\fxtp{f(\xp,t')}
\def\psxt{\ps(\x,t)}
\def\phxp{\ph(\xp)}
\def\psxp{\ps(\xp)}
\def\Ak{A(k)}
\def\uzt{U(z,t)}
\def\Ex{\E(\x)}
\def\Ext{\E(\x,t)}
\def\Eo{\E(\om)}
\def\Exo{\E(\x,\om)}
\def\Eko{\E(\k,\om)}
\def\Dx{\D(\x)}
\def\Cx{\C(\x)}
\def\Dxt{\D(\x,t)}
\def\Dxo{\D(\x,\om)}
\def\Px{\P(\x)}
\def\Pxt{\P(\x,t)}
\def\Pxo{\P(\x,\om)}
\def\Po{\P(\om)}
\def\Ax{\A(\x)}
\def\Axt{\A(\x,t)}
\def\Ako{\A(\k,\om)}
\def\Bx{\B(\x)}
\def\Bxt{\B(\x,t)}
\def\Bxo{\B(\x,\om)}
\def\Bko{\B(\k,\om)}
\def\Hx{\H(\x)}
\def\Hxt{\H(\x,t)}
\def\Mx{\M(\x)}
\def\Mxt{\M(\x,t)}
\def\Mxp{\M(\xp)}
\def\Jx{\J(\x)}
\def\Jxt{\J(\x,t)}
\def\Jko{\J(\k,\om)}
\def\Jxo{\J(\x,\om)}
\def\Jxop{\J(\x,\om')}
\def\Jxp{\J(\xp)}
\def\chx{\ch(\x)}
\def\chxt{\ch(\x,t)}
\def\v{{\bf v}}
\def\l{{\bf l}}
\def\b{{\bf b}}
\def\c{{\bf c}}
\def\u{{\bf u}}
\def\vfxt{{\bf f}(\x,t)}
\def\mm{{\bf \mu}}
\def\N{{\bf N}}
\def\eq#1{Eq.~(#1)}
\def\rhx{\rh(\x)}
\def\rhr{\rh(\vr)}
\def\rhxt{\rh(\x,t)}
\def\rhko{\rh(\k,\om)}
\def\rhxp{\rh(\xp)}
\def\rhxo{\rh(\x,\om)}
\def\rhxop{\rh(\x,\om')}
\def\leb{\left[}
\def\rib{\right]}
\def\six{\si(\x)}
\def\sixp{\si(\xp)}
\def\gxxp{G(\x,\xp)}
\def\gxt{G(\x,t;\xp,t')}
\def\gko{g(\k,\om)}
\def\dexxp{\de(\x-\xp)}
\def\dettp{\de(t-t')}
\def\dekkp{\de(\k-\k')}
\def\deoop{\de(\om-\om')}
\def\fxxp{F(\x,\xp)}
\def\lapp{\nabla'^2}
\def\xxpi{\frac{1}{|\x-\xp|}}
\def\xxpa{|\x-\xp|}
\def\xxp{(\x-\xp)}
\def\dvl{d{\bf l}}
\def\gradp{\nabla'}
\def\nnp{{\bf n}'}
\def\hsp#1{\hspace*{#1 in}}
\def\vsp#1{\vspace*{#1 in}}
\def\lel{\left|}
\def\ril{\right|}
\def\zh{\hat{{\bf z}}}
\def\xh{\hat{{\bf x}}}
\def\yh{\hat{{\bf y}}}
\def\p{{\bf p}}
\def\r{{\bf r}}
\def\s{{\bf s}}
\def\rhat{\hat{{\bf r}}}
\def\eqs#1#2{Eqs.~(#1) and (#2)}
\def\eqss#1#2#3{Eqs.~(#1),~(#2), and~(#3)}
\def\sde#1#2{\frac{d^2#1}{d#2^2}}
\def\csde#1#2#3{\frac{d^2#1}{d#2 d#3}}
\def\y{{\bf y}}
\def\yp{{\bf y'}}
\def\unet{U_n(\et)}
\def\umet{U_m(\et)}
\def\unets{U_n^*(\et)}
\def\umets{U_m^*(\et)}
\def\inabet{\int_a^bd\et}
\def\Ps{\Psi}
\def\sth{\sin\th}
\def\cth{\cos\th}
\def\sph{\sin\ph}
\def\cph{\cos\ph}
\def\inoou{\int_{-1}^1du}
\def\lec{\left\{}
\def\ric{\right\}}
\def\ylm{Y_{l,m}(\th,\ph)}
\def\ylms{Y_{l,m}^*(\th,\ph)}
\def\ylmsp{Y_{l,m}^*(\th',\ph')}
\def\ylmp{Y_{l'm',}(\th,\ph)}
\def\ylmps{Y_{l'm'}^*(\th,\ph)}
\def\inom{\int d\Om}
\def\les{\left/}
\def\led{\left.}
\def\rid{\right.}
\def\ris{\right/}
\def\Ga{\Gamma}
\def\De{\Delta}
\def\Th{\Theta}
\def\jnx{J_\nu(x)}
\def\nnx{N_\nu(x)}
\def\jnxy{J_\nu(x_{\nu n}y)}
\def\jxy{J_\nu(xy)}
\def\jxpy{J_\nu(x'y)}
\def\xnn{x_{\nu n}}
\def\jnxpy{J_\nu(x_{\nu n'}y)}
\def\xnnp{x_{\nu n'}}
\def\nlb{\nolinebreak}
\def\edm{{\bf p}}
\def\mdm{{\bf m}}
\def\Pxp{{\bf P}(\xp)}
\def\rhh{\hat{{\bf r}}}
\def\thh{\hat{{\bf \th}}}
\def\phh{\hat{{\bf \ph}}}
\def\vereq#1#2{\lower3pt\vbox{\baselineskip1.5pt \lineskip1.5pt
\ialign{$#1\hfill##\hfil$\crcr#2\crcr\sim\crcr}}}
\def\alt{\mathrel{\mathpalette\vereq<}}
\def\agt{\mathrel{\mathpalette\vereq>}}
\def\SE{{\Sc E}}
\def\SL{{\Sc L}}
\def\j{{\bf j}}
\def\i{{\bf i}}
\def\jd{{\dot{\bf j}}}
\def\La{\Lambda}
\def\hb{\hbar}
\def\hbo{\hb \omega }
\def\hod{\hb \omega _D}
\def\kpr{k ^{\prime }}
\def\thk{\th _k}
\def\thkp{\th _{k^{\prime }}}

%% file: main.bbl
\begin{thebibliography}{227}%
\makeatletter
\providecommand \@ifxundefined [1]{%
 \@ifx{#1\undefined}
}%
\providecommand \@ifnum [1]{%
 \ifnum #1\expandafter \@firstoftwo
 \else \expandafter \@secondoftwo
 \fi
}%
\providecommand \@ifx [1]{%
 \ifx #1\expandafter \@firstoftwo
 \else \expandafter \@secondoftwo
 \fi
}%
\providecommand \natexlab [1]{#1}%
\providecommand \enquote  [1]{``#1''}%
\providecommand \bibnamefont  [1]{#1}%
\providecommand \bibfnamefont [1]{#1}%
\providecommand \citenamefont [1]{#1}%
\providecommand \href@noop [0]{\@secondoftwo}%
\providecommand \href [0]{\begingroup \@sanitize@url \@href}%
\providecommand \@href[1]{\@@startlink{#1}\@@href}%
\providecommand \@@href[1]{\endgroup#1\@@endlink}%
\providecommand \@sanitize@url [0]{\catcode `\\12\catcode `\$12\catcode
  `\&12\catcode `\#12\catcode `\^12\catcode `\_12\catcode `\%12\relax}%
\providecommand \@@startlink[1]{}%
\providecommand \@@endlink[0]{}%
\providecommand \url  [0]{\begingroup\@sanitize@url \@url }%
\providecommand \@url [1]{\endgroup\@href {#1}{\urlprefix }}%
\providecommand \urlprefix  [0]{URL }%
\providecommand \Eprint [0]{\href }%
\providecommand \doibase [0]{http://dx.doi.org/}%
\providecommand \selectlanguage [0]{\@gobble}%
\providecommand \bibinfo  [0]{\@secondoftwo}%
\providecommand \bibfield  [0]{\@secondoftwo}%
\providecommand \translation [1]{[#1]}%
\providecommand \BibitemOpen [0]{}%
\providecommand \bibitemStop [0]{}%
\providecommand \bibitemNoStop [0]{.\EOS\space}%
\providecommand \EOS [0]{\spacefactor3000\relax}%
\providecommand \BibitemShut  [1]{\csname bibitem#1\endcsname}%
\let\auto@bib@innerbib\@empty
\bibitem [{new(2012)}]{newyorktimes}%
  \BibitemOpen
  \ (\bibinfo  {publisher} {New York Times},\ \bibinfo {year}
  {2012})\BibitemShut {NoStop}%
\bibitem [{gre(2011)}]{greenpeace}%
  \BibitemOpen
  \ (\bibinfo  {publisher} {Greenpeace International},\ \bibinfo {year}
  {2011})\BibitemShut {NoStop}%
\bibitem [{\citenamefont {Imada}\ \emph {et~al.}(1998)\citenamefont {Imada},
  \citenamefont {Fujimori},\ and\ \citenamefont {Tokura}}]{imada_mit}%
  \BibitemOpen
  \bibfield  {author} {\bibinfo {author} {\bibfnamefont {M.}~\bibnamefont
  {Imada}}, \bibinfo {author} {\bibfnamefont {A.}~\bibnamefont {Fujimori}}, \
  and\ \bibinfo {author} {\bibfnamefont {Y.}~\bibnamefont {Tokura}},\ }\href
  {\doibase 10.1103/RevModPhys.70.1039} {\bibfield  {journal} {\bibinfo
  {journal} {Rev. Mod. Phys.}\ }\textbf {\bibinfo {volume} {70}},\ \bibinfo
  {pages} {1039} (\bibinfo {year} {1998})}\BibitemShut {NoStop}%
\bibitem [{\citenamefont {Mott}(1968)}]{n_mott_68}%
  \BibitemOpen
  \bibfield  {author} {\bibinfo {author} {\bibfnamefont {N.~F.}\ \bibnamefont
  {Mott}},\ }\href {\doibase 10.1103/RevModPhys.40.677} {\bibfield  {journal}
  {\bibinfo  {journal} {Rev. Mod. Phys.}\ }\textbf {\bibinfo {volume} {40}},\
  \bibinfo {pages} {677} (\bibinfo {year} {1968})}\BibitemShut {NoStop}%
\bibitem [{\citenamefont {Belitz}\ and\ \citenamefont
  {Kirkpatrick}(1994)}]{d_belitz_94}%
  \BibitemOpen
  \bibfield  {author} {\bibinfo {author} {\bibfnamefont {D.}~\bibnamefont
  {Belitz}}\ and\ \bibinfo {author} {\bibfnamefont {T.~R.}\ \bibnamefont
  {Kirkpatrick}},\ }\href {\doibase 10.1103/RevModPhys.66.261} {\bibfield
  {journal} {\bibinfo  {journal} {Rev. Mod. Phys.}\ }\textbf {\bibinfo {volume}
  {66}},\ \bibinfo {pages} {261} (\bibinfo {year} {1994})}\BibitemShut
  {NoStop}%
\bibitem [{\citenamefont {Evers}\ and\ \citenamefont
  {Mirlin}(2008)}]{f_evers_08}%
  \BibitemOpen
  \bibfield  {author} {\bibinfo {author} {\bibfnamefont {F.}~\bibnamefont
  {Evers}}\ and\ \bibinfo {author} {\bibfnamefont {A.~D.}\ \bibnamefont
  {Mirlin}},\ }\href {\doibase 10.1103/RevModPhys.80.1355} {\bibfield
  {journal} {\bibinfo  {journal} {Rev. Mod. Phys.}\ }\textbf {\bibinfo {volume}
  {80}},\ \bibinfo {pages} {1355} (\bibinfo {year} {2008})}\BibitemShut
  {NoStop}%
\bibitem [{\citenamefont {V.~Dobrosavljević}\ and\ \citenamefont
  {J.~M.~Valles}(2012)}]{dobrosavljevic_book}%
  \BibitemOpen
  \bibfield  {author} {\bibinfo {author} {\bibfnamefont {N.~T.}\ \bibnamefont
  {V.~Dobrosavljević}}\ and\ \bibinfo {author} {\bibfnamefont
  {J.}~\bibnamefont {J.~M.~Valles}},\ }\href@noop {} {\emph {\bibinfo {title}
  {Conductor Insulator Quantum Phase Transitions}}}\ (\bibinfo  {publisher}
  {Oxford University Press},\ \bibinfo {year} {2012})\BibitemShut {NoStop}%
\bibitem [{\citenamefont {Jarrell}\ and\ \citenamefont
  {Krishnamurthy}(2001)}]{m_jarrell_01a}%
  \BibitemOpen
  \bibfield  {author} {\bibinfo {author} {\bibfnamefont {M.}~\bibnamefont
  {Jarrell}}\ and\ \bibinfo {author} {\bibfnamefont {H.~R.}\ \bibnamefont
  {Krishnamurthy}},\ }\href {\doibase 10.1103/PhysRevB.63.125102} {\bibfield
  {journal} {\bibinfo  {journal} {Phys. Rev. B}\ }\textbf {\bibinfo {volume}
  {63}},\ \bibinfo {pages} {125102} (\bibinfo {year} {2001})}\BibitemShut
  {NoStop}%
\bibitem [{\citenamefont {Dobrosavljevi\'{c}}\ \emph
  {et~al.}(2003)\citenamefont {Dobrosavljevi\'{c}}, \citenamefont {Pastor},\
  and\ \citenamefont {Nikoli\'{c}}}]{v_dobrosavljevic_03}%
  \BibitemOpen
  \bibfield  {author} {\bibinfo {author} {\bibfnamefont {V.}~\bibnamefont
  {Dobrosavljevi\'{c}}}, \bibinfo {author} {\bibfnamefont {A.~A.}\ \bibnamefont
  {Pastor}}, \ and\ \bibinfo {author} {\bibfnamefont {B.~K.}\ \bibnamefont
  {Nikoli\'{c}}},\ }\href@noop {} {\bibfield  {journal} {\bibinfo  {journal}
  {EPL}\ }\textbf {\bibinfo {volume} {62}},\ \bibinfo {pages} {76} (\bibinfo
  {year} {2003})}\BibitemShut {NoStop}%
\bibitem [{\citenamefont {Ekuma}\ \emph
  {et~al.}(2015{\natexlab{a}})\citenamefont {Ekuma}, \citenamefont {Moore},
  \citenamefont {Terletska}, \citenamefont {Tam}, \citenamefont {Moreno},
  \citenamefont {Jarrell},\ and\ \citenamefont {Vidhyadhiraja}}]{c_ekuma_15b}%
  \BibitemOpen
  \bibfield  {author} {\bibinfo {author} {\bibfnamefont {C.~E.}\ \bibnamefont
  {Ekuma}}, \bibinfo {author} {\bibfnamefont {C.}~\bibnamefont {Moore}},
  \bibinfo {author} {\bibfnamefont {H.}~\bibnamefont {Terletska}}, \bibinfo
  {author} {\bibfnamefont {K.-M.}\ \bibnamefont {Tam}}, \bibinfo {author}
  {\bibfnamefont {J.}~\bibnamefont {Moreno}}, \bibinfo {author} {\bibfnamefont
  {M.}~\bibnamefont {Jarrell}}, \ and\ \bibinfo {author} {\bibfnamefont
  {N.~S.}\ \bibnamefont {Vidhyadhiraja}},\ }\href {\doibase
  10.1103/PhysRevB.92.014209} {\bibfield  {journal} {\bibinfo  {journal} {Phys.
  Rev. B}\ }\textbf {\bibinfo {volume} {92}},\ \bibinfo {pages} {014209}
  (\bibinfo {year} {2015}{\natexlab{a}})}\BibitemShut {NoStop}%
\bibitem [{\citenamefont {Ekuma}\ \emph
  {et~al.}(2015{\natexlab{b}})\citenamefont {Ekuma}, \citenamefont {Yang},
  \citenamefont {Terletska}, \citenamefont {Tam}, \citenamefont
  {Vidhyadhiraja}, \citenamefont {Moreno},\ and\ \citenamefont
  {Jarrell}}]{c_ekuma_15c}%
  \BibitemOpen
  \bibfield  {author} {\bibinfo {author} {\bibfnamefont {C.~E.}\ \bibnamefont
  {Ekuma}}, \bibinfo {author} {\bibfnamefont {S.-X.}\ \bibnamefont {Yang}},
  \bibinfo {author} {\bibfnamefont {H.}~\bibnamefont {Terletska}}, \bibinfo
  {author} {\bibfnamefont {K.-M.}\ \bibnamefont {Tam}}, \bibinfo {author}
  {\bibfnamefont {N.~S.}\ \bibnamefont {Vidhyadhiraja}}, \bibinfo {author}
  {\bibfnamefont {J.}~\bibnamefont {Moreno}}, \ and\ \bibinfo {author}
  {\bibfnamefont {M.}~\bibnamefont {Jarrell}},\ }\href {\doibase
  10.1103/PhysRevB.92.201114} {\bibfield  {journal} {\bibinfo  {journal} {Phys.
  Rev. B}\ }\textbf {\bibinfo {volume} {92}},\ \bibinfo {pages} {201114}
  (\bibinfo {year} {2015}{\natexlab{b}})}\BibitemShut {NoStop}%
\bibitem [{\citenamefont {Zhang}\ \emph
  {et~al.}(2015{\natexlab{a}})\citenamefont {Zhang}, \citenamefont {Terletska},
  \citenamefont {Moore}, \citenamefont {Ekuma}, \citenamefont {Tam},
  \citenamefont {Berlijn}, \citenamefont {Ku}, \citenamefont {Moreno},\ and\
  \citenamefont {Jarrell}}]{y_zhang_15a}%
  \BibitemOpen
  \bibfield  {author} {\bibinfo {author} {\bibfnamefont {Y.}~\bibnamefont
  {Zhang}}, \bibinfo {author} {\bibfnamefont {H.}~\bibnamefont {Terletska}},
  \bibinfo {author} {\bibfnamefont {C.}~\bibnamefont {Moore}}, \bibinfo
  {author} {\bibfnamefont {C.}~\bibnamefont {Ekuma}}, \bibinfo {author}
  {\bibfnamefont {K.-M.}\ \bibnamefont {Tam}}, \bibinfo {author} {\bibfnamefont
  {T.}~\bibnamefont {Berlijn}}, \bibinfo {author} {\bibfnamefont
  {W.}~\bibnamefont {Ku}}, \bibinfo {author} {\bibfnamefont {J.}~\bibnamefont
  {Moreno}}, \ and\ \bibinfo {author} {\bibfnamefont {M.}~\bibnamefont
  {Jarrell}},\ }\href {\doibase 10.1103/PhysRevB.92.205111} {\bibfield
  {journal} {\bibinfo  {journal} {Phys. Rev. B}\ }\textbf {\bibinfo {volume}
  {92}},\ \bibinfo {pages} {205111} (\bibinfo {year}
  {2015}{\natexlab{a}})}\BibitemShut {NoStop}%
\bibitem [{\citenamefont {Metzner}\ and\ \citenamefont
  {Vollhardt}(1989{\natexlab{a}})}]{w_metzner_89a}%
  \BibitemOpen
  \bibfield  {author} {\bibinfo {author} {\bibfnamefont {W.}~\bibnamefont
  {Metzner}}\ and\ \bibinfo {author} {\bibfnamefont {D.}~\bibnamefont
  {Vollhardt}},\ }\href@noop {} {\bibfield  {journal} {\bibinfo  {journal}
  {Phys.~Rev.~Lett.}\ }\textbf {\bibinfo {volume} {62}},\ \bibinfo {pages}
  {324} (\bibinfo {year} {1989}{\natexlab{a}})}\BibitemShut {NoStop}%
\bibitem [{\citenamefont
  {M\"{u}ller-Hartmann}(1989{\natexlab{a}})}]{e_mullerhartmann_89a}%
  \BibitemOpen
  \bibfield  {author} {\bibinfo {author} {\bibfnamefont {E.}~\bibnamefont
  {M\"{u}ller-Hartmann}},\ }\href@noop {} {\bibfield  {journal} {\bibinfo
  {journal} {Z. Phys. B (Condensed Matter)}\ }\textbf {\bibinfo {volume}
  {74}},\ \bibinfo {pages} {507} (\bibinfo {year}
  {1989}{\natexlab{a}})}\BibitemShut {NoStop}%
\bibitem [{\citenamefont
  {M\"{u}ller-Hartmann}(1989{\natexlab{b}})}]{e_mullerhartmann_89b}%
  \BibitemOpen
  \bibfield  {author} {\bibinfo {author} {\bibfnamefont {E.}~\bibnamefont
  {M\"{u}ller-Hartmann}},\ }\href@noop {} {\bibfield  {journal} {\bibinfo
  {journal} {Z. Phys. B (Condensed Matter)}\ }\textbf {\bibinfo {volume}
  {76}},\ \bibinfo {pages} {211} (\bibinfo {year}
  {1989}{\natexlab{b}})}\BibitemShut {NoStop}%
\bibitem [{\citenamefont {Georges}\ and\ \citenamefont
  {Kotliar}(1992)}]{a_georges_92a}%
  \BibitemOpen
  \bibfield  {author} {\bibinfo {author} {\bibfnamefont {A.}~\bibnamefont
  {Georges}}\ and\ \bibinfo {author} {\bibfnamefont {G.}~\bibnamefont
  {Kotliar}},\ }\href@noop {} {\bibfield  {journal} {\bibinfo  {journal}
  {Phys.~Rev.~B}\ }\textbf {\bibinfo {volume} {45}},\ \bibinfo {pages} {6479}
  (\bibinfo {year} {1992})}\BibitemShut {NoStop}%
\bibitem [{\citenamefont {Jarrell}(1992)}]{m_jarrell_92a}%
  \BibitemOpen
  \bibfield  {author} {\bibinfo {author} {\bibfnamefont {M.}~\bibnamefont
  {Jarrell}},\ }\href@noop {} {\bibfield  {journal} {\bibinfo  {journal} {Phys.
  Rev. Lett.}\ }\textbf {\bibinfo {volume} {69}},\ \bibinfo {pages} {168}
  (\bibinfo {year} {1992})}\BibitemShut {NoStop}%
\bibitem [{\citenamefont {Pruschke}\ \emph {et~al.}(1995)\citenamefont
  {Pruschke}, \citenamefont {Jarrell},\ and\ \citenamefont
  {Freericks}}]{t_pruschke_95}%
  \BibitemOpen
  \bibfield  {author} {\bibinfo {author} {\bibfnamefont {T.}~\bibnamefont
  {Pruschke}}, \bibinfo {author} {\bibfnamefont {M.}~\bibnamefont {Jarrell}}, \
  and\ \bibinfo {author} {\bibfnamefont {J.}~\bibnamefont {Freericks}},\
  }\href@noop {} {\bibfield  {journal} {\bibinfo  {journal} {Adv. Phys.}\
  }\textbf {\bibinfo {volume} {44}},\ \bibinfo {pages} {187} (\bibinfo {year}
  {1995})}\BibitemShut {NoStop}%
\bibitem [{\citenamefont {Georges}\ \emph {et~al.}(1996)\citenamefont
  {Georges}, \citenamefont {Kotliar}, \citenamefont {Krauth},\ and\
  \citenamefont {Rozenberg}}]{a_georges_96a}%
  \BibitemOpen
  \bibfield  {author} {\bibinfo {author} {\bibfnamefont {A.}~\bibnamefont
  {Georges}}, \bibinfo {author} {\bibfnamefont {G.}~\bibnamefont {Kotliar}},
  \bibinfo {author} {\bibfnamefont {W.}~\bibnamefont {Krauth}}, \ and\ \bibinfo
  {author} {\bibfnamefont {M.}~\bibnamefont {Rozenberg}},\ }\href@noop {}
  {\bibfield  {journal} {\bibinfo  {journal} {Rev. Mod. Phys.}\ }\textbf
  {\bibinfo {volume} {68}},\ \bibinfo {pages} {13} (\bibinfo {year}
  {1996})}\BibitemShut {NoStop}%
\bibitem [{\citenamefont {Soven}(1967)}]{soven_cpa}%
  \BibitemOpen
  \bibfield  {author} {\bibinfo {author} {\bibfnamefont {P.}~\bibnamefont
  {Soven}},\ }\href {\doibase 10.1103/PhysRev.156.809} {\bibfield  {journal}
  {\bibinfo  {journal} {Phys. Rev.}\ }\textbf {\bibinfo {volume} {156}},\
  \bibinfo {pages} {809} (\bibinfo {year} {1967})}\BibitemShut {NoStop}%
\bibitem [{\citenamefont {Velick\'y}\ \emph {et~al.}(1968)\citenamefont
  {Velick\'y}, \citenamefont {Kirkpatrick},\ and\ \citenamefont
  {Ehrenreich}}]{velicky_cpa}%
  \BibitemOpen
  \bibfield  {author} {\bibinfo {author} {\bibfnamefont {B.}~\bibnamefont
  {Velick\'y}}, \bibinfo {author} {\bibfnamefont {S.}~\bibnamefont
  {Kirkpatrick}}, \ and\ \bibinfo {author} {\bibfnamefont {H.}~\bibnamefont
  {Ehrenreich}},\ }\href {\doibase 10.1103/PhysRev.175.747} {\bibfield
  {journal} {\bibinfo  {journal} {Phys. Rev.}\ }\textbf {\bibinfo {volume}
  {175}},\ \bibinfo {pages} {747} (\bibinfo {year} {1968})}\BibitemShut
  {NoStop}%
\bibitem [{\citenamefont {Elliott}\ \emph {et~al.}(1974)\citenamefont
  {Elliott}, \citenamefont {Krumhansl},\ and\ \citenamefont
  {Leath}}]{r_elliott_74}%
  \BibitemOpen
  \bibfield  {author} {\bibinfo {author} {\bibfnamefont {R.~J.}\ \bibnamefont
  {Elliott}}, \bibinfo {author} {\bibfnamefont {J.~A.}\ \bibnamefont
  {Krumhansl}}, \ and\ \bibinfo {author} {\bibfnamefont {P.~L.}\ \bibnamefont
  {Leath}},\ }\href {\doibase 10.1103/RevModPhys.46.465} {\bibfield  {journal}
  {\bibinfo  {journal} {Rev. Mod. Phys.}\ }\textbf {\bibinfo {volume} {46}},\
  \bibinfo {pages} {465} (\bibinfo {year} {1974})}\BibitemShut {NoStop}%
\bibitem [{\citenamefont {Hettler}\ \emph {et~al.}(1998)\citenamefont
  {Hettler}, \citenamefont {Tahvildar-Zadeh}, \citenamefont {Jarrell},
  \citenamefont {Pruschke},\ and\ \citenamefont
  {Krishnamurthy}}]{m_hettler_98a}%
  \BibitemOpen
  \bibfield  {author} {\bibinfo {author} {\bibfnamefont {M.}~\bibnamefont
  {Hettler}}, \bibinfo {author} {\bibfnamefont {A.}~\bibnamefont
  {Tahvildar-Zadeh}}, \bibinfo {author} {\bibfnamefont {M.}~\bibnamefont
  {Jarrell}}, \bibinfo {author} {\bibfnamefont {T.}~\bibnamefont {Pruschke}}, \
  and\ \bibinfo {author} {\bibfnamefont {H.}~\bibnamefont {Krishnamurthy}},\
  }\href@noop {} {\bibfield  {journal} {\bibinfo  {journal} {Phys.~Rev.~B}\
  }\textbf {\bibinfo {volume} {58}},\ \bibinfo {pages} {7475} (\bibinfo {year}
  {1998})}\BibitemShut {NoStop}%
\bibitem [{\citenamefont {Hettler}\ \emph {et~al.}(2000)\citenamefont
  {Hettler}, \citenamefont {Mukherjee}, \citenamefont {Jarrell},\ and\
  \citenamefont {Krishnamurthy}}]{m_hettler_00a}%
  \BibitemOpen
  \bibfield  {author} {\bibinfo {author} {\bibfnamefont {M.}~\bibnamefont
  {Hettler}}, \bibinfo {author} {\bibfnamefont {M.}~\bibnamefont {Mukherjee}},
  \bibinfo {author} {\bibfnamefont {M.}~\bibnamefont {Jarrell}}, \ and\
  \bibinfo {author} {\bibfnamefont {H.}~\bibnamefont {Krishnamurthy}},\ }\href
  {\doibase 10.1103/PhysRevB.61.12739} {\bibfield  {journal} {\bibinfo
  {journal} {Phys. Rev. B}\ }\textbf {\bibinfo {volume} {61}},\ \bibinfo
  {pages} {12739} (\bibinfo {year} {2000})}\BibitemShut {NoStop}%
\bibitem [{\citenamefont {{Anderson}}(1958)}]{p_anderson_58}%
  \BibitemOpen
  \bibfield  {author} {\bibinfo {author} {\bibfnamefont {P.~W.}\ \bibnamefont
  {{Anderson}}},\ }\href {\doibase 10.1103/PhysRev.109.1492} {\bibfield
  {journal} {\bibinfo  {journal} {Phys. Rev.}\ }\textbf {\bibinfo {volume}
  {109}},\ \bibinfo {pages} {1492} (\bibinfo {year} {1958})}\BibitemShut
  {NoStop}%
\bibitem [{\citenamefont {{Dagotto}}(2005)}]{e_dagotto_05}%
  \BibitemOpen
  \bibfield  {author} {\bibinfo {author} {\bibfnamefont {E.}~\bibnamefont
  {{Dagotto}}},\ }\href@noop {} {\bibfield  {journal} {\bibinfo  {journal}
  {Science}\ }\textbf {\bibinfo {volume} {309}},\ \bibinfo {pages} {257}
  (\bibinfo {year} {2005})}\BibitemShut {NoStop}%
\bibitem [{\citenamefont {Rokhinson}\ \emph {et~al.}(2007)\citenamefont
  {Rokhinson}, \citenamefont {Lyanda-Geller}, \citenamefont {Ge}, \citenamefont
  {Shen}, \citenamefont {Liu}, \citenamefont {Dobrowolska},\ and\ \citenamefont
  {Furdyna}}]{l_rokhinson_07}%
  \BibitemOpen
  \bibfield  {author} {\bibinfo {author} {\bibfnamefont {L.~P.}\ \bibnamefont
  {Rokhinson}}, \bibinfo {author} {\bibfnamefont {Y.}~\bibnamefont
  {Lyanda-Geller}}, \bibinfo {author} {\bibfnamefont {Z.}~\bibnamefont {Ge}},
  \bibinfo {author} {\bibfnamefont {S.}~\bibnamefont {Shen}}, \bibinfo {author}
  {\bibfnamefont {X.}~\bibnamefont {Liu}}, \bibinfo {author} {\bibfnamefont
  {M.}~\bibnamefont {Dobrowolska}}, \ and\ \bibinfo {author} {\bibfnamefont
  {J.~K.}\ \bibnamefont {Furdyna}},\ }\href {\doibase
  10.1103/PhysRevB.76.161201} {\bibfield  {journal} {\bibinfo  {journal} {Phys.
  Rev. B}\ }\textbf {\bibinfo {volume} {76}},\ \bibinfo {pages} {161201}
  (\bibinfo {year} {2007})}\BibitemShut {NoStop}%
\bibitem [{\citenamefont {Dobrowolska}\ \emph {et~al.}(2012)\citenamefont
  {Dobrowolska}, \citenamefont {Tivakornsasithorn}, \citenamefont {Liu},
  \citenamefont {Furdyna}, \citenamefont {Berciu}, \citenamefont {Yu},\ and\
  \citenamefont {Walukiewicz}}]{Dobrowolska12}%
  \BibitemOpen
  \bibfield  {author} {\bibinfo {author} {\bibfnamefont {M.}~\bibnamefont
  {Dobrowolska}}, \bibinfo {author} {\bibfnamefont {K.}~\bibnamefont
  {Tivakornsasithorn}}, \bibinfo {author} {\bibfnamefont {X.}~\bibnamefont
  {Liu}}, \bibinfo {author} {\bibfnamefont {J.~K.}\ \bibnamefont {Furdyna}},
  \bibinfo {author} {\bibfnamefont {M.}~\bibnamefont {Berciu}}, \bibinfo
  {author} {\bibfnamefont {K.~M.}\ \bibnamefont {Yu}}, \ and\ \bibinfo {author}
  {\bibfnamefont {W.}~\bibnamefont {Walukiewicz}},\ }\href@noop {} {\bibfield
  {journal} {\bibinfo  {journal} {Nat. Mater.}\ }\textbf {\bibinfo {volume}
  {11}},\ \bibinfo {pages} {444} (\bibinfo {year} {2012})}\BibitemShut
  {NoStop}%
\bibitem [{\citenamefont {Sawicki}\ \emph {et~al.}(2010)\citenamefont
  {Sawicki}, \citenamefont {Chiba}, \citenamefont {Korbecka}, \citenamefont
  {Nishitani}, \citenamefont {Majewski}, \citenamefont {Matsukura},
  \citenamefont {Dielt},\ and\ \citenamefont {Ohno}}]{m_sawicki_10a}%
  \BibitemOpen
  \bibfield  {author} {\bibinfo {author} {\bibfnamefont {M.}~\bibnamefont
  {Sawicki}}, \bibinfo {author} {\bibfnamefont {D.}~\bibnamefont {Chiba}},
  \bibinfo {author} {\bibfnamefont {A.}~\bibnamefont {Korbecka}}, \bibinfo
  {author} {\bibfnamefont {Y.}~\bibnamefont {Nishitani}}, \bibinfo {author}
  {\bibfnamefont {J.}~\bibnamefont {Majewski}}, \bibinfo {author}
  {\bibfnamefont {F.}~\bibnamefont {Matsukura}}, \bibinfo {author}
  {\bibfnamefont {T.}~\bibnamefont {Dielt}}, \ and\ \bibinfo {author}
  {\bibfnamefont {H.}~\bibnamefont {Ohno}},\ }\href@noop {} {\bibfield
  {journal} {\bibinfo  {journal} {Nat. Phys.}\ }\textbf {\bibinfo {volume}
  {6}},\ \bibinfo {pages} {22} (\bibinfo {year} {2010})}\BibitemShut {NoStop}%
\bibitem [{\citenamefont {Flatte}(2011)}]{m_flatte_11}%
  \BibitemOpen
  \bibfield  {author} {\bibinfo {author} {\bibfnamefont {M.~E.}\ \bibnamefont
  {Flatte}},\ }\href {\doibase 10.1038/nphys1971} {\bibfield  {journal}
  {\bibinfo  {journal} {Nat. Phys.}\ }\textbf {\bibinfo {volume} {7}},\
  \bibinfo {pages} {285} (\bibinfo {year} {2011})}\BibitemShut {NoStop}%
\bibitem [{\citenamefont {Samarth}(2012)}]{n_samarth_12}%
  \BibitemOpen
  \bibfield  {author} {\bibinfo {author} {\bibfnamefont {N.}~\bibnamefont
  {Samarth}},\ }\href@noop {} {\bibfield  {journal} {\bibinfo  {journal} {Nat.
  Mater.}\ }\textbf {\bibinfo {volume} {11}},\ \bibinfo {pages} {360} (\bibinfo
  {year} {2012})}\BibitemShut {NoStop}%
\bibitem [{\citenamefont {Luque}\ and\ \citenamefont
  {Martí}(2001)}]{a_luque_01}%
  \BibitemOpen
  \bibfield  {author} {\bibinfo {author} {\bibfnamefont {A.}~\bibnamefont
  {Luque}}\ and\ \bibinfo {author} {\bibfnamefont {A.}~\bibnamefont {Martí}},\
  }\href {\doibase 10.1002/pip.354} {\bibfield  {journal} {\bibinfo  {journal}
  {Prog. Photovolt: Res Appl}\ }\textbf {\bibinfo {volume} {9}},\ \bibinfo
  {pages} {73} (\bibinfo {year} {2001})}\BibitemShut {NoStop}%
\bibitem [{\citenamefont {Okada}\ \emph {et~al.}(2015)\citenamefont {Okada},
  \citenamefont {Ekins-Daukes}, \citenamefont {Kita}, \citenamefont {Tamaki},
  \citenamefont {Yoshida}, \citenamefont {Pusch}, \citenamefont {Hess},
  \citenamefont {Phillips}, \citenamefont {Farrell}, \citenamefont {Yoshida},
  \citenamefont {Ahsan}, \citenamefont {Shoji}, \citenamefont {Sogabe},\ and\
  \citenamefont {Guillemoles}}]{y_okada_15}%
  \BibitemOpen
  \bibfield  {author} {\bibinfo {author} {\bibfnamefont {Y.}~\bibnamefont
  {Okada}}, \bibinfo {author} {\bibfnamefont {N.~J.}\ \bibnamefont
  {Ekins-Daukes}}, \bibinfo {author} {\bibfnamefont {T.}~\bibnamefont {Kita}},
  \bibinfo {author} {\bibfnamefont {R.}~\bibnamefont {Tamaki}}, \bibinfo
  {author} {\bibfnamefont {M.}~\bibnamefont {Yoshida}}, \bibinfo {author}
  {\bibfnamefont {A.}~\bibnamefont {Pusch}}, \bibinfo {author} {\bibfnamefont
  {O.}~\bibnamefont {Hess}}, \bibinfo {author} {\bibfnamefont {C.~C.}\
  \bibnamefont {Phillips}}, \bibinfo {author} {\bibfnamefont {D.~J.}\
  \bibnamefont {Farrell}}, \bibinfo {author} {\bibfnamefont {K.}~\bibnamefont
  {Yoshida}}, \bibinfo {author} {\bibfnamefont {N.}~\bibnamefont {Ahsan}},
  \bibinfo {author} {\bibfnamefont {Y.}~\bibnamefont {Shoji}}, \bibinfo
  {author} {\bibfnamefont {T.}~\bibnamefont {Sogabe}}, \ and\ \bibinfo {author}
  {\bibfnamefont {J.-F.}\ \bibnamefont {Guillemoles}},\ }\href {\doibase
  http://dx.doi.org/10.1063/1.4916561} {\bibfield  {journal} {\bibinfo
  {journal} {Appl. Phys. Rev.}\ }\textbf {\bibinfo {volume} {2}},\ \bibinfo
  {eid} {021302} (\bibinfo {year} {2015}),\
  http://dx.doi.org/10.1063/1.4916561}\BibitemShut {NoStop}%
\bibitem [{\citenamefont {Zhang}\ \emph
  {et~al.}(2015{\natexlab{b}})\citenamefont {Zhang}, \citenamefont {He},\ and\
  \citenamefont {Pan}}]{j_zhang_15}%
  \BibitemOpen
  \bibfield  {author} {\bibinfo {author} {\bibfnamefont {J.}~\bibnamefont
  {Zhang}}, \bibinfo {author} {\bibfnamefont {H.}~\bibnamefont {He}}, \ and\
  \bibinfo {author} {\bibfnamefont {B.}~\bibnamefont {Pan}},\ }\href@noop {}
  {\bibfield  {journal} {\bibinfo  {journal} {Nanotechnology}\ }\textbf
  {\bibinfo {volume} {26}},\ \bibinfo {pages} {195401} (\bibinfo {year}
  {2015}{\natexlab{b}})}\BibitemShut {NoStop}%
\bibitem [{\citenamefont {Manley}\ \emph {et~al.}(2014)\citenamefont {Manley},
  \citenamefont {Lynn}, \citenamefont {Abernathy}, \citenamefont {Specht},
  \citenamefont {Delaire}, \citenamefont {Bishop}, \citenamefont {Sahul},\ and\
  \citenamefont {Budai}}]{m_manley_14}%
  \BibitemOpen
  \bibfield  {author} {\bibinfo {author} {\bibfnamefont {M.}~\bibnamefont
  {Manley}}, \bibinfo {author} {\bibfnamefont {J.}~\bibnamefont {Lynn}},
  \bibinfo {author} {\bibfnamefont {D.}~\bibnamefont {Abernathy}}, \bibinfo
  {author} {\bibfnamefont {E.}~\bibnamefont {Specht}}, \bibinfo {author}
  {\bibfnamefont {O.}~\bibnamefont {Delaire}}, \bibinfo {author} {\bibfnamefont
  {A.}~\bibnamefont {Bishop}}, \bibinfo {author} {\bibfnamefont
  {R.}~\bibnamefont {Sahul}}, \ and\ \bibinfo {author} {\bibfnamefont
  {J.}~\bibnamefont {Budai}},\ }\href@noop {} {\bibfield  {journal} {\bibinfo
  {journal} {Nat. Commun.}\ }\textbf {\bibinfo {volume} {5}} (\bibinfo {year}
  {2014})}\BibitemShut {NoStop}%
\bibitem [{\citenamefont {Anderson}\ \emph {et~al.}(1977)\citenamefont
  {Anderson}, \citenamefont {Mott},\ and\ \citenamefont {van
  Vleck}}]{1977Nobel}%
  \BibitemOpen
  \bibfield  {author} {\bibinfo {author} {\bibfnamefont {P.~W.}\ \bibnamefont
  {Anderson}}, \bibinfo {author} {\bibfnamefont {N.~F.}\ \bibnamefont {Mott}},
  \ and\ \bibinfo {author} {\bibfnamefont {J.~H.}\ \bibnamefont {van Vleck}},\
  }\href@noop {} {}\bibinfo {howpublished}
  {\url{https://www.nobelprize.org/nobel_prizes/physics/laureates/1977/}}
  (\bibinfo {year} {1977})\BibitemShut {NoStop}%
\bibitem [{\citenamefont {Lagendijk}\ \emph {et~al.}(2009)\citenamefont
  {Lagendijk}, \citenamefont {van Tiggelen},\ and\ \citenamefont
  {Wiersma}}]{a_lagendijk_09}%
  \BibitemOpen
  \bibfield  {author} {\bibinfo {author} {\bibfnamefont {A.}~\bibnamefont
  {Lagendijk}}, \bibinfo {author} {\bibfnamefont {B.}~\bibnamefont {van
  Tiggelen}}, \ and\ \bibinfo {author} {\bibfnamefont {D.~S.}\ \bibnamefont
  {Wiersma}},\ }\href@noop {} {\bibfield  {journal} {\bibinfo  {journal} {Phys.
  Today}\ }\textbf {\bibinfo {volume} {62}},\ \bibinfo {pages} {24} (\bibinfo
  {year} {2009})}\BibitemShut {NoStop}%
\bibitem [{\citenamefont {Abrahams}(2010)}]{e_abrahams_10}%
  \BibitemOpen
  \bibinfo {editor} {\bibfnamefont {E.}~\bibnamefont {Abrahams}},\ ed.,\
  \href@noop {} {\emph {\bibinfo {title} {{50 Years of Anderson
  Localization}}}}\ (\bibinfo  {publisher} {World Scientific},\ \bibinfo {year}
  {2010})\BibitemShut {NoStop}%
\bibitem [{\citenamefont {Kramer}\ \emph {et~al.}(2010)\citenamefont {Kramer},
  \citenamefont {MacKinnon}, \citenamefont {Ohtsuki}, ,\ and\ \citenamefont
  {Slevin}}]{Kramer_etal_2010}%
  \BibitemOpen
  \bibfield  {author} {\bibinfo {author} {\bibfnamefont {B.}~\bibnamefont
  {Kramer}}, \bibinfo {author} {\bibfnamefont {A.}~\bibnamefont {MacKinnon}},
  \bibinfo {author} {\bibfnamefont {T.}~\bibnamefont {Ohtsuki}}, , \ and\
  \bibinfo {author} {\bibfnamefont {K.}~\bibnamefont {Slevin}},\ }\href
  {\doibase 10.1142/S0217979210064630} {\bibfield  {journal} {\bibinfo
  {journal} {Int. J. Mod. Phys. B}\ }\textbf {\bibinfo {volume} {24}},\
  \bibinfo {pages} {1841} (\bibinfo {year} {2010})}\BibitemShut {NoStop}%
\bibitem [{\citenamefont {Markos}(2006)}]{Markos_2006}%
  \BibitemOpen
  \bibfield  {author} {\bibinfo {author} {\bibfnamefont {P.}~\bibnamefont
  {Markos}},\ }\href {\doibase 10.2478/v10155-010-0081-0} {\bibfield  {journal}
  {\bibinfo  {journal} {Acta Phys. Slovaca}\ }\textbf {\bibinfo {volume}
  {56}},\ \bibinfo {pages} {561} (\bibinfo {year} {2006})}\BibitemShut
  {NoStop}%
\bibitem [{\citenamefont {Kramer}\ and\ \citenamefont
  {MacKinnon}(1993)}]{Kramer_MacKinnon_1993}%
  \BibitemOpen
  \bibfield  {author} {\bibinfo {author} {\bibfnamefont {B.}~\bibnamefont
  {Kramer}}\ and\ \bibinfo {author} {\bibfnamefont {A.}~\bibnamefont
  {MacKinnon}},\ }\href {http://stacks.iop.org/0034-4885/56/i=12/a=001}
  {\bibfield  {journal} {\bibinfo  {journal} {Rep. Prog. Phys.}\ }\textbf
  {\bibinfo {volume} {56}},\ \bibinfo {pages} {1469} (\bibinfo {year}
  {1993})}\BibitemShut {NoStop}%
\bibitem [{\citenamefont {Wei\ss{}e}\ \emph {et~al.}(2006)\citenamefont
  {Wei\ss{}e}, \citenamefont {Wellein}, \citenamefont {Alvermann},\ and\
  \citenamefont {Fehske}}]{a_weisse_06}%
  \BibitemOpen
  \bibfield  {author} {\bibinfo {author} {\bibfnamefont {A.}~\bibnamefont
  {Wei\ss{}e}}, \bibinfo {author} {\bibfnamefont {G.}~\bibnamefont {Wellein}},
  \bibinfo {author} {\bibfnamefont {A.}~\bibnamefont {Alvermann}}, \ and\
  \bibinfo {author} {\bibfnamefont {H.}~\bibnamefont {Fehske}},\ }\href
  {\doibase 10.1103/RevModPhys.78.275} {\bibfield  {journal} {\bibinfo
  {journal} {Rev. Mod. Phys.}\ }\textbf {\bibinfo {volume} {78}},\ \bibinfo
  {pages} {275} (\bibinfo {year} {2006})}\BibitemShut {NoStop}%
\bibitem [{\citenamefont {Vollhardt}(2010)}]{Vollhardt_2010}%
  \BibitemOpen
  \bibfield  {author} {\bibinfo {author} {\bibfnamefont {D.}~\bibnamefont
  {Vollhardt}},\ }\href {\doibase 10.1063/1.3518901} {\bibfield  {journal}
  {\bibinfo  {journal} {AIP Conf. Proc.}\ }\textbf {\bibinfo {volume} {1297}},\
  \bibinfo {pages} {339} (\bibinfo {year} {2010})}\BibitemShut {NoStop}%
\bibitem [{\citenamefont {Jarrell}\ \emph {et~al.}(1993)\citenamefont
  {Jarrell}, \citenamefont {Akhlaghpour},\ and\ \citenamefont
  {Pruschke}}]{m_jarrell_93b}%
  \BibitemOpen
  \bibfield  {author} {\bibinfo {author} {\bibfnamefont {M.}~\bibnamefont
  {Jarrell}}, \bibinfo {author} {\bibfnamefont {H.}~\bibnamefont
  {Akhlaghpour}}, \ and\ \bibinfo {author} {\bibfnamefont {T.}~\bibnamefont
  {Pruschke}},\ }\href@noop {} {\bibfield  {journal} {\bibinfo  {journal}
  {Phys. Rev. Lett.}\ }\textbf {\bibinfo {volume} {70}},\ \bibinfo {pages}
  {1670} (\bibinfo {year} {1993})}\BibitemShut {NoStop}%
\bibitem [{\citenamefont {Freericks}\ \emph {et~al.}(1993)\citenamefont
  {Freericks}, \citenamefont {Jarrell},\ and\ \citenamefont
  {Scalapino}}]{j_freericks_93b}%
  \BibitemOpen
  \bibfield  {author} {\bibinfo {author} {\bibfnamefont {J.~K.}\ \bibnamefont
  {Freericks}}, \bibinfo {author} {\bibfnamefont {M.}~\bibnamefont {Jarrell}},
  \ and\ \bibinfo {author} {\bibfnamefont {D.~J.}\ \bibnamefont {Scalapino}},\
  }\href {\doibase 10.1103/PhysRevB.48.6302} {\bibfield  {journal} {\bibinfo
  {journal} {Phys. Rev. B}\ }\textbf {\bibinfo {volume} {48}},\ \bibinfo
  {pages} {6302} (\bibinfo {year} {1993})}\BibitemShut {NoStop}%
\bibitem [{\citenamefont {Maier}\ \emph
  {et~al.}(2005{\natexlab{a}})\citenamefont {Maier}, \citenamefont {Jarrell},
  \citenamefont {Pruschke},\ and\ \citenamefont {Hettler}}]{th_maier_05a}%
  \BibitemOpen
  \bibfield  {author} {\bibinfo {author} {\bibfnamefont {T.}~\bibnamefont
  {Maier}}, \bibinfo {author} {\bibfnamefont {M.}~\bibnamefont {Jarrell}},
  \bibinfo {author} {\bibfnamefont {T.}~\bibnamefont {Pruschke}}, \ and\
  \bibinfo {author} {\bibfnamefont {M.~H.}\ \bibnamefont {Hettler}},\ }\href
  {\doibase 10.1103/RevModPhys.77.1027} {\bibfield  {journal} {\bibinfo
  {journal} {Rev. Mod. Phys.}\ }\textbf {\bibinfo {volume} {77}},\ \bibinfo
  {pages} {1027} (\bibinfo {year} {2005}{\natexlab{a}})}\BibitemShut {NoStop}%
\bibitem [{\citenamefont {Kotliar}\ \emph {et~al.}(2001)\citenamefont
  {Kotliar}, \citenamefont {Savrasov}, \citenamefont {Palsson},\ and\
  \citenamefont {Biroli}}]{g_kotliar_01}%
  \BibitemOpen
  \bibfield  {author} {\bibinfo {author} {\bibfnamefont {G.}~\bibnamefont
  {Kotliar}}, \bibinfo {author} {\bibfnamefont {S.}~\bibnamefont {Savrasov}},
  \bibinfo {author} {\bibfnamefont {G.}~\bibnamefont {Palsson}}, \ and\
  \bibinfo {author} {\bibfnamefont {G.}~\bibnamefont {Biroli}},\ }\href@noop {}
  {\bibfield  {journal} {\bibinfo  {journal} {Phys. Rev. Lett.}\ }\textbf
  {\bibinfo {volume} {87}},\ \bibinfo {pages} {186401} (\bibinfo {year}
  {2001})}\BibitemShut {NoStop}%
\bibitem [{\citenamefont {Jarrell}\ and\ \citenamefont
  {Pruschke}(1993)}]{m_jarrell_93a}%
  \BibitemOpen
  \bibfield  {author} {\bibinfo {author} {\bibfnamefont {M.}~\bibnamefont
  {Jarrell}}\ and\ \bibinfo {author} {\bibfnamefont {T.}~\bibnamefont
  {Pruschke}},\ }\href@noop {} {\bibfield  {journal} {\bibinfo  {journal} {Z.
  Phys. B.}\ }\textbf {\bibinfo {volume} {90}},\ \bibinfo {pages} {187}
  (\bibinfo {year} {1993})}\BibitemShut {NoStop}%
\bibitem [{\citenamefont {Maier}\ \emph
  {et~al.}(2005{\natexlab{b}})\citenamefont {Maier}, \citenamefont {Jarrell},
  \citenamefont {Schulthess}, \citenamefont {Kent},\ and\ \citenamefont
  {White}}]{th_maier_05b}%
  \BibitemOpen
  \bibfield  {author} {\bibinfo {author} {\bibfnamefont {T.}~\bibnamefont
  {Maier}}, \bibinfo {author} {\bibfnamefont {M.}~\bibnamefont {Jarrell}},
  \bibinfo {author} {\bibfnamefont {T.}~\bibnamefont {Schulthess}}, \bibinfo
  {author} {\bibfnamefont {P.}~\bibnamefont {Kent}}, \ and\ \bibinfo {author}
  {\bibfnamefont {J.}~\bibnamefont {White}},\ }\href@noop {} {\bibfield
  {journal} {\bibinfo  {journal} {Phys. Rev. Lett.}\ }\textbf {\bibinfo
  {volume} {95}},\ \bibinfo {pages} {237001} (\bibinfo {year}
  {2005}{\natexlab{b}})}\BibitemShut {NoStop}%
\bibitem [{\citenamefont {Yonezawa}\ and\ \citenamefont
  {Morigaki}(1973)}]{yo.mo.73}%
  \BibitemOpen
  \bibfield  {author} {\bibinfo {author} {\bibfnamefont {F.}~\bibnamefont
  {Yonezawa}}\ and\ \bibinfo {author} {\bibfnamefont {K.}~\bibnamefont
  {Morigaki}},\ }\href@noop {} {\bibfield  {journal} {\bibinfo  {journal}
  {Suppl. Prog. Theor. Phys.}\ }\textbf {\bibinfo {volume} {53}},\ \bibinfo
  {pages} {1} (\bibinfo {year} {1973})}\BibitemShut {NoStop}%
\bibitem [{\citenamefont {Ziman}(1979)}]{ziman_79}%
  \BibitemOpen
  \bibfield  {author} {\bibinfo {author} {\bibfnamefont {J.~M.}\ \bibnamefont
  {Ziman}},\ }\href@noop {} {\emph {\bibinfo {title} {Models of disorder}}}\
  (\bibinfo  {publisher} {Cambridge University Press},\ \bibinfo {year}
  {1979})\BibitemShut {NoStop}%
\bibitem [{\citenamefont {Taylor}(1967)}]{d_taylor_67}%
  \BibitemOpen
  \bibfield  {author} {\bibinfo {author} {\bibfnamefont {D.}~\bibnamefont
  {Taylor}},\ }\href@noop {} {\bibfield  {journal} {\bibinfo  {journal} {Phys.
  Rev.}\ }\textbf {\bibinfo {volume} {156}},\ \bibinfo {pages} {1017} (\bibinfo
  {year} {1967})}\BibitemShut {NoStop}%
\bibitem [{\citenamefont {Gy\"{o}rffy}(1972)}]{Gyorffy72}%
  \BibitemOpen
  \bibfield  {author} {\bibinfo {author} {\bibfnamefont {B.}~\bibnamefont
  {Gy\"{o}rffy}},\ }\href@noop {} {\bibfield  {journal} {\bibinfo  {journal}
  {Phys. Rev. B}\ }\textbf {\bibinfo {volume} {5}},\ \bibinfo {pages} {2382}
  (\bibinfo {year} {1972})}\BibitemShut {NoStop}%
\bibitem [{\citenamefont {Johnson}\ \emph {et~al.}(1986)\citenamefont
  {Johnson}, \citenamefont {Nicholson}, \citenamefont {Pinski}, \citenamefont
  {Gyorffy},\ and\ \citenamefont {Stocks}}]{Johnson_1986a}%
  \BibitemOpen
  \bibfield  {author} {\bibinfo {author} {\bibfnamefont {D.}~\bibnamefont
  {Johnson}}, \bibinfo {author} {\bibfnamefont {D.}~\bibnamefont {Nicholson}},
  \bibinfo {author} {\bibfnamefont {F.}~\bibnamefont {Pinski}}, \bibinfo
  {author} {\bibfnamefont {B.}~\bibnamefont {Gyorffy}}, \ and\ \bibinfo
  {author} {\bibfnamefont {G.}~\bibnamefont {Stocks}},\ }\href@noop {}
  {\bibfield  {journal} {\bibinfo  {journal} {Phys. Rev. Lett.}\ }\textbf
  {\bibinfo {volume} {56}},\ \bibinfo {pages} {2088} (\bibinfo {year}
  {1986})}\BibitemShut {NoStop}%
\bibitem [{\citenamefont {Vitos}\ \emph {et~al.}(2001)\citenamefont {Vitos},
  \citenamefont {Abrikosov},\ and\ \citenamefont {Johansson}}]{vi.ab.01}%
  \BibitemOpen
  \bibfield  {author} {\bibinfo {author} {\bibfnamefont {L.}~\bibnamefont
  {Vitos}}, \bibinfo {author} {\bibfnamefont {I.~A.}\ \bibnamefont
  {Abrikosov}}, \ and\ \bibinfo {author} {\bibfnamefont {B.}~\bibnamefont
  {Johansson}},\ }\href {\doibase 10.1103/PhysRevLett.87.156401} {\bibfield
  {journal} {\bibinfo  {journal} {Phys. Rev. Lett.}\ }\textbf {\bibinfo
  {volume} {87}},\ \bibinfo {pages} {156401} (\bibinfo {year}
  {2001})}\BibitemShut {NoStop}%
\bibitem [{\citenamefont {Singh}\ \emph {et~al.}(1993)\citenamefont {Singh},
  \citenamefont {Gonis},\ and\ \citenamefont {Turchi}}]{si.go.93}%
  \BibitemOpen
  \bibfield  {author} {\bibinfo {author} {\bibfnamefont {P.~P.}\ \bibnamefont
  {Singh}}, \bibinfo {author} {\bibfnamefont {A.}~\bibnamefont {Gonis}}, \ and\
  \bibinfo {author} {\bibfnamefont {P.~E.~A.}\ \bibnamefont {Turchi}},\
  }\href@noop {} {\bibfield  {journal} {\bibinfo  {journal} {Phys. Rev. Lett.}\
  }\textbf {\bibinfo {volume} {71}},\ \bibinfo {pages} {1605} (\bibinfo {year}
  {1993})}\BibitemShut {NoStop}%
\bibitem [{\citenamefont {Faulkner}(1982)}]{faul.82}%
  \BibitemOpen
  \bibfield  {author} {\bibinfo {author} {\bibfnamefont {J.~S.}\ \bibnamefont
  {Faulkner}},\ }\href@noop {} {\bibfield  {journal} {\bibinfo  {journal}
  {Prog. Mater. Sci.}\ }\textbf {\bibinfo {volume} {27}},\ \bibinfo {pages} {1}
  (\bibinfo {year} {1982})}\BibitemShut {NoStop}%
\bibitem [{\citenamefont {Johnson}\ and\ \citenamefont
  {Pinski}(1993)}]{jo.pi.93}%
  \BibitemOpen
  \bibfield  {author} {\bibinfo {author} {\bibfnamefont {D.~D.}\ \bibnamefont
  {Johnson}}\ and\ \bibinfo {author} {\bibfnamefont {F.~J.}\ \bibnamefont
  {Pinski}},\ }\href@noop {} {\bibfield  {journal} {\bibinfo  {journal} {Phys.
  Rev. B}\ }\textbf {\bibinfo {volume} {48}},\ \bibinfo {pages} {11553}
  (\bibinfo {year} {1993})}\BibitemShut {NoStop}%
\bibitem [{\citenamefont {Korzhavyi}\ \emph {et~al.}(1995)\citenamefont
  {Korzhavyi}, \citenamefont {Ruban}, \citenamefont {Abrikosov},\ and\
  \citenamefont {Skriver}}]{ko.ru.95}%
  \BibitemOpen
  \bibfield  {author} {\bibinfo {author} {\bibfnamefont {P.~A.}\ \bibnamefont
  {Korzhavyi}}, \bibinfo {author} {\bibfnamefont {A.~V.}\ \bibnamefont
  {Ruban}}, \bibinfo {author} {\bibfnamefont {I.~A.}\ \bibnamefont
  {Abrikosov}}, \ and\ \bibinfo {author} {\bibfnamefont {H.~L.}\ \bibnamefont
  {Skriver}},\ }\href@noop {} {\bibfield  {journal} {\bibinfo  {journal} {Phys.
  Rev. B}\ }\textbf {\bibinfo {volume} {51}},\ \bibinfo {pages} {5773}
  (\bibinfo {year} {1995})}\BibitemShut {NoStop}%
\bibitem [{\citenamefont {Ruban}\ \emph {et~al.}(1995)\citenamefont {Ruban},
  \citenamefont {Abrikosov},\ and\ \citenamefont {Skriver}}]{ru.ab.95}%
  \BibitemOpen
  \bibfield  {author} {\bibinfo {author} {\bibfnamefont {A.~V.}\ \bibnamefont
  {Ruban}}, \bibinfo {author} {\bibfnamefont {I.~A.}\ \bibnamefont
  {Abrikosov}}, \ and\ \bibinfo {author} {\bibfnamefont {H.~L.}\ \bibnamefont
  {Skriver}},\ }\href@noop {} {\bibfield  {journal} {\bibinfo  {journal} {Phys.
  Rev. B}\ }\textbf {\bibinfo {volume} {51}},\ \bibinfo {pages} {12958}
  (\bibinfo {year} {1995})}\BibitemShut {NoStop}%
\bibitem [{\citenamefont {Gy\"orffy}\ and\ \citenamefont
  {Stocks}(1983)}]{gy.st.83}%
  \BibitemOpen
  \bibfield  {author} {\bibinfo {author} {\bibfnamefont {B.~L.}\ \bibnamefont
  {Gy\"orffy}}\ and\ \bibinfo {author} {\bibfnamefont {G.~M.}\ \bibnamefont
  {Stocks}},\ }\href@noop {} {\bibfield  {journal} {\bibinfo  {journal} {Phys.
  Rev. Lett.}\ }\textbf {\bibinfo {volume} {50}},\ \bibinfo {pages} {374}
  (\bibinfo {year} {1983})}\BibitemShut {NoStop}%
\bibitem [{\citenamefont {Althoff}\ \emph {et~al.}(1995)\citenamefont
  {Althoff}, \citenamefont {Johnson},\ and\ \citenamefont {Pinski}}]{al.jo.95}%
  \BibitemOpen
  \bibfield  {author} {\bibinfo {author} {\bibfnamefont {J.~D.}\ \bibnamefont
  {Althoff}}, \bibinfo {author} {\bibfnamefont {D.~D.}\ \bibnamefont
  {Johnson}}, \ and\ \bibinfo {author} {\bibfnamefont {F.~J.}\ \bibnamefont
  {Pinski}},\ }\href@noop {} {\bibfield  {journal} {\bibinfo  {journal} {Phys.
  Rev. Lett.}\ }\textbf {\bibinfo {volume} {74}},\ \bibinfo {pages} {138}
  (\bibinfo {year} {1995})}\BibitemShut {NoStop}%
\bibitem [{\citenamefont {Abrikosov}\ \emph {et~al.}(1993)\citenamefont
  {Abrikosov}, \citenamefont {Ruban}, \citenamefont {Kats},\ and\ \citenamefont
  {Vekilov}}]{ab.ru.93}%
  \BibitemOpen
  \bibfield  {author} {\bibinfo {author} {\bibfnamefont {I.~A.}\ \bibnamefont
  {Abrikosov}}, \bibinfo {author} {\bibfnamefont {A.~V.}\ \bibnamefont
  {Ruban}}, \bibinfo {author} {\bibfnamefont {D.~Y.}\ \bibnamefont {Kats}}, \
  and\ \bibinfo {author} {\bibfnamefont {Y.~H.}\ \bibnamefont {Vekilov}},\
  }\href@noop {} {\bibfield  {journal} {\bibinfo  {journal} {J. Phys.: Condens.
  Matter}\ }\textbf {\bibinfo {volume} {5}},\ \bibinfo {pages} {1271} (\bibinfo
  {year} {1993})}\BibitemShut {NoStop}%
\bibitem [{\citenamefont {Vitos}(2007)}]{vito.07}%
  \BibitemOpen
  \bibfield  {author} {\bibinfo {author} {\bibfnamefont {L.}~\bibnamefont
  {Vitos}},\ }\href@noop {} {\emph {\bibinfo {title} {Computational Quantum
  Mechanics for Materials Engineers}}}\ (\bibinfo  {publisher} {Springer},\
  \bibinfo {year} {2007})\BibitemShut {NoStop}%
\bibitem [{\citenamefont {Akai}\ and\ \citenamefont
  {Dederichs}(1993)}]{ak.de.93}%
  \BibitemOpen
  \bibfield  {author} {\bibinfo {author} {\bibfnamefont {H.}~\bibnamefont
  {Akai}}\ and\ \bibinfo {author} {\bibfnamefont {P.~H.}\ \bibnamefont
  {Dederichs}},\ }\href@noop {} {\bibfield  {journal} {\bibinfo  {journal}
  {Phys. Rev. B}\ }\textbf {\bibinfo {volume} {47}},\ \bibinfo {pages} {8739}
  (\bibinfo {year} {1993})}\BibitemShut {NoStop}%
\bibitem [{\citenamefont {Turek}\ \emph {et~al.}(1994)\citenamefont {Turek},
  \citenamefont {Kudrnovsk\'y}, \citenamefont {Drchal},\ and\ \citenamefont
  {Weinberger}}]{tu.ku.94}%
  \BibitemOpen
  \bibfield  {author} {\bibinfo {author} {\bibfnamefont {I.}~\bibnamefont
  {Turek}}, \bibinfo {author} {\bibfnamefont {J.}~\bibnamefont {Kudrnovsk\'y}},
  \bibinfo {author} {\bibfnamefont {V.}~\bibnamefont {Drchal}}, \ and\ \bibinfo
  {author} {\bibfnamefont {P.}~\bibnamefont {Weinberger}},\ }\href@noop {}
  {\bibfield  {journal} {\bibinfo  {journal} {Phys. Rev. B}\ }\textbf {\bibinfo
  {volume} {49}},\ \bibinfo {pages} {3352} (\bibinfo {year}
  {1994})}\BibitemShut {NoStop}%
\bibitem [{\citenamefont {Abrikosov}\ \emph {et~al.}(1995)\citenamefont
  {Abrikosov}, \citenamefont {Eriksson}, \citenamefont {S\"oderlind},
  \citenamefont {Skriver},\ and\ \citenamefont {Johansson}}]{ab.er.95}%
  \BibitemOpen
  \bibfield  {author} {\bibinfo {author} {\bibfnamefont {I.~A.}\ \bibnamefont
  {Abrikosov}}, \bibinfo {author} {\bibfnamefont {O.}~\bibnamefont {Eriksson}},
  \bibinfo {author} {\bibfnamefont {P.}~\bibnamefont {S\"oderlind}}, \bibinfo
  {author} {\bibfnamefont {H.~L.}\ \bibnamefont {Skriver}}, \ and\ \bibinfo
  {author} {\bibfnamefont {B.}~\bibnamefont {Johansson}},\ }\href@noop {}
  {\bibfield  {journal} {\bibinfo  {journal} {Phys. Rev. B}\ }\textbf {\bibinfo
  {volume} {51}},\ \bibinfo {pages} {1058} (\bibinfo {year}
  {1995})}\BibitemShut {NoStop}%
\bibitem [{\citenamefont {Kudrnovsk\'y}\ \emph {et~al.}(1992)\citenamefont
  {Kudrnovsk\'y}, \citenamefont {Turek}, \citenamefont {Drchal}, \citenamefont
  {Weinberger}, \citenamefont {Christensen},\ and\ \citenamefont
  {Bose}}]{ku.tu.92}%
  \BibitemOpen
  \bibfield  {author} {\bibinfo {author} {\bibfnamefont {J.}~\bibnamefont
  {Kudrnovsk\'y}}, \bibinfo {author} {\bibfnamefont {I.}~\bibnamefont {Turek}},
  \bibinfo {author} {\bibfnamefont {V.}~\bibnamefont {Drchal}}, \bibinfo
  {author} {\bibfnamefont {P.}~\bibnamefont {Weinberger}}, \bibinfo {author}
  {\bibfnamefont {N.~E.}\ \bibnamefont {Christensen}}, \ and\ \bibinfo {author}
  {\bibfnamefont {S.~K.}\ \bibnamefont {Bose}},\ }\href@noop {} {\bibfield
  {journal} {\bibinfo  {journal} {Phys. Rev. B}\ }\textbf {\bibinfo {volume}
  {46}},\ \bibinfo {pages} {4222} (\bibinfo {year} {1992})}\BibitemShut
  {NoStop}%
\bibitem [{\citenamefont {MacLaren}\ \emph {et~al.}(1992)\citenamefont
  {MacLaren}, \citenamefont {Gonis},\ and\ \citenamefont
  {Schadler}}]{ma.go.92}%
  \BibitemOpen
  \bibfield  {author} {\bibinfo {author} {\bibfnamefont {J.~M.}\ \bibnamefont
  {MacLaren}}, \bibinfo {author} {\bibfnamefont {A.}~\bibnamefont {Gonis}}, \
  and\ \bibinfo {author} {\bibfnamefont {G.}~\bibnamefont {Schadler}},\
  }\href@noop {} {\bibfield  {journal} {\bibinfo  {journal} {Phys. Rev. B}\
  }\textbf {\bibinfo {volume} {45}},\ \bibinfo {pages} {14392} (\bibinfo {year}
  {1992})}\BibitemShut {NoStop}%
\bibitem [{\citenamefont {Abrikosov}\ and\ \citenamefont
  {Skriver}(1993)}]{ab.sk.93}%
  \BibitemOpen
  \bibfield  {author} {\bibinfo {author} {\bibfnamefont {I.~A.}\ \bibnamefont
  {Abrikosov}}\ and\ \bibinfo {author} {\bibfnamefont {H.~L.}\ \bibnamefont
  {Skriver}},\ }\href@noop {} {\bibfield  {journal} {\bibinfo  {journal} {Phys.
  Rev. B}\ }\textbf {\bibinfo {volume} {47}},\ \bibinfo {pages} {16532}
  (\bibinfo {year} {1993})}\BibitemShut {NoStop}%
\bibitem [{\citenamefont {Ruban}\ \emph {et~al.}(1994)\citenamefont {Ruban},
  \citenamefont {Abrikosov}, \citenamefont {Kats}, \citenamefont {Gorelikov},
  \citenamefont {Jacobsen},\ and\ \citenamefont {Skriver}}]{ru.ab.94}%
  \BibitemOpen
  \bibfield  {author} {\bibinfo {author} {\bibfnamefont {A.~V.}\ \bibnamefont
  {Ruban}}, \bibinfo {author} {\bibfnamefont {I.~A.}\ \bibnamefont
  {Abrikosov}}, \bibinfo {author} {\bibfnamefont {D.~Y.}\ \bibnamefont {Kats}},
  \bibinfo {author} {\bibfnamefont {D.}~\bibnamefont {Gorelikov}}, \bibinfo
  {author} {\bibfnamefont {K.~W.}\ \bibnamefont {Jacobsen}}, \ and\ \bibinfo
  {author} {\bibfnamefont {H.~L.}\ \bibnamefont {Skriver}},\ }\href@noop {}
  {\bibfield  {journal} {\bibinfo  {journal} {Phys. Rev. B}\ }\textbf {\bibinfo
  {volume} {49}},\ \bibinfo {pages} {11383} (\bibinfo {year}
  {1994})}\BibitemShut {NoStop}%
\bibitem [{\citenamefont {Pasturel}\ \emph {et~al.}(1993)\citenamefont
  {Pasturel}, \citenamefont {Drchal}, \citenamefont {Kudrnovsk\'y},\ and\
  \citenamefont {Weinberger}}]{pa.dr.93}%
  \BibitemOpen
  \bibfield  {author} {\bibinfo {author} {\bibfnamefont {A.}~\bibnamefont
  {Pasturel}}, \bibinfo {author} {\bibfnamefont {V.}~\bibnamefont {Drchal}},
  \bibinfo {author} {\bibfnamefont {J.}~\bibnamefont {Kudrnovsk\'y}}, \ and\
  \bibinfo {author} {\bibfnamefont {P.}~\bibnamefont {Weinberger}},\
  }\href@noop {} {\bibfield  {journal} {\bibinfo  {journal} {Phys. Rev. B}\
  }\textbf {\bibinfo {volume} {48}},\ \bibinfo {pages} {2704} (\bibinfo {year}
  {1993})}\BibitemShut {NoStop}%
\bibitem [{\citenamefont {Min\'ar}\ \emph {et~al.}(2017)\citenamefont
  {Min\'ar}, \citenamefont {Ebert},\ and\ \citenamefont
  {Chioncel}}]{j_minar_17}%
  \BibitemOpen
  \bibfield  {author} {\bibinfo {author} {\bibfnamefont {J.}~\bibnamefont
  {Min\'ar}}, \bibinfo {author} {\bibfnamefont {H.}~\bibnamefont {Ebert}}, \
  and\ \bibinfo {author} {\bibfnamefont {L.}~\bibnamefont {Chioncel}},\ }\href
  {\doibase 10.1140/epjst/e2017-70047-5} {\bibfield  {journal} {\bibinfo
  {journal} {Eur. Phys. J. Special Topics}\ }\textbf {\bibinfo {volume}
  {226}},\ \bibinfo {pages} {2477} (\bibinfo {year} {2017})}\BibitemShut
  {NoStop}%
\bibitem [{\citenamefont {Marzari}\ and\ \citenamefont
  {Vanderbilt}(1997)}]{marzari_1997}%
  \BibitemOpen
  \bibfield  {author} {\bibinfo {author} {\bibfnamefont {N.}~\bibnamefont
  {Marzari}}\ and\ \bibinfo {author} {\bibfnamefont {D.}~\bibnamefont
  {Vanderbilt}},\ }\href {\doibase 10.1103/PhysRevB.56.12847} {\bibfield
  {journal} {\bibinfo  {journal} {Phys. Rev. B}\ }\textbf {\bibinfo {volume}
  {56}},\ \bibinfo {pages} {12847} (\bibinfo {year} {1997})}\BibitemShut
  {NoStop}%
\bibitem [{\citenamefont {Ku}\ \emph {et~al.}(2002)\citenamefont {Ku},
  \citenamefont {Rosner}, \citenamefont {Pickett},\ and\ \citenamefont
  {Scalettar}}]{w_ku_02}%
  \BibitemOpen
  \bibfield  {author} {\bibinfo {author} {\bibfnamefont {W.}~\bibnamefont
  {Ku}}, \bibinfo {author} {\bibfnamefont {H.}~\bibnamefont {Rosner}}, \bibinfo
  {author} {\bibfnamefont {W.~E.}\ \bibnamefont {Pickett}}, \ and\ \bibinfo
  {author} {\bibfnamefont {R.~T.}\ \bibnamefont {Scalettar}},\ }\href {\doibase
  10.1103/PhysRevLett.89.167204} {\bibfield  {journal} {\bibinfo  {journal}
  {Phys. Rev. Lett.}\ }\textbf {\bibinfo {volume} {89}},\ \bibinfo {pages}
  {167204} (\bibinfo {year} {2002})}\BibitemShut {NoStop}%
\bibitem [{\citenamefont {Anisimov}\ \emph {et~al.}(2005)\citenamefont
  {Anisimov}, \citenamefont {Kondakov}, \citenamefont {Kozhevnikov},
  \citenamefont {Nekrasov}, \citenamefont {Pchelkina}, \citenamefont {Allen},
  \citenamefont {Mo}, \citenamefont {Kim}, \citenamefont {Metcalf},
  \citenamefont {Suga}, \citenamefont {Sekiyama}, \citenamefont {Keller},
  \citenamefont {Leonov}, \citenamefont {Ren},\ and\ \citenamefont
  {Vollhardt}}]{Anisimov:2005ix}%
  \BibitemOpen
  \bibfield  {author} {\bibinfo {author} {\bibfnamefont {V.~I.}\ \bibnamefont
  {Anisimov}}, \bibinfo {author} {\bibfnamefont {D.}~\bibnamefont {Kondakov}},
  \bibinfo {author} {\bibfnamefont {A.~V.}\ \bibnamefont {Kozhevnikov}},
  \bibinfo {author} {\bibfnamefont {I.}~\bibnamefont {Nekrasov}}, \bibinfo
  {author} {\bibfnamefont {Z.}~\bibnamefont {Pchelkina}}, \bibinfo {author}
  {\bibfnamefont {J.}~\bibnamefont {Allen}}, \bibinfo {author} {\bibfnamefont
  {S.-K.}\ \bibnamefont {Mo}}, \bibinfo {author} {\bibfnamefont
  {H.}~\bibnamefont {Kim}}, \bibinfo {author} {\bibfnamefont {P.}~\bibnamefont
  {Metcalf}}, \bibinfo {author} {\bibfnamefont {S.}~\bibnamefont {Suga}},
  \bibinfo {author} {\bibfnamefont {A.}~\bibnamefont {Sekiyama}}, \bibinfo
  {author} {\bibfnamefont {G.}~\bibnamefont {Keller}}, \bibinfo {author}
  {\bibfnamefont {I.}~\bibnamefont {Leonov}}, \bibinfo {author} {\bibfnamefont
  {X.}~\bibnamefont {Ren}}, \ and\ \bibinfo {author} {\bibfnamefont
  {D.}~\bibnamefont {Vollhardt}},\ }\href@noop {} {\bibfield  {journal}
  {\bibinfo  {journal} {Phys. Rev. B}\ }\textbf {\bibinfo {volume} {71}},\
  \bibinfo {pages} {125119} (\bibinfo {year} {2005})}\BibitemShut {NoStop}%
\bibitem [{\citenamefont {Gonis}(1992)}]{a_gonis_92}%
  \BibitemOpen
  \bibfield  {author} {\bibinfo {author} {\bibfnamefont {A.}~\bibnamefont
  {Gonis}},\ }\href@noop {} {\emph {\bibinfo {title} {{Green functions for
  ordered and disordered systems}}}}\ (\bibinfo  {publisher} {North-Holland
  Amsterdam},\ \bibinfo {year} {1992})\BibitemShut {NoStop}%
\bibitem [{\citenamefont {Kotliar}\ \emph {et~al.}(2006)\citenamefont
  {Kotliar}, \citenamefont {Savrasov}, \citenamefont {Haule}, \citenamefont
  {Oudovenko}, \citenamefont {Parcollet},\ and\ \citenamefont
  {Marianetti}}]{kotliar_lda_dmft}%
  \BibitemOpen
  \bibfield  {author} {\bibinfo {author} {\bibfnamefont {G.}~\bibnamefont
  {Kotliar}}, \bibinfo {author} {\bibfnamefont {S.~Y.}\ \bibnamefont
  {Savrasov}}, \bibinfo {author} {\bibfnamefont {K.}~\bibnamefont {Haule}},
  \bibinfo {author} {\bibfnamefont {V.~S.}\ \bibnamefont {Oudovenko}}, \bibinfo
  {author} {\bibfnamefont {O.}~\bibnamefont {Parcollet}}, \ and\ \bibinfo
  {author} {\bibfnamefont {C.~A.}\ \bibnamefont {Marianetti}},\ }\href
  {\doibase 10.1103/RevModPhys.78.865} {\bibfield  {journal} {\bibinfo
  {journal} {Rev. Mod. Phys.}\ }\textbf {\bibinfo {volume} {78}},\ \bibinfo
  {pages} {865} (\bibinfo {year} {2006})}\BibitemShut {NoStop}%
\bibitem [{\citenamefont {Bergmann}(1984)}]{Bergmann_1984}%
  \BibitemOpen
  \bibfield  {author} {\bibinfo {author} {\bibfnamefont {G.}~\bibnamefont
  {Bergmann}},\ }\href {\doibase https://doi.org/10.1016/0370-1573(84)90103-0}
  {\bibfield  {journal} {\bibinfo  {journal} {Phys. Rep.}\ }\textbf {\bibinfo
  {volume} {107}},\ \bibinfo {pages} {1 } (\bibinfo {year} {1984})}\BibitemShut
  {NoStop}%
\bibitem [{\citenamefont {Langer}(1960)}]{Langer_1960}%
  \BibitemOpen
  \bibfield  {author} {\bibinfo {author} {\bibfnamefont {J.~S.}\ \bibnamefont
  {Langer}},\ }\href {\doibase 10.1103/PhysRev.120.714} {\bibfield  {journal}
  {\bibinfo  {journal} {Phys. Rev.}\ }\textbf {\bibinfo {volume} {120}},\
  \bibinfo {pages} {714} (\bibinfo {year} {1960})}\BibitemShut {NoStop}%
\bibitem [{\citenamefont {Langer}\ and\ \citenamefont
  {Neal}(1966)}]{Langer_Neal_1966}%
  \BibitemOpen
  \bibfield  {author} {\bibinfo {author} {\bibfnamefont {J.~S.}\ \bibnamefont
  {Langer}}\ and\ \bibinfo {author} {\bibfnamefont {T.}~\bibnamefont {Neal}},\
  }\href {\doibase 10.1103/PhysRevLett.16.984} {\bibfield  {journal} {\bibinfo
  {journal} {Phys. Rev. Lett.}\ }\textbf {\bibinfo {volume} {16}},\ \bibinfo
  {pages} {984} (\bibinfo {year} {1966})}\BibitemShut {NoStop}%
\bibitem [{\citenamefont {Dobrosavljevi\'{c}}(2010)}]{v_dobrosavljevic_10}%
  \BibitemOpen
  \bibfield  {author} {\bibinfo {author} {\bibfnamefont {V.}~\bibnamefont
  {Dobrosavljevi\'{c}}},\ }\href@noop {} {\bibfield  {journal} {\bibinfo
  {journal} {Int. J. Mod. Phys. B}\ }\textbf {\bibinfo {volume} {24}},\
  \bibinfo {pages} {1680} (\bibinfo {year} {2010})}\BibitemShut {NoStop}%
\bibitem [{\citenamefont {Mott}(1967)}]{Mott_1967}%
  \BibitemOpen
  \bibfield  {author} {\bibinfo {author} {\bibfnamefont {N.}~\bibnamefont
  {Mott}},\ }\href {\doibase 10.1080/00018736700101265} {\bibfield  {journal}
  {\bibinfo  {journal} {Adv. Phys.}\ }\textbf {\bibinfo {volume} {16}},\
  \bibinfo {pages} {49} (\bibinfo {year} {1967})}\BibitemShut {NoStop}%
\bibitem [{\citenamefont {Cohen}\ \emph {et~al.}(1969)\citenamefont {Cohen},
  \citenamefont {Fritzsche},\ and\ \citenamefont
  {Ovshinsky}}]{Cohen_Fritzsche_Ovshinsky_1969}%
  \BibitemOpen
  \bibfield  {author} {\bibinfo {author} {\bibfnamefont {M.~H.}\ \bibnamefont
  {Cohen}}, \bibinfo {author} {\bibfnamefont {H.}~\bibnamefont {Fritzsche}}, \
  and\ \bibinfo {author} {\bibfnamefont {S.~R.}\ \bibnamefont {Ovshinsky}},\
  }\href {\doibase 10.1103/PhysRevLett.22.1065} {\bibfield  {journal} {\bibinfo
   {journal} {Phys. Rev. Lett.}\ }\textbf {\bibinfo {volume} {22}},\ \bibinfo
  {pages} {1065} (\bibinfo {year} {1969})}\BibitemShut {NoStop}%
\bibitem [{\citenamefont {Economou}\ and\ \citenamefont
  {Cohen}(1972)}]{Economou_Cohen_1972}%
  \BibitemOpen
  \bibfield  {author} {\bibinfo {author} {\bibfnamefont {E.~N.}\ \bibnamefont
  {Economou}}\ and\ \bibinfo {author} {\bibfnamefont {M.~H.}\ \bibnamefont
  {Cohen}},\ }\href {\doibase 10.1103/PhysRevB.5.2931} {\bibfield  {journal}
  {\bibinfo  {journal} {Phys. Rev. B}\ }\textbf {\bibinfo {volume} {5}},\
  \bibinfo {pages} {2931} (\bibinfo {year} {1972})}\BibitemShut {NoStop}%
\bibitem [{\citenamefont {Edwards}\ and\ \citenamefont
  {Thouless}(1972)}]{Edwards_Thouless_1972}%
  \BibitemOpen
  \bibfield  {author} {\bibinfo {author} {\bibfnamefont {J.~T.}\ \bibnamefont
  {Edwards}}\ and\ \bibinfo {author} {\bibfnamefont {D.~J.}\ \bibnamefont
  {Thouless}},\ }\href {http://stacks.iop.org/0022-3719/5/i=8/a=007} {\bibfield
   {journal} {\bibinfo  {journal} {J. Phys. C: Solid State Physics}\ }\textbf
  {\bibinfo {volume} {5}},\ \bibinfo {pages} {807} (\bibinfo {year}
  {1972})}\BibitemShut {NoStop}%
\bibitem [{\citenamefont {Licciardello}\ and\ \citenamefont
  {Thouless}(1975)}]{Licciardello_Thouless_1974}%
  \BibitemOpen
  \bibfield  {author} {\bibinfo {author} {\bibfnamefont {D.~C.}\ \bibnamefont
  {Licciardello}}\ and\ \bibinfo {author} {\bibfnamefont {D.~J.}\ \bibnamefont
  {Thouless}},\ }\href {\doibase 10.1103/PhysRevLett.35.1475} {\bibfield
  {journal} {\bibinfo  {journal} {Phys. Rev. Lett.}\ }\textbf {\bibinfo
  {volume} {35}},\ \bibinfo {pages} {1475} (\bibinfo {year}
  {1975})}\BibitemShut {NoStop}%
\bibitem [{\citenamefont {Abrahams}\ \emph {et~al.}(1979)\citenamefont
  {Abrahams}, \citenamefont {Anderson}, \citenamefont {Licciardello},\ and\
  \citenamefont {Ramakrishnan}}]{e_abrahams_79}%
  \BibitemOpen
  \bibfield  {author} {\bibinfo {author} {\bibfnamefont {E.}~\bibnamefont
  {Abrahams}}, \bibinfo {author} {\bibfnamefont {P.~W.}\ \bibnamefont
  {Anderson}}, \bibinfo {author} {\bibfnamefont {D.~C.}\ \bibnamefont
  {Licciardello}}, \ and\ \bibinfo {author} {\bibfnamefont {T.~V.}\
  \bibnamefont {Ramakrishnan}},\ }\href {\doibase 10.1103/PhysRevLett.42.673}
  {\bibfield  {journal} {\bibinfo  {journal} {Phys. Rev. Lett.}\ }\textbf
  {\bibinfo {volume} {42}},\ \bibinfo {pages} {673} (\bibinfo {year}
  {1979})}\BibitemShut {NoStop}%
\bibitem [{\citenamefont {Gor'kov}\ \emph {et~al.}(1979)\citenamefont
  {Gor'kov}, \citenamefont {Larkin},\ and\ \citenamefont
  {Khmel'nitskii}}]{Gorkov_Larkin_Khmelnitskii_1979}%
  \BibitemOpen
  \bibfield  {author} {\bibinfo {author} {\bibfnamefont {L.~P.}\ \bibnamefont
  {Gor'kov}}, \bibinfo {author} {\bibfnamefont {A.~I.}\ \bibnamefont {Larkin}},
  \ and\ \bibinfo {author} {\bibfnamefont {D.~E.}\ \bibnamefont
  {Khmel'nitskii}},\ }\href
  {http://www.jetpletters.ac.ru/ps/1364/article_20629.shtml} {\bibfield
  {journal} {\bibinfo  {journal} {JETP}\ }\textbf {\bibinfo {volume} {30}},\
  \bibinfo {pages} {248} (\bibinfo {year} {1979})}\BibitemShut {NoStop}%
\bibitem [{\citenamefont {Aharony}\ and\ \citenamefont
  {Imry}(1977)}]{Aharony_Imry_1977}%
  \BibitemOpen
  \bibfield  {author} {\bibinfo {author} {\bibfnamefont {A.}~\bibnamefont
  {Aharony}}\ and\ \bibinfo {author} {\bibfnamefont {Y.}~\bibnamefont {Imry}},\
  }\href {http://stacks.iop.org/0022-3719/10/i=17/a=005} {\bibfield  {journal}
  {\bibinfo  {journal} {Journal of Physics C: Solid State Physics}\ }\textbf
  {\bibinfo {volume} {10}},\ \bibinfo {pages} {L487} (\bibinfo {year}
  {1977})}\BibitemShut {NoStop}%
\bibitem [{\citenamefont {Wegner}(1979)}]{Wegner_1979}%
  \BibitemOpen
  \bibfield  {author} {\bibinfo {author} {\bibfnamefont {F.}~\bibnamefont
  {Wegner}},\ }\href {\doibase 10.1007/BF01319839} {\bibfield  {journal}
  {\bibinfo  {journal} {Z. Phys. B Condensed Matter}\ }\textbf {\bibinfo
  {volume} {35}},\ \bibinfo {pages} {207} (\bibinfo {year} {1979})}\BibitemShut
  {NoStop}%
\bibitem [{\citenamefont {Wegner}(1980)}]{Wegner_1980}%
  \BibitemOpen
  \bibfield  {author} {\bibinfo {author} {\bibfnamefont {F.}~\bibnamefont
  {Wegner}},\ }\href {\doibase 10.1007/BF01325284} {\bibfield  {journal}
  {\bibinfo  {journal} {Z. Phys. B Condensed Matter}\ }\textbf {\bibinfo
  {volume} {36}},\ \bibinfo {pages} {209} (\bibinfo {year} {1980})}\BibitemShut
  {NoStop}%
\bibitem [{\citenamefont {Sch{\"a}fer}\ and\ \citenamefont
  {Wegner}(1980)}]{Schafer_Wegner_1980}%
  \BibitemOpen
  \bibfield  {author} {\bibinfo {author} {\bibfnamefont {L.}~\bibnamefont
  {Sch{\"a}fer}}\ and\ \bibinfo {author} {\bibfnamefont {F.}~\bibnamefont
  {Wegner}},\ }\href {\doibase 10.1007/BF01598751} {\bibfield  {journal}
  {\bibinfo  {journal} {Z. Phys. B Condensed Matter}\ }\textbf {\bibinfo
  {volume} {38}},\ \bibinfo {pages} {113} (\bibinfo {year} {1980})}\BibitemShut
  {NoStop}%
\bibitem [{\citenamefont {Castellani}\ and\ \citenamefont
  {Peliti}(1986)}]{Castellani_Peliti_1986}%
  \BibitemOpen
  \bibfield  {author} {\bibinfo {author} {\bibfnamefont {C.}~\bibnamefont
  {Castellani}}\ and\ \bibinfo {author} {\bibfnamefont {L.}~\bibnamefont
  {Peliti}},\ }\href {http://stacks.iop.org/0305-4470/19/i=8/a=004} {\bibfield
  {journal} {\bibinfo  {journal} {J. Phys. A: Mathematical and General}\
  }\textbf {\bibinfo {volume} {19}},\ \bibinfo {pages} {L429} (\bibinfo {year}
  {1986})}\BibitemShut {NoStop}%
\bibitem [{\citenamefont {Hikami}\ \emph {et~al.}(1980)\citenamefont {Hikami},
  \citenamefont {Larkin},\ and\ \citenamefont
  {Nagaoka}}]{Hikami_Larkin_Nagaoka_1980}%
  \BibitemOpen
  \bibfield  {author} {\bibinfo {author} {\bibfnamefont {S.}~\bibnamefont
  {Hikami}}, \bibinfo {author} {\bibfnamefont {A.~I.}\ \bibnamefont {Larkin}},
  \ and\ \bibinfo {author} {\bibfnamefont {Y.}~\bibnamefont {Nagaoka}},\ }\href
  {\doibase 10.1143/PTP.63.707} {\bibfield  {journal} {\bibinfo  {journal}
  {Prog. Theor. Phys.}\ }\textbf {\bibinfo {volume} {63}},\ \bibinfo {pages}
  {707} (\bibinfo {year} {1980})}\BibitemShut {NoStop}%
\bibitem [{\citenamefont {Efetov}(1982)}]{Efetov_1982}%
  \BibitemOpen
  \bibfield  {author} {\bibinfo {author} {\bibfnamefont {K.~B.}\ \bibnamefont
  {Efetov}},\ }\href {http://www.jetpletters.ac.ru/ps/1364/article_20629.shtml}
  {\bibfield  {journal} {\bibinfo  {journal} {JETP}\ }\textbf {\bibinfo
  {volume} {82}},\ \bibinfo {pages} {872} (\bibinfo {year} {1982})}\BibitemShut
  {NoStop}%
\bibitem [{\citenamefont {Vollhardt}\ and\ \citenamefont
  {W\"olfle}(1980)}]{Vollhardt_Wolfle_1980}%
  \BibitemOpen
  \bibfield  {author} {\bibinfo {author} {\bibfnamefont {D.}~\bibnamefont
  {Vollhardt}}\ and\ \bibinfo {author} {\bibfnamefont {P.}~\bibnamefont
  {W\"olfle}},\ }\href {\doibase 10.1103/PhysRevB.22.4666} {\bibfield
  {journal} {\bibinfo  {journal} {Phys. Rev. B}\ }\textbf {\bibinfo {volume}
  {22}},\ \bibinfo {pages} {4666} (\bibinfo {year} {1980})}\BibitemShut
  {NoStop}%
\bibitem [{\citenamefont {Vollhardt}\ and\ \citenamefont
  {W{\"o}lfle}(1992)}]{Vollhardt_Wolfle_1992}%
  \BibitemOpen
  \bibfield  {author} {\bibinfo {author} {\bibfnamefont {D.}~\bibnamefont
  {Vollhardt}}\ and\ \bibinfo {author} {\bibfnamefont {P.}~\bibnamefont
  {W{\"o}lfle}},\ }in\ \href {\doibase
  https://doi.org/10.1016/B978-0-444-88885-3.50006-8} {\emph {\bibinfo
  {booktitle} {Electronic Phase Transitions}}},\ \bibinfo {series} {Modern
  Problems in Condensed Matter Sciences}, Vol.~\bibinfo {volume} {32},\
  \bibinfo {editor} {edited by\ \bibinfo {editor} {\bibfnamefont
  {W.}~\bibnamefont {HANKE}}\ and\ \bibinfo {editor} {\bibfnamefont
  {Y.}~\bibnamefont {KOPAEV}}}\ (\bibinfo  {publisher} {Elsevier},\ \bibinfo
  {year} {1992})\ pp.\ \bibinfo {pages} {1 -- 78}\BibitemShut {NoStop}%
\bibitem [{\citenamefont {Vollhardt}\ and\ \citenamefont
  {W\"olfle}(1982)}]{Vollhardt_Wolfle_1982}%
  \BibitemOpen
  \bibfield  {author} {\bibinfo {author} {\bibfnamefont {D.}~\bibnamefont
  {Vollhardt}}\ and\ \bibinfo {author} {\bibfnamefont {P.}~\bibnamefont
  {W\"olfle}},\ }\href {\doibase 10.1103/PhysRevLett.48.699} {\bibfield
  {journal} {\bibinfo  {journal} {Phys. Rev. Lett.}\ }\textbf {\bibinfo
  {volume} {48}},\ \bibinfo {pages} {699} (\bibinfo {year} {1982})}\BibitemShut
  {NoStop}%
\bibitem [{\citenamefont {Altland}\ and\ \citenamefont
  {Zirnbauer}(1997)}]{Atland_Zirnbauer_1997}%
  \BibitemOpen
  \bibfield  {author} {\bibinfo {author} {\bibfnamefont {A.}~\bibnamefont
  {Altland}}\ and\ \bibinfo {author} {\bibfnamefont {M.~R.}\ \bibnamefont
  {Zirnbauer}},\ }\href {\doibase 10.1103/PhysRevB.55.1142} {\bibfield
  {journal} {\bibinfo  {journal} {Phys. Rev. B}\ }\textbf {\bibinfo {volume}
  {55}},\ \bibinfo {pages} {1142} (\bibinfo {year} {1997})}\BibitemShut
  {NoStop}%
\bibitem [{\citenamefont {Schnyder}\ \emph {et~al.}(2008)\citenamefont
  {Schnyder}, \citenamefont {Ryu}, \citenamefont {Furusaki},\ and\
  \citenamefont {Ludwig}}]{Schnyder_etal_2009}%
  \BibitemOpen
  \bibfield  {author} {\bibinfo {author} {\bibfnamefont {A.~P.}\ \bibnamefont
  {Schnyder}}, \bibinfo {author} {\bibfnamefont {S.}~\bibnamefont {Ryu}},
  \bibinfo {author} {\bibfnamefont {A.}~\bibnamefont {Furusaki}}, \ and\
  \bibinfo {author} {\bibfnamefont {A.~W.~W.}\ \bibnamefont {Ludwig}},\ }\href
  {\doibase 10.1103/PhysRevB.78.195125} {\bibfield  {journal} {\bibinfo
  {journal} {Phys. Rev. B}\ }\textbf {\bibinfo {volume} {78}},\ \bibinfo
  {pages} {195125} (\bibinfo {year} {2008})}\BibitemShut {NoStop}%
\bibitem [{\citenamefont {Chiu}\ \emph {et~al.}(2016)\citenamefont {Chiu},
  \citenamefont {Teo}, \citenamefont {Schnyder},\ and\ \citenamefont
  {Ryu}}]{Chiu_etal_2016}%
  \BibitemOpen
  \bibfield  {author} {\bibinfo {author} {\bibfnamefont {C.-K.}\ \bibnamefont
  {Chiu}}, \bibinfo {author} {\bibfnamefont {J.~C.~Y.}\ \bibnamefont {Teo}},
  \bibinfo {author} {\bibfnamefont {A.~P.}\ \bibnamefont {Schnyder}}, \ and\
  \bibinfo {author} {\bibfnamefont {S.}~\bibnamefont {Ryu}},\ }\href {\doibase
  10.1103/RevModPhys.88.035005} {\bibfield  {journal} {\bibinfo  {journal}
  {Rev. Mod. Phys.}\ }\textbf {\bibinfo {volume} {88}},\ \bibinfo {pages}
  {035005} (\bibinfo {year} {2016})}\BibitemShut {NoStop}%
\bibitem [{\citenamefont {Zirnbauer}(1986)}]{Zirnbauer_1986}%
  \BibitemOpen
  \bibfield  {author} {\bibinfo {author} {\bibfnamefont {M.~R.}\ \bibnamefont
  {Zirnbauer}},\ }\href {\doibase 10.1103/PhysRevB.34.6394} {\bibfield
  {journal} {\bibinfo  {journal} {Phys. Rev. B}\ }\textbf {\bibinfo {volume}
  {34}},\ \bibinfo {pages} {6394} (\bibinfo {year} {1986})}\BibitemShut
  {NoStop}%
\bibitem [{\citenamefont {Efetov}(1987)}]{Efetov_1987}%
  \BibitemOpen
  \bibfield  {author} {\bibinfo {author} {\bibfnamefont {K.~B.}\ \bibnamefont
  {Efetov}},\ }\href {http://www.jetp.ac.ru/cgi-bin/e/index/e/65/2/p360?a=list}
  {\bibfield  {journal} {\bibinfo  {journal} {JETP}\ }\textbf {\bibinfo
  {volume} {65}},\ \bibinfo {pages} {360} (\bibinfo {year} {1987})}\BibitemShut
  {NoStop}%
\bibitem [{\citenamefont {Mirlin}\ and\ \citenamefont
  {Fyodorov}(1994)}]{Mirlin_Fyodorov_1994}%
  \BibitemOpen
  \bibfield  {author} {\bibinfo {author} {\bibfnamefont {A.~D.}\ \bibnamefont
  {Mirlin}}\ and\ \bibinfo {author} {\bibfnamefont {Y.~V.}\ \bibnamefont
  {Fyodorov}},\ }\href {\doibase 10.1103/PhysRevLett.72.526} {\bibfield
  {journal} {\bibinfo  {journal} {Phys. Rev. Lett.}\ }\textbf {\bibinfo
  {volume} {72}},\ \bibinfo {pages} {526} (\bibinfo {year} {1994})}\BibitemShut
  {NoStop}%
\bibitem [{\citenamefont {Schubert}\ \emph {et~al.}(2010)\citenamefont
  {Schubert}, \citenamefont {Schleede}, \citenamefont {Byczuk}, \citenamefont
  {Fehske},\ and\ \citenamefont {Vollhardt}}]{g_schubert_10}%
  \BibitemOpen
  \bibfield  {author} {\bibinfo {author} {\bibfnamefont {G.}~\bibnamefont
  {Schubert}}, \bibinfo {author} {\bibfnamefont {J.}~\bibnamefont {Schleede}},
  \bibinfo {author} {\bibfnamefont {K.}~\bibnamefont {Byczuk}}, \bibinfo
  {author} {\bibfnamefont {H.}~\bibnamefont {Fehske}}, \ and\ \bibinfo {author}
  {\bibfnamefont {D.}~\bibnamefont {Vollhardt}},\ }\href {\doibase
  10.1103/PhysRevB.81.155106} {\bibfield  {journal} {\bibinfo  {journal} {Phys.
  Rev. B}\ }\textbf {\bibinfo {volume} {81}},\ \bibinfo {pages} {155106}
  (\bibinfo {year} {2010})}\BibitemShut {NoStop}%
\bibitem [{\citenamefont {Thomas}(1927)}]{Thomas_1927}%
  \BibitemOpen
  \bibfield  {author} {\bibinfo {author} {\bibfnamefont {L.~H.}\ \bibnamefont
  {Thomas}},\ }\href {\doibase 10.1017/S0305004100011683} {\bibfield  {journal}
  {\bibinfo  {journal} {Math. Proc. Camb. Philos. Soc.}\ }\textbf {\bibinfo
  {volume} {23}},\ \bibinfo {pages} {542} (\bibinfo {year} {1927})}\BibitemShut
  {NoStop}%
\bibitem [{\citenamefont {Fermi}(1928)}]{Fermi_1928}%
  \BibitemOpen
  \bibfield  {author} {\bibinfo {author} {\bibfnamefont {E.}~\bibnamefont
  {Fermi}},\ }\href {\doibase 10.1007/BF01351576} {\bibfield  {journal}
  {\bibinfo  {journal} {Z. Phys.}\ }\textbf {\bibinfo {volume} {48}},\ \bibinfo
  {pages} {73} (\bibinfo {year} {1928})}\BibitemShut {NoStop}%
\bibitem [{\citenamefont {Dirac}(1930)}]{Dirac_1930}%
  \BibitemOpen
  \bibfield  {author} {\bibinfo {author} {\bibfnamefont {P.~A.~M.}\
  \bibnamefont {Dirac}},\ }\href {\doibase 10.1017/S0305004100016108}
  {\bibfield  {journal} {\bibinfo  {journal} {Math. Proc. Camb. Philos. Soc.}\
  }\textbf {\bibinfo {volume} {26}},\ \bibinfo {pages} {376–385} (\bibinfo
  {year} {1930})}\BibitemShut {NoStop}%
\bibitem [{\citenamefont {Ibach}\ and\ \citenamefont
  {L{\"u}th}(2009)}]{h_ibach_09}%
  \BibitemOpen
  \bibfield  {author} {\bibinfo {author} {\bibfnamefont {H.}~\bibnamefont
  {Ibach}}\ and\ \bibinfo {author} {\bibfnamefont {H.}~\bibnamefont
  {L{\"u}th}},\ }\href {https://books.google.com/books?id=qjxv68JFe3gC} {\emph
  {\bibinfo {title} {Solid-State Physics: An Introduction to Principles of
  Materials Science}}},\ Advanced texts in physics\ (\bibinfo  {publisher}
  {Springer Berlin Heidelberg},\ \bibinfo {year} {2009})\BibitemShut {NoStop}%
\bibitem [{\citenamefont {Mott}(1949)}]{n_mott_49}%
  \BibitemOpen
  \bibfield  {author} {\bibinfo {author} {\bibfnamefont {N.~F.}\ \bibnamefont
  {Mott}},\ }\href@noop {} {\bibfield  {journal} {\bibinfo  {journal} {Proc.
  Phys. Soc., Sect. A}\ }\textbf {\bibinfo {volume} {62}},\ \bibinfo {pages}
  {416} (\bibinfo {year} {1949})}\BibitemShut {NoStop}%
\bibitem [{\citenamefont {Li}\ \emph {et~al.}(2006)\citenamefont {Li},
  \citenamefont {Luo},\ and\ \citenamefont {Kr{\"o}ger}}]{y_li_06}%
  \BibitemOpen
  \bibfield  {author} {\bibinfo {author} {\bibfnamefont {Y.}~\bibnamefont
  {Li}}, \bibinfo {author} {\bibfnamefont {X.}~\bibnamefont {Luo}}, \ and\
  \bibinfo {author} {\bibfnamefont {H.}~\bibnamefont {Kr{\"o}ger}},\ }\href
  {\doibase 10.1007/s11433-004-0020-5} {\bibfield  {journal} {\bibinfo
  {journal} {Sci. China, Ser. G}\ }\textbf {\bibinfo {volume} {49}},\ \bibinfo
  {pages} {60} (\bibinfo {year} {2006})}\BibitemShut {NoStop}%
\bibitem [{\citenamefont {{Pergament}}\ \emph {et~al.}(2014)\citenamefont
  {{Pergament}}, \citenamefont {{Stefanovich}},\ and\ \citenamefont
  {{Markova}}}]{a_pergament_14}%
  \BibitemOpen
  \bibfield  {author} {\bibinfo {author} {\bibfnamefont {A.}~\bibnamefont
  {{Pergament}}}, \bibinfo {author} {\bibfnamefont {G.}~\bibnamefont
  {{Stefanovich}}}, \ and\ \bibinfo {author} {\bibfnamefont {N.}~\bibnamefont
  {{Markova}}},\ }\href@noop {} {\bibfield  {journal} {\bibinfo  {journal}
  {ArXiv e-prints}\ } (\bibinfo {year} {2014})},\ \Eprint
  {http://arxiv.org/abs/1411.4372} {arXiv:1411.4372 [cond-mat.str-el]}
  \BibitemShut {NoStop}%
\bibitem [{\citenamefont {Kravchenko}\ \emph {et~al.}(1995)\citenamefont
  {Kravchenko}, \citenamefont {Mason}, \citenamefont {Bowker}, \citenamefont
  {Furneaux}, \citenamefont {Pudalov},\ and\ \citenamefont
  {D'Iorio}}]{Kravchenko_etal_1995}%
  \BibitemOpen
  \bibfield  {author} {\bibinfo {author} {\bibfnamefont {S.~V.}\ \bibnamefont
  {Kravchenko}}, \bibinfo {author} {\bibfnamefont {W.~E.}\ \bibnamefont
  {Mason}}, \bibinfo {author} {\bibfnamefont {G.~E.}\ \bibnamefont {Bowker}},
  \bibinfo {author} {\bibfnamefont {J.~E.}\ \bibnamefont {Furneaux}}, \bibinfo
  {author} {\bibfnamefont {V.~M.}\ \bibnamefont {Pudalov}}, \ and\ \bibinfo
  {author} {\bibfnamefont {M.}~\bibnamefont {D'Iorio}},\ }\href {\doibase
  10.1103/PhysRevB.51.7038} {\bibfield  {journal} {\bibinfo  {journal} {Phys.
  Rev. B}\ }\textbf {\bibinfo {volume} {51}},\ \bibinfo {pages} {7038}
  (\bibinfo {year} {1995})}\BibitemShut {NoStop}%
\bibitem [{\citenamefont {Kravchenko}\ \emph {et~al.}(1994)\citenamefont
  {Kravchenko}, \citenamefont {Kravchenko}, \citenamefont {Furneaux},
  \citenamefont {Pudalov},\ and\ \citenamefont
  {D'Iorio}}]{Kravchenko_etal_1994}%
  \BibitemOpen
  \bibfield  {author} {\bibinfo {author} {\bibfnamefont {S.~V.}\ \bibnamefont
  {Kravchenko}}, \bibinfo {author} {\bibfnamefont {G.~V.}\ \bibnamefont
  {Kravchenko}}, \bibinfo {author} {\bibfnamefont {J.~E.}\ \bibnamefont
  {Furneaux}}, \bibinfo {author} {\bibfnamefont {V.~M.}\ \bibnamefont
  {Pudalov}}, \ and\ \bibinfo {author} {\bibfnamefont {M.}~\bibnamefont
  {D'Iorio}},\ }\href {\doibase 10.1103/PhysRevB.50.8039} {\bibfield  {journal}
  {\bibinfo  {journal} {Phys. Rev. B}\ }\textbf {\bibinfo {volume} {50}},\
  \bibinfo {pages} {8039} (\bibinfo {year} {1994})}\BibitemShut {NoStop}%
\bibitem [{\citenamefont {Radonji\ifmmode~\acute{c}\else \'{c}\fi{}}\ \emph
  {et~al.}(2010)\citenamefont {Radonji\ifmmode~\acute{c}\else \'{c}\fi{}},
  \citenamefont {Tanaskovi\ifmmode~\acute{c}\else \'{c}\fi{}}, \citenamefont
  {Dobrosavljevi\ifmmode~\acute{c}\else \'{c}\fi{}},\ and\ \citenamefont
  {Haule}}]{dobrosavljevic_mosfet}%
  \BibitemOpen
  \bibfield  {author} {\bibinfo {author} {\bibfnamefont {M.~c. v.~M.}\
  \bibnamefont {Radonji\ifmmode~\acute{c}\else \'{c}\fi{}}}, \bibinfo {author}
  {\bibfnamefont {D.}~\bibnamefont {Tanaskovi\ifmmode~\acute{c}\else
  \'{c}\fi{}}}, \bibinfo {author} {\bibfnamefont {V.}~\bibnamefont
  {Dobrosavljevi\ifmmode~\acute{c}\else \'{c}\fi{}}}, \ and\ \bibinfo {author}
  {\bibfnamefont {K.}~\bibnamefont {Haule}},\ }\href {\doibase
  10.1103/PhysRevB.81.075118} {\bibfield  {journal} {\bibinfo  {journal} {Phys.
  Rev. B}\ }\textbf {\bibinfo {volume} {81}},\ \bibinfo {pages} {075118}
  (\bibinfo {year} {2010})}\BibitemShut {NoStop}%
\bibitem [{\citenamefont {Radonji\ifmmode~\acute{c}\else \'{c}\fi{}}\ \emph
  {et~al.}(2012)\citenamefont {Radonji\ifmmode~\acute{c}\else \'{c}\fi{}},
  \citenamefont {Tanaskovi\ifmmode~\acute{c}\else \'{c}\fi{}}, \citenamefont
  {Dobrosavljevi\ifmmode~\acute{c}\else \'{c}\fi{}}, \citenamefont {Haule},\
  and\ \citenamefont {Kotliar}}]{wigner_mott}%
  \BibitemOpen
  \bibfield  {author} {\bibinfo {author} {\bibfnamefont {M.~M.}\ \bibnamefont
  {Radonji\ifmmode~\acute{c}\else \'{c}\fi{}}}, \bibinfo {author}
  {\bibfnamefont {D.}~\bibnamefont {Tanaskovi\ifmmode~\acute{c}\else
  \'{c}\fi{}}}, \bibinfo {author} {\bibfnamefont {V.}~\bibnamefont
  {Dobrosavljevi\ifmmode~\acute{c}\else \'{c}\fi{}}}, \bibinfo {author}
  {\bibfnamefont {K.}~\bibnamefont {Haule}}, \ and\ \bibinfo {author}
  {\bibfnamefont {G.}~\bibnamefont {Kotliar}},\ }\href {\doibase
  10.1103/PhysRevB.85.085133} {\bibfield  {journal} {\bibinfo  {journal} {Phys.
  Rev. B}\ }\textbf {\bibinfo {volume} {85}},\ \bibinfo {pages} {085133}
  (\bibinfo {year} {2012})}\BibitemShut {NoStop}%
\bibitem [{\citenamefont
  {Finkel'stein}(1984{\natexlab{a}})}]{Finkelstein_1984a}%
  \BibitemOpen
  \bibfield  {author} {\bibinfo {author} {\bibfnamefont {A.~M.}\ \bibnamefont
  {Finkel'stein}},\ }\href@noop {} {\bibfield  {journal} {\bibinfo  {journal}
  {JETP Lett.}\ }\textbf {\bibinfo {volume} {40}},\ \bibinfo {pages} {796}
  (\bibinfo {year} {1984}{\natexlab{a}})}\BibitemShut {NoStop}%
\bibitem [{\citenamefont
  {Finkel'stein}(1984{\natexlab{b}})}]{Finkelstein_1984b}%
  \BibitemOpen
  \bibfield  {author} {\bibinfo {author} {\bibfnamefont {A.~M.}\ \bibnamefont
  {Finkel'stein}},\ }\href@noop {} {\bibfield  {journal} {\bibinfo  {journal}
  {JETP}\ }\textbf {\bibinfo {volume} {59}},\ \bibinfo {pages} {212} (\bibinfo
  {year} {1984}{\natexlab{b}})}\BibitemShut {NoStop}%
\bibitem [{\citenamefont
  {Finkel'stein}(1984{\natexlab{c}})}]{Finkelstein_1984c}%
  \BibitemOpen
  \bibfield  {author} {\bibinfo {author} {\bibfnamefont {A.~M.}\ \bibnamefont
  {Finkel'stein}},\ }\href@noop {} {\bibfield  {journal} {\bibinfo  {journal}
  {Z. Phys. B Condensed Matter}\ }\textbf {\bibinfo {volume} {56}},\ \bibinfo
  {pages} {189} (\bibinfo {year} {1984}{\natexlab{c}})}\BibitemShut {NoStop}%
\bibitem [{\citenamefont {Finkl'stein}(1983)}]{Finkelstein_1983}%
  \BibitemOpen
  \bibfield  {author} {\bibinfo {author} {\bibfnamefont {A.~M.}\ \bibnamefont
  {Finkl'stein}},\ }\href@noop {} {\bibfield  {journal} {\bibinfo  {journal}
  {JETP}\ }\textbf {\bibinfo {volume} {57}},\ \bibinfo {pages} {97} (\bibinfo
  {year} {1983})}\BibitemShut {NoStop}%
\bibitem [{\citenamefont {Lee}\ and\ \citenamefont
  {Ramakrishnan}(1985)}]{p_lee_85}%
  \BibitemOpen
  \bibfield  {author} {\bibinfo {author} {\bibfnamefont {P.~A.}\ \bibnamefont
  {Lee}}\ and\ \bibinfo {author} {\bibfnamefont {T.~V.}\ \bibnamefont
  {Ramakrishnan}},\ }\href@noop {} {\bibfield  {journal} {\bibinfo  {journal}
  {Rev. Mod. Phys.}\ }\textbf {\bibinfo {volume} {57}},\ \bibinfo {pages} {287}
  (\bibinfo {year} {1985})}\BibitemShut {NoStop}%
\bibitem [{\citenamefont {Belitz}\ and\ \citenamefont
  {Kirkpatrick}(1993)}]{Belitz_Kirkpatrick_1993}%
  \BibitemOpen
  \bibfield  {author} {\bibinfo {author} {\bibfnamefont {D.}~\bibnamefont
  {Belitz}}\ and\ \bibinfo {author} {\bibfnamefont {T.~R.}\ \bibnamefont
  {Kirkpatrick}},\ }\href {\doibase 10.1103/PhysRevB.48.14072} {\bibfield
  {journal} {\bibinfo  {journal} {Phys. Rev. B}\ }\textbf {\bibinfo {volume}
  {48}},\ \bibinfo {pages} {14072} (\bibinfo {year} {1993})}\BibitemShut
  {NoStop}%
\bibitem [{\citenamefont {Castellani}\ \emph {et~al.}(1984)\citenamefont
  {Castellani}, \citenamefont {Di~Castro}, \citenamefont {Lee},\ and\
  \citenamefont {Ma}}]{Castellani_etal_1984a}%
  \BibitemOpen
  \bibfield  {author} {\bibinfo {author} {\bibfnamefont {C.}~\bibnamefont
  {Castellani}}, \bibinfo {author} {\bibfnamefont {C.}~\bibnamefont
  {Di~Castro}}, \bibinfo {author} {\bibfnamefont {P.~A.}\ \bibnamefont {Lee}},
  \ and\ \bibinfo {author} {\bibfnamefont {M.}~\bibnamefont {Ma}},\ }\href
  {\doibase 10.1103/PhysRevB.30.527} {\bibfield  {journal} {\bibinfo  {journal}
  {Phys. Rev. B}\ }\textbf {\bibinfo {volume} {30}},\ \bibinfo {pages} {527}
  (\bibinfo {year} {1984})}\BibitemShut {NoStop}%
\bibitem [{\citenamefont {Aguiar}\ \emph {et~al.}(2009)\citenamefont {Aguiar},
  \citenamefont {Dobrosavljevi\ifmmode~\acute{c}\else \'{c}\fi{}},
  \citenamefont {Abrahams},\ and\ \citenamefont {Kotliar}}]{Vlad_tmt_critical}%
  \BibitemOpen
  \bibfield  {author} {\bibinfo {author} {\bibfnamefont {M.~C.~O.}\
  \bibnamefont {Aguiar}}, \bibinfo {author} {\bibfnamefont {V.}~\bibnamefont
  {Dobrosavljevi\ifmmode~\acute{c}\else \'{c}\fi{}}}, \bibinfo {author}
  {\bibfnamefont {E.}~\bibnamefont {Abrahams}}, \ and\ \bibinfo {author}
  {\bibfnamefont {G.}~\bibnamefont {Kotliar}},\ }\href {\doibase
  10.1103/PhysRevLett.102.156402} {\bibfield  {journal} {\bibinfo  {journal}
  {Phys. Rev. Lett.}\ }\textbf {\bibinfo {volume} {102}},\ \bibinfo {pages}
  {156402} (\bibinfo {year} {2009})}\BibitemShut {NoStop}%
\bibitem [{\citenamefont {Byczuk}\ \emph {et~al.}(2005)\citenamefont {Byczuk},
  \citenamefont {Hofstetter},\ and\ \citenamefont {Vollhardt}}]{k_byczuk_05}%
  \BibitemOpen
  \bibfield  {author} {\bibinfo {author} {\bibfnamefont {K.}~\bibnamefont
  {Byczuk}}, \bibinfo {author} {\bibfnamefont {W.}~\bibnamefont {Hofstetter}},
  \ and\ \bibinfo {author} {\bibfnamefont {D.}~\bibnamefont {Vollhardt}},\
  }\href {\doibase 10.1103/PhysRevLett.94.056404} {\bibfield  {journal}
  {\bibinfo  {journal} {Phys. Rev. Lett.}\ }\textbf {\bibinfo {volume} {94}},\
  \bibinfo {pages} {056404} (\bibinfo {year} {2005})}\BibitemShut {NoStop}%
\bibitem [{\citenamefont {Byczuk}\ \emph {et~al.}(2009)\citenamefont {Byczuk},
  \citenamefont {Hofsletter}, \citenamefont {Yu},\ and\ \citenamefont
  {Vollhardt}}]{k_byczuk_09}%
  \BibitemOpen
  \bibfield  {author} {\bibinfo {author} {\bibfnamefont {K.}~\bibnamefont
  {Byczuk}}, \bibinfo {author} {\bibfnamefont {W.}~\bibnamefont {Hofsletter}},
  \bibinfo {author} {\bibfnamefont {U.}~\bibnamefont {Yu}}, \ and\ \bibinfo
  {author} {\bibfnamefont {D.}~\bibnamefont {Vollhardt}},\ }\href@noop {}
  {\bibfield  {journal} {\bibinfo  {journal} {The European Physics Journal
  Special Topics}\ }\textbf {\bibinfo {volume} {180}},\ \bibinfo {pages} {135}
  (\bibinfo {year} {2009})}\BibitemShut {NoStop}%
\bibitem [{\citenamefont {Derrida}\ \emph {et~al.}(1984)\citenamefont
  {Derrida}, ,\ and\ \citenamefont {Gardner}}]{Derrida_Gardner_1984}%
  \BibitemOpen
  \bibfield  {author} {\bibinfo {author} {\bibfnamefont {B.}~\bibnamefont
  {Derrida}}, , \ and\ \bibinfo {author} {\bibfnamefont {E.}~\bibnamefont
  {Gardner}},\ }\href {\doibase 10.1051/jphys:019840045080128300} {\bibfield
  {journal} {\bibinfo  {journal} {J. Phys. France}\ }\textbf {\bibinfo {volume}
  {45}},\ \bibinfo {pages} {1283} (\bibinfo {year} {1984})}\BibitemShut
  {NoStop}%
\bibitem [{\citenamefont {Oseledets}(1968)}]{Oseledets_1968}%
  \BibitemOpen
  \bibfield  {author} {\bibinfo {author} {\bibfnamefont {V.~I.}\ \bibnamefont
  {Oseledets}},\ }\href
  {http://www.mathnet.ru/php/archive.phtml?wshow=paper&jrnid=mmo&paperid=214&option_lang=eng}
  {\bibfield  {journal} {\bibinfo  {journal} {Tr. Mosk. Mat. Obs.}\ }\textbf
  {\bibinfo {volume} {19}},\ \bibinfo {pages} {179} (\bibinfo {year}
  {1968})}\BibitemShut {NoStop}%
\bibitem [{\citenamefont {Pichard}(1986)}]{Pichard_1986}%
  \BibitemOpen
  \bibfield  {author} {\bibinfo {author} {\bibfnamefont {J.~L.}\ \bibnamefont
  {Pichard}},\ }\href {http://stacks.iop.org/0022-3719/19/i=10/a=009}
  {\bibfield  {journal} {\bibinfo  {journal} {J. Phys. C: Solid State Physics}\
  }\textbf {\bibinfo {volume} {19}},\ \bibinfo {pages} {1519} (\bibinfo {year}
  {1986})}\BibitemShut {NoStop}%
\bibitem [{\citenamefont {Furstenberg}(1963)}]{Furstenberg_1963}%
  \BibitemOpen
  \bibfield  {author} {\bibinfo {author} {\bibfnamefont {H.}~\bibnamefont
  {Furstenberg}},\ }\href {\doibase 10.1090/S0002-9947-1963-0163345-0}
  {\bibfield  {journal} {\bibinfo  {journal} {Trans. Amer. Math. Soc.}\
  }\textbf {\bibinfo {volume} {108}},\ \bibinfo {pages} {377} (\bibinfo {year}
  {1963})}\BibitemShut {NoStop}%
\bibitem [{\citenamefont {MacKinnon}\ and\ \citenamefont
  {Kramer}(1983)}]{MacKinnon_Kramer_1983}%
  \BibitemOpen
  \bibfield  {author} {\bibinfo {author} {\bibfnamefont {A.}~\bibnamefont
  {MacKinnon}}\ and\ \bibinfo {author} {\bibfnamefont {B.}~\bibnamefont
  {Kramer}},\ }\href {\doibase 10.1007/BF01578242} {\bibfield  {journal}
  {\bibinfo  {journal} {Z. Phys. B Condensed Matter}\ }\textbf {\bibinfo
  {volume} {53}},\ \bibinfo {pages} {1} (\bibinfo {year} {1983})}\BibitemShut
  {NoStop}%
\bibitem [{\citenamefont {Furstenberg}\ and\ \citenamefont
  {Kesten}(1960)}]{Furstenberg_Kesten_1960}%
  \BibitemOpen
  \bibfield  {author} {\bibinfo {author} {\bibfnamefont {H.}~\bibnamefont
  {Furstenberg}}\ and\ \bibinfo {author} {\bibfnamefont {H.}~\bibnamefont
  {Kesten}},\ }\href {\doibase doi:10.1214/aoms/1177705909} {\bibfield
  {journal} {\bibinfo  {journal} {Ann. Math. Statist.}\ }\textbf {\bibinfo
  {volume} {31}},\ \bibinfo {pages} {457} (\bibinfo {year} {1960})}\BibitemShut
  {NoStop}%
\bibitem [{\citenamefont {Pichard}\ and\ \citenamefont
  {Sarma}(1981)}]{Pichard_Sarma_1981}%
  \BibitemOpen
  \bibfield  {author} {\bibinfo {author} {\bibfnamefont {J.~L.}\ \bibnamefont
  {Pichard}}\ and\ \bibinfo {author} {\bibfnamefont {G.}~\bibnamefont
  {Sarma}},\ }\href {http://stacks.iop.org/0022-3719/14/i=6/a=003} {\bibfield
  {journal} {\bibinfo  {journal} {J. Phys. C: Solid State Physics}\ }\textbf
  {\bibinfo {volume} {14}},\ \bibinfo {pages} {L127} (\bibinfo {year}
  {1981})}\BibitemShut {NoStop}%
\bibitem [{\citenamefont {Wang}(1994)}]{Wang_1994}%
  \BibitemOpen
  \bibfield  {author} {\bibinfo {author} {\bibfnamefont {L.-W.}\ \bibnamefont
  {Wang}},\ }\href {\doibase 10.1103/PhysRevB.49.10154} {\bibfield  {journal}
  {\bibinfo  {journal} {Phys. Rev. B}\ }\textbf {\bibinfo {volume} {49}},\
  \bibinfo {pages} {10154} (\bibinfo {year} {1994})}\BibitemShut {NoStop}%
\bibitem [{\citenamefont {Silver}\ and\ \citenamefont
  {R\"oder}(1994)}]{Silver_Roder_1994}%
  \BibitemOpen
  \bibfield  {author} {\bibinfo {author} {\bibfnamefont {R.}~\bibnamefont
  {Silver}}\ and\ \bibinfo {author} {\bibfnamefont {H.}~\bibnamefont
  {R\"oder}},\ }\href {\doibase 10.1142/S0129183194000842} {\bibfield
  {journal} {\bibinfo  {journal} {Int. J. Mod. Phys. C}\ }\textbf {\bibinfo
  {volume} {05}},\ \bibinfo {pages} {735} (\bibinfo {year} {1994})}\BibitemShut
  {NoStop}%
\bibitem [{\citenamefont {Silver}\ \emph {et~al.}(1996)\citenamefont {Silver},
  \citenamefont {Roeder}, \citenamefont {Voter},\ and\ \citenamefont
  {Kress}}]{Silver_roeder_Voter_etal_1996}%
  \BibitemOpen
  \bibfield  {author} {\bibinfo {author} {\bibfnamefont {R.}~\bibnamefont
  {Silver}}, \bibinfo {author} {\bibfnamefont {H.}~\bibnamefont {Roeder}},
  \bibinfo {author} {\bibfnamefont {A.}~\bibnamefont {Voter}}, \ and\ \bibinfo
  {author} {\bibfnamefont {J.}~\bibnamefont {Kress}},\ }\href {\doibase
  https://doi.org/10.1006/jcph.1996.0048} {\bibfield  {journal} {\bibinfo
  {journal} {J. Comput. Phys.}\ }\textbf {\bibinfo {volume} {124}},\ \bibinfo
  {pages} {115 } (\bibinfo {year} {1996})}\BibitemShut {NoStop}%
\bibitem [{\citenamefont {Silver}\ and\ \citenamefont
  {R\"oder}(1997)}]{Silver_Roeder_1997}%
  \BibitemOpen
  \bibfield  {author} {\bibinfo {author} {\bibfnamefont {R.~N.}\ \bibnamefont
  {Silver}}\ and\ \bibinfo {author} {\bibfnamefont {H.}~\bibnamefont
  {R\"oder}},\ }\href {\doibase 10.1103/PhysRevE.56.4822} {\bibfield  {journal}
  {\bibinfo  {journal} {Phys. Rev. E}\ }\textbf {\bibinfo {volume} {56}},\
  \bibinfo {pages} {4822} (\bibinfo {year} {1997})}\BibitemShut {NoStop}%
\bibitem [{\citenamefont {Jackson}(1930)}]{Jackson_1930}%
  \BibitemOpen
  \bibfield  {author} {\bibinfo {author} {\bibfnamefont {D.}~\bibnamefont
  {Jackson}},\ }\href {\doibase 10.1002/zamm.19310110117} {\bibfield  {journal}
  {\bibinfo  {journal} {J. Appl. Math. Mech.}\ }\textbf {\bibinfo {volume}
  {11}},\ \bibinfo {pages} {77} (\bibinfo {year} {1930})}\BibitemShut {NoStop}%
\bibitem [{\citenamefont {Lancoz}(1950)}]{Lanczos_1950}%
  \BibitemOpen
  \bibfield  {author} {\bibinfo {author} {\bibfnamefont {C.}~\bibnamefont
  {Lancoz}},\ }\href@noop {} {\bibfield  {journal} {\bibinfo  {journal} {J.
  Res. Nat’l Bur. Std.}\ }\textbf {\bibinfo {volume} {45}},\ \bibinfo {pages}
  {255} (\bibinfo {year} {1950})}\BibitemShut {NoStop}%
\bibitem [{\citenamefont {Arnoldi}(1951)}]{Arnoldi_1951}%
  \BibitemOpen
  \bibfield  {author} {\bibinfo {author} {\bibfnamefont {W.~E.}\ \bibnamefont
  {Arnoldi}},\ }\href@noop {} {\bibfield  {journal} {\bibinfo  {journal}
  {Quart. Appl. Math.}\ }\textbf {\bibinfo {volume} {9}} (\bibinfo {year}
  {1951})}\BibitemShut {NoStop}%
\bibitem [{\citenamefont {Lin}\ and\ \citenamefont
  {Gubernatis}(1993)}]{Lin_etal_1993}%
  \BibitemOpen
  \bibfield  {author} {\bibinfo {author} {\bibfnamefont {H.}~\bibnamefont
  {Lin}}\ and\ \bibinfo {author} {\bibfnamefont {J.}~\bibnamefont
  {Gubernatis}},\ }\href {\doibase 10.1063/1.4823192} {\bibfield  {journal}
  {\bibinfo  {journal} {Comput. Phys.}\ }\textbf {\bibinfo {volume} {7}},\
  \bibinfo {pages} {400} (\bibinfo {year} {1993})}\BibitemShut {NoStop}%
\bibitem [{\citenamefont {Wei{\ss}e}\ and\ \citenamefont
  {Fehske}(2008)}]{Weisse_Fehske_2008}%
  \BibitemOpen
  \bibfield  {author} {\bibinfo {author} {\bibfnamefont {A.}~\bibnamefont
  {Wei{\ss}e}}\ and\ \bibinfo {author} {\bibfnamefont {H.}~\bibnamefont
  {Fehske}},\ }\enquote {\bibinfo {title} {Exact diagonalization techniques},}\
  in\ \href {\doibase 10.1007/978-3-540-74686-7_18} {\emph {\bibinfo
  {booktitle} {Computational Many-Particle Physics}}},\ \bibinfo {editor}
  {edited by\ \bibinfo {editor} {\bibfnamefont {H.}~\bibnamefont {Fehske}},
  \bibinfo {editor} {\bibfnamefont {R.}~\bibnamefont {Schneider}}, \ and\
  \bibinfo {editor} {\bibfnamefont {A.}~\bibnamefont {Wei{\ss}e}}}\ (\bibinfo
  {publisher} {Springer Berlin Heidelberg},\ \bibinfo {address} {Berlin,
  Heidelberg},\ \bibinfo {year} {2008})\ pp.\ \bibinfo {pages}
  {529--544}\BibitemShut {NoStop}%
\bibitem [{\citenamefont {Noack}\ and\ \citenamefont
  {Manmana}(2005)}]{Noack_Manmana_2005}%
  \BibitemOpen
  \bibfield  {author} {\bibinfo {author} {\bibfnamefont {R.~M.}\ \bibnamefont
  {Noack}}\ and\ \bibinfo {author} {\bibfnamefont {S.~R.}\ \bibnamefont
  {Manmana}},\ }\href {\doibase 10.1063/1.2080349} {\bibfield  {journal}
  {\bibinfo  {journal} {AIP Conference Proceedings}\ }\textbf {\bibinfo
  {volume} {789}},\ \bibinfo {pages} {93} (\bibinfo {year} {2005})},\ \Eprint
  {http://arxiv.org/abs/https://aip.scitation.org/doi/pdf/10.1063/1.2080349}
  {https://aip.scitation.org/doi/pdf/10.1063/1.2080349} \BibitemShut {NoStop}%
\bibitem [{\citenamefont {Ericsson}\ and\ \citenamefont
  {Ruhe}(1980)}]{Ericsson_Ruhe_1980}%
  \BibitemOpen
  \bibfield  {author} {\bibinfo {author} {\bibfnamefont {T.}~\bibnamefont
  {Ericsson}}\ and\ \bibinfo {author} {\bibfnamefont {A.}~\bibnamefont
  {Ruhe}},\ }\href {\doibase https://doi.org/10.1090/S0025-5718-1980-0583502-2}
  {\bibfield  {journal} {\bibinfo  {journal} {Math. Comp. 35}\ ,\ \bibinfo
  {pages} {1251}} (\bibinfo {year} {1980})}\BibitemShut {NoStop}%
\bibitem [{\citenamefont {Kawamura}\ \emph {et~al.}(2017)\citenamefont
  {Kawamura}, \citenamefont {Yoshimi}, \citenamefont {Misawa}, \citenamefont
  {Yamaji}, \citenamefont {Todo},\ and\ \citenamefont
  {Kawashima}}]{Kawamura_etal_2017}%
  \BibitemOpen
  \bibfield  {author} {\bibinfo {author} {\bibfnamefont {M.}~\bibnamefont
  {Kawamura}}, \bibinfo {author} {\bibfnamefont {K.}~\bibnamefont {Yoshimi}},
  \bibinfo {author} {\bibfnamefont {T.}~\bibnamefont {Misawa}}, \bibinfo
  {author} {\bibfnamefont {Y.}~\bibnamefont {Yamaji}}, \bibinfo {author}
  {\bibfnamefont {S.}~\bibnamefont {Todo}}, \ and\ \bibinfo {author}
  {\bibfnamefont {N.}~\bibnamefont {Kawashima}},\ }\href {\doibase
  https://doi.org/10.1016/j.cpc.2017.04.006} {\bibfield  {journal} {\bibinfo
  {journal} {Comput. Phys. Commun.}\ }\textbf {\bibinfo {volume} {217}},\
  \bibinfo {pages} {180 } (\bibinfo {year} {2017})}\BibitemShut {NoStop}%
\bibitem [{\citenamefont {Davidson}(1975)}]{Davidson_1975}%
  \BibitemOpen
  \bibfield  {author} {\bibinfo {author} {\bibfnamefont {E.~R.}\ \bibnamefont
  {Davidson}},\ }\href {\doibase https://doi.org/10.1016/0021-9991(75)90065-0}
  {\bibfield  {journal} {\bibinfo  {journal} {J. Comput. Phys.}\ }\textbf
  {\bibinfo {volume} {17}},\ \bibinfo {pages} {87 } (\bibinfo {year}
  {1975})}\BibitemShut {NoStop}%
\bibitem [{\citenamefont {Dupont}\ \emph {et~al.}(1968)\citenamefont {Dupont},
  \citenamefont {Kendall},\ and\ \citenamefont
  {Rachford}}]{Dupont_Kendall_Rachford_1968}%
  \BibitemOpen
  \bibfield  {author} {\bibinfo {author} {\bibfnamefont {T.}~\bibnamefont
  {Dupont}}, \bibinfo {author} {\bibfnamefont {R.}~\bibnamefont {Kendall}}, \
  and\ \bibinfo {author} {\bibfnamefont {H.}~\bibnamefont {Rachford},
  \bibfnamefont {Jr.}},\ }\href {\doibase 10.1137/0705045} {\bibfield
  {journal} {\bibinfo  {journal} {SIAM J. Numer. Anal.}\ }\textbf {\bibinfo
  {volume} {5}},\ \bibinfo {pages} {559} (\bibinfo {year} {1968})}\BibitemShut
  {NoStop}%
\bibitem [{\citenamefont {Meijerink}\ and\ \citenamefont {van~der
  Vorst}(1977)}]{Meijerink_Vorst_1977}%
  \BibitemOpen
  \bibfield  {author} {\bibinfo {author} {\bibfnamefont {J.~A.}\ \bibnamefont
  {Meijerink}}\ and\ \bibinfo {author} {\bibfnamefont {H.~A.}\ \bibnamefont
  {van~der Vorst}},\ }\href {http://www.jstor.org/stable/2005786} {\bibfield
  {journal} {\bibinfo  {journal} {Math. Comput.}\ }\textbf {\bibinfo {volume}
  {31}},\ \bibinfo {pages} {148} (\bibinfo {year} {1977})}\BibitemShut
  {NoStop}%
\bibitem [{\citenamefont {Bollh\"ofer}\ and\ \citenamefont
  {Notay}(2007)}]{Bollhofer_Notay_2007}%
  \BibitemOpen
  \bibfield  {author} {\bibinfo {author} {\bibfnamefont {M.}~\bibnamefont
  {Bollh\"ofer}}\ and\ \bibinfo {author} {\bibfnamefont {Y.}~\bibnamefont
  {Notay}},\ }\href {\doibase https://doi.org/10.1016/j.cpc.2007.08.004}
  {\bibfield  {journal} {\bibinfo  {journal} {Comput. Phys. Commun.}\ }\textbf
  {\bibinfo {volume} {177}},\ \bibinfo {pages} {951 } (\bibinfo {year}
  {2007})}\BibitemShut {NoStop}%
\bibitem [{\citenamefont {Rodriguez}\ \emph {et~al.}(2010)\citenamefont
  {Rodriguez}, \citenamefont {Vasquez}, \citenamefont {Slevin},\ and\
  \citenamefont {R\"omer}}]{Rodriguez_etal_2010}%
  \BibitemOpen
  \bibfield  {author} {\bibinfo {author} {\bibfnamefont {A.}~\bibnamefont
  {Rodriguez}}, \bibinfo {author} {\bibfnamefont {L.~J.}\ \bibnamefont
  {Vasquez}}, \bibinfo {author} {\bibfnamefont {K.}~\bibnamefont {Slevin}}, \
  and\ \bibinfo {author} {\bibfnamefont {R.~A.}\ \bibnamefont {R\"omer}},\
  }\href {\doibase 10.1103/PhysRevLett.105.046403} {\bibfield  {journal}
  {\bibinfo  {journal} {Phys. Rev. Lett.}\ }\textbf {\bibinfo {volume} {105}},\
  \bibinfo {pages} {046403} (\bibinfo {year} {2010})}\BibitemShut {NoStop}%
\bibitem [{\citenamefont {Ujfalusi}\ and\ \citenamefont
  {Varga}(2015)}]{Ujfalusi_Varga_2015}%
  \BibitemOpen
  \bibfield  {author} {\bibinfo {author} {\bibfnamefont {L.}~\bibnamefont
  {Ujfalusi}}\ and\ \bibinfo {author} {\bibfnamefont {I.}~\bibnamefont
  {Varga}},\ }\href {\doibase 10.1103/PhysRevB.91.184206} {\bibfield  {journal}
  {\bibinfo  {journal} {Phys. Rev. B}\ }\textbf {\bibinfo {volume} {91}},\
  \bibinfo {pages} {184206} (\bibinfo {year} {2015})}\BibitemShut {NoStop}%
\bibitem [{\citenamefont {Hubbard}(1963)}]{j_hubbard_63}%
  \BibitemOpen
  \bibfield  {author} {\bibinfo {author} {\bibfnamefont {J.}~\bibnamefont
  {Hubbard}},\ }\href {\doibase 10.1098/rspa.1963.0204} {\bibfield  {journal}
  {\bibinfo  {journal} {Proc. Royal Soc. A}\ }\textbf {\bibinfo {volume}
  {276}},\ \bibinfo {pages} {238} (\bibinfo {year} {1963})}\BibitemShut
  {NoStop}%
\bibitem [{\citenamefont {Anderson}(2006)}]{p_anderson_06}%
  \BibitemOpen
  \bibfield  {author} {\bibinfo {author} {\bibfnamefont {P.~W.}\ \bibnamefont
  {Anderson}},\ }\href@noop {} {\bibfield  {journal} {\bibinfo  {journal} {Low
  Temp. Phys.}\ }\textbf {\bibinfo {volume} {32}},\ \bibinfo {pages} {282}
  (\bibinfo {year} {2006})}\BibitemShut {NoStop}%
\bibitem [{\citenamefont {Metzner}\ and\ \citenamefont
  {Vollhardt}(1989{\natexlab{b}})}]{w_metzner_89b}%
  \BibitemOpen
  \bibfield  {author} {\bibinfo {author} {\bibfnamefont {W.}~\bibnamefont
  {Metzner}}\ and\ \bibinfo {author} {\bibfnamefont {D.}~\bibnamefont
  {Vollhardt}},\ }\href@noop {} {\bibfield  {journal} {\bibinfo  {journal}
  {Phys. Rev. B}\ }\textbf {\bibinfo {volume} {39}},\ \bibinfo {pages} {4462}
  (\bibinfo {year} {1989}{\natexlab{b}})}\BibitemShut {NoStop}%
\bibitem [{\citenamefont {Jarrell}\ \emph {et~al.}(1996)\citenamefont
  {Jarrell}, \citenamefont {Pang}, \citenamefont {Cox},\ and\ \citenamefont
  {Luk}}]{m_jarrell_96b}%
  \BibitemOpen
  \bibfield  {author} {\bibinfo {author} {\bibfnamefont {M.}~\bibnamefont
  {Jarrell}}, \bibinfo {author} {\bibfnamefont {H.}~\bibnamefont {Pang}},
  \bibinfo {author} {\bibfnamefont {D.}~\bibnamefont {Cox}}, \ and\ \bibinfo
  {author} {\bibfnamefont {K.}~\bibnamefont {Luk}},\ }\href@noop {} {\bibfield
  {journal} {\bibinfo  {journal} {Phys. Rev. Lett.}\ }\textbf {\bibinfo
  {volume} {77}},\ \bibinfo {pages} {1612} (\bibinfo {year}
  {1996})}\BibitemShut {NoStop}%
\bibitem [{\citenamefont {Weik}(2001)}]{m_weik_01}%
  \BibitemOpen
  \bibfield  {author} {\bibinfo {author} {\bibfnamefont {M.~H.}\ \bibnamefont
  {Weik}},\ }\enquote {\bibinfo {title} {Nyquist theorem},}\ in\ \href
  {\doibase 10.1007/1-4020-0613-6_12654} {\emph {\bibinfo {booktitle} {Computer
  Science and Communications Dictionary}}}\ (\bibinfo  {publisher} {Springer
  US},\ \bibinfo {address} {Boston, MA},\ \bibinfo {year} {2001})\ pp.\
  \bibinfo {pages} {1127--1127}\BibitemShut {NoStop}%
\bibitem [{\citenamefont {Jarrell}\ \emph {et~al.}(2001)\citenamefont
  {Jarrell}, \citenamefont {Maier}, \citenamefont {Huscroft},\ and\
  \citenamefont {Moukouri}}]{m_jarrell_01c}%
  \BibitemOpen
  \bibfield  {author} {\bibinfo {author} {\bibfnamefont {M.}~\bibnamefont
  {Jarrell}}, \bibinfo {author} {\bibfnamefont {T.}~\bibnamefont {Maier}},
  \bibinfo {author} {\bibfnamefont {C.}~\bibnamefont {Huscroft}}, \ and\
  \bibinfo {author} {\bibfnamefont {S.}~\bibnamefont {Moukouri}},\ }\href
  {\doibase 10.1103/PhysRevB.64.195130} {\bibfield  {journal} {\bibinfo
  {journal} {Phys. Rev. B}\ }\textbf {\bibinfo {volume} {64}},\ \bibinfo
  {pages} {195130} (\bibinfo {year} {2001})}\BibitemShut {NoStop}%
\bibitem [{\citenamefont {Zlatic}\ and\ \citenamefont
  {Horvatic}(1990)}]{v_zlatic_90}%
  \BibitemOpen
  \bibfield  {author} {\bibinfo {author} {\bibfnamefont {V.}~\bibnamefont
  {Zlatic}}\ and\ \bibinfo {author} {\bibfnamefont {B.}~\bibnamefont
  {Horvatic}},\ }\href {\doibase https://doi.org/10.1016/0038-1098(90)90282-G}
  {\bibfield  {journal} {\bibinfo  {journal} {Solid State Commun.}\ }\textbf
  {\bibinfo {volume} {75}},\ \bibinfo {pages} {263 } (\bibinfo {year}
  {1990})}\BibitemShut {NoStop}%
\bibitem [{\citenamefont {Baym}\ and\ \citenamefont
  {Kadanoff}(1961)}]{g_baym_61}%
  \BibitemOpen
  \bibfield  {author} {\bibinfo {author} {\bibfnamefont {G.}~\bibnamefont
  {Baym}}\ and\ \bibinfo {author} {\bibfnamefont {L.~P.}\ \bibnamefont
  {Kadanoff}},\ }\href {\doibase 10.1103/PhysRev.124.287} {\bibfield  {journal}
  {\bibinfo  {journal} {Phys. Rev.}\ }\textbf {\bibinfo {volume} {124}},\
  \bibinfo {pages} {287} (\bibinfo {year} {1961})}\BibitemShut {NoStop}%
\bibitem [{\citenamefont {Baym}(1962)}]{g_baym_62}%
  \BibitemOpen
  \bibfield  {author} {\bibinfo {author} {\bibfnamefont {G.}~\bibnamefont
  {Baym}},\ }\href {\doibase 10.1103/PhysRev.127.1391} {\bibfield  {journal}
  {\bibinfo  {journal} {Phys. Rev.}\ }\textbf {\bibinfo {volume} {127}},\
  \bibinfo {pages} {1391} (\bibinfo {year} {1962})}\BibitemShut {NoStop}%
\bibitem [{\citenamefont {Jarrell}\ and\ \citenamefont
  {Gubernatis}(1996)}]{m_jarrell_96a}%
  \BibitemOpen
  \bibfield  {author} {\bibinfo {author} {\bibfnamefont {M.}~\bibnamefont
  {Jarrell}}\ and\ \bibinfo {author} {\bibfnamefont {J.}~\bibnamefont
  {Gubernatis}},\ }\href@noop {} {\bibfield  {journal} {\bibinfo  {journal}
  {Phys. Rep.}\ }\textbf {\bibinfo {volume} {269}},\ \bibinfo {pages} {133}
  (\bibinfo {year} {1996})}\BibitemShut {NoStop}%
\bibitem [{\citenamefont {Terletska}\ \emph
  {et~al.}(2013{\natexlab{a}})\citenamefont {Terletska}, \citenamefont {Yang},
  \citenamefont {Meng}, \citenamefont {Moreno},\ and\ \citenamefont
  {Jarrell}}]{h_terletska_13}%
  \BibitemOpen
  \bibfield  {author} {\bibinfo {author} {\bibfnamefont {H.}~\bibnamefont
  {Terletska}}, \bibinfo {author} {\bibfnamefont {S.-X.}\ \bibnamefont {Yang}},
  \bibinfo {author} {\bibfnamefont {Z.~Y.}\ \bibnamefont {Meng}}, \bibinfo
  {author} {\bibfnamefont {J.}~\bibnamefont {Moreno}}, \ and\ \bibinfo {author}
  {\bibfnamefont {M.}~\bibnamefont {Jarrell}},\ }\href {\doibase
  10.1103/PhysRevB.87.134208} {\bibfield  {journal} {\bibinfo  {journal} {Phys.
  Rev. B}\ }\textbf {\bibinfo {volume} {87}},\ \bibinfo {pages} {134208}
  (\bibinfo {year} {2013}{\natexlab{a}})}\BibitemShut {NoStop}%
\bibitem [{\citenamefont {Rammer}\ and\ \citenamefont
  {Smith}(1986)}]{j_rammer_86}%
  \BibitemOpen
  \bibfield  {author} {\bibinfo {author} {\bibfnamefont {J.}~\bibnamefont
  {Rammer}}\ and\ \bibinfo {author} {\bibfnamefont {H.}~\bibnamefont {Smith}},\
  }\href {\doibase 10.1103/RevModPhys.58.323} {\bibfield  {journal} {\bibinfo
  {journal} {Rev. Mod. Phys.}\ }\textbf {\bibinfo {volume} {58}},\ \bibinfo
  {pages} {323} (\bibinfo {year} {1986})}\BibitemShut {NoStop}%
\bibitem [{\citenamefont {Keldysh}(1965)}]{Keldysh_1965}%
  \BibitemOpen
  \bibfield  {author} {\bibinfo {author} {\bibfnamefont {L.~V.}\ \bibnamefont
  {Keldysh}},\ }\href@noop {} {\bibfield  {journal} {\bibinfo  {journal}
  {JETP}\ }\textbf {\bibinfo {volume} {20}},\ \bibinfo {pages} {1018} (\bibinfo
  {year} {1965})}\BibitemShut {NoStop}%
\bibitem [{\citenamefont {Wagner}(1991)}]{m_wagner_91}%
  \BibitemOpen
  \bibfield  {author} {\bibinfo {author} {\bibfnamefont {M.}~\bibnamefont
  {Wagner}},\ }\href {\doibase 10.1103/PhysRevB.44.6104} {\bibfield  {journal}
  {\bibinfo  {journal} {Phys. Rev. B}\ }\textbf {\bibinfo {volume} {44}},\
  \bibinfo {pages} {6104} (\bibinfo {year} {1991})}\BibitemShut {NoStop}%
\bibitem [{\citenamefont {Edwards}\ and\ \citenamefont
  {Anderson}(1975)}]{Edwards_Anderson_1975}%
  \BibitemOpen
  \bibfield  {author} {\bibinfo {author} {\bibfnamefont {S.~F.}\ \bibnamefont
  {Edwards}}\ and\ \bibinfo {author} {\bibfnamefont {P.~W.}\ \bibnamefont
  {Anderson}},\ }\href {http://stacks.iop.org/0305-4608/5/i=5/a=017} {\bibfield
   {journal} {\bibinfo  {journal} {J. Phys. F: Metal Physics}\ }\textbf
  {\bibinfo {volume} {5}},\ \bibinfo {pages} {965} (\bibinfo {year}
  {1975})}\BibitemShut {NoStop}%
\bibitem [{\citenamefont {Terletska}\ \emph
  {et~al.}(2013{\natexlab{b}})\citenamefont {Terletska}, \citenamefont {Yang},
  \citenamefont {Meng}, \citenamefont {Moreno},\ and\ \citenamefont
  {Jarrell}}]{h_terletska_df}%
  \BibitemOpen
  \bibfield  {author} {\bibinfo {author} {\bibfnamefont {H.}~\bibnamefont
  {Terletska}}, \bibinfo {author} {\bibfnamefont {S.-X.}\ \bibnamefont {Yang}},
  \bibinfo {author} {\bibfnamefont {Z.~Y.}\ \bibnamefont {Meng}}, \bibinfo
  {author} {\bibfnamefont {J.}~\bibnamefont {Moreno}}, \ and\ \bibinfo {author}
  {\bibfnamefont {M.}~\bibnamefont {Jarrell}},\ }\href {\doibase
  10.1103/PhysRevB.87.134208} {\bibfield  {journal} {\bibinfo  {journal} {Phys.
  Rev. B}\ }\textbf {\bibinfo {volume} {87}},\ \bibinfo {pages} {134208}
  (\bibinfo {year} {2013}{\natexlab{b}})}\BibitemShut {NoStop}%
\bibitem [{\citenamefont {Betts}\ and\ \citenamefont
  {Stewart}(1997)}]{d_betts_97}%
  \BibitemOpen
  \bibfield  {author} {\bibinfo {author} {\bibfnamefont {D.~D.}\ \bibnamefont
  {Betts}}\ and\ \bibinfo {author} {\bibfnamefont {G.~E.}\ \bibnamefont
  {Stewart}},\ }\href@noop {} {\bibfield  {journal} {\bibinfo  {journal} {Can.
  J. Phys.}\ }\textbf {\bibinfo {volume} {75}},\ \bibinfo {pages} {47}
  (\bibinfo {year} {1997})}\BibitemShut {NoStop}%
\bibitem [{\citenamefont {Betts}\ \emph {et~al.}(1999)\citenamefont {Betts},
  \citenamefont {Lin},\ and\ \citenamefont {Flynn}}]{d_betts_99a}%
  \BibitemOpen
  \bibfield  {author} {\bibinfo {author} {\bibfnamefont {D.~D.}\ \bibnamefont
  {Betts}}, \bibinfo {author} {\bibfnamefont {H.~Q.}\ \bibnamefont {Lin}}, \
  and\ \bibinfo {author} {\bibfnamefont {J.~S.}\ \bibnamefont {Flynn}},\
  }\href@noop {} {\bibfield  {journal} {\bibinfo  {journal} {Can. J. Phys.}\
  }\textbf {\bibinfo {volume} {77}},\ \bibinfo {pages} {353} (\bibinfo {year}
  {1999})}\BibitemShut {NoStop}%
\bibitem [{\citenamefont {Kent}\ \emph {et~al.}(2005)\citenamefont {Kent},
  \citenamefont {Jarrell}, \citenamefont {Maier},\ and\ \citenamefont
  {Pruschke}}]{p_kent_05}%
  \BibitemOpen
  \bibfield  {author} {\bibinfo {author} {\bibfnamefont {P.~R.~C.}\
  \bibnamefont {Kent}}, \bibinfo {author} {\bibfnamefont {M.}~\bibnamefont
  {Jarrell}}, \bibinfo {author} {\bibfnamefont {T.~A.}\ \bibnamefont {Maier}},
  \ and\ \bibinfo {author} {\bibfnamefont {T.}~\bibnamefont {Pruschke}},\
  }\href@noop {} {\bibfield  {journal} {\bibinfo  {journal} {Phys. Rev. B}\
  }\textbf {\bibinfo {volume} {72}},\ \bibinfo {pages} {060411} (\bibinfo
  {year} {2005})}\BibitemShut {NoStop}%
\bibitem [{\citenamefont {Ekuma}\ \emph
  {et~al.}(2014{\natexlab{a}})\citenamefont {Ekuma}, \citenamefont {Terletska},
  \citenamefont {Meng}, \citenamefont {Moreno}, \citenamefont {Jarrell},
  \citenamefont {Mahmoudian},\ and\ \citenamefont
  {Dobrosavljevic}}]{c_ekuma_14a}%
  \BibitemOpen
  \bibfield  {author} {\bibinfo {author} {\bibfnamefont {C.~E.}\ \bibnamefont
  {Ekuma}}, \bibinfo {author} {\bibfnamefont {H.}~\bibnamefont {Terletska}},
  \bibinfo {author} {\bibfnamefont {Z.~Y.}\ \bibnamefont {Meng}}, \bibinfo
  {author} {\bibfnamefont {J.}~\bibnamefont {Moreno}}, \bibinfo {author}
  {\bibfnamefont {M.}~\bibnamefont {Jarrell}}, \bibinfo {author} {\bibfnamefont
  {S.}~\bibnamefont {Mahmoudian}}, \ and\ \bibinfo {author} {\bibfnamefont
  {V.}~\bibnamefont {Dobrosavljevic}},\ }\href@noop {} {\bibfield  {journal}
  {\bibinfo  {journal} {J. Phys. Condens. Matter}\ }\textbf {\bibinfo {volume}
  {26}},\ \bibinfo {pages} {274209} (\bibinfo {year}
  {2014}{\natexlab{a}})}\BibitemShut {NoStop}%
\bibitem [{\citenamefont {Zhang}\ \emph {et~al.}(2016)\citenamefont {Zhang},
  \citenamefont {Nelson}, \citenamefont {Siddiqui}, \citenamefont {Tam},
  \citenamefont {Yu}, \citenamefont {Berlijn}, \citenamefont {Ku},
  \citenamefont {Vidhyadhiraja}, \citenamefont {Moreno},\ and\ \citenamefont
  {Jarrell}}]{y_zhang_16}%
  \BibitemOpen
  \bibfield  {author} {\bibinfo {author} {\bibfnamefont {Y.}~\bibnamefont
  {Zhang}}, \bibinfo {author} {\bibfnamefont {R.}~\bibnamefont {Nelson}},
  \bibinfo {author} {\bibfnamefont {E.}~\bibnamefont {Siddiqui}}, \bibinfo
  {author} {\bibfnamefont {K.-M.}\ \bibnamefont {Tam}}, \bibinfo {author}
  {\bibfnamefont {U.}~\bibnamefont {Yu}}, \bibinfo {author} {\bibfnamefont
  {T.}~\bibnamefont {Berlijn}}, \bibinfo {author} {\bibfnamefont
  {W.}~\bibnamefont {Ku}}, \bibinfo {author} {\bibfnamefont {N.~S.}\
  \bibnamefont {Vidhyadhiraja}}, \bibinfo {author} {\bibfnamefont
  {J.}~\bibnamefont {Moreno}}, \ and\ \bibinfo {author} {\bibfnamefont
  {M.}~\bibnamefont {Jarrell}},\ }\href {\doibase 10.1103/PhysRevB.94.224208}
  {\bibfield  {journal} {\bibinfo  {journal} {Phys. Rev. B}\ }\textbf {\bibinfo
  {volume} {94}},\ \bibinfo {pages} {224208} (\bibinfo {year}
  {2016})}\BibitemShut {NoStop}%
\bibitem [{\citenamefont {Blackman}\ \emph {et~al.}(1971)\citenamefont
  {Blackman}, \citenamefont {Esterling},\ and\ \citenamefont
  {Berk}}]{Blackman_1917}%
  \BibitemOpen
  \bibfield  {author} {\bibinfo {author} {\bibfnamefont {J.~A.}\ \bibnamefont
  {Blackman}}, \bibinfo {author} {\bibfnamefont {D.~M.}\ \bibnamefont
  {Esterling}}, \ and\ \bibinfo {author} {\bibfnamefont {N.~F.}\ \bibnamefont
  {Berk}},\ }\href {\doibase 10.1103/PhysRevB.4.2412} {\bibfield  {journal}
  {\bibinfo  {journal} {Phys. Rev. B}\ }\textbf {\bibinfo {volume} {4}},\
  \bibinfo {pages} {2412} (\bibinfo {year} {1971})}\BibitemShut {NoStop}%
\bibitem [{\citenamefont {Ekuma}\ \emph
  {et~al.}(2014{\natexlab{b}})\citenamefont {Ekuma}, \citenamefont {Terletska},
  \citenamefont {Tam}, \citenamefont {Meng}, \citenamefont {Moreno},\ and\
  \citenamefont {Jarrell}}]{c_ekuma_14b}%
  \BibitemOpen
  \bibfield  {author} {\bibinfo {author} {\bibfnamefont {C.~E.}\ \bibnamefont
  {Ekuma}}, \bibinfo {author} {\bibfnamefont {H.}~\bibnamefont {Terletska}},
  \bibinfo {author} {\bibfnamefont {K.-M.}\ \bibnamefont {Tam}}, \bibinfo
  {author} {\bibfnamefont {Z.-Y.}\ \bibnamefont {Meng}}, \bibinfo {author}
  {\bibfnamefont {J.}~\bibnamefont {Moreno}}, \ and\ \bibinfo {author}
  {\bibfnamefont {M.}~\bibnamefont {Jarrell}},\ }\href {\doibase
  10.1103/PhysRevB.89.081107} {\bibfield  {journal} {\bibinfo  {journal} {Phys.
  Rev. B}\ }\textbf {\bibinfo {volume} {89}},\ \bibinfo {pages} {081107}
  (\bibinfo {year} {2014}{\natexlab{b}})}\BibitemShut {NoStop}%
\bibitem [{\citenamefont {Altshuler}\ and\ \citenamefont
  {Aronov}(1979)}]{Altshuler_Aronov_1979}%
  \BibitemOpen
  \bibfield  {author} {\bibinfo {author} {\bibfnamefont {B.}~\bibnamefont
  {Altshuler}}\ and\ \bibinfo {author} {\bibfnamefont {A.}~\bibnamefont
  {Aronov}},\ }\href {\doibase https://doi.org/10.1016/0038-1098(79)90967-0}
  {\bibfield  {journal} {\bibinfo  {journal} {Solid State Commun.}\ }\textbf
  {\bibinfo {volume} {30}},\ \bibinfo {pages} {115 } (\bibinfo {year}
  {1979})}\BibitemShut {NoStop}%
\bibitem [{\citenamefont {Efros}\ and\ \citenamefont
  {Shklovskii}(1975)}]{Efros_Shklovskii_1975}%
  \BibitemOpen
  \bibfield  {author} {\bibinfo {author} {\bibfnamefont {A.~L.}\ \bibnamefont
  {Efros}}\ and\ \bibinfo {author} {\bibfnamefont {B.~I.}\ \bibnamefont
  {Shklovskii}},\ }\href {http://stacks.iop.org/0022-3719/8/i=4/a=003}
  {\bibfield  {journal} {\bibinfo  {journal} {J. Phys. C: Solid State Physics}\
  }\textbf {\bibinfo {volume} {8}},\ \bibinfo {pages} {L49} (\bibinfo {year}
  {1975})}\BibitemShut {NoStop}%
\bibitem [{\citenamefont {Dobrosavljevi\ifmmode~\acute{c}\else \'{c}\fi{}}\
  \emph {et~al.}(1997)\citenamefont {Dobrosavljevi\ifmmode~\acute{c}\else
  \'{c}\fi{}}, \citenamefont {Abrahams}, \citenamefont {Miranda},\ and\
  \citenamefont {Chakravarty}}]{v_dobrosavljevic_97}%
  \BibitemOpen
  \bibfield  {author} {\bibinfo {author} {\bibfnamefont {V.}~\bibnamefont
  {Dobrosavljevi\ifmmode~\acute{c}\else \'{c}\fi{}}}, \bibinfo {author}
  {\bibfnamefont {E.}~\bibnamefont {Abrahams}}, \bibinfo {author}
  {\bibfnamefont {E.}~\bibnamefont {Miranda}}, \ and\ \bibinfo {author}
  {\bibfnamefont {S.}~\bibnamefont {Chakravarty}},\ }\href {\doibase
  10.1103/PhysRevLett.79.455} {\bibfield  {journal} {\bibinfo  {journal} {Phys.
  Rev. Lett.}\ }\textbf {\bibinfo {volume} {79}},\ \bibinfo {pages} {455}
  (\bibinfo {year} {1997})}\BibitemShut {NoStop}%
\bibitem [{\citenamefont {Atland}\ and\ \citenamefont {Simons}(2010)}]{atland}%
  \BibitemOpen
  \bibfield  {author} {\bibinfo {author} {\bibfnamefont {A.}~\bibnamefont
  {Atland}}\ and\ \bibinfo {author} {\bibfnamefont {B.}~\bibnamefont
  {Simons}},\ }\href@noop {} {\emph {\bibinfo {title} {{Condensed Matter Field
  Thory}}}}\ (\bibinfo  {publisher} {Cambridge University Press},\ \bibinfo
  {year} {2010})\BibitemShut {NoStop}%
\bibitem [{\citenamefont {Miranda}\ and\ \citenamefont
  {Dobrosavljevi{\'c}}(2013)}]{e_miranda_12}%
  \BibitemOpen
  \bibfield  {author} {\bibinfo {author} {\bibfnamefont {E.}~\bibnamefont
  {Miranda}}\ and\ \bibinfo {author} {\bibfnamefont {V.}~\bibnamefont
  {Dobrosavljevi{\'c}}},\ }in\ \href@noop {} {\emph {\bibinfo {booktitle}
  {Conductor-Insulator Quantum Phase Transitions}}},\ \bibinfo {editor} {edited
  by\ \bibinfo {editor} {\bibfnamefont {V.}~\bibnamefont {Dobrosavljevic}},
  \bibinfo {editor} {\bibfnamefont {N.}~\bibnamefont {Trivedi}}, \ and\
  \bibinfo {editor} {\bibfnamefont {J.}~\bibnamefont {Valles}}}\ (\bibinfo
  {publisher} {Oxford University Press},\ \bibinfo {year} {2013})\ pp.\
  \bibinfo {pages} {161--243}\BibitemShut {NoStop}%
\bibitem [{\citenamefont {Galpin}\ \emph {et~al.}(2009)\citenamefont {Galpin},
  \citenamefont {Gilbert},\ and\ \citenamefont {Logan}}]{m_galpin_09}%
  \BibitemOpen
  \bibfield  {author} {\bibinfo {author} {\bibfnamefont {M.~R.}\ \bibnamefont
  {Galpin}}, \bibinfo {author} {\bibfnamefont {A.~B.}\ \bibnamefont {Gilbert}},
  \ and\ \bibinfo {author} {\bibfnamefont {D.~E.}\ \bibnamefont {Logan}},\
  }\href {http://stacks.iop.org/0953-8984/21/i=37/a=375602} {\bibfield
  {journal} {\bibinfo  {journal} {J. Phys. Condens. Matter}\ }\textbf {\bibinfo
  {volume} {21}},\ \bibinfo {pages} {375602} (\bibinfo {year}
  {2009})}\BibitemShut {NoStop}%
\bibitem [{\citenamefont {Bulla}\ \emph {et~al.}(2008)\citenamefont {Bulla},
  \citenamefont {Costi},\ and\ \citenamefont {Pruschke}}]{r_bulla_08}%
  \BibitemOpen
  \bibfield  {author} {\bibinfo {author} {\bibfnamefont {R.}~\bibnamefont
  {Bulla}}, \bibinfo {author} {\bibfnamefont {T.~A.}\ \bibnamefont {Costi}}, \
  and\ \bibinfo {author} {\bibfnamefont {T.}~\bibnamefont {Pruschke}},\ }\href
  {\doibase 10.1103/RevModPhys.80.395} {\bibfield  {journal} {\bibinfo
  {journal} {Rev. Mod. Phys.}\ }\textbf {\bibinfo {volume} {80}},\ \bibinfo
  {pages} {395} (\bibinfo {year} {2008})}\BibitemShut {NoStop}%
\bibitem [{\citenamefont {Gull}\ \emph {et~al.}(2011)\citenamefont {Gull},
  \citenamefont {Millis}, \citenamefont {Lichtenstein}, \citenamefont
  {Rubtsov}, \citenamefont {Troyer},\ and\ \citenamefont {Werner}}]{e_gull_11}%
  \BibitemOpen
  \bibfield  {author} {\bibinfo {author} {\bibfnamefont {E.}~\bibnamefont
  {Gull}}, \bibinfo {author} {\bibfnamefont {A.~J.}\ \bibnamefont {Millis}},
  \bibinfo {author} {\bibfnamefont {A.~I.}\ \bibnamefont {Lichtenstein}},
  \bibinfo {author} {\bibfnamefont {A.~N.}\ \bibnamefont {Rubtsov}}, \bibinfo
  {author} {\bibfnamefont {M.}~\bibnamefont {Troyer}}, \ and\ \bibinfo {author}
  {\bibfnamefont {P.}~\bibnamefont {Werner}},\ }\href {\doibase
  10.1103/RevModPhys.83.349} {\bibfield  {journal} {\bibinfo  {journal} {Rev.
  Mod. Phys.}\ }\textbf {\bibinfo {volume} {83}},\ \bibinfo {pages} {349}
  (\bibinfo {year} {2011})}\BibitemShut {NoStop}%
\bibitem [{\citenamefont {Assaad}(2014)}]{Assaad_2014}%
  \BibitemOpen
  \bibfield  {author} {\bibinfo {author} {\bibfnamefont {F.~F.}\ \bibnamefont
  {Assaad}},\ }\enquote {\bibinfo {title} {Exact diagonalization techniques},}\
  in\ \href {https://www.cond-mat.de/events/correl14/manuscripts/} {\emph
  {\bibinfo {booktitle} {DMFT at 25: Infinite Dimensions}}},\ \bibinfo {editor}
  {edited by\ \bibinfo {editor} {\bibfnamefont {E.}~\bibnamefont {Pavarini}},
  \bibinfo {editor} {\bibfnamefont {E.}~\bibnamefont {Koch}}, \bibinfo {editor}
  {\bibfnamefont {D.}~\bibnamefont {Vollhardt}}, \ and\ \bibinfo {editor}
  {\bibfnamefont {A.}~\bibnamefont {Lichtenstein}}}\ (\bibinfo  {publisher}
  {Verlag des Forschungszentrum Jülich},\ \bibinfo {address} {Jurlich},\
  \bibinfo {year} {2014})\BibitemShut {NoStop}%
\bibitem [{\citenamefont {Miranda}\ \emph {et~al.}(1996)\citenamefont
  {Miranda}, \citenamefont {Dobrosavljevi{\'c}},\ and\ \citenamefont
  {Kotliar}}]{Miranda_Kondo}%
  \BibitemOpen
  \bibfield  {author} {\bibinfo {author} {\bibfnamefont {E.}~\bibnamefont
  {Miranda}}, \bibinfo {author} {\bibnamefont {Dobrosavljevi{\'c}}}, \ and\
  \bibinfo {author} {\bibfnamefont {G.}~\bibnamefont {Kotliar}},\ }\href@noop
  {} {\bibfield  {journal} {\bibinfo  {journal} {J. Phys.: Condens. Matter}\
  }\textbf {\bibinfo {volume} {9871}} (\bibinfo {year} {1996})}\BibitemShut
  {NoStop}%
\bibitem [{\citenamefont {Miranda}\ \emph {et~al.}(1997)\citenamefont
  {Miranda}, \citenamefont {Dobrosavljevi\ifmmode~\acute{c}\else \'{c}\fi{}},\
  and\ \citenamefont {Kotliar}}]{Miranda_Kondo_prb}%
  \BibitemOpen
  \bibfield  {author} {\bibinfo {author} {\bibfnamefont {E.}~\bibnamefont
  {Miranda}}, \bibinfo {author} {\bibfnamefont {V.}~\bibnamefont
  {Dobrosavljevi\ifmmode~\acute{c}\else \'{c}\fi{}}}, \ and\ \bibinfo {author}
  {\bibfnamefont {G.}~\bibnamefont {Kotliar}},\ }\href {\doibase
  10.1103/PhysRevLett.78.290} {\bibfield  {journal} {\bibinfo  {journal} {Phys.
  Rev. Lett.}\ }\textbf {\bibinfo {volume} {78}},\ \bibinfo {pages} {290}
  (\bibinfo {year} {1997})}\BibitemShut {NoStop}%
\bibitem [{\citenamefont {Chattopadhyay}\ \emph {et~al.}(1998)\citenamefont
  {Chattopadhyay}, \citenamefont {Jarrell}, \citenamefont {Krishnamurthy},
  \citenamefont {Ng}, \citenamefont {Sarrao},\ and\ \citenamefont
  {Fisk}}]{Fisk_Kondo}%
  \BibitemOpen
  \bibfield  {author} {\bibinfo {author} {\bibfnamefont {A.}~\bibnamefont
  {Chattopadhyay}}, \bibinfo {author} {\bibfnamefont {M.}~\bibnamefont
  {Jarrell}}, \bibinfo {author} {\bibfnamefont {H.~R.}\ \bibnamefont
  {Krishnamurthy}}, \bibinfo {author} {\bibfnamefont {H.~K.}\ \bibnamefont
  {Ng}}, \bibinfo {author} {\bibfnamefont {J.}~\bibnamefont {Sarrao}}, \ and\
  \bibinfo {author} {\bibfnamefont {Z.}~\bibnamefont {Fisk}},\ }\href@noop {}
  {\  (\bibinfo {year} {1998})},\ \Eprint
  {http://arxiv.org/abs/arXiv:cond-mat/9805127} {arXiv:cond-mat/9805127}
  \BibitemShut {NoStop}%
\bibitem [{\citenamefont {Sen}\ \emph {et~al.}(2016)\citenamefont {Sen},
  \citenamefont {Terletska}, \citenamefont {Moreno}, \citenamefont
  {Vidhyadhiraja},\ and\ \citenamefont {Jarrell}}]{s_sen_16a}%
  \BibitemOpen
  \bibfield  {author} {\bibinfo {author} {\bibfnamefont {S.}~\bibnamefont
  {Sen}}, \bibinfo {author} {\bibfnamefont {H.}~\bibnamefont {Terletska}},
  \bibinfo {author} {\bibfnamefont {J.}~\bibnamefont {Moreno}}, \bibinfo
  {author} {\bibfnamefont {N.~S.}\ \bibnamefont {Vidhyadhiraja}}, \ and\
  \bibinfo {author} {\bibfnamefont {M.}~\bibnamefont {Jarrell}},\ }\href
  {\doibase 10.1103/PhysRevB.94.235104} {\bibfield  {journal} {\bibinfo
  {journal} {Phys. Rev. B}\ }\textbf {\bibinfo {volume} {94}},\ \bibinfo
  {pages} {235104} (\bibinfo {year} {2016})}\BibitemShut {NoStop}%
\bibitem [{\citenamefont {Aguiar}\ \emph {et~al.}(2006)\citenamefont {Aguiar},
  \citenamefont {Dobrosavljevi\'{c}}, \citenamefont {Abrahams},\ and\
  \citenamefont {Kotliar}}]{Vlad_scaling_tmt}%
  \BibitemOpen
  \bibfield  {author} {\bibinfo {author} {\bibfnamefont {M.~C.~O.}\
  \bibnamefont {Aguiar}}, \bibinfo {author} {\bibfnamefont {V.}~\bibnamefont
  {Dobrosavljevi\'{c}}}, \bibinfo {author} {\bibfnamefont {E.}~\bibnamefont
  {Abrahams}}, \ and\ \bibinfo {author} {\bibfnamefont {G.}~\bibnamefont
  {Kotliar}},\ }\href {\doibase 10.1103/PhysRevB.73.115117} {\bibfield
  {journal} {\bibinfo  {journal} {Phys. Rev. B}\ }\textbf {\bibinfo {volume}
  {73}},\ \bibinfo {pages} {115117} (\bibinfo {year} {2006})}\BibitemShut
  {NoStop}%
\bibitem [{\citenamefont {Sen}\ \emph {et~al.}(2018)\citenamefont {Sen},
  \citenamefont {Vidhyadhiraja},\ and\ \citenamefont
  {Jarrell}}]{Sen_etal_2018}%
  \BibitemOpen
  \bibfield  {author} {\bibinfo {author} {\bibfnamefont {S.}~\bibnamefont
  {Sen}}, \bibinfo {author} {\bibfnamefont {N.~S.}\ \bibnamefont
  {Vidhyadhiraja}}, \ and\ \bibinfo {author} {\bibfnamefont {M.}~\bibnamefont
  {Jarrell}},\ }\href {\doibase 10.1103/PhysRevB.98.075112} {\bibfield
  {journal} {\bibinfo  {journal} {Phys. Rev. B}\ }\textbf {\bibinfo {volume}
  {98}},\ \bibinfo {pages} {075112} (\bibinfo {year} {2018})}\BibitemShut
  {NoStop}%
\bibitem [{\citenamefont {Zhang}\ \emph {et~al.}(2017)\citenamefont {Zhang},
  \citenamefont {Zhang}, \citenamefont {Yang}, \citenamefont {Tam},
  \citenamefont {Vidhyadhiraja},\ and\ \citenamefont {Jarrell}}]{y_zhang_17}%
  \BibitemOpen
  \bibfield  {author} {\bibinfo {author} {\bibfnamefont {Y.}~\bibnamefont
  {Zhang}}, \bibinfo {author} {\bibfnamefont {Y.~F.}\ \bibnamefont {Zhang}},
  \bibinfo {author} {\bibfnamefont {S.~X.}\ \bibnamefont {Yang}}, \bibinfo
  {author} {\bibfnamefont {K.-M.}\ \bibnamefont {Tam}}, \bibinfo {author}
  {\bibfnamefont {N.~S.}\ \bibnamefont {Vidhyadhiraja}}, \ and\ \bibinfo
  {author} {\bibfnamefont {M.}~\bibnamefont {Jarrell}},\ }\href {\doibase
  10.1103/PhysRevB.95.144208} {\bibfield  {journal} {\bibinfo  {journal} {Phys.
  Rev. B}\ }\textbf {\bibinfo {volume} {95}},\ \bibinfo {pages} {144208}
  (\bibinfo {year} {2017})}\BibitemShut {NoStop}%
\bibitem [{\citenamefont {Terletska}\ \emph {et~al.}(2017)\citenamefont
  {Terletska}, \citenamefont {Zhang}, \citenamefont {Chioncel}, \citenamefont
  {Vollhardt},\ and\ \citenamefont {Jarrell}}]{h_terletska_17}%
  \BibitemOpen
  \bibfield  {author} {\bibinfo {author} {\bibfnamefont {H.}~\bibnamefont
  {Terletska}}, \bibinfo {author} {\bibfnamefont {Y.}~\bibnamefont {Zhang}},
  \bibinfo {author} {\bibfnamefont {L.}~\bibnamefont {Chioncel}}, \bibinfo
  {author} {\bibfnamefont {D.}~\bibnamefont {Vollhardt}}, \ and\ \bibinfo
  {author} {\bibfnamefont {M.}~\bibnamefont {Jarrell}},\ }\href {\doibase
  10.1103/PhysRevB.95.134204} {\bibfield  {journal} {\bibinfo  {journal} {Phys.
  Rev. B}\ }\textbf {\bibinfo {volume} {95}},\ \bibinfo {pages} {134204}
  (\bibinfo {year} {2017})}\BibitemShut {NoStop}%
\bibitem [{\citenamefont {Berlijn}\ \emph {et~al.}(2011)\citenamefont
  {Berlijn}, \citenamefont {Volja},\ and\ \citenamefont {Ku}}]{naxcoo2}%
  \BibitemOpen
  \bibfield  {author} {\bibinfo {author} {\bibfnamefont {T.}~\bibnamefont
  {Berlijn}}, \bibinfo {author} {\bibfnamefont {D.}~\bibnamefont {Volja}}, \
  and\ \bibinfo {author} {\bibfnamefont {W.}~\bibnamefont {Ku}},\ }\href
  {\doibase 10.1103/PhysRevLett.106.077005} {\bibfield  {journal} {\bibinfo
  {journal} {Phys. Rev. Lett.}\ }\textbf {\bibinfo {volume} {106}},\ \bibinfo
  {pages} {077005} (\bibinfo {year} {2011})}\BibitemShut {NoStop}%
\bibitem [{\citenamefont {Soler}\ \emph {et~al.}(2002)\citenamefont {Soler},
  \citenamefont {Artacho}, \citenamefont {Gale}, \citenamefont {García},
  \citenamefont {Junquera}, \citenamefont {Ordejón},\ and\ \citenamefont
  {Sánchez-Portal}}]{soler_2002}%
  \BibitemOpen
  \bibfield  {author} {\bibinfo {author} {\bibfnamefont {J.~M.}\ \bibnamefont
  {Soler}}, \bibinfo {author} {\bibfnamefont {E.}~\bibnamefont {Artacho}},
  \bibinfo {author} {\bibfnamefont {J.~D.}\ \bibnamefont {Gale}}, \bibinfo
  {author} {\bibfnamefont {A.}~\bibnamefont {García}}, \bibinfo {author}
  {\bibfnamefont {J.}~\bibnamefont {Junquera}}, \bibinfo {author}
  {\bibfnamefont {P.}~\bibnamefont {Ordejón}}, \ and\ \bibinfo {author}
  {\bibfnamefont {D.}~\bibnamefont {Sánchez-Portal}},\ }\href
  {http://stacks.iop.org/0953-8984/14/i=11/a=302} {\bibfield  {journal}
  {\bibinfo  {journal} {J. Phys. Condens. Matter}\ }\textbf {\bibinfo {volume}
  {14}},\ \bibinfo {pages} {2745} (\bibinfo {year} {2002})}\BibitemShut
  {NoStop}%
\bibitem [{\citenamefont {Junquera}\ \emph {et~al.}(2001)\citenamefont
  {Junquera}, \citenamefont {Paz}, \citenamefont {S\'anchez-Portal},\ and\
  \citenamefont {Artacho}}]{nao_2001}%
  \BibitemOpen
  \bibfield  {author} {\bibinfo {author} {\bibfnamefont {J.}~\bibnamefont
  {Junquera}}, \bibinfo {author} {\bibfnamefont {O.}~\bibnamefont {Paz}},
  \bibinfo {author} {\bibfnamefont {D.}~\bibnamefont {S\'anchez-Portal}}, \
  and\ \bibinfo {author} {\bibfnamefont {E.}~\bibnamefont {Artacho}},\ }\href
  {\doibase 10.1103/PhysRevB.64.235111} {\bibfield  {journal} {\bibinfo
  {journal} {Phys. Rev. B}\ }\textbf {\bibinfo {volume} {64}},\ \bibinfo
  {pages} {235111} (\bibinfo {year} {2001})}\BibitemShut {NoStop}%
\bibitem [{\citenamefont {Blum}\ \emph {et~al.}(2009)\citenamefont {Blum},
  \citenamefont {Gehrke}, \citenamefont {Hanke}, \citenamefont {Havu},
  \citenamefont {Havu}, \citenamefont {Ren}, \citenamefont {Reuter},\ and\
  \citenamefont {Scheffler}}]{blum_2009}%
  \BibitemOpen
  \bibfield  {author} {\bibinfo {author} {\bibfnamefont {V.}~\bibnamefont
  {Blum}}, \bibinfo {author} {\bibfnamefont {R.}~\bibnamefont {Gehrke}},
  \bibinfo {author} {\bibfnamefont {F.}~\bibnamefont {Hanke}}, \bibinfo
  {author} {\bibfnamefont {P.}~\bibnamefont {Havu}}, \bibinfo {author}
  {\bibfnamefont {V.}~\bibnamefont {Havu}}, \bibinfo {author} {\bibfnamefont
  {X.}~\bibnamefont {Ren}}, \bibinfo {author} {\bibfnamefont {K.}~\bibnamefont
  {Reuter}}, \ and\ \bibinfo {author} {\bibfnamefont {M.}~\bibnamefont
  {Scheffler}},\ }\href {\doibase https://doi.org/10.1016/j.cpc.2009.06.022}
  {\bibfield  {journal} {\bibinfo  {journal} {Comput. Phys. Commun.}\ }\textbf
  {\bibinfo {volume} {180}},\ \bibinfo {pages} {2175 } (\bibinfo {year}
  {2009})}\BibitemShut {NoStop}%
\bibitem [{\citenamefont {Koskinen}\ and\ \citenamefont
  {Mäkinen}(2009)}]{dftb_2009}%
  \BibitemOpen
  \bibfield  {author} {\bibinfo {author} {\bibfnamefont {P.}~\bibnamefont
  {Koskinen}}\ and\ \bibinfo {author} {\bibfnamefont {V.}~\bibnamefont
  {Mäkinen}},\ }\href {\doibase
  https://doi.org/10.1016/j.commatsci.2009.07.013} {\bibfield  {journal}
  {\bibinfo  {journal} {Comput. Mater. Sci.}\ }\textbf {\bibinfo {volume}
  {47}},\ \bibinfo {pages} {237 } (\bibinfo {year} {2009})}\BibitemShut
  {NoStop}%
\bibitem [{\citenamefont {Van~der Ven}\ \emph {et~al.}(1998)\citenamefont
  {Van~der Ven}, \citenamefont {Aydinol}, \citenamefont {Ceder}, \citenamefont
  {Kresse},\ and\ \citenamefont {Hafner}}]{ceder_1998}%
  \BibitemOpen
  \bibfield  {author} {\bibinfo {author} {\bibfnamefont {A.}~\bibnamefont
  {Van~der Ven}}, \bibinfo {author} {\bibfnamefont {M.~K.}\ \bibnamefont
  {Aydinol}}, \bibinfo {author} {\bibfnamefont {G.}~\bibnamefont {Ceder}},
  \bibinfo {author} {\bibfnamefont {G.}~\bibnamefont {Kresse}}, \ and\ \bibinfo
  {author} {\bibfnamefont {J.}~\bibnamefont {Hafner}},\ }\href {\doibase
  10.1103/PhysRevB.58.2975} {\bibfield  {journal} {\bibinfo  {journal} {Phys.
  Rev. B}\ }\textbf {\bibinfo {volume} {58}},\ \bibinfo {pages} {2975}
  (\bibinfo {year} {1998})}\BibitemShut {NoStop}%
\bibitem [{\citenamefont {Berlijn}\ \emph
  {et~al.}(2012{\natexlab{a}})\citenamefont {Berlijn}, \citenamefont {Lin},
  \citenamefont {Garber},\ and\ \citenamefont {Ku}}]{tm122}%
  \BibitemOpen
  \bibfield  {author} {\bibinfo {author} {\bibfnamefont {T.}~\bibnamefont
  {Berlijn}}, \bibinfo {author} {\bibfnamefont {C.-H.}\ \bibnamefont {Lin}},
  \bibinfo {author} {\bibfnamefont {W.}~\bibnamefont {Garber}}, \ and\ \bibinfo
  {author} {\bibfnamefont {W.}~\bibnamefont {Ku}},\ }\href {\doibase
  10.1103/PhysRevLett.108.207003} {\bibfield  {journal} {\bibinfo  {journal}
  {Phys. Rev. Lett.}\ }\textbf {\bibinfo {volume} {108}},\ \bibinfo {pages}
  {207003} (\bibinfo {year} {2012}{\natexlab{a}})}\BibitemShut {NoStop}%
\bibitem [{\citenamefont {Berlijn}\ \emph
  {et~al.}(2012{\natexlab{b}})\citenamefont {Berlijn}, \citenamefont
  {Hirschfeld},\ and\ \citenamefont {Ku}}]{kfe2se2}%
  \BibitemOpen
  \bibfield  {author} {\bibinfo {author} {\bibfnamefont {T.}~\bibnamefont
  {Berlijn}}, \bibinfo {author} {\bibfnamefont {P.~J.}\ \bibnamefont
  {Hirschfeld}}, \ and\ \bibinfo {author} {\bibfnamefont {W.}~\bibnamefont
  {Ku}},\ }\href {\doibase 10.1103/PhysRevLett.109.147003} {\bibfield
  {journal} {\bibinfo  {journal} {Phys. Rev. Lett.}\ }\textbf {\bibinfo
  {volume} {109}},\ \bibinfo {pages} {147003} (\bibinfo {year}
  {2012}{\natexlab{b}})}\BibitemShut {NoStop}%
\bibitem [{\citenamefont {Wang}\ \emph {et~al.}(2013)\citenamefont {Wang},
  \citenamefont {Berlijn}, \citenamefont {Wang}, \citenamefont {Lin},
  \citenamefont {Hirschfeld},\ and\ \citenamefont {Ku}}]{ru122}%
  \BibitemOpen
  \bibfield  {author} {\bibinfo {author} {\bibfnamefont {L.}~\bibnamefont
  {Wang}}, \bibinfo {author} {\bibfnamefont {T.}~\bibnamefont {Berlijn}},
  \bibinfo {author} {\bibfnamefont {Y.}~\bibnamefont {Wang}}, \bibinfo {author}
  {\bibfnamefont {C.-H.}\ \bibnamefont {Lin}}, \bibinfo {author} {\bibfnamefont
  {P.~J.}\ \bibnamefont {Hirschfeld}}, \ and\ \bibinfo {author} {\bibfnamefont
  {W.}~\bibnamefont {Ku}},\ }\href {\doibase 10.1103/PhysRevLett.110.037001}
  {\bibfield  {journal} {\bibinfo  {journal} {Phys. Rev. Lett.}\ }\textbf
  {\bibinfo {volume} {110}},\ \bibinfo {pages} {037001} (\bibinfo {year}
  {2013})}\BibitemShut {NoStop}%
\bibitem [{\citenamefont {Berlijn}\ \emph {et~al.}(2014)\citenamefont
  {Berlijn}, \citenamefont {Cheng}, \citenamefont {Hirschfeld},\ and\
  \citenamefont {Ku}}]{11sto}%
  \BibitemOpen
  \bibfield  {author} {\bibinfo {author} {\bibfnamefont {T.}~\bibnamefont
  {Berlijn}}, \bibinfo {author} {\bibfnamefont {H.-P.}\ \bibnamefont {Cheng}},
  \bibinfo {author} {\bibfnamefont {P.~J.}\ \bibnamefont {Hirschfeld}}, \ and\
  \bibinfo {author} {\bibfnamefont {W.}~\bibnamefont {Ku}},\ }\href {\doibase
  10.1103/PhysRevB.89.020501} {\bibfield  {journal} {\bibinfo  {journal} {Phys.
  Rev. B}\ }\textbf {\bibinfo {volume} {89}},\ \bibinfo {pages} {020501}
  (\bibinfo {year} {2014})}\BibitemShut {NoStop}%
\bibitem [{\citenamefont {Berlijn}(2011)}]{thesistberlijn}%
  \BibitemOpen
  \bibfield  {author} {\bibinfo {author} {\bibfnamefont {T.}~\bibnamefont
  {Berlijn}},\ }\emph {\bibinfo {title} {Effects of Disordered Dopants on the
  Electronic Structure of Functional Materials: Wannier Function-Based First
  Principles Methods for Disordered Systems}},\ \href@noop {} {Ph.D. thesis},\
  \bibinfo  {school} {Stony Brook University} (\bibinfo {year}
  {2011})\BibitemShut {NoStop}%
\bibitem [{\citenamefont {Anisimov}\ and\ \citenamefont
  {Gunnarsson}(1991)}]{tbanisimov}%
  \BibitemOpen
  \bibfield  {author} {\bibinfo {author} {\bibfnamefont {V.~I.}\ \bibnamefont
  {Anisimov}}\ and\ \bibinfo {author} {\bibfnamefont {O.}~\bibnamefont
  {Gunnarsson}},\ }\href {\doibase 10.1103/PhysRevB.43.7570} {\bibfield
  {journal} {\bibinfo  {journal} {Phys. Rev. B}\ }\textbf {\bibinfo {volume}
  {43}},\ \bibinfo {pages} {7570} (\bibinfo {year} {1991})}\BibitemShut
  {NoStop}%
\bibitem [{\citenamefont {Cococcioni}(2012)}]{tbcococcioni}%
  \BibitemOpen
  \bibfield  {author} {\bibinfo {author} {\bibfnamefont {M.}~\bibnamefont
  {Cococcioni}},\ }\enquote {\bibinfo {title} {Correlated electrons: From
  models to materials modeling and simulation},}\ \ (\bibinfo  {publisher}
  {Verlag des Forschungszentrum J\"ulich},\ \bibinfo {address} {J\"ulich},\
  \bibinfo {year} {2012})\ Chap.\ \bibinfo {chapter} {The LDA+U Approach: A
  Simple Hubbard Correction for Correlated Ground States}\BibitemShut {NoStop}%
\bibitem [{\citenamefont {Nelson}\ \emph {et~al.}(2015)\citenamefont {Nelson},
  \citenamefont {Berlijn}, \citenamefont {Moreno}, \citenamefont {Jarrell},\
  and\ \citenamefont {Ku}}]{rykynelson}%
  \BibitemOpen
  \bibfield  {author} {\bibinfo {author} {\bibfnamefont {R.}~\bibnamefont
  {Nelson}}, \bibinfo {author} {\bibfnamefont {T.}~\bibnamefont {Berlijn}},
  \bibinfo {author} {\bibfnamefont {J.}~\bibnamefont {Moreno}}, \bibinfo
  {author} {\bibfnamefont {M.}~\bibnamefont {Jarrell}}, \ and\ \bibinfo
  {author} {\bibfnamefont {W.}~\bibnamefont {Ku}},\ }\href {\doibase
  10.1103/PhysRevLett.115.197203} {\bibfield  {journal} {\bibinfo  {journal}
  {Phys. Rev. Lett.}\ }\textbf {\bibinfo {volume} {115}},\ \bibinfo {pages}
  {197203} (\bibinfo {year} {2015})}\BibitemShut {NoStop}%
\bibitem [{\citenamefont {Aryasetiawan}\ \emph {et~al.}(2004)\citenamefont
  {Aryasetiawan}, \citenamefont {Imada}, \citenamefont {Georges}, \citenamefont
  {Kotliar}, \citenamefont {Biermann},\ and\ \citenamefont
  {Lichtenstein}}]{crpa}%
  \BibitemOpen
  \bibfield  {author} {\bibinfo {author} {\bibfnamefont {F.}~\bibnamefont
  {Aryasetiawan}}, \bibinfo {author} {\bibfnamefont {M.}~\bibnamefont {Imada}},
  \bibinfo {author} {\bibfnamefont {A.}~\bibnamefont {Georges}}, \bibinfo
  {author} {\bibfnamefont {G.}~\bibnamefont {Kotliar}}, \bibinfo {author}
  {\bibfnamefont {S.}~\bibnamefont {Biermann}}, \ and\ \bibinfo {author}
  {\bibfnamefont {A.~I.}\ \bibnamefont {Lichtenstein}},\ }\href {\doibase
  10.1103/PhysRevB.70.195104} {\bibfield  {journal} {\bibinfo  {journal} {Phys.
  Rev. B}\ }\textbf {\bibinfo {volume} {70}},\ \bibinfo {pages} {195104}
  (\bibinfo {year} {2004})}\BibitemShut {NoStop}%
\bibitem [{\citenamefont {Chandrasekharan}\ and\ \citenamefont
  {Wiese}(1999)}]{tbchandrasekharan}%
  \BibitemOpen
  \bibfield  {author} {\bibinfo {author} {\bibfnamefont {S.}~\bibnamefont
  {Chandrasekharan}}\ and\ \bibinfo {author} {\bibfnamefont {U.-J.}\
  \bibnamefont {Wiese}},\ }\href {\doibase 10.1103/PhysRevLett.83.3116}
  {\bibfield  {journal} {\bibinfo  {journal} {Phys. Rev. Lett.}\ }\textbf
  {\bibinfo {volume} {83}},\ \bibinfo {pages} {3116} (\bibinfo {year}
  {1999})}\BibitemShut {NoStop}%
\bibitem [{\citenamefont {Delaire}\ \emph {et~al.}(2015)\citenamefont
  {Delaire}, \citenamefont {Al-Qasir}, \citenamefont {May}, \citenamefont {Li},
  \citenamefont {Sales}, \citenamefont {Niedziela}, \citenamefont {Ma},
  \citenamefont {Matsuda}, \citenamefont {Abernathy},\ and\ \citenamefont
  {Berlijn}}]{fesi}%
  \BibitemOpen
  \bibfield  {author} {\bibinfo {author} {\bibfnamefont {O.}~\bibnamefont
  {Delaire}}, \bibinfo {author} {\bibfnamefont {I.~I.}\ \bibnamefont
  {Al-Qasir}}, \bibinfo {author} {\bibfnamefont {A.~F.}\ \bibnamefont {May}},
  \bibinfo {author} {\bibfnamefont {C.~W.}\ \bibnamefont {Li}}, \bibinfo
  {author} {\bibfnamefont {B.~C.}\ \bibnamefont {Sales}}, \bibinfo {author}
  {\bibfnamefont {J.~L.}\ \bibnamefont {Niedziela}}, \bibinfo {author}
  {\bibfnamefont {J.}~\bibnamefont {Ma}}, \bibinfo {author} {\bibfnamefont
  {M.}~\bibnamefont {Matsuda}}, \bibinfo {author} {\bibfnamefont {D.~L.}\
  \bibnamefont {Abernathy}}, \ and\ \bibinfo {author} {\bibfnamefont
  {T.}~\bibnamefont {Berlijn}},\ }\href {\doibase 10.1103/PhysRevB.91.094307}
  {\bibfield  {journal} {\bibinfo  {journal} {Phys. Rev. B}\ }\textbf {\bibinfo
  {volume} {91}},\ \bibinfo {pages} {094307} (\bibinfo {year}
  {2015})}\BibitemShut {NoStop}%
\bibitem [{\citenamefont {Bulka}\ \emph {et~al.}(1985)\citenamefont {Bulka},
  \citenamefont {Kramer},\ and\ \citenamefont {MacKinnon}}]{b_bulka_85}%
  \BibitemOpen
  \bibfield  {author} {\bibinfo {author} {\bibfnamefont {B.}~\bibnamefont
  {Bulka}}, \bibinfo {author} {\bibfnamefont {B.}~\bibnamefont {Kramer}}, \
  and\ \bibinfo {author} {\bibfnamefont {A.}~\bibnamefont {MacKinnon}},\ }\href
  {\doibase 10.1007/BF01312638} {\bibfield  {journal} {\bibinfo  {journal} {Z.
  Phys. B Condensed Matter}\ }\textbf {\bibinfo {volume} {60}},\ \bibinfo
  {pages} {13} (\bibinfo {year} {1985})}\BibitemShut {NoStop}%
\bibitem [{\citenamefont {Slevin}\ and\ \citenamefont
  {Ohtsuki}(2014)}]{k_slevin_14}%
  \BibitemOpen
  \bibfield  {author} {\bibinfo {author} {\bibfnamefont {K.}~\bibnamefont
  {Slevin}}\ and\ \bibinfo {author} {\bibfnamefont {T.}~\bibnamefont
  {Ohtsuki}},\ }\href {http://stacks.iop.org/1367-2630/16/i=1/a=015012}
  {\bibfield  {journal} {\bibinfo  {journal} {New J. Phys.}\ }\textbf {\bibinfo
  {volume} {16}},\ \bibinfo {pages} {015012} (\bibinfo {year}
  {2014})}\BibitemShut {NoStop}%
\bibitem [{\citenamefont {Terletska}\ \emph {et~al.}(2014)\citenamefont
  {Terletska}, \citenamefont {Ekuma}, \citenamefont {Moore}, \citenamefont
  {Tam}, \citenamefont {Moreno},\ and\ \citenamefont
  {Jarrell}}]{h_terletska_14a}%
  \BibitemOpen
  \bibfield  {author} {\bibinfo {author} {\bibfnamefont {H.}~\bibnamefont
  {Terletska}}, \bibinfo {author} {\bibfnamefont {C.~E.}\ \bibnamefont
  {Ekuma}}, \bibinfo {author} {\bibfnamefont {C.}~\bibnamefont {Moore}},
  \bibinfo {author} {\bibfnamefont {K.-M.}\ \bibnamefont {Tam}}, \bibinfo
  {author} {\bibfnamefont {J.}~\bibnamefont {Moreno}}, \ and\ \bibinfo {author}
  {\bibfnamefont {M.}~\bibnamefont {Jarrell}},\ }\href {\doibase
  10.1103/PhysRevB.90.094208} {\bibfield  {journal} {\bibinfo  {journal} {Phys.
  Rev. B}\ }\textbf {\bibinfo {volume} {90}},\ \bibinfo {pages} {094208}
  (\bibinfo {year} {2014})}\BibitemShut {NoStop}%
\bibitem [{\citenamefont {Wei\ss{}e}(2004)}]{Weisse_2004}%
  \BibitemOpen
  \bibfield  {author} {\bibinfo {author} {\bibfnamefont {A.}~\bibnamefont
  {Wei\ss{}e}},\ }\href {\doibase 10.1140/epjb/e2004-00250-6} {\bibfield
  {journal} {\bibinfo  {journal} {Eur. Phys. J. B}\ }\textbf {\bibinfo {volume}
  {40}},\ \bibinfo {pages} {125} (\bibinfo {year} {2004})}\BibitemShut
  {NoStop}%
\bibitem [{\citenamefont {Leconte}\ \emph {et~al.}(2016)\citenamefont
  {Leconte}, \citenamefont {Ferreira},\ and\ \citenamefont
  {Jung}}]{Ferreira_2016}%
  \BibitemOpen
  \bibfield  {author} {\bibinfo {author} {\bibfnamefont {N.}~\bibnamefont
  {Leconte}}, \bibinfo {author} {\bibfnamefont {A.}~\bibnamefont {Ferreira}}, \
  and\ \bibinfo {author} {\bibfnamefont {J.}~\bibnamefont {Jung}},\ }in\ \href
  {\doibase http://dx.doi.org/10.1016/bs.semsem.2016.04.002} {\emph {\bibinfo
  {booktitle} {2D Materials}}},\ \bibinfo {series} {Semiconductors and
  Semimetals}, Vol.~\bibinfo {volume} {95},\ \bibinfo {editor} {edited by\
  \bibinfo {editor} {\bibfnamefont {J.~J.~B.}\ \bibnamefont
  {Francesca~Iacopi}}\ and\ \bibinfo {editor} {\bibfnamefont {C.}~\bibnamefont
  {Jagadish}}}\ (\bibinfo  {publisher} {Elsevier},\ \bibinfo {address}
  {Burlington},\ \bibinfo {year} {2016})\ pp.\ \bibinfo {pages} {35 --
  99}\BibitemShut {NoStop}%
\bibitem [{\citenamefont {Garcia}\ \emph {et~al.}(2015)\citenamefont {Garcia},
  \citenamefont {Covaci},\ and\ \citenamefont {Rappoport}}]{Garcia_2015}%
  \BibitemOpen
  \bibfield  {author} {\bibinfo {author} {\bibfnamefont {J.~H.}\ \bibnamefont
  {Garcia}}, \bibinfo {author} {\bibfnamefont {L.}~\bibnamefont {Covaci}}, \
  and\ \bibinfo {author} {\bibfnamefont {T.~G.}\ \bibnamefont {Rappoport}},\
  }\href {\doibase 10.1103/PhysRevLett.114.116602} {\bibfield  {journal}
  {\bibinfo  {journal} {Phys. Rev. Lett.}\ }\textbf {\bibinfo {volume} {114}},\
  \bibinfo {pages} {116602} (\bibinfo {year} {2015})}\BibitemShut {NoStop}%
\bibitem [{\citenamefont {Tanaskovi\ifmmode~\acute{c}\else \'{c}\fi{}}\ \emph
  {et~al.}(2003)\citenamefont {Tanaskovi\ifmmode~\acute{c}\else \'{c}\fi{}},
  \citenamefont {Dobrosavljevi\ifmmode~\acute{c}\else \'{c}\fi{}},
  \citenamefont {Abrahams},\ and\ \citenamefont {Kotliar}}]{d_tanaskovic_03}%
  \BibitemOpen
  \bibfield  {author} {\bibinfo {author} {\bibfnamefont {D.}~\bibnamefont
  {Tanaskovi\ifmmode~\acute{c}\else \'{c}\fi{}}}, \bibinfo {author}
  {\bibfnamefont {V.}~\bibnamefont {Dobrosavljevi\ifmmode~\acute{c}\else
  \'{c}\fi{}}}, \bibinfo {author} {\bibfnamefont {E.}~\bibnamefont {Abrahams}},
  \ and\ \bibinfo {author} {\bibfnamefont {G.}~\bibnamefont {Kotliar}},\ }\href
  {\doibase 10.1103/PhysRevLett.91.066603} {\bibfield  {journal} {\bibinfo
  {journal} {Phys. Rev. Lett.}\ }\textbf {\bibinfo {volume} {91}},\ \bibinfo
  {pages} {066603} (\bibinfo {year} {2003})}\BibitemShut {NoStop}%
\bibitem [{\citenamefont {Guo}\ \emph {et~al.}(2010{\natexlab{a}})\citenamefont
  {Guo}, \citenamefont {Jin}, \citenamefont {Wang}, \citenamefont {Wang},
  \citenamefont {Zhu}, \citenamefont {Zhou}, \citenamefont {He},\ and\
  \citenamefont {Chen}}]{j_guo_10}%
  \BibitemOpen
  \bibfield  {author} {\bibinfo {author} {\bibfnamefont {J.}~\bibnamefont
  {Guo}}, \bibinfo {author} {\bibfnamefont {S.}~\bibnamefont {Jin}}, \bibinfo
  {author} {\bibfnamefont {G.}~\bibnamefont {Wang}}, \bibinfo {author}
  {\bibfnamefont {S.}~\bibnamefont {Wang}}, \bibinfo {author} {\bibfnamefont
  {K.}~\bibnamefont {Zhu}}, \bibinfo {author} {\bibfnamefont {T.}~\bibnamefont
  {Zhou}}, \bibinfo {author} {\bibfnamefont {M.}~\bibnamefont {He}}, \ and\
  \bibinfo {author} {\bibfnamefont {X.}~\bibnamefont {Chen}},\ }\href {\doibase
  10.1103/PhysRevB.82.180520} {\bibfield  {journal} {\bibinfo  {journal} {Phys.
  Rev. B}\ }\textbf {\bibinfo {volume} {82}},\ \bibinfo {pages} {180520}
  (\bibinfo {year} {2010}{\natexlab{a}})}\BibitemShut {NoStop}%
\bibitem [{\citenamefont {Wei}\ \emph {et~al.}(2011)\citenamefont {Wei},
  \citenamefont {Qing-Zhen}, \citenamefont {Gen-Fu}, \citenamefont {Green},
  \citenamefont {Du-Ming}, \citenamefont {Jun-Bao},\ and\ \citenamefont
  {Yi-Ming}}]{Wei_2011}%
  \BibitemOpen
  \bibfield  {author} {\bibinfo {author} {\bibfnamefont {B.}~\bibnamefont
  {Wei}}, \bibinfo {author} {\bibfnamefont {H.}~\bibnamefont {Qing-Zhen}},
  \bibinfo {author} {\bibfnamefont {C.}~\bibnamefont {Gen-Fu}}, \bibinfo
  {author} {\bibfnamefont {M.~A.}\ \bibnamefont {Green}}, \bibinfo {author}
  {\bibfnamefont {W.}~\bibnamefont {Du-Ming}}, \bibinfo {author} {\bibfnamefont
  {H.}~\bibnamefont {Jun-Bao}}, \ and\ \bibinfo {author} {\bibfnamefont
  {Q.}~\bibnamefont {Yi-Ming}},\ }\href
  {http://stacks.iop.org/0256-307X/28/i=8/a=086104} {\bibfield  {journal}
  {\bibinfo  {journal} {Chin. Phys. Lett.}\ }\textbf {\bibinfo {volume} {28}},\
  \bibinfo {pages} {086104} (\bibinfo {year} {2011})}\BibitemShut {NoStop}%
\bibitem [{\citenamefont {Jungwirth}\ \emph {et~al.}(2006)\citenamefont
  {Jungwirth}, \citenamefont {Sinova}, \citenamefont {Ma\ifmmode~\check{s}\else
  \v{s}\fi{}ek}, \citenamefont {Ku\ifmmode~\check{c}\else \v{c}\fi{}era},\ and\
  \citenamefont {MacDonald}}]{jungwirth_rmp_2006}%
  \BibitemOpen
  \bibfield  {author} {\bibinfo {author} {\bibfnamefont {T.}~\bibnamefont
  {Jungwirth}}, \bibinfo {author} {\bibfnamefont {J.}~\bibnamefont {Sinova}},
  \bibinfo {author} {\bibfnamefont {J.}~\bibnamefont {Ma\ifmmode~\check{s}\else
  \v{s}\fi{}ek}}, \bibinfo {author} {\bibfnamefont {J.}~\bibnamefont
  {Ku\ifmmode~\check{c}\else \v{c}\fi{}era}}, \ and\ \bibinfo {author}
  {\bibfnamefont {A.~H.}\ \bibnamefont {MacDonald}},\ }\href {\doibase
  10.1103/RevModPhys.78.809} {\bibfield  {journal} {\bibinfo  {journal} {Rev.
  Mod. Phys.}\ }\textbf {\bibinfo {volume} {78}},\ \bibinfo {pages} {809}
  (\bibinfo {year} {2006})}\BibitemShut {NoStop}%
\bibitem [{\citenamefont {Dietl}\ \emph {et~al.}(2001)\citenamefont {Dietl},
  \citenamefont {Ohno},\ and\ \citenamefont {Matsukura}}]{t_dietl_01a}%
  \BibitemOpen
  \bibfield  {author} {\bibinfo {author} {\bibfnamefont {T.}~\bibnamefont
  {Dietl}}, \bibinfo {author} {\bibfnamefont {H.}~\bibnamefont {Ohno}}, \ and\
  \bibinfo {author} {\bibfnamefont {F.}~\bibnamefont {Matsukura}},\ }\href@noop
  {} {\bibfield  {journal} {\bibinfo  {journal} {Phys. Rev. B}\ }\textbf
  {\bibinfo {volume} {63}},\ \bibinfo {pages} {195205} (\bibinfo {year}
  {2001})}\BibitemShut {NoStop}%
\bibitem [{\citenamefont {Zajac}\ \emph {et~al.}(2001)\citenamefont {Zajac},
  \citenamefont {Gosk}, \citenamefont {Kami{\'n}ska}, \citenamefont
  {Twardowski}, \citenamefont {Szyszko},\ and\ \citenamefont
  {Podsiadlo}}]{m_zajac_01}%
  \BibitemOpen
  \bibfield  {author} {\bibinfo {author} {\bibfnamefont {M.}~\bibnamefont
  {Zajac}}, \bibinfo {author} {\bibfnamefont {J.}~\bibnamefont {Gosk}},
  \bibinfo {author} {\bibfnamefont {M.}~\bibnamefont {Kami{\'n}ska}}, \bibinfo
  {author} {\bibfnamefont {A.}~\bibnamefont {Twardowski}}, \bibinfo {author}
  {\bibfnamefont {T.}~\bibnamefont {Szyszko}}, \ and\ \bibinfo {author}
  {\bibfnamefont {S.}~\bibnamefont {Podsiadlo}},\ }\href@noop {} {\bibfield
  {journal} {\bibinfo  {journal} {Appl. Phys. Lett.}\ }\textbf {\bibinfo
  {volume} {79}},\ \bibinfo {pages} {2432} (\bibinfo {year}
  {2001})}\BibitemShut {NoStop}%
\bibitem [{\citenamefont {Dhar}\ \emph {et~al.}(2003)\citenamefont {Dhar},
  \citenamefont {Brandt}, \citenamefont {Trampert}, \citenamefont {Friedland},
  \citenamefont {Sun},\ and\ \citenamefont {Ploog}}]{s_dhar_03}%
  \BibitemOpen
  \bibfield  {author} {\bibinfo {author} {\bibfnamefont {S.}~\bibnamefont
  {Dhar}}, \bibinfo {author} {\bibfnamefont {O.}~\bibnamefont {Brandt}},
  \bibinfo {author} {\bibfnamefont {A.}~\bibnamefont {Trampert}}, \bibinfo
  {author} {\bibfnamefont {K.~J.}\ \bibnamefont {Friedland}}, \bibinfo {author}
  {\bibfnamefont {Y.~J.}\ \bibnamefont {Sun}}, \ and\ \bibinfo {author}
  {\bibfnamefont {K.~H.}\ \bibnamefont {Ploog}},\ }\href@noop {} {\bibfield
  {journal} {\bibinfo  {journal} {Phys. Rev. B}\ }\textbf {\bibinfo {volume}
  {67}},\ \bibinfo {pages} {165205} (\bibinfo {year} {2003})}\BibitemShut
  {NoStop}%
\bibitem [{\citenamefont {Overberg}\ \emph {et~al.}(2001)\citenamefont
  {Overberg}, \citenamefont {Abernathy}, \citenamefont {Pearton}, \citenamefont
  {Theodoropoulou}, \citenamefont {McCarthy},\ and\ \citenamefont
  {Hebard}}]{m_overberg_01}%
  \BibitemOpen
  \bibfield  {author} {\bibinfo {author} {\bibfnamefont {M.~E.}\ \bibnamefont
  {Overberg}}, \bibinfo {author} {\bibfnamefont {C.~R.}\ \bibnamefont
  {Abernathy}}, \bibinfo {author} {\bibfnamefont {S.~J.}\ \bibnamefont
  {Pearton}}, \bibinfo {author} {\bibfnamefont {N.~A.}\ \bibnamefont
  {Theodoropoulou}}, \bibinfo {author} {\bibfnamefont {K.~T.}\ \bibnamefont
  {McCarthy}}, \ and\ \bibinfo {author} {\bibfnamefont {A.~F.}\ \bibnamefont
  {Hebard}},\ }\href@noop {} {\bibfield  {journal} {\bibinfo  {journal} {Appl.
  Phys. Lett.}\ }\textbf {\bibinfo {volume} {79}},\ \bibinfo {pages} {1312}
  (\bibinfo {year} {2001})}\BibitemShut {NoStop}%
\bibitem [{\citenamefont {Stefanowicz}\ \emph {et~al.}(2013)\citenamefont
  {Stefanowicz}, \citenamefont {Kunert}, \citenamefont {Simserides},
  \citenamefont {Majewski}, \citenamefont {Stefanowicz}, \citenamefont {Kruse},
  \citenamefont {Figge}, \citenamefont {Li}, \citenamefont {Jakie\l{}a},
  \citenamefont {Trohidou}, \citenamefont {Bonanni}, \citenamefont {Hommel},
  \citenamefont {Sawicki},\ and\ \citenamefont {Dietl}}]{s_stefanowicz_13}%
  \BibitemOpen
  \bibfield  {author} {\bibinfo {author} {\bibfnamefont {S.}~\bibnamefont
  {Stefanowicz}}, \bibinfo {author} {\bibfnamefont {G.}~\bibnamefont {Kunert}},
  \bibinfo {author} {\bibfnamefont {C.}~\bibnamefont {Simserides}}, \bibinfo
  {author} {\bibfnamefont {J.~A.}\ \bibnamefont {Majewski}}, \bibinfo {author}
  {\bibfnamefont {W.}~\bibnamefont {Stefanowicz}}, \bibinfo {author}
  {\bibfnamefont {C.}~\bibnamefont {Kruse}}, \bibinfo {author} {\bibfnamefont
  {S.}~\bibnamefont {Figge}}, \bibinfo {author} {\bibfnamefont
  {T.}~\bibnamefont {Li}}, \bibinfo {author} {\bibfnamefont {R.}~\bibnamefont
  {Jakie\l{}a}}, \bibinfo {author} {\bibfnamefont {K.~N.}\ \bibnamefont
  {Trohidou}}, \bibinfo {author} {\bibfnamefont {A.}~\bibnamefont {Bonanni}},
  \bibinfo {author} {\bibfnamefont {D.}~\bibnamefont {Hommel}}, \bibinfo
  {author} {\bibfnamefont {M.}~\bibnamefont {Sawicki}}, \ and\ \bibinfo
  {author} {\bibfnamefont {T.}~\bibnamefont {Dietl}},\ }\href {\doibase
  10.1103/PhysRevB.88.081201} {\bibfield  {journal} {\bibinfo  {journal} {Phys.
  Rev. B}\ }\textbf {\bibinfo {volume} {88}},\ \bibinfo {pages} {081201}
  (\bibinfo {year} {2013})}\BibitemShut {NoStop}%
\bibitem [{\citenamefont {Sasaki}\ \emph {et~al.}(2002)\citenamefont {Sasaki},
  \citenamefont {Sonoda}, \citenamefont {Yamamoto}, \citenamefont {Suga},
  \citenamefont {Shimizu}, \citenamefont {Kindo},\ and\ \citenamefont
  {Hori}}]{t_sasaki_02}%
  \BibitemOpen
  \bibfield  {author} {\bibinfo {author} {\bibfnamefont {T.}~\bibnamefont
  {Sasaki}}, \bibinfo {author} {\bibfnamefont {S.}~\bibnamefont {Sonoda}},
  \bibinfo {author} {\bibfnamefont {Y.}~\bibnamefont {Yamamoto}}, \bibinfo
  {author} {\bibfnamefont {K.-I.}\ \bibnamefont {Suga}}, \bibinfo {author}
  {\bibfnamefont {S.}~\bibnamefont {Shimizu}}, \bibinfo {author} {\bibfnamefont
  {K.}~\bibnamefont {Kindo}}, \ and\ \bibinfo {author} {\bibfnamefont
  {H.}~\bibnamefont {Hori}},\ }\href@noop {} {\bibfield  {journal} {\bibinfo
  {journal} {J. Appl. Phys.}\ }\textbf {\bibinfo {volume} {91}},\ \bibinfo
  {pages} {7911} (\bibinfo {year} {2002})}\BibitemShut {NoStop}%
\bibitem [{\citenamefont {Li}\ \emph {et~al.}(2009)\citenamefont {Li},
  \citenamefont {Chu}, \citenamefont {Jain},\ and\ \citenamefont
  {Shen}}]{Jian_Li_2009}%
  \BibitemOpen
  \bibfield  {author} {\bibinfo {author} {\bibfnamefont {J.}~\bibnamefont
  {Li}}, \bibinfo {author} {\bibfnamefont {R.-L.}\ \bibnamefont {Chu}},
  \bibinfo {author} {\bibfnamefont {J.~K.}\ \bibnamefont {Jain}}, \ and\
  \bibinfo {author} {\bibfnamefont {S.-Q.}\ \bibnamefont {Shen}},\ }\href
  {\doibase 10.1103/PhysRevLett.102.136806} {\bibfield  {journal} {\bibinfo
  {journal} {Phys. Rev. Lett.}\ }\textbf {\bibinfo {volume} {102}},\ \bibinfo
  {pages} {136806} (\bibinfo {year} {2009})}\BibitemShut {NoStop}%
\bibitem [{\citenamefont {Guo}\ \emph {et~al.}(2010{\natexlab{b}})\citenamefont
  {Guo}, \citenamefont {Rosenberg}, \citenamefont {Refael},\ and\ \citenamefont
  {Franz}}]{H_Guo_2010}%
  \BibitemOpen
  \bibfield  {author} {\bibinfo {author} {\bibfnamefont {H.-M.}\ \bibnamefont
  {Guo}}, \bibinfo {author} {\bibfnamefont {G.}~\bibnamefont {Rosenberg}},
  \bibinfo {author} {\bibfnamefont {G.}~\bibnamefont {Refael}}, \ and\ \bibinfo
  {author} {\bibfnamefont {M.}~\bibnamefont {Franz}},\ }\href {\doibase
  10.1103/PhysRevLett.105.216601} {\bibfield  {journal} {\bibinfo  {journal}
  {Phys. Rev. Lett.}\ }\textbf {\bibinfo {volume} {105}},\ \bibinfo {pages}
  {216601} (\bibinfo {year} {2010}{\natexlab{b}})}\BibitemShut {NoStop}%
\end{thebibliography}%
